\documentclass[fleqn,usenatbib]{mnras}

\usepackage[T1]{fontenc}

\DeclareRobustCommand{\VAN}[3]{#2}
\let\VANthebibliography\thebibliography
\def\thebibliography{\DeclareRobustCommand{\VAN}[3]{##3}\VANthebibliography}

\usepackage{graphicx}	
\usepackage{amsmath}	
\usepackage{amssymb}	
\usepackage{natbib}
\usepackage{multicol}
\usepackage{subfig}
\usepackage{appendix}
\usepackage{multirow}
\usepackage{pdflscape}	
\usepackage{footnote}
\usepackage{threeparttable}
\usepackage{hyperref}
\usepackage{newtxtext,newtxmath}

\title[Galaxy evolution in the Spiderweb protocluster]{Signs of environmental effects on star-forming galaxies in the Spiderweb protocluster at z=2.16}

\author[J. M. Perez-Martinez et al.]{J. M. P\'erez-Mart\'inez,$^{1,2,3}$\thanks{E-mail: jm.perez@astr.tohoku.ac.jp}
H. Dannerbauer,$^{2,3}$
T. Kodama,$^{1}$
Y. Koyama,$^{4}$
R. Shimakawa,$^{5}$
T. L. Suzuki,$^{6}$
\newauthor R. Calvi,$^{2,3}$
Z. Chen,$^{2,3,7}$
K. Daikuhara,$^{1}$
N. A. Hatch,$^{8}$
A. Laza-Ramos,$^{2,3}$
D. Sobral,$^{9}$
J. P. Stott,$^{10}$
I. Tanaka$^{4}$
\\
$^{1}$ Astronomical Institute, Tohoku University, 6-3, Aramaki, Aoba, Sendai, Miyagi, 980-8578, Japan.\\
$^{2}$ Instituto de Astrof\'isica de Canarias (IAC), E-38205, La Laguna, Tenerife, Spain.\\
$^{3}$ Universidad de La Laguna, Dpto. Astrof\'isica, E-38206, La Laguna, Tenerife, Spain.\\
$^{4}$ Subaru Telescope, National Astronomical Observatory of Japan, National Institutes of Natural Sciences, 650 North A’ohoku Place, Hilo, HI 96720, USA.\\
$^{5}$ National Astronomical Observatory of Japan, 2-21-1, Osawa, Mitaka, Tokyo, 181-8588, Japan.\\
$^{6}$ Kavli Institute for the Physics and Mathematics of the Universe (WPI), University of Tokyo, Kashiwa, Chiba, 277-8583, Japan.\\
$^{7}$ School of Astronomy and Space Science, Nanjing University, Nanjing 210093, PR China\\
$^{8}$ School of Physics and Astronomy, University of Nottingham, Nottingham, NG7 2RD, UK\\
$^{9}$ Departamento de Física, Faculdade de Ciências, Universidade de Lisboa, Edifício C8, Campo Grande, PT1749-016 Lisbon, Portugal.\\
$^{10}$ Department of Physics, Lancaster University, Lancaster, LA1 4YB, UK.\\
}

\date{Accepted XXX. Received YYY; in original form ZZZ}

\pubyear{2022}

\begin{document}
\label{firstpage}
\pagerange{\pageref{firstpage}--\pageref{lastpage}}
\maketitle

\begin{abstract}
{We use multi-object near-infrared (NIR) spectroscopy with VLT/KMOS to investigate the role of the environment in the evolution of the ionized gas properties of narrow-band selected H$\alpha$ emitters (HAEs) in the Spiderweb protocluster at $z=2.16$. Based on rest-frame optical emission lines, H$\alpha$ and [N{\sc{ii}}]$\lambda$6584, we confirm the cluster membership of 39 of our targets (i.e. 93\% success rate), and measure their star-formation rates (SFR), gas-phase oxygen abundances and effective radius. We parametrize the environment where our targets reside by using local and global density indicators based on previous samples of spectroscopic and narrow-band cluster members. We find that star-forming galaxies embedded in the Spiderweb protocluster display SFRs compatible with those of the main sequence and morphologies comparable to those of late-type galaxies at $z=2.2$ in the field. We also report a mild gas-phase metallicity enhancement ($0.6\pm0.3$ dex) at intermediate stellar-masses. Furthermore, we identify two UVJ-selected quiescent galaxies with residual H$\alpha$-based star formation and find signs of extreme dust obscuration in a small sample of SMGs based on their FIR and H$\alpha$ emission. Interestingly, the spatial distribution of these objects differs from the rest of HAEs, avoiding the protocluster core. Finally, we explore the gas fraction-gas metallicity diagram for 7 galaxies with molecular gas masses measured by ATCA using CO(1-0). In the context of the gas-regulator model, our objects are consistent with relatively low mass-loading factors, suggesting lower outflow activity than field samples at the cosmic noon and thus, hinting at the onset of environmental effects in this massive protocluster.}

\end{abstract}




\section{Introduction}
\label{S:Intro}

Local galaxy clusters host a very distinct distribution of galaxy populations with respect to the field. While the latter is dominated by galaxies that are blue in color, have relatively high star formation rates (SFRs) and disk-like morphologies, the former mainly host redder objects with low levels of star formation and triaxial shapes (e.g. \citealt{Dressler80}; \citealt{Balogh98}; \citealt{Hogg03}). These discrepancies in terms of galaxy populations strongly correlate with the cluster total mass and with the distance to their centers, suggesting that the impact of the environment is controlled by both the accretion history of the clusters (\citealt{Dekel06}) and the spatial and temporal trajectory that galaxies follow in them (e.g. \citealt{Muzzin14}; \citealt{Haines15}; \citealt{Rhee17}).

This qualitative description has been extensively studied from a quantitative point of view over the last decades (\citealt{Barsanti21}) crystallizing into several correlations that expose the gradual effect of the environment over the galaxies' physical properties. Among them, we find the morphology-density relation (\citealt{Dressler80}), the color-density relation (\citealt{Hogg03}) and the star-formation density relation (\citealt{Balogh98}; \citealt{Lewis02}) which describe the discrepancies between the cluster and field outlined above. Furthermore, signs of enhanced metal enrichment have also been found in the densest regions of clusters, specially for relatively low-mass galaxies (e.g. \citealt{Petropoulou11, Petropoulou12}; \citealt{Paulino-Afonso18}; \citealt{Ciocan20}). The simultaneous manifestation of all these correlations demonstrate the influence of the environment in the accelerated transformation of galaxy properties. However, the exact mechanism behind these changes is still a matter of debate since we need to find cluster-specific processes that account for the dynamical transformation of cluster galaxies (\citealt{Mortlock13}; \citealt{Swinbank17}) as well as for the evolution of their interstellar medium (ISM, \citealt{Peng14}; \citealt{Maiolino19}). 

Several possibilities have been proposed to tackle this conundrum, with mergers and repeated tidal interactions (i.e. harassment, \citealt{Moore96}) likely driving the dynamical transformation while ram-pressure stripping (RPS, \citealt{Gunn72}, see also \citealt{Jaffe15}) and strangulation (\citealt{Peng15}) being responsible for the ISM evolution. In this scenario, cluster galaxies are first detached from the cosmic web once they enter into the cluster halo, restricting the amount of cold gas inflows they get from the cosmic web (\citealt{Dekel09a}). Thus, cluster galaxies start depleting their gas reservoir in the outskirts of the cluster while they keep enriching their ISM through successive generations of stars. On top of this, the outside pressure of the intracluster medium (ICM) would push back the outflowing gas from feedback processes, forcing the galaxy to recycle the already processed gas before being fully quenched due to RPS and gas exhaustion in the cluster core (\citealt{Wetzel13}). This description of galaxy evolution in massive clusters holds until $z\sim1$. 

At earlier epochs, an increasing fraction of overdense regions are still in the processs of being fully assembled, with most of them commonly referred to as protoclusters beyond $z=2$ (see \citealt{Overzier16} for a review). Interestingly, the progenitors of the quiescent galaxies that populate the inner cores of local clusters formed the bulk of their stellar budget more than $\gtrsim$10 Gyrs ago, at the same time that protoclusters were being assembled. Thus, the so-called "cosmic-noon" (i.e. $z\approx1.5-3$, \citealt{Madau14}) is a crucial epoch to understand the assembly of the main components of massive clusters as well as to explore the onset of the first environmental effects and their influence over the early evolution of protocluster galaxies. Over the last years, several attempts have been made to examine the star-formation activity and the gas-phase metallicity of protocluster galaxies compared to the field. The star-forming population is dominant in these structures and even starburst galaxies are relatively common within them (\citealt{Dannerbauer14}; \citealt{Popesso15}; \citealt{Casey17}). However, larger samples display a big diversity of behaviors regarding star-formation. Several studies have shown that star formation is enhanced relative to the field in dense environments at high redshift (\citealt{Alberts14}; \citealt{Shimakawa18a}; \citealt{Lemaux20}; \citealt{Monson21}), suggesting protocluster galaxies may have undergone accelerated mass assembly compared to their field counterparts. However, there are also some of protoclusters where no such differences are seen as compared to the field (e.g. \citealt{Toshikawa14}; \citealt{Cucciati14}; \citealt{Shi21}; \citealt{Sattari21}). The discrepancies between works may arise from the diverse galaxy populations studied, the lack of statistics due to small sample size, and the different evolutionary stages in which these protoclusters are observed (\citealt{Overzier08}; \citealt{Toshikawa14}).
 
 In addition, the early metal enrichment of protocluster galaxies is still a matter of debate. Some authors found evidence of enhanced gas-phase metallicity in low-mass cluster and protocluster galaxies at $z=1.5-3$ (e.g. \citealt{Kulas13}; \citealt{Shimakawa15}; \citealt{Maier19}), which can be explained by the shut down of pristine gas inflows in the cluster environment due to the early onset of the ICM. However, recent works have reported various levels of metallicity deficiency compared to field galaxies at the same redshift (\citealt{Valentino15}; \citealt{Chartab21}; \citealt{Sattari21}). To explain these results, it has been proposed that the dark matter haloes of young and not yet massive ($\mathrm{\log M_*/M_{\odot}<13.5}$) protoclusters at high-z would still be on a phase of powerful cold stream accretion (\citealt{Dekel06}). This extra supply of gas on protocluster galaxies would dilute their current metallicities with respect to the general field. Conversely, some other studies have not observed significant environmental dependence of the mass-metallicity relation during this epoch (e.g. \citealt{Tran15}; \citealt{Kacprzak15}; \citealt{Namiki19}). While the total mass of the overdensity seems to play a key role on explaining these metallicity discrepancies between protoclusters in terms of their accretion mode (cold vs hot, \citealt{Dekel09a}), there are still other potential biases that should be taken into account, such as the presence and relevance of the AGN fraction within these forming structures (see \citealt{Macuga19} and \citealt{Monson21} for some examples) and the different selection criteria for the parent galaxy samples under scrutiny. Furthermore, the molecular gas properties of protocluster at $z>2$, which are key to understand the gas feeding and consumption processes that fuels star formation, remain largely unexplored except in a few cases and over small sample sizes (\citealt{Dannerbauer17};  \citealt{Wang18}; \citealt{Tadaki19}; \citealt{Zavala19}; \citealt{Champagne21}; \citealt{Aoyama22}). Only very recently there has been some attempts to map the CO emission of galaxies in protoclusters (e.g. \citealt{Jin21}) providing a new window to investigate galaxy evolution in these forming large scale structures. Thus, in order to shed light into the early stages of galaxy formation and evolution in protoclusters at the cosmic noon we must trace the star-formation activity, metal enrichment and gas reservoir of the individual objects that belong to these large scale structures in formation.

Among the many assembling clusters detected during the last years, PKS1138-262 at $z=2.16$ (hereafter the Spiderweb protocluster) stands as the one of the most massive ($\mathrm{M_{200}>2\times10^{14}\,M_\odot}$, \citealt{Shimakawa14}) and best studied systems beyond $z=2$ both in terms of its galaxy populations and large-scale structure. This protocluster was first discovered by \cite{Kurk00} using narrow and broad-band photometry to identify an overdensity of Lyman-$\alpha$ emitters (LAEs) around the radio galaxy MRC1138-262 or Spiderweb galaxy (\citealt{Roettgering94}; \citealt{Pentericci97}). These initial reports about the presence of an overdensity of LAEs at $z=2.16$ were later followed up by \cite{Pentericci00} who spectroscopically confirmed the cluster membership of 15 LAEs. From that moment, the Spiderweb protocluster has been subject to exhaustive spectrophotometric campaigns to unveil the properties and distribution of its galaxy populations. This includes the characterization of the central radio galaxy (\citealt{Pentericci00}; \citealt{Carilli02}; \citealt{Miley06}; \citealt{Hatch08,Hatch09}; \citealt{Emonts16,Emonts18}: \citealt{DeBreuck22}, \citealt{Carilli22}), the location of several X-ray emitters (\citealt{Pentericci02}; \citealt{Croft05}; \citealt{Tozzi22}), the observation of an emerging red sequence within the cluster core (\citealt{Kurk04}; \citealt{Kodama07}; \citealt{Zirm08}; \citealt{Tanaka10, Tanaka13}), the discovery of a network of starbursty submillimeter galaxies (SMGs, \citealt{Dannerbauer14, Dannerbauer17}), the location of a rich population of H$\alpha$ emitters (HAEs) which represent the bulk of the known cluster members up to date (\citealt{Kuiper11}; \citealt{Hatch11}; \citealt{Koyama13}; \citealt{Shimakawa14, Shimakawa15, Shimakawa18b}), and the use of (sub-)millimeter observations to trace the dust content and gas reservoirs of several protocluster members (\citealt{Emonts18}; \citealt{Tadaki19}) as well as the mapping of the protocluster large scale structure in CO(1-0) using ATCA (\citealt{Jin21}). 

In this work, we investigate the environmental imprints of galaxy evolution in the Spiderweb protocluster focusing on the star formation, gas-phase metallicity, stellar-disk size, and molecular gas properties of the protocluster members. This manuscript is structured in the following way: $\text{Sect.}$\,\ref{S:Data} describes our new KMOS spectroscopic observations in the Spiderweb protocluster and the wealth of archival data available within this field. $\text{Sect.}$\,\ref{S:Methods} outlines the methods used to analyze the physical properties of our targets and the environmental parameters measured to them. $\text{Sect.}$\,\ref{S:Results} and \ref{S:Discussion} present our main results and the discussion of their physical interpretation in the context of galaxy evolution respectively. Finally, $\text{Sect.}$\,\ref{S:Conclusions} outlines the major conclusions of this study. Throughout this article we assume a \citet{Chabrier03} initial mass function (IMF), and adopt a flat cosmology with $\Omega_{\Lambda}$=0.7, $\Omega_{m}$=0.3, and $H_{0}$=70 km\,s$^{-1}$Mpc$^{-1}$. All magnitudes quoted in this paper are in the AB system (\citealt{Oke83}).

\section{Observations}
\label{S:Data}

In this section, we describe the main characteristics of the datasets we used in our analysis. In particular, we combine a new set of NIR spectroscopic observation with previous multiwavelength photometry covering a significant fraction of the Spiderweb protocluster field.

\subsection{KMOS spectroscopy}
\label{SS:KMOS} 

\begin{figure}
\includegraphics[width=\linewidth]{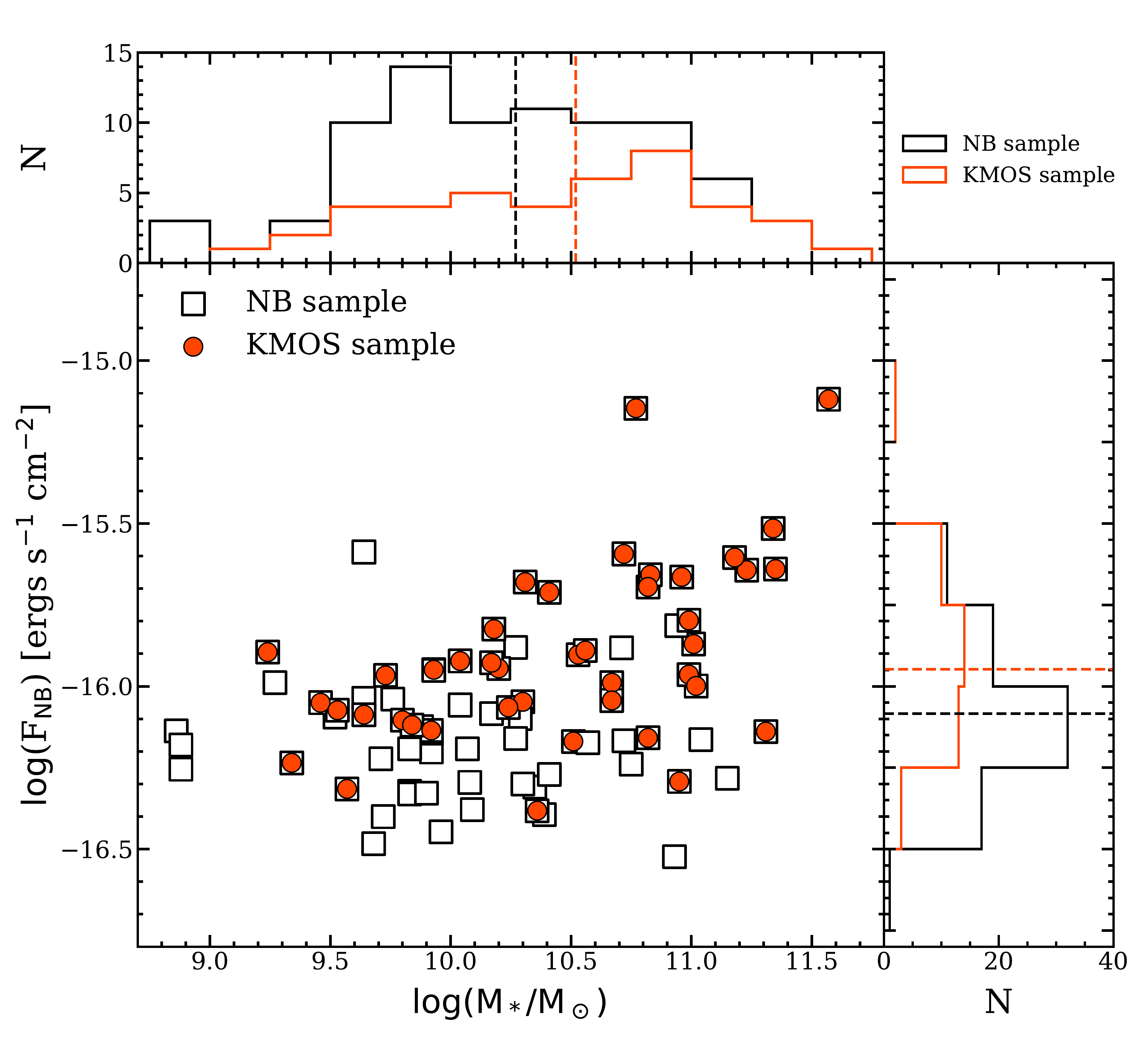}
\caption{Stellar-mass and narrow-band flux properties of both our spectroscopic targets and their parent sample of narrow-band emitters from \protect\cite{Koyama13}. Main diagram: Red circles represent the KMOS spectroscopic sample. Empty squares display the parent sample. Side panels: Distribution (histograms) and median values (dashed lines) of the KMOS sample and its parent sample. Colors follow the same scheme applied in the main diagram.}
\label{F:SEL}
\end{figure}

\begin{figure*}
 \centering
      \includegraphics[width=\linewidth]{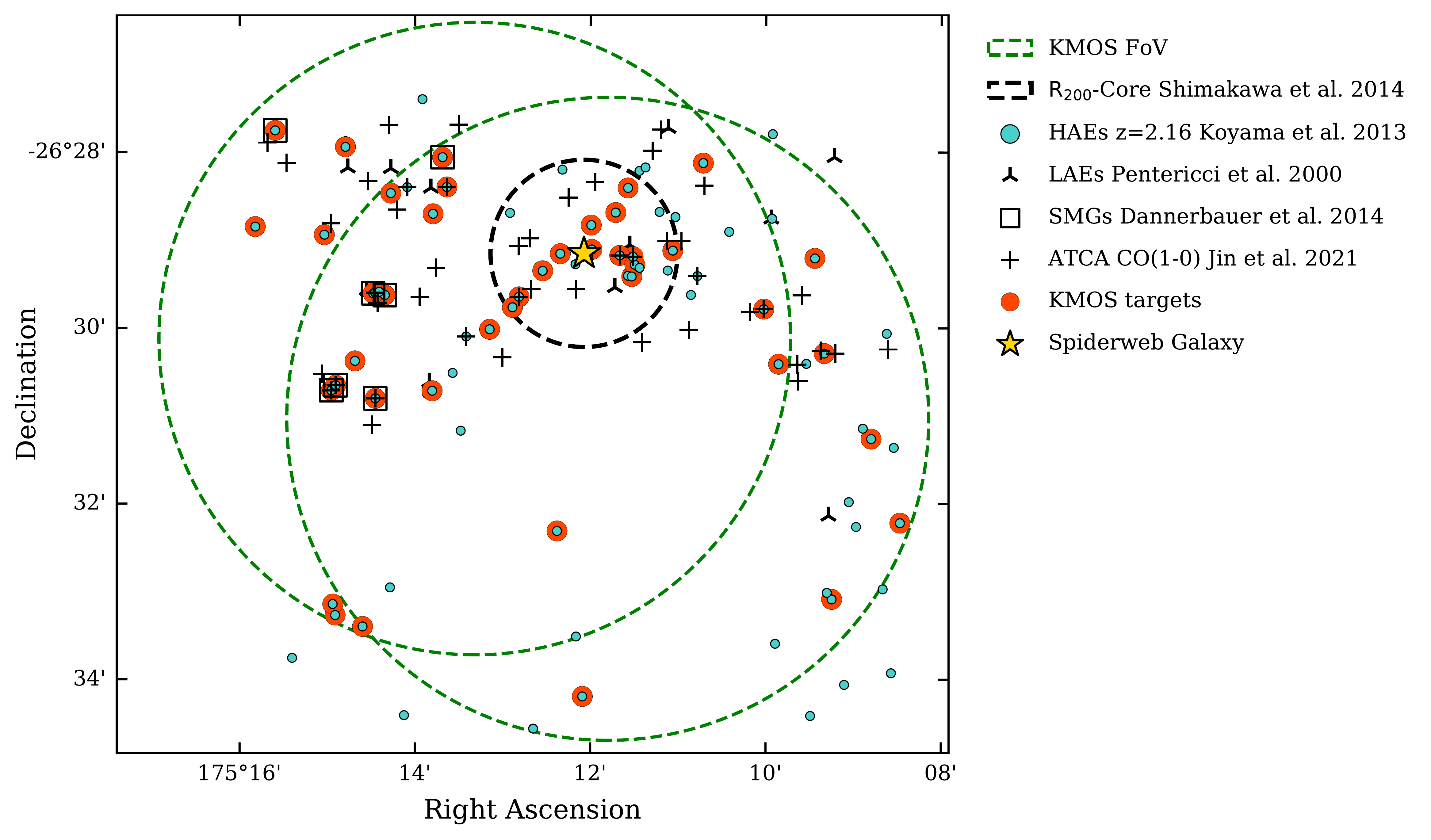}\par 
      \caption{Distribution of spectroscopic and narrow-band selected cluster members in the field of the Spiderweb protocluster. Red circles show the sample of KMOS-HAEs studied in this work. Empty circles depict the parent sample of narrow-band detected HAEs from \protect\cite{Koyama13}. Empty squares highlight a small overlapping sample of SMGs from \protect\cite{Dannerbauer14}. Purple crosses display the position of LAEs from \protect\cite{Pentericci00}. Black crosses show the positions of ATCA CO(1-0) emitters detected by \protect\cite{Jin21} in this field. The yellow star shows the position of the radio galaxy MRC $1138-262$ (\protect\citealt{Roettgering94}. The black dashed circle encloses $\mathrm{R_{200}}$ according to the mass estimation performed by \protect\cite{Shimakawa14} for the protocluster core. Finally, the two dashed green circles show the FoV of the two KMOS configurations used to observe our targets.}
         \label{F:Map}
      \end{figure*}

We carried out multi-object integral field spectroscopy observations of a sample of 42 narrow-band selected HAEs in the Spiderweb protocluster to obtain H$\alpha$ and [N{\sc{ii}}]$\lambda6584$ emission-line fluxes and study the galaxies' star-formation activity and gas-phase metallicities as a function of several environmental indicators. In Fig.\,\ref{F:SEL} we put in context our spectroscopic targets (hereafter KMOS sample) with their parent sample of narrow-band (NB) emitters from \cite{Koyama13}. The KMOS sample was selected to have narrow-band fluxes $\mathrm{F_{NB}\gtrsim4\times10^{-17}}$ erg\,s$^{-1}$\,cm$^{-2}$ and a similar stellar mass range than its parent sample. The distribution and median values (dashed lines) of both the parent and the KMOS sample show that the latter is slightly biased towards massive galaxies, and that at a fixed stellar mass it includes objects with relatively high narrow-band fluxes. In addition, our targets are selected to trace both the inner core and some of the filaments around the large scale structure of the protocluster simultaneously (see Fig.\,\ref{F:Map}). As a result of the narrow-band selection criteria, our targets span a very narrow range in redshift space ($\langle z\rangle=2.159\pm0.008$) though they encompass a broad range of cluster-centric distances (up to $R=5R_{200}$). 

The observations were carried out using the K-band Multi-Object Spectrograph (KMOS, \citealt{Sharples13}) installed at the Very Large Telescope (VLT) in Cerro Paranal, Chile (Program ID: 095.A-0500(A), PI: Y. Koyama). Our program employed two different arm configurations targeting 42 HAEs identified by \cite{Koyama13}. The observations where executed in service mode between April and June 2015 under average seeing conditions of $0.6\arcsec$, airmass value of 1.3, and no moonlight contamination (i.e., dark time). The observations were carried out in the K-band (19300-24600\,\AA), which captures H$\alpha$ and [N{\sc{ii}}]$\lambda6584$ at $z=2.16$ with a nominal spectral resolving power of R$\sim$4200 around its central wavelength, which translates to $ \mathrm{\sigma_{ins} \approx 35}$ km\,s$^{-1}$ in velocity space. Each KMOS arm carries a small IFU ($2.8\arcsec\times2.8\arcsec$) with spatial resolution of $0.2\arcsec$ per spaxel. For each configuration, one arm was used to take sky exposures while two additional arms were fixed to monitor stars. In summary, 21 arms were used for science targets per pointing. The observing time of each configuration was divided into $\sim$1h observing blocks (OB) made of 5 on-source subexposures of 450 seconds each plus overheads. The total on-source time per configuration is 3h. The spectroscopic data reduction was carried out using the ESO-Reflex workflow pipeline (\citealt{Davies13}). The main reduction steps were bias subtraction, flat-field normalization, sky subtraction, wavelength and flux calibration, and frame stacking. The resulting datacubes conserve the spatial and spectral properties previously outlined in this paragraph.  

\subsection{Archival Data}
\label{SS:Photometry} 

In addition to our spectroscopic campaign, we used abundant optical to NIR complimentary archival imaging data in this field. This includes the Subaru/Suprime-Cam B and z'-band, and the Subaru/MOIRCS J and $\mathrm{K_s}$-band from a previous MAHALO-Subaru project publication (\citealt{Koyama13}). The NIR campaign carried out by VLT/HAWK-I to obtain Y, H, $\mathrm{K_s}$ deep imaging (PI: A. Kurk, program IDs 088.A-0754, 091.A-0106, 094.A-0104, see \citealt{Dannerbauer17} and \citealt{Shimakawa18b}). Furthermore, we use the Post-BCD (PBCD) products from the Spitzer data archive library to obtain IRAC broad band imaging at 3.6 and 4.5 $\micron$ (PI: D. Stern, campaign IDs 736 and 793, see \citealt{Seymour07}). Finally, we use the reduced Hubble Space Telescope (HST) ACS/WFC data in filters F475W and F814W from the Hubble Legacy Archive (PI: H. Ford, proposal ID 10327, see \citealt{Miley06}). The exposure times and seeing conditions of the retrieved coadded mosaic images are shown in Table\,\ref{T:imaging}. Finally, we also make use of the ATCA CO(1-0) map presented in \cite{Jin21}, which belongs to the COALAS project (large program ID: C3181, PI: H. Dannerbauer). The coordinates, redshifts, and general properties of our final protocluster galaxy sample are summarized in the appendix.

\begin{table*}
\caption{Properties of the optical to NIR photometric bands used in this work. Ellipsis are used to avoid repetitions in the instrument and reference columns.}
\centering
\begin{tabular}{llcccccc}
\hline
\noalign{\vskip 0.1cm}
Instrument &  Filter   & Exp. Time &  FWHM & Pixel-size & Reference \\ 
           &           & (s)       &   ($\arcsec$) & ($\arcsec$) &  \\ \hline 
\noalign{\vskip 0.1cm}
HST/ACS       & F475W         &  20670   &  0.11 & 0.05 & \citealt{Miley06} \\  
\ldots        & F814W         &  23004   &  0.11 & 0.05 & \ldots \\
Suprime-Cam/Subaru & B        &  6300    &  1.15 & 0.20 & \citealt{Shimakawa18b}\\ 
\ldots        & z'            &  4500    &  0.70 & 0.20 & \citealt{Koyama13} \\
MOIRCS/Subaru & J             &  9060    &  0.69 & 0.12 & \ldots \\ 
\ldots        & $\mathrm{K_s}$            &  3300    &  0.63 & 0.12 & \ldots \\  
\ldots        & NB2071        &  11160   &  0.63 & 0.12 & \ldots \\
HAWKI/VLT     & Y             &  26880   &  0.37 & 0.11 & \citealt{Dannerbauer17}\\
\ldots        & H             &  14832   &  0.49 & 0.11 & \ldots\\
\ldots        & $\mathrm{K_s}$            &  9228    &  0.38 & 0.11 & \ldots\\
Spitzer/IRAC  & 3.6$\mu m$    &  3000    &  1.80 & 0.40 & \citealt{Seymour07}\\
\ldots        & 4.5$\mu m$    &  3000    &  1.80 & 0.40 & \ldots \\ \hline

\end{tabular}
\label{T:imaging}
\end{table*}

\section{Methods}
\label{S:Methods}

\subsection{Emission-line fitting}
\label{SS:EL}

In order to confirm the cluster membership of our targets and to measure their star formation activities and metallicities we need to carry out a line-fitting procedure. To carry out this task, we developed a self-written code in {\sc{python}} with the Astropy libray \citep{Astropy13, Astropy18}. In the next paragraphs we describe the most important steps that we followed during the emission line detection and fitting.

First, we visually inspect each datacube with the software QFitsView, which is a publicly available IFS visual tool developed by Thomas Ott at the MPE. We search for emission lines in the wavelength range defined by the width of the narrow-band filter used to classify our targets as HAEs (see \citealt{Koyama13}). At least one emission line is clearly detected in 40 out of the 42 original KMOS targets. In these cases, we integrate the spectral axis of each datacube over the wavelength limits of the detected emission line and define the brightest spaxel of the light distribution as the spatial center of each target within the IFU. Next, we place $1.4\arcsec\times1.4\arcsec$ (i.e. $7\times7$ spaxels) squared apertures (red squares in Fig.\,\ref{F:IFU}) there and extract a 1d spectrum for each galaxy. The same procedure and aperture position is applied to the noise datacube. 

After the extraction of the 1D spectra, we fit and subtract a local continuum around the H$\alpha$ line consisting of several windows free from skyline contamination that cover a few hundred angstroms in both directions around the H$\alpha$ observed wavelength. Finally, we perform a triple gaussian fit for H$\alpha$ and the [N{\sc{ii}}]$\lambda6584$ and $\lambda6548$ lines allowing for small variations in the center, amplitude and width of the first two. However, the flux ratio [N{\sc{ii}}]$\lambda6584$/[N{\sc{ii}}]$\lambda6548$ is fixed to 1/3. Furthermore, their widths are tied and fixed to be equal or smaller than the one from H$\alpha$. The reason for this is that Balmer lines may exhibit broad components when the ionization mechanism is (at least partly) related to AGN activity or skewed profiles when inflows and outflows are present. These effects may broaden the width of the single component fit applied to the H$\alpha$ emission line, and thus, this value is taken as the upper bound for the [N{\sc{ii}}] line. Two objects (IDs 911 and 647) display extremely broad H$\alpha$ profiles ($\sigma>700$ km\,s$^{-1}$), indicative of AGN type 1 activity. These sources are also identified as X-ray emitters in \cite{Croft05} and \cite{Tozzi22}, confirming their AGN nature. In these cases, we expand our method by including one additional component to trace the H$\alpha$ broad emission. One object (ID 897) display three emission lines within the inspected spectral window. These lines are consistent with [O{\sc{iii}}]$\lambda$5007, [O\,{\sc{iii}}]$\lambda$4959, and H$\beta$ emission from a background galaxy at $z=3.166$ and thus, this object is removed from the subsequent analyses of the protocluster sample. In total, we report 39 H$\alpha$ spectroscopic detections with 12 of them being new sources unknown to previous spectroscopic studies in the Spiderweb protocluster (e.g. \citealt{Pentericci00}; \citealt{Kurk04}; \citealt{Croft05}; \citealt{Doherty10}; \citealt{Kuiper11}; \citealt{Tanaka13}, \citealt{Shimakawa14, Shimakawa18b}; \citealt{Tadaki19}; \citealt{Jin21})

In order to take into account the uncertainties of both the fitting procedure and the datacube noise we conduct a Montecarlo approach where the fitting method outlined above is repeated 1000 times allowing for random gaussian variations on every single spectral data point with a maximum amplitude equal to the value found in the 1D noise spectra. After this, we take the mean value of this parameters and compute the total flux for the emission lines under scrutiny. The flux error is estimated by the standard deviation of the gaussian fit parameters over the 1000 realizations. We consider a line to be detected if the ratio between the total flux and the flux error measured in the way described above is higher than two. This constrain apply both to H$\alpha$ and [N{\sc{ii}}]. In Fig.\,\ref{F:IFU} we show the 2D image of the collapsed datacube around H$\alpha$ and the extracted 1D spectrum for each target.

\begin{figure*}
 \centering
 \begin{multicols}{2}
      \includegraphics[width=\linewidth]{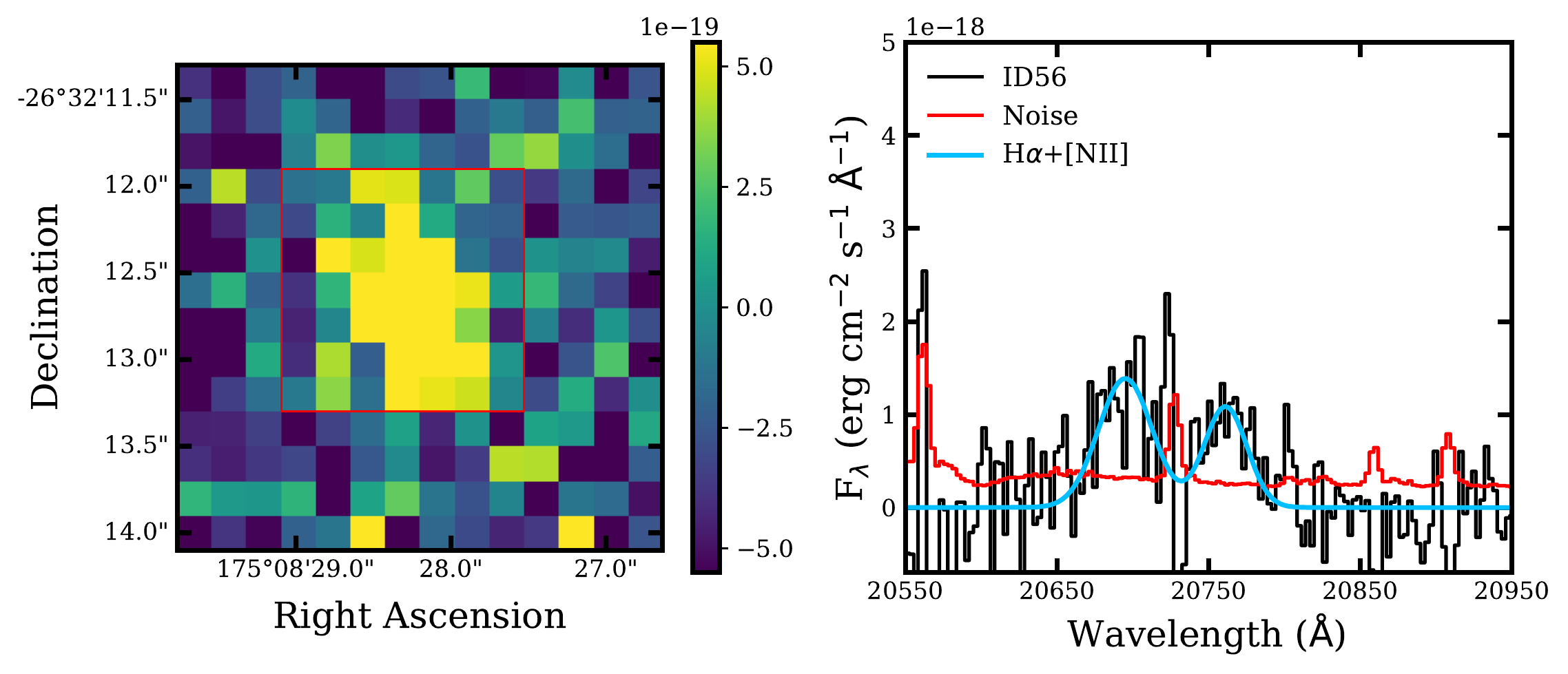}\par
      \includegraphics[width=\linewidth]{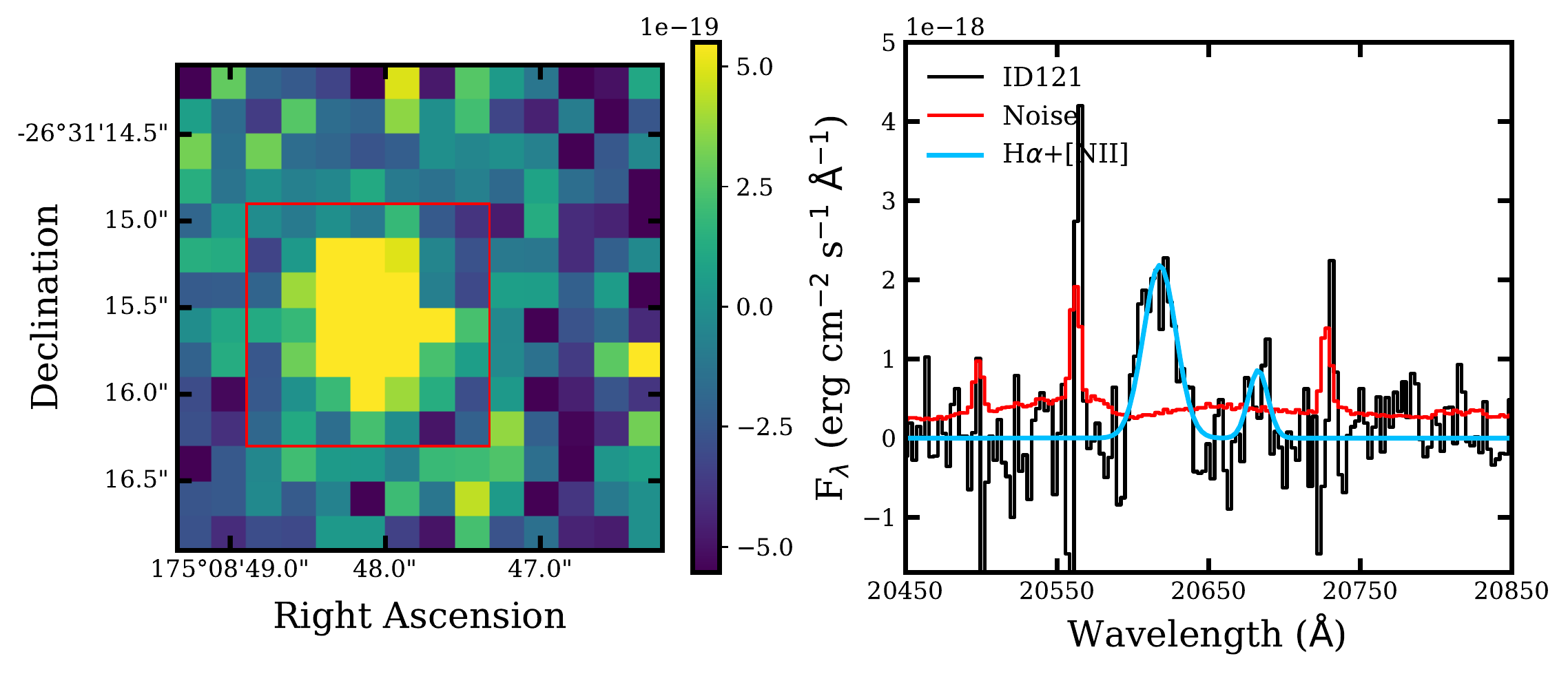}\par
      \includegraphics[width=\linewidth]{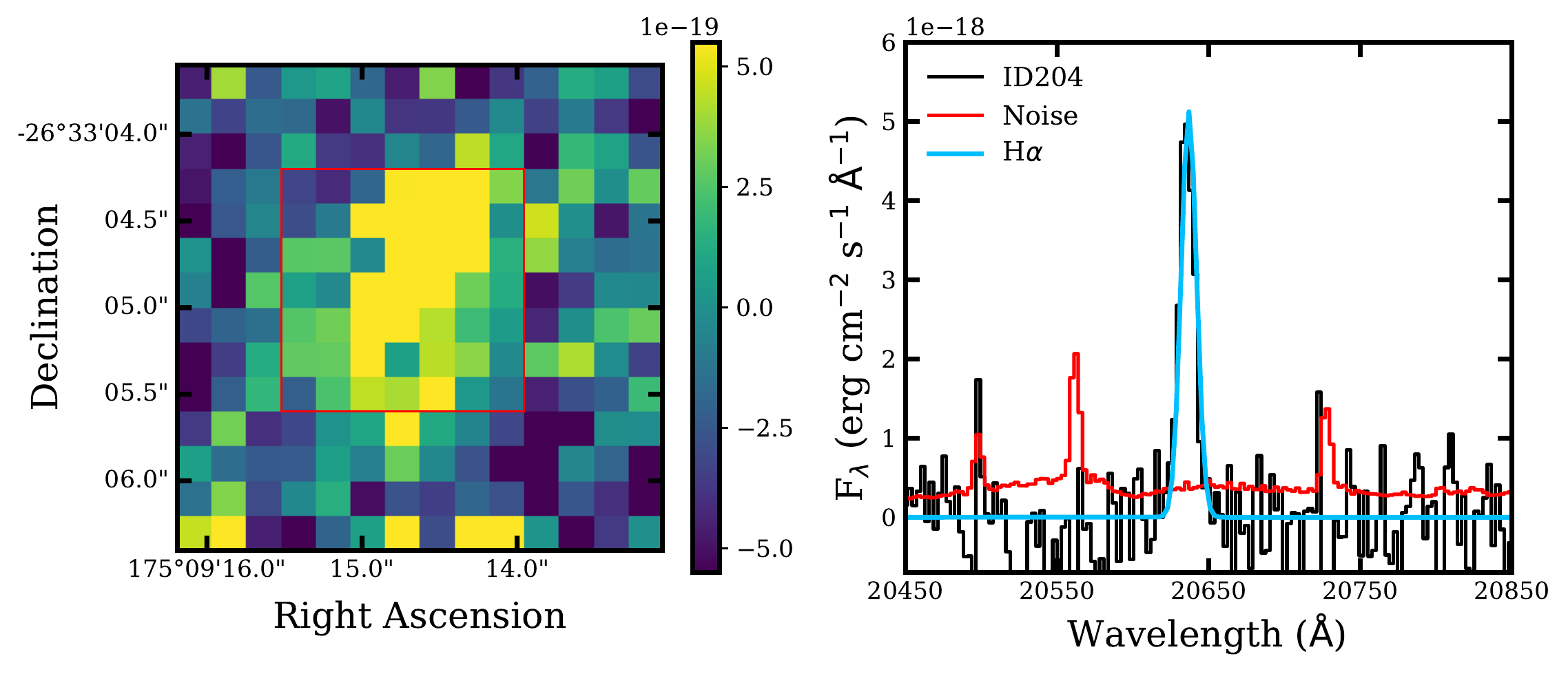}\par
      \includegraphics[width=\linewidth]{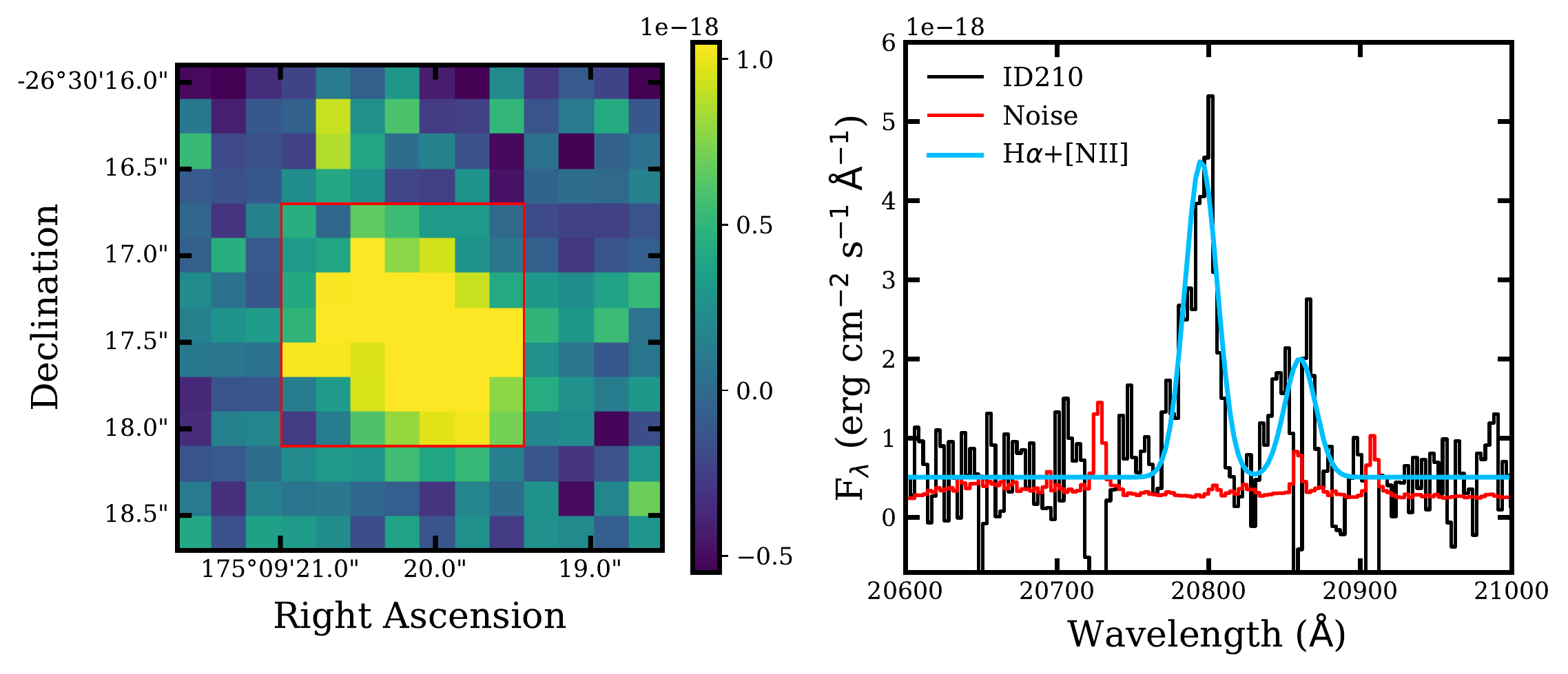}\par
      \includegraphics[width=\linewidth]{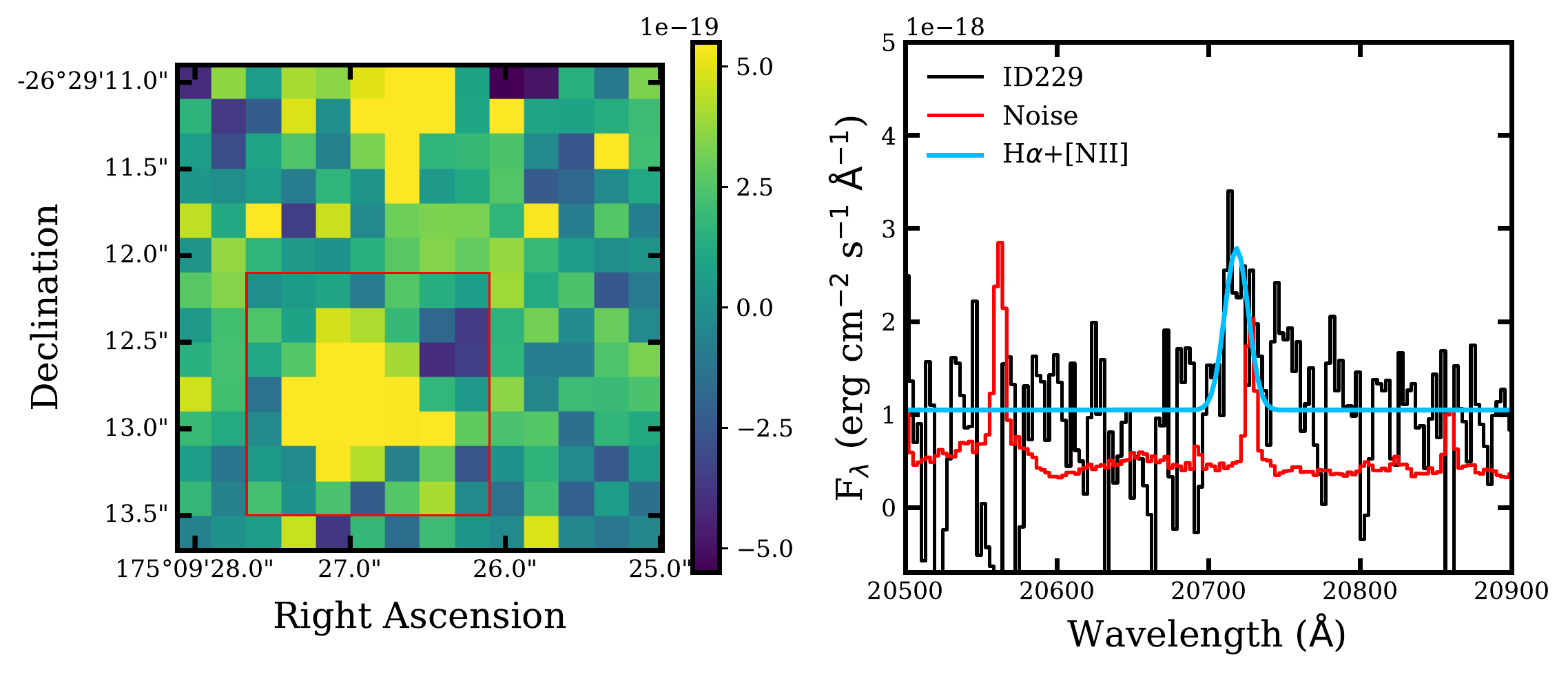}\par
      \includegraphics[width=\linewidth]{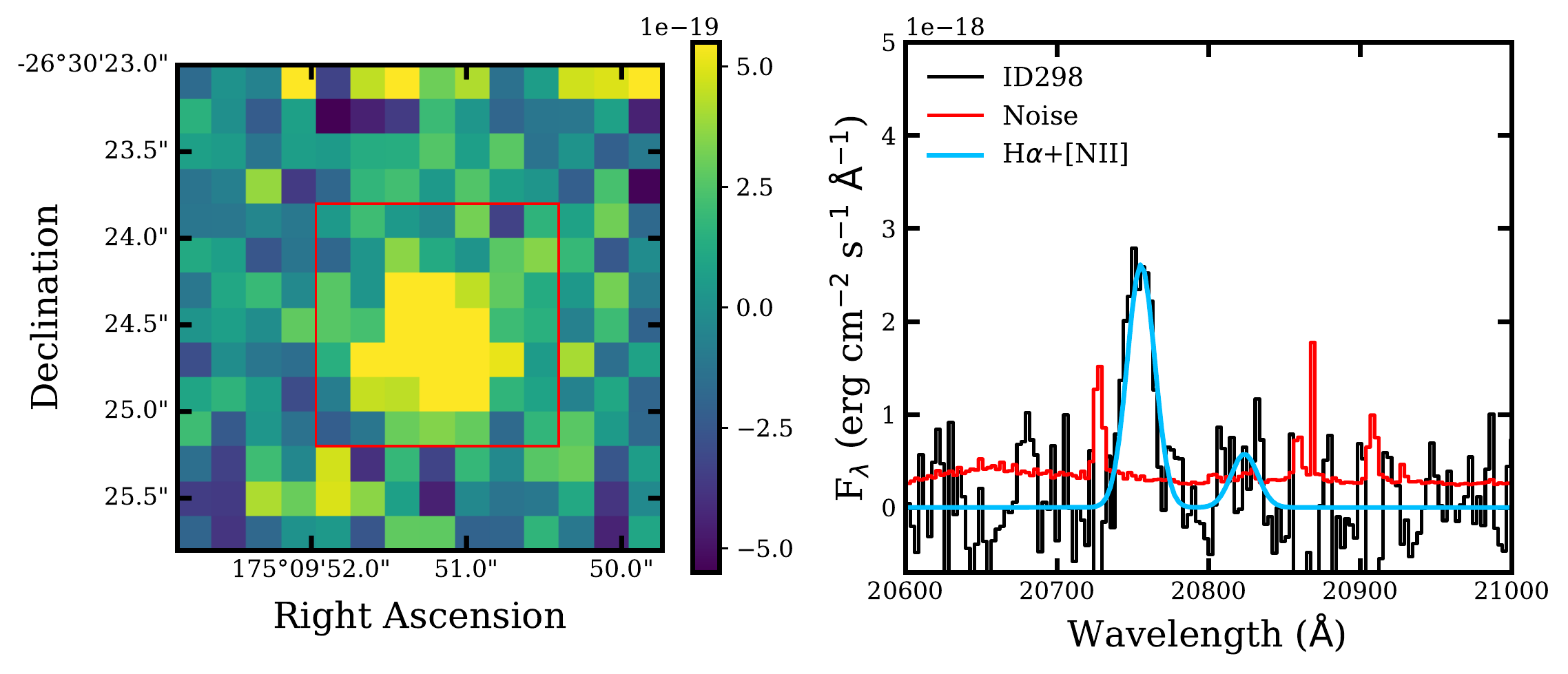}\par
      \includegraphics[width=\linewidth]{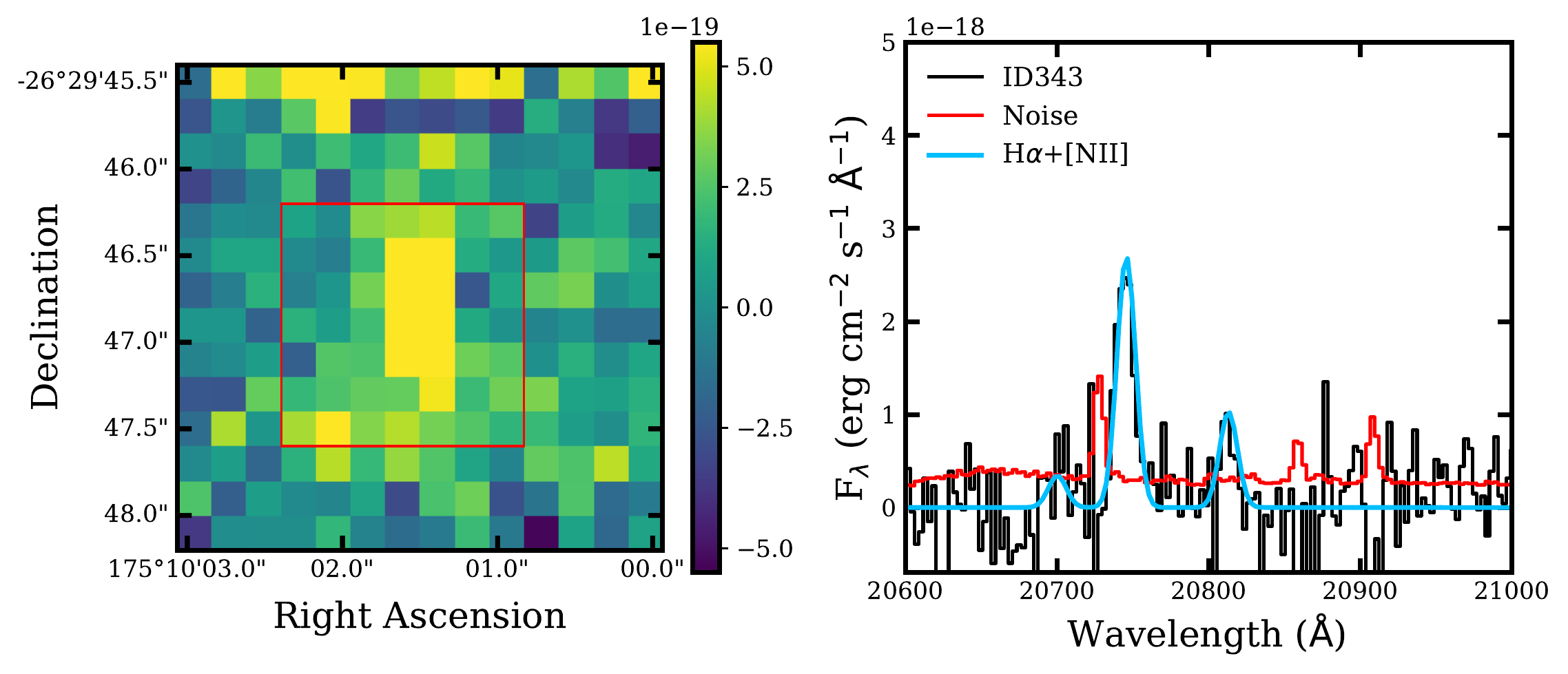}\par
      \includegraphics[width=\linewidth]{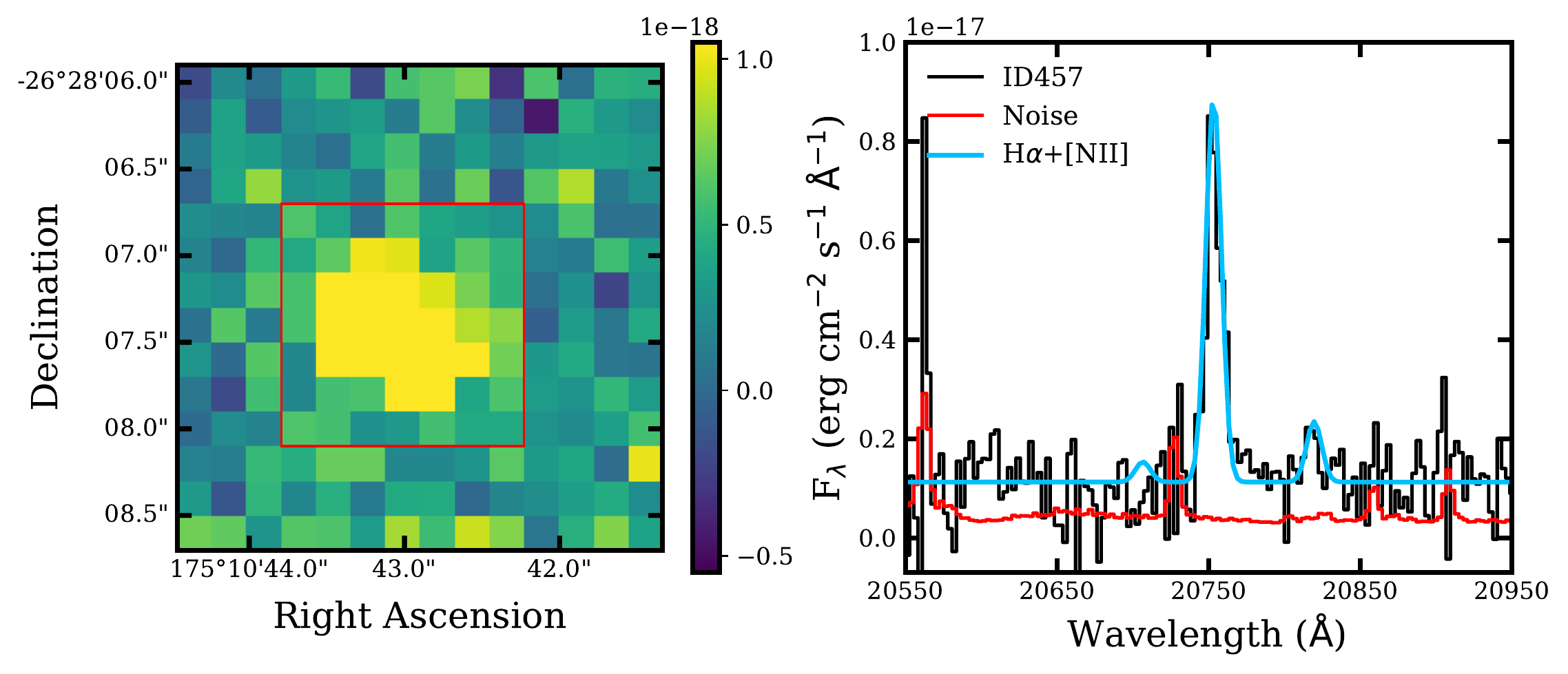}\par
      \includegraphics[width=\linewidth]{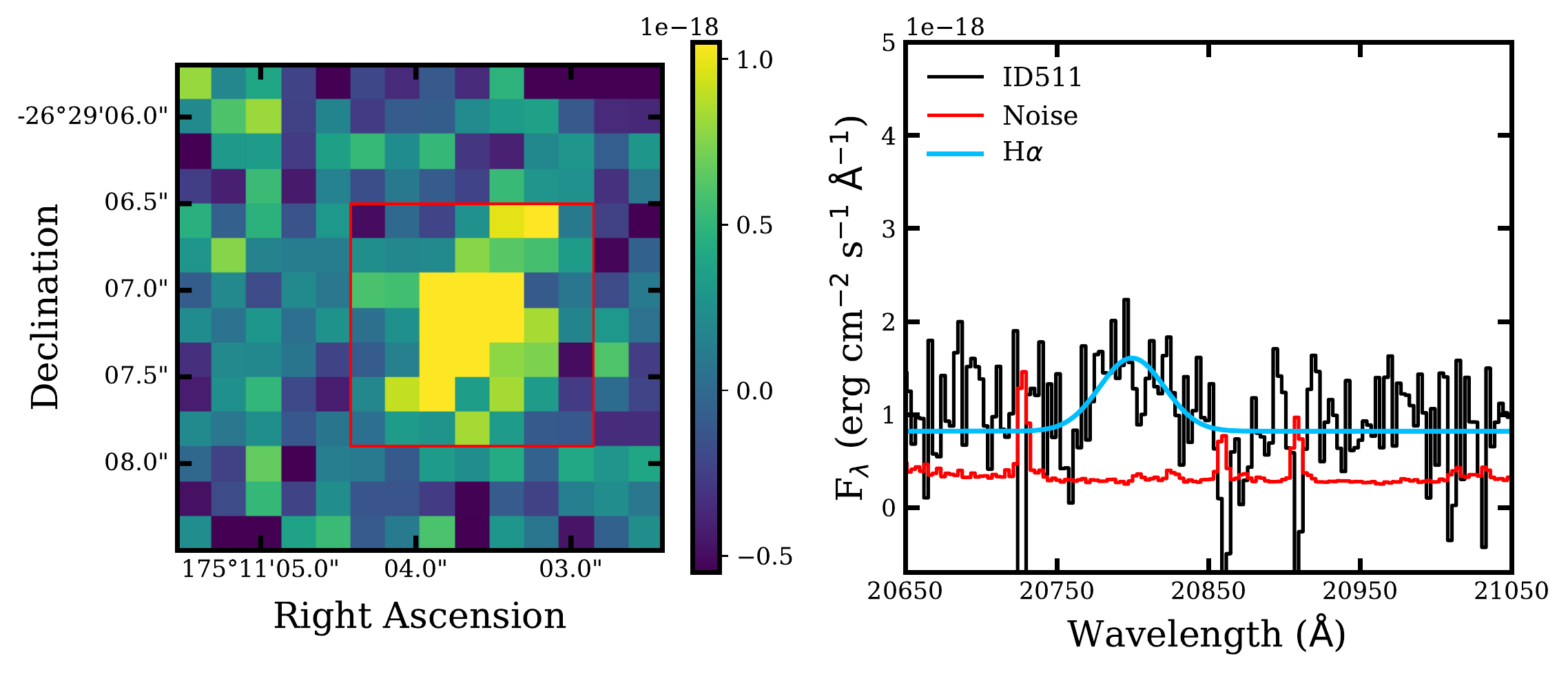}\par
      \includegraphics[width=\linewidth]{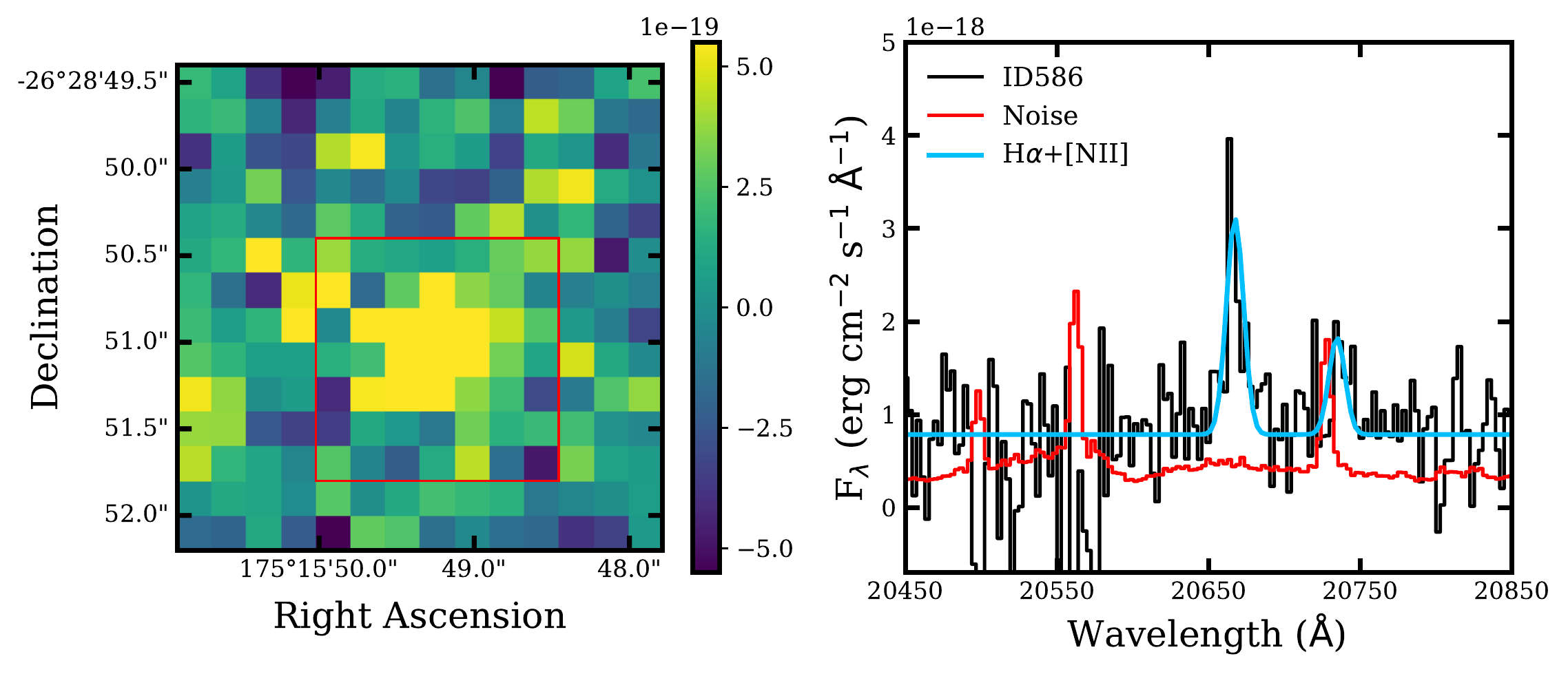}\par
      \includegraphics[width=\linewidth]{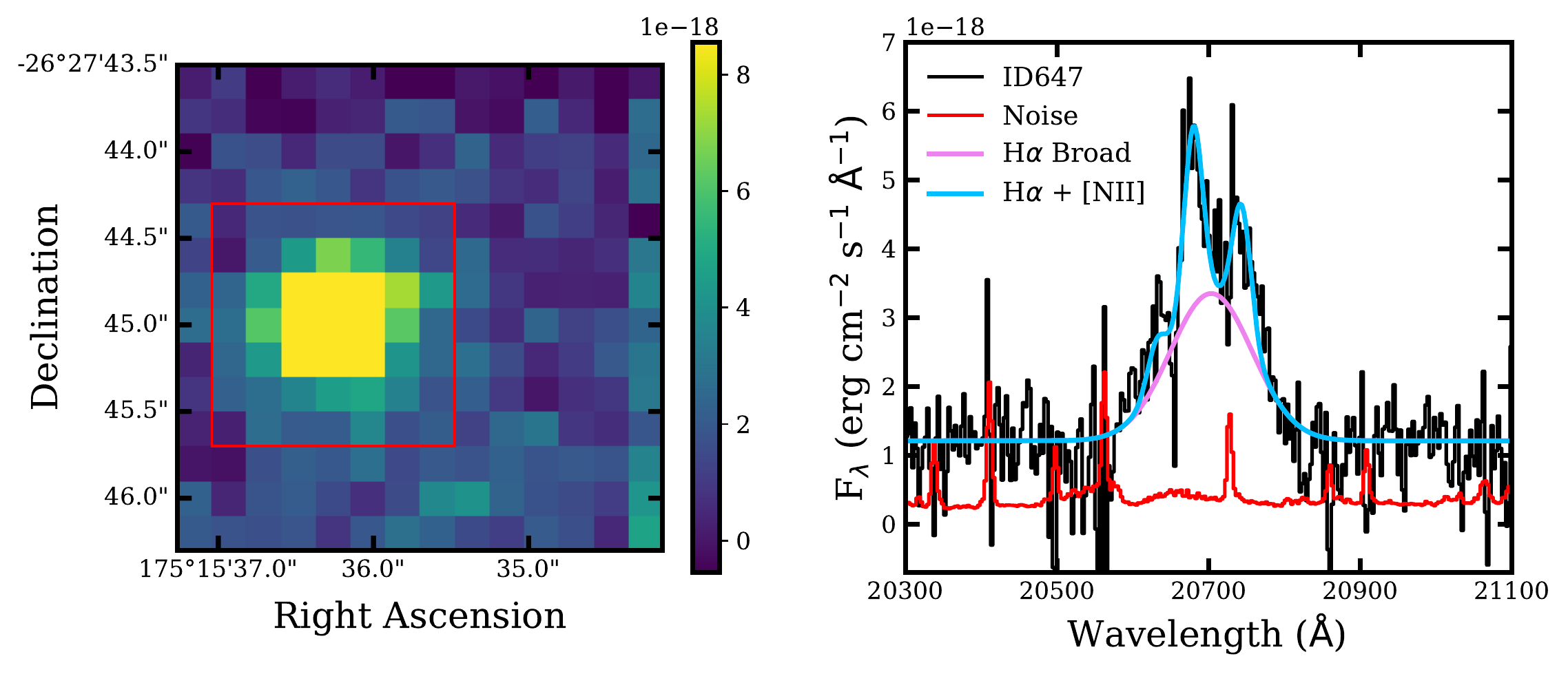}\par
      \includegraphics[width=\linewidth]{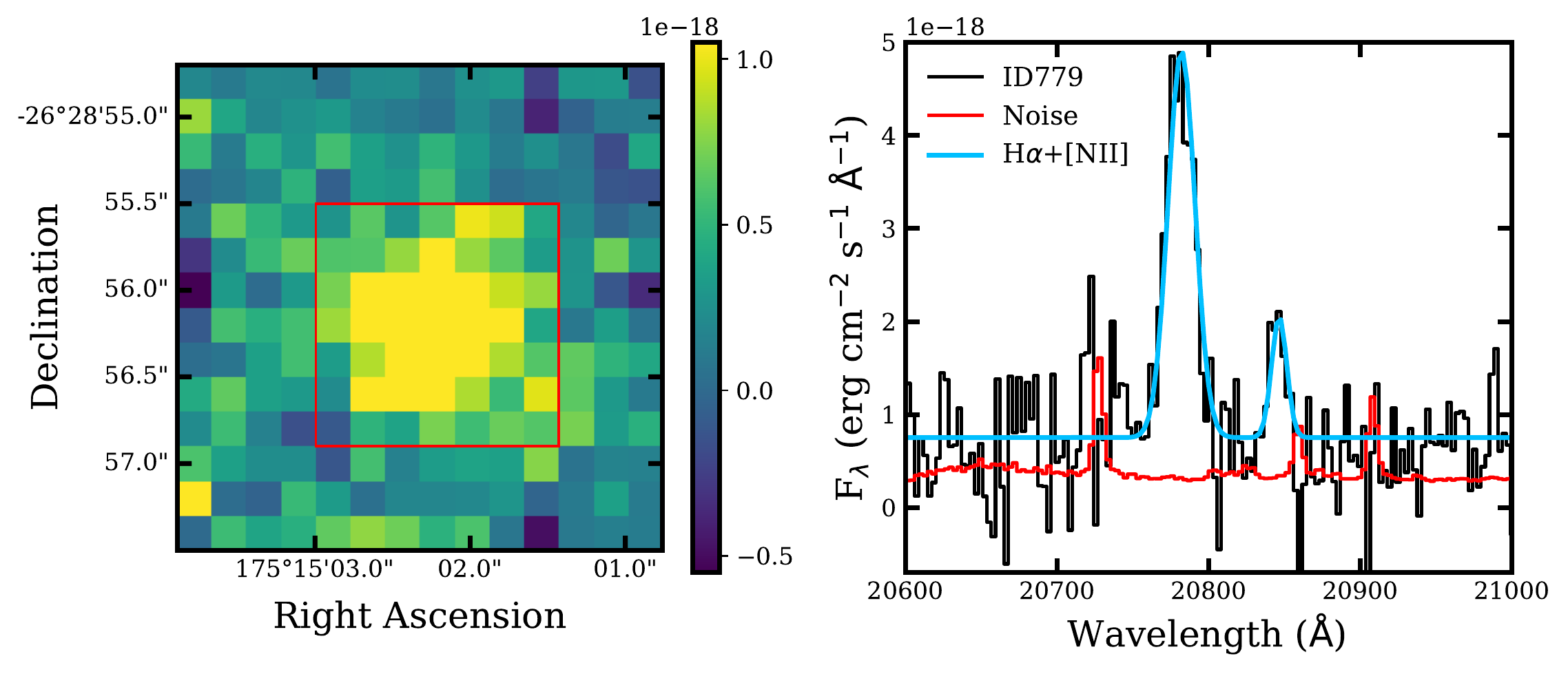}\par
      \end{multicols}
      \caption{Left: Image of the H$\alpha$-collapsed datacube. The red square shows the $1.4\arcsec$ squared aperture used to extract the 1D spectrum of each object. Right: Aperture extracted signal spectrum (black), noise spectrum (red) and fitted spectrum (blue) to the H$\alpha$ and [N{\sc{ii}}] emission lines. The used IDs follow the numbering given by \protect\cite{Koyama13}.}
         \label{F:IFU}
\end{figure*}

\begin{figure*}
 \centering
 \begin{multicols}{2}
      \includegraphics[width=\linewidth]{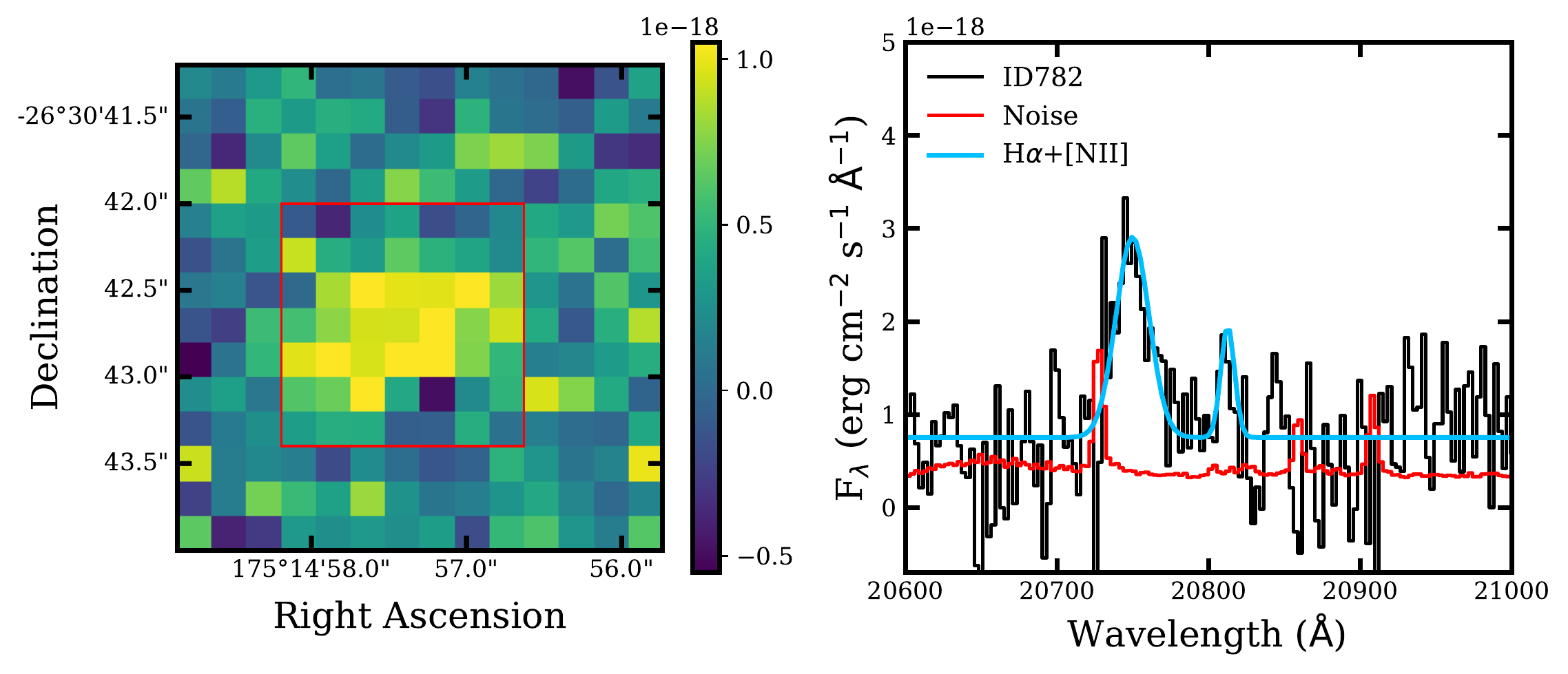}\par
      \includegraphics[width=\linewidth]{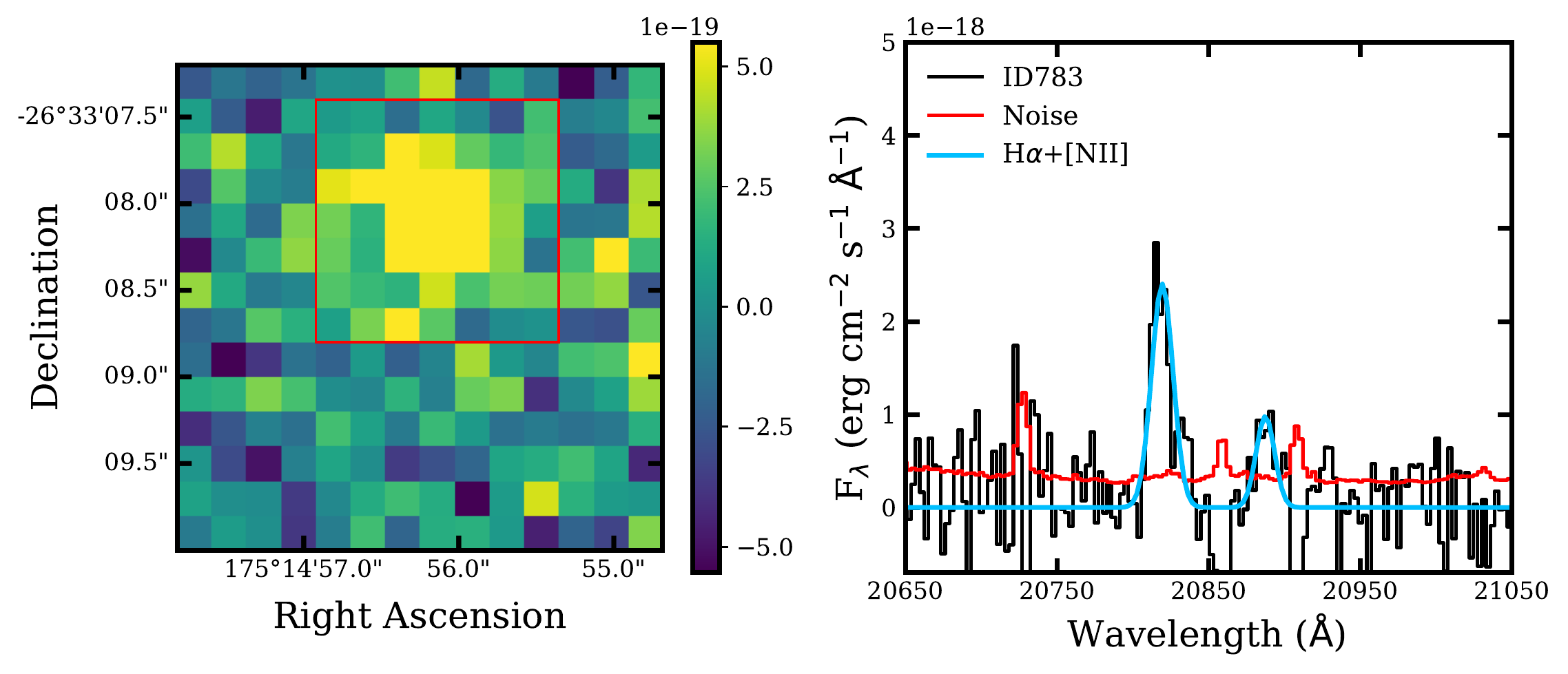}\par
      \includegraphics[width=\linewidth]{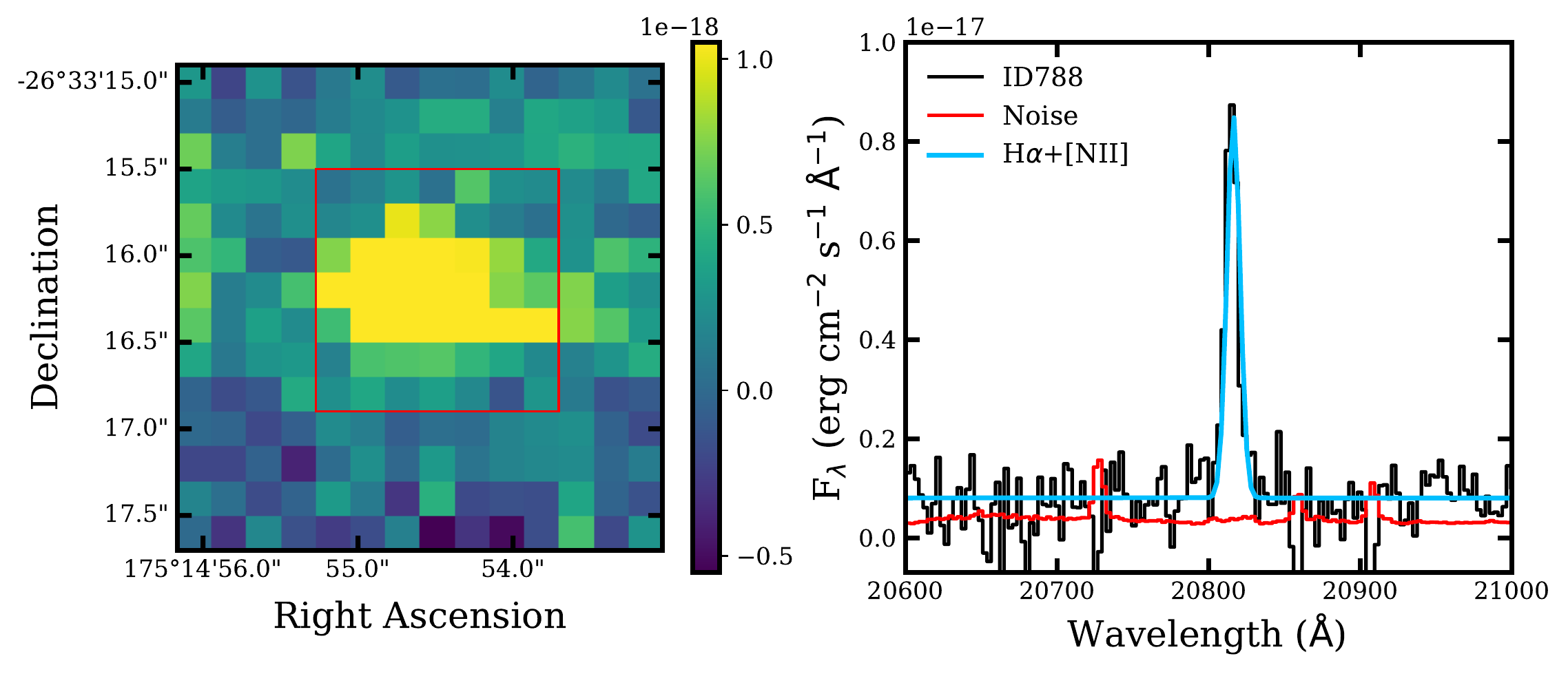}\par
      \includegraphics[width=\linewidth]{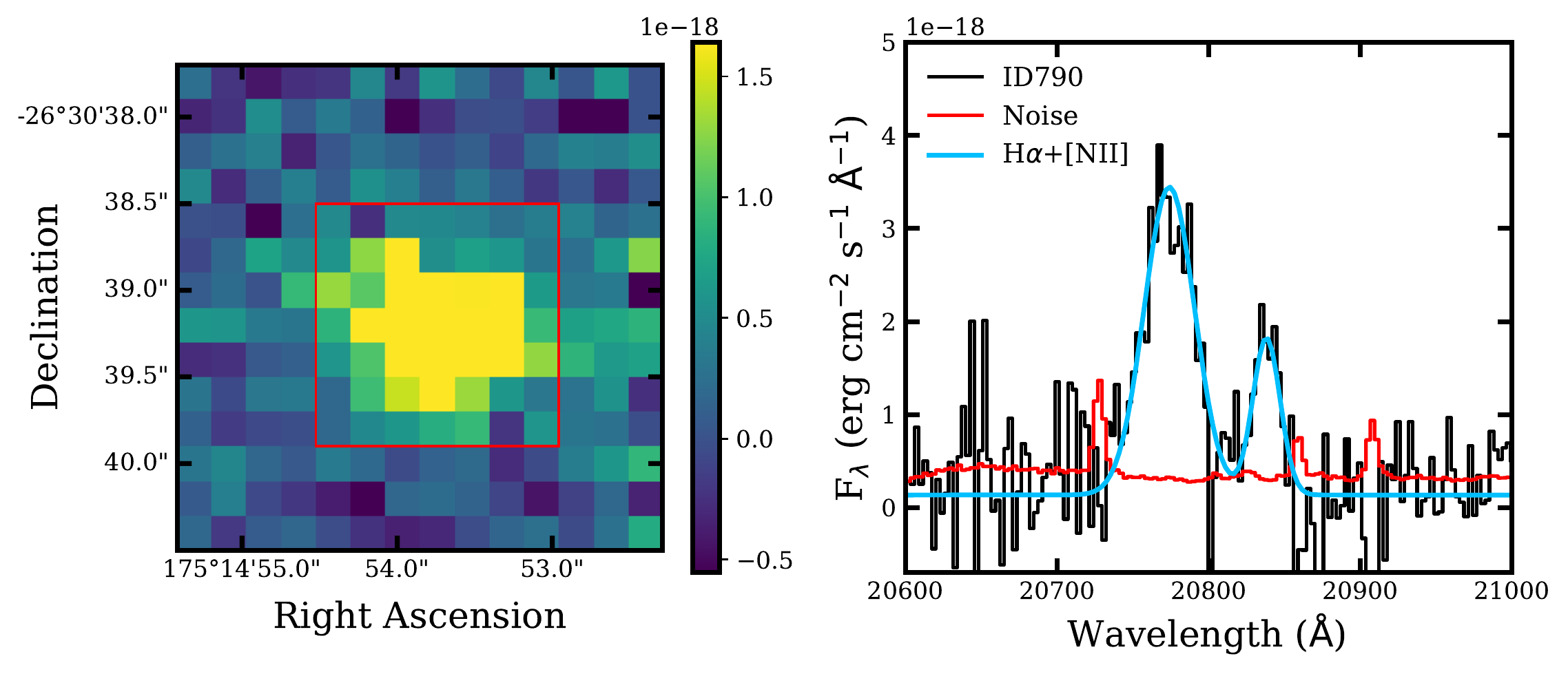}\par
      \includegraphics[width=\linewidth]{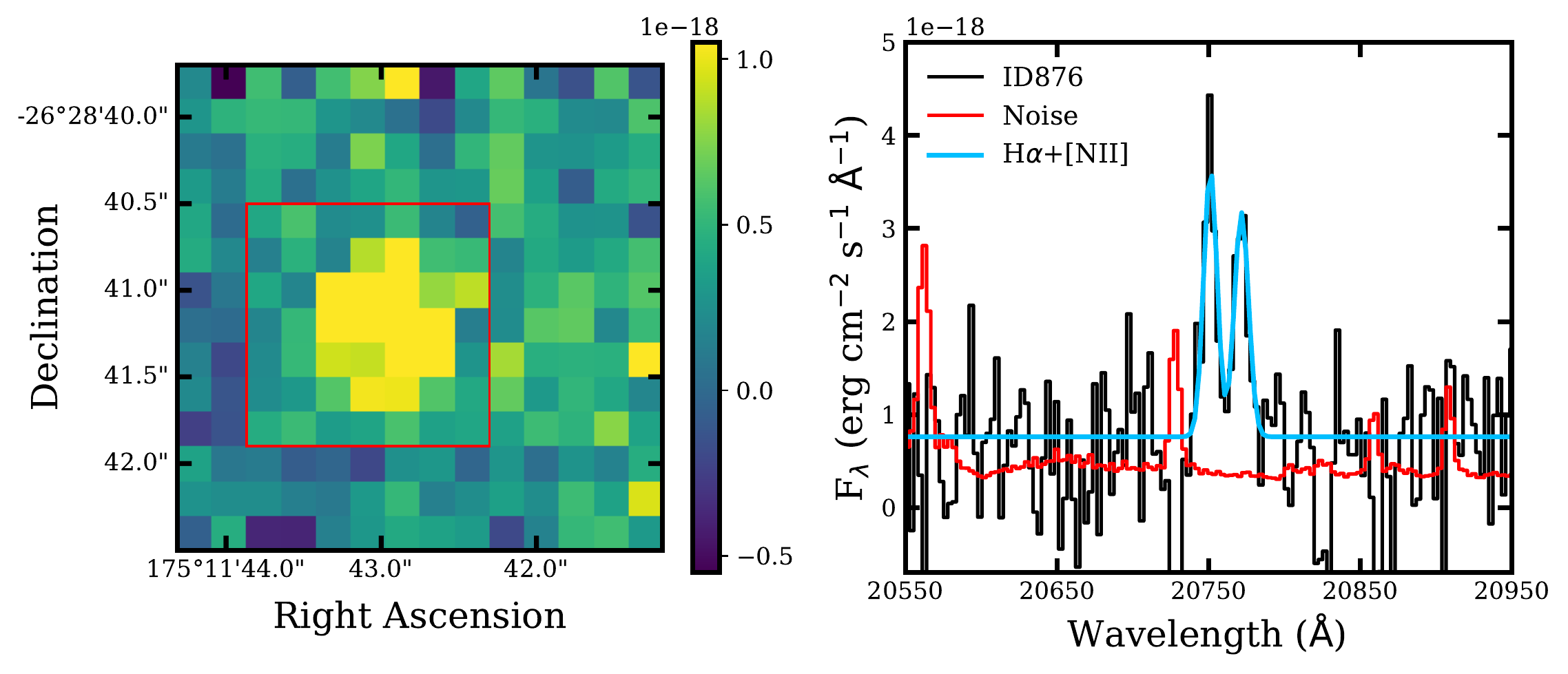}\par
      \includegraphics[width=\linewidth]{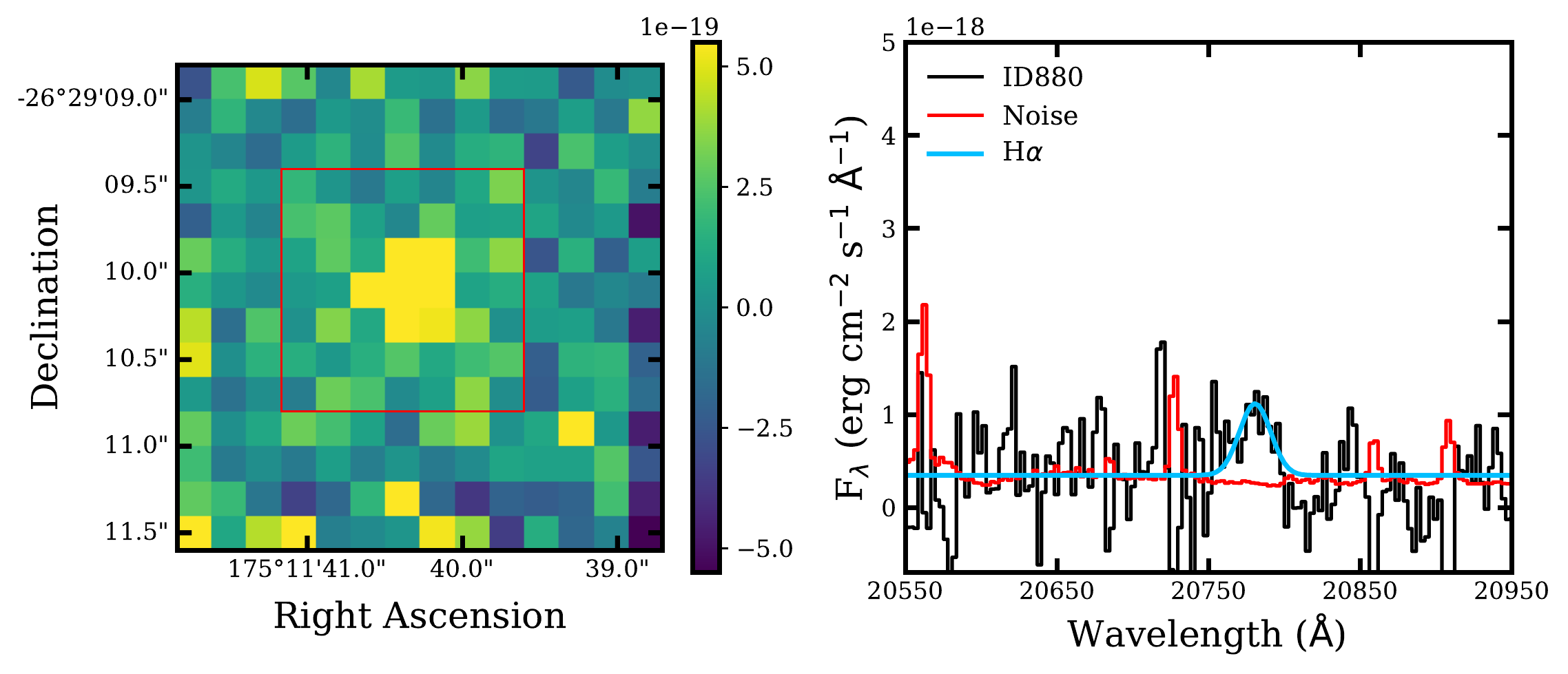}\par
      \includegraphics[width=\linewidth]{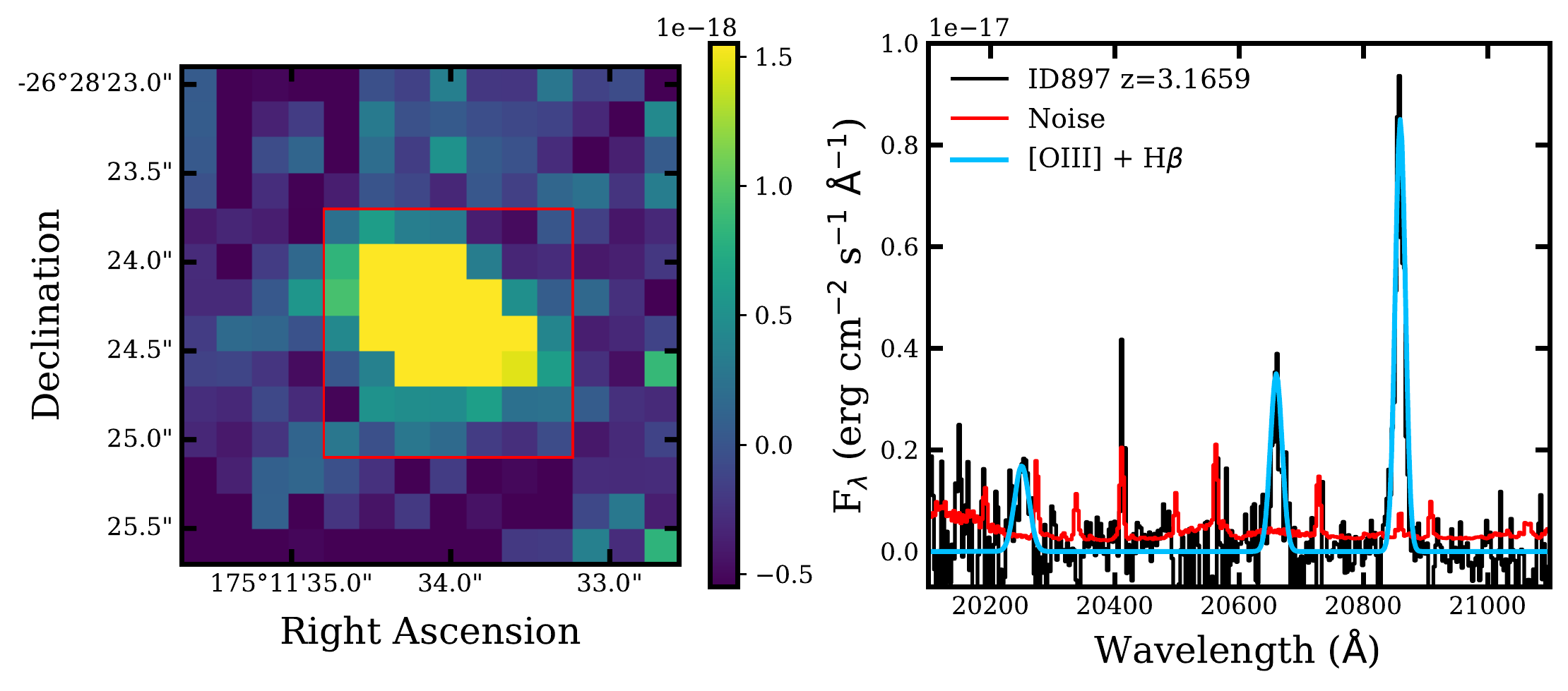}\par
      \includegraphics[width=\linewidth]{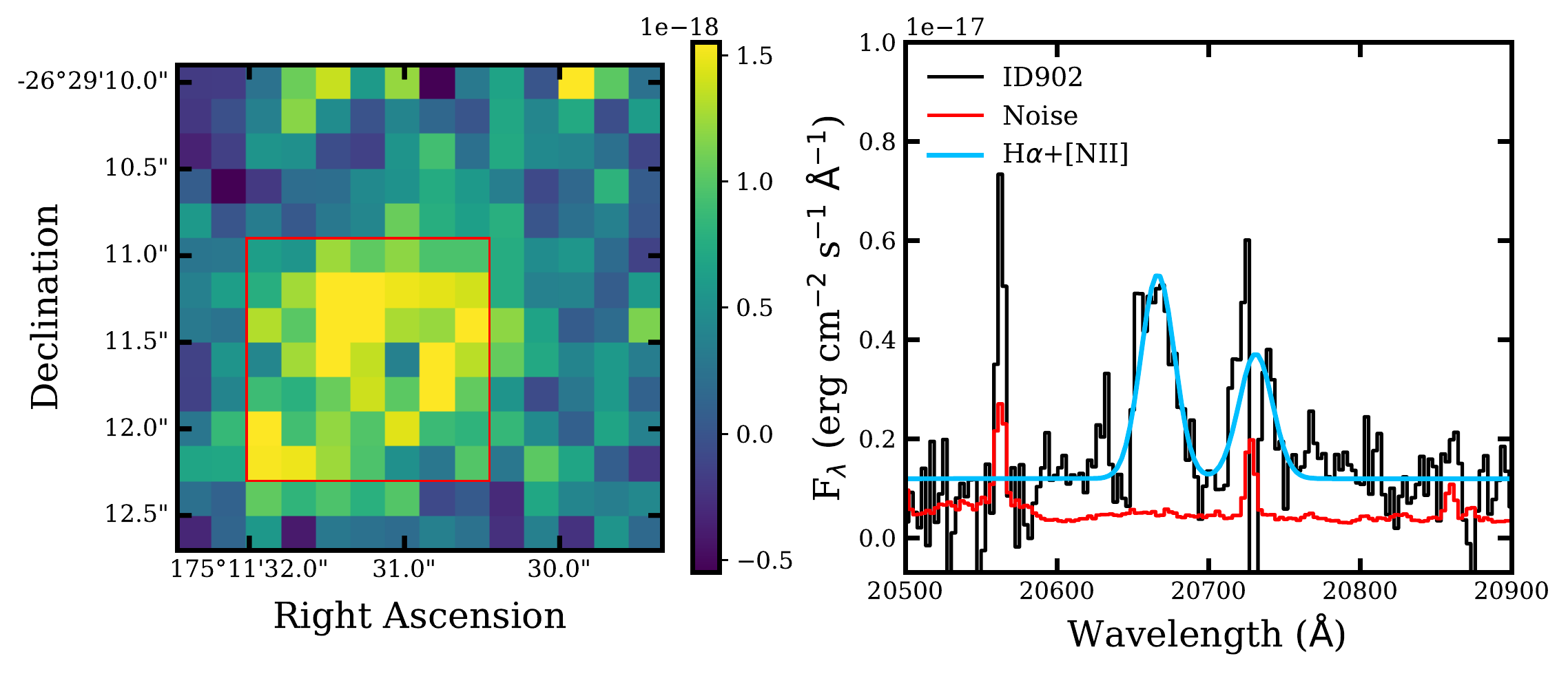}\par
      \includegraphics[width=\linewidth]{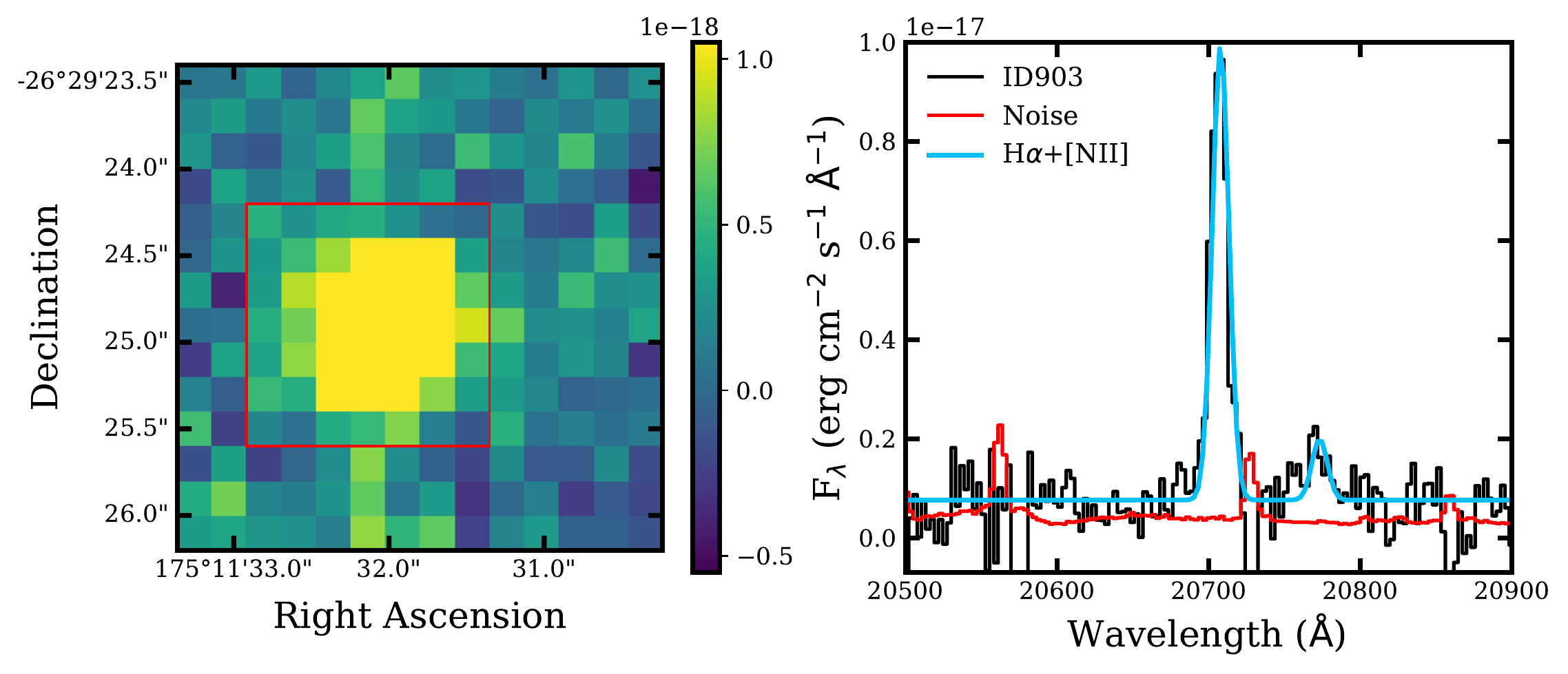}\par
      \includegraphics[width=\linewidth]{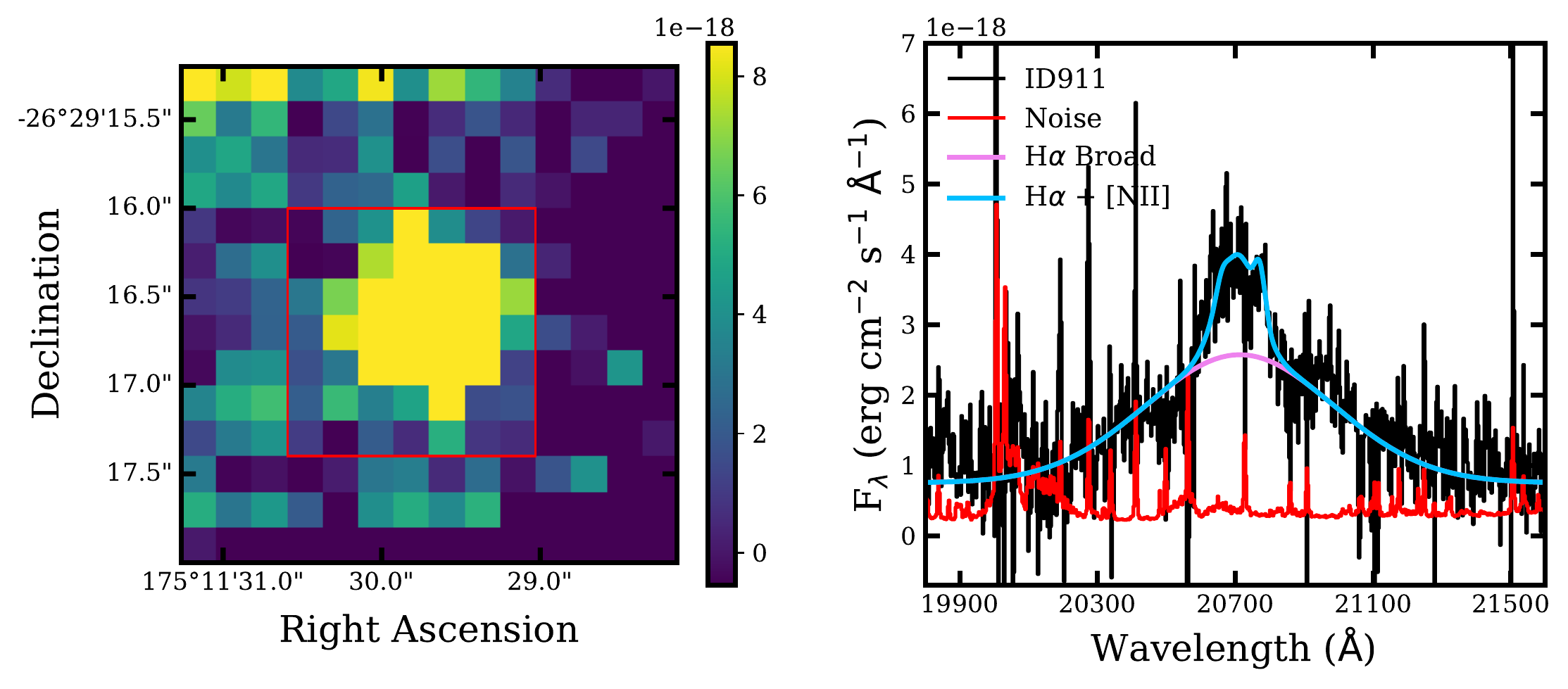}\par
      \includegraphics[width=\linewidth]{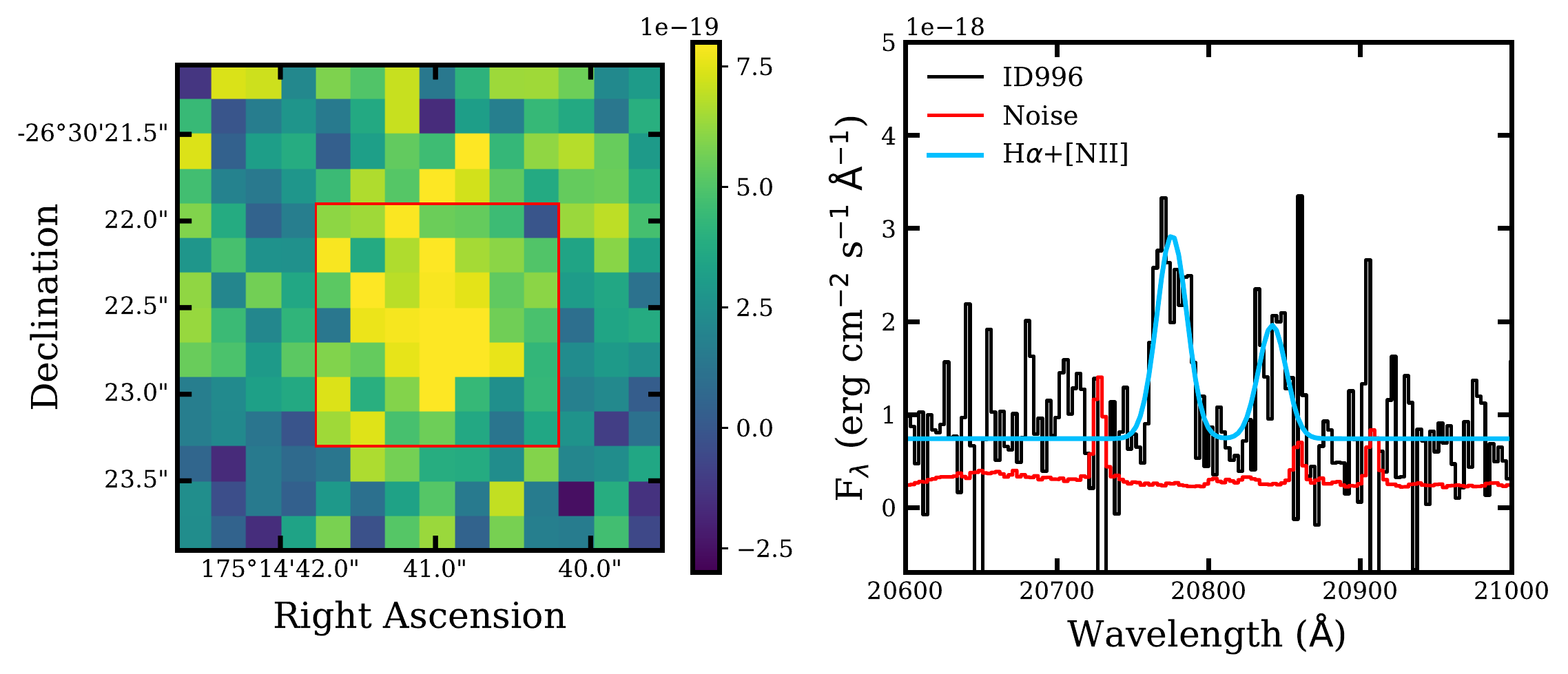}\par
      \includegraphics[width=\linewidth]{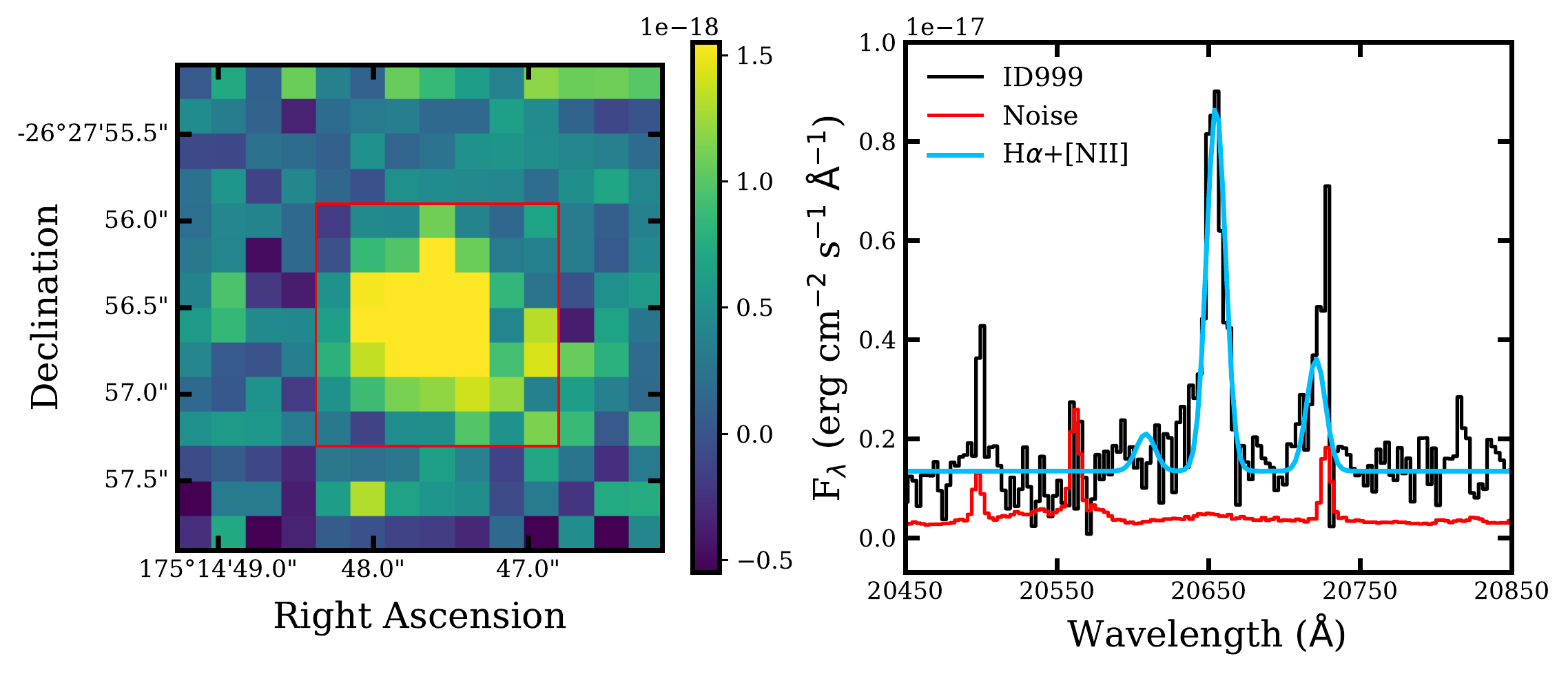}\par
      \end{multicols}
      \contcaption{}
         \label{F:SFR}
\end{figure*}

\begin{figure*}
 \centering
 \begin{multicols}{2}
      \includegraphics[width=\linewidth]{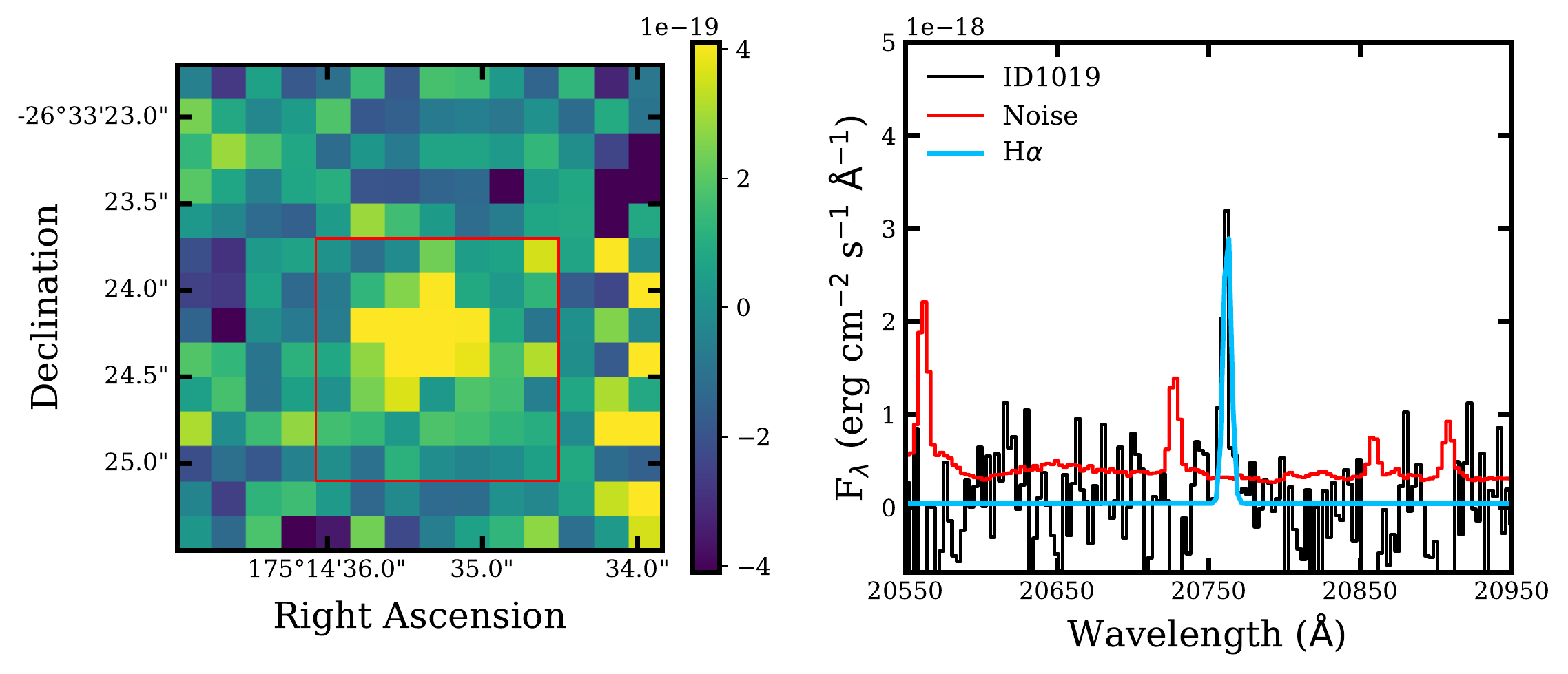}\par
      \includegraphics[width=\linewidth]{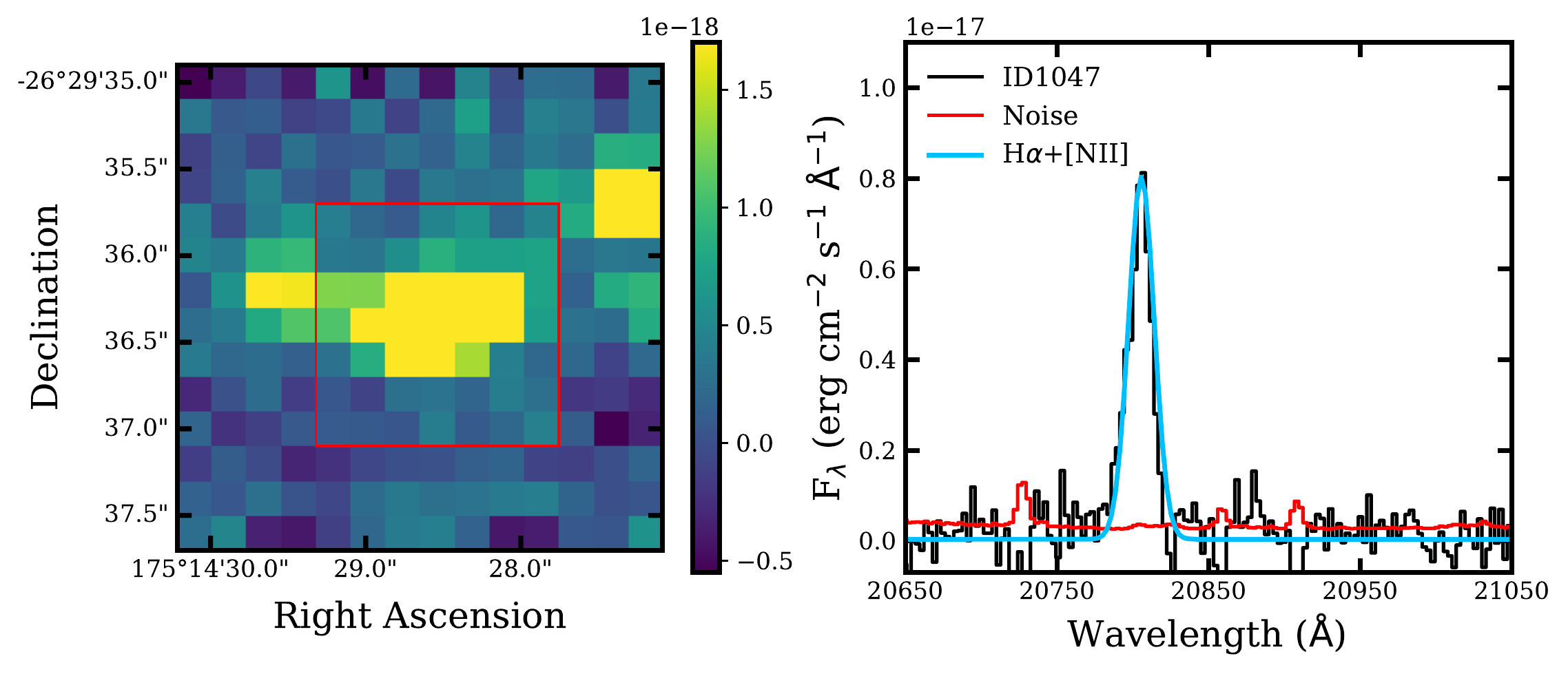}\par
      \includegraphics[width=\linewidth]{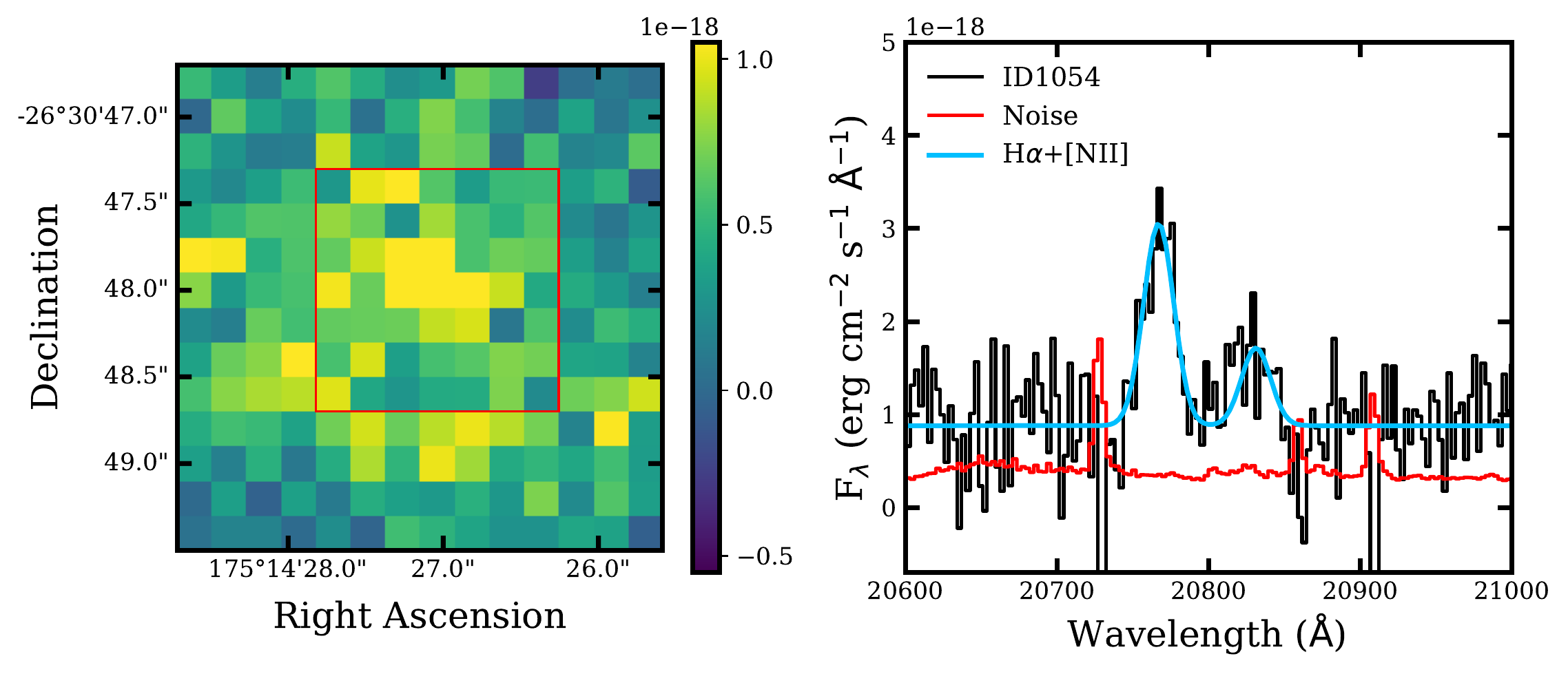}\par
      \includegraphics[width=\linewidth]{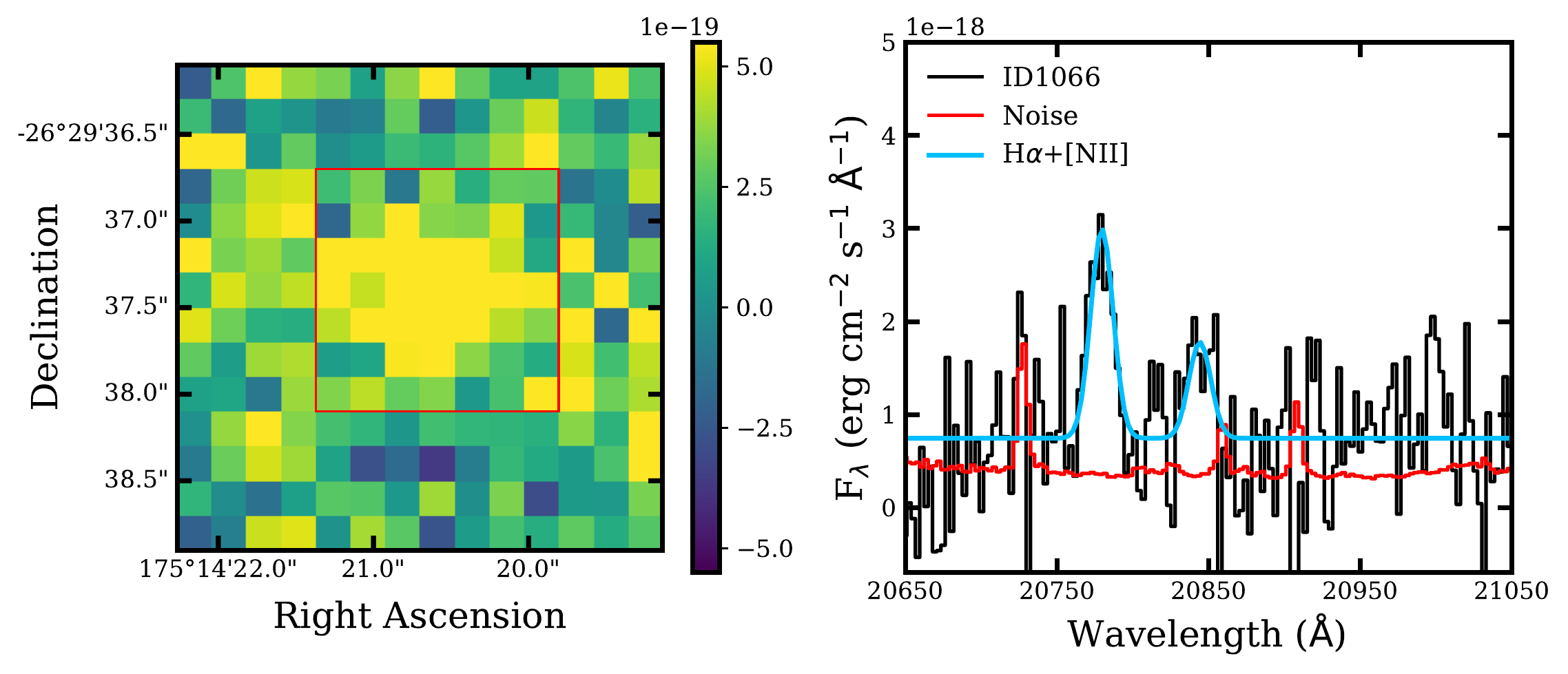}\par
      \includegraphics[width=\linewidth]{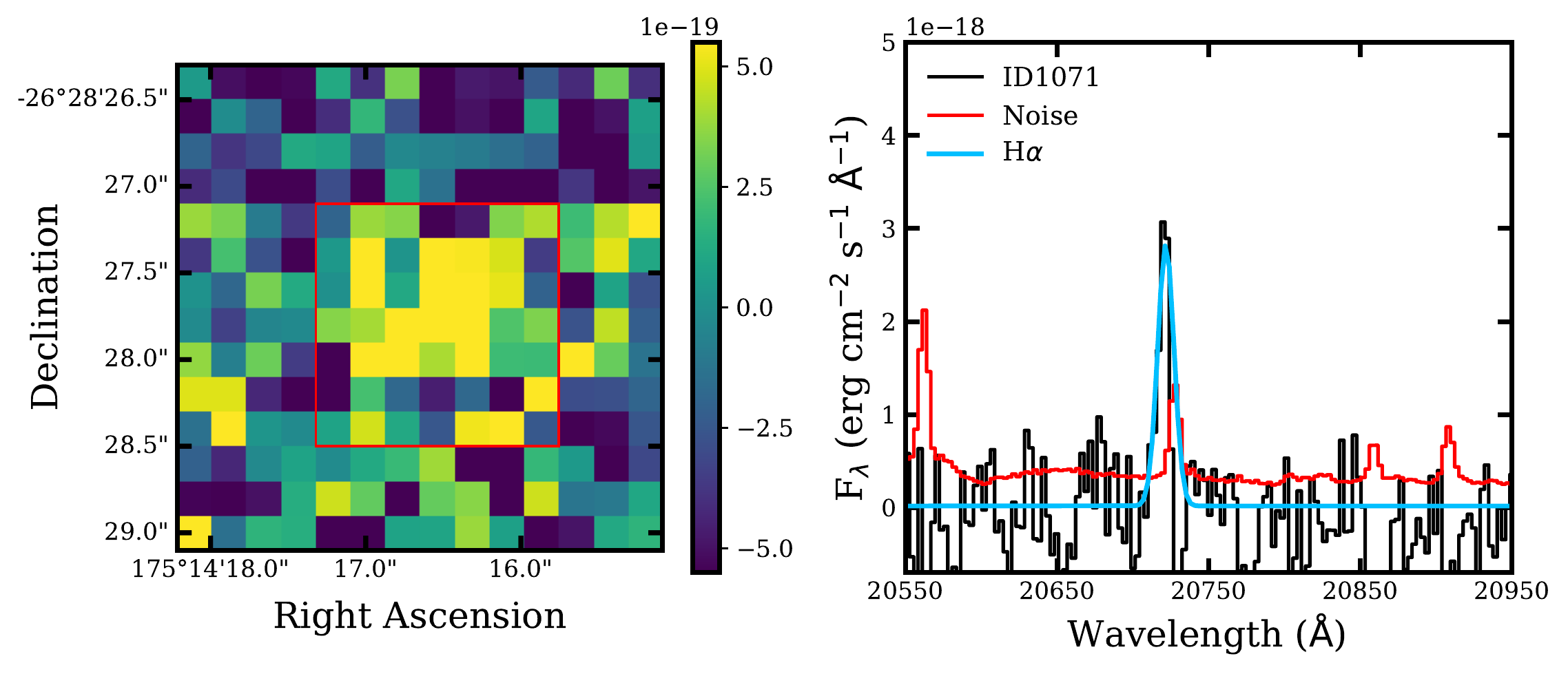}\par
      \includegraphics[width=\linewidth]{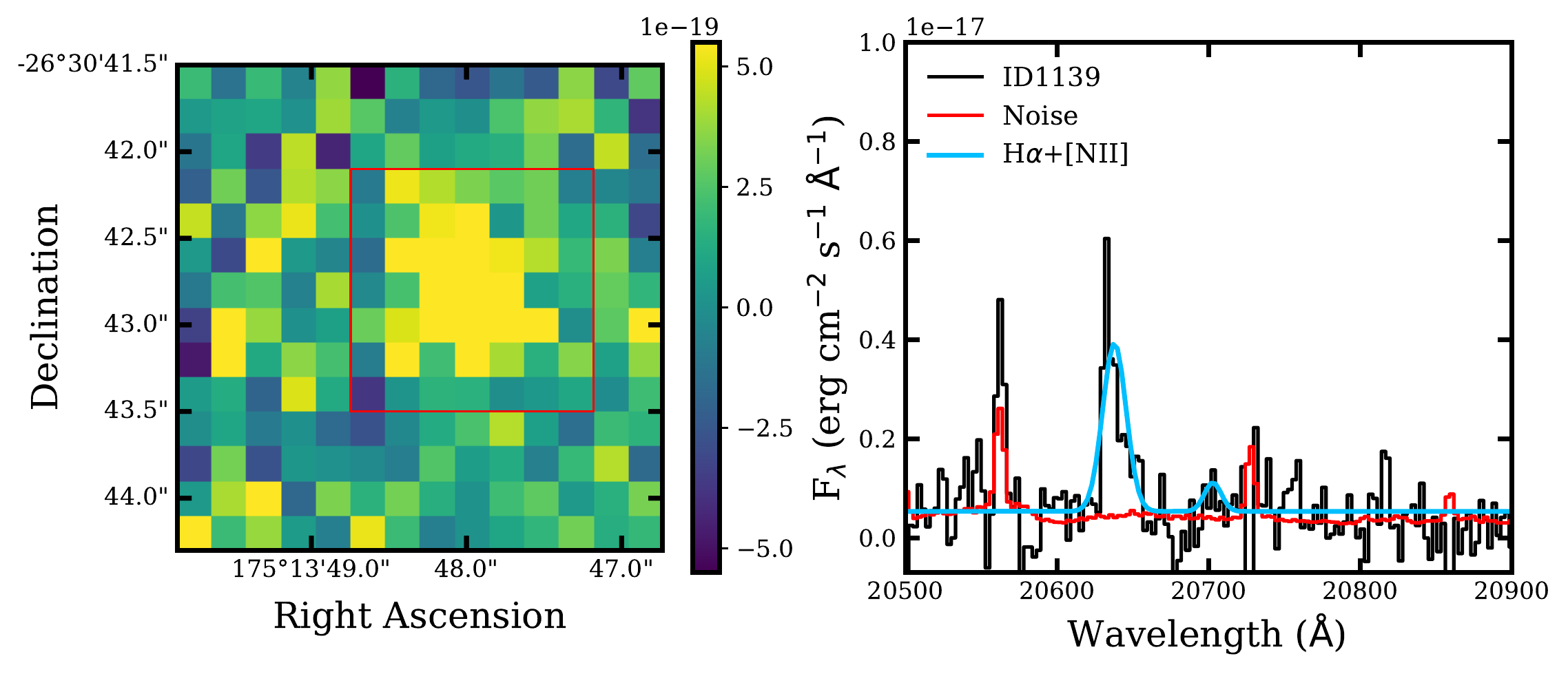}\par
      \includegraphics[width=\linewidth]{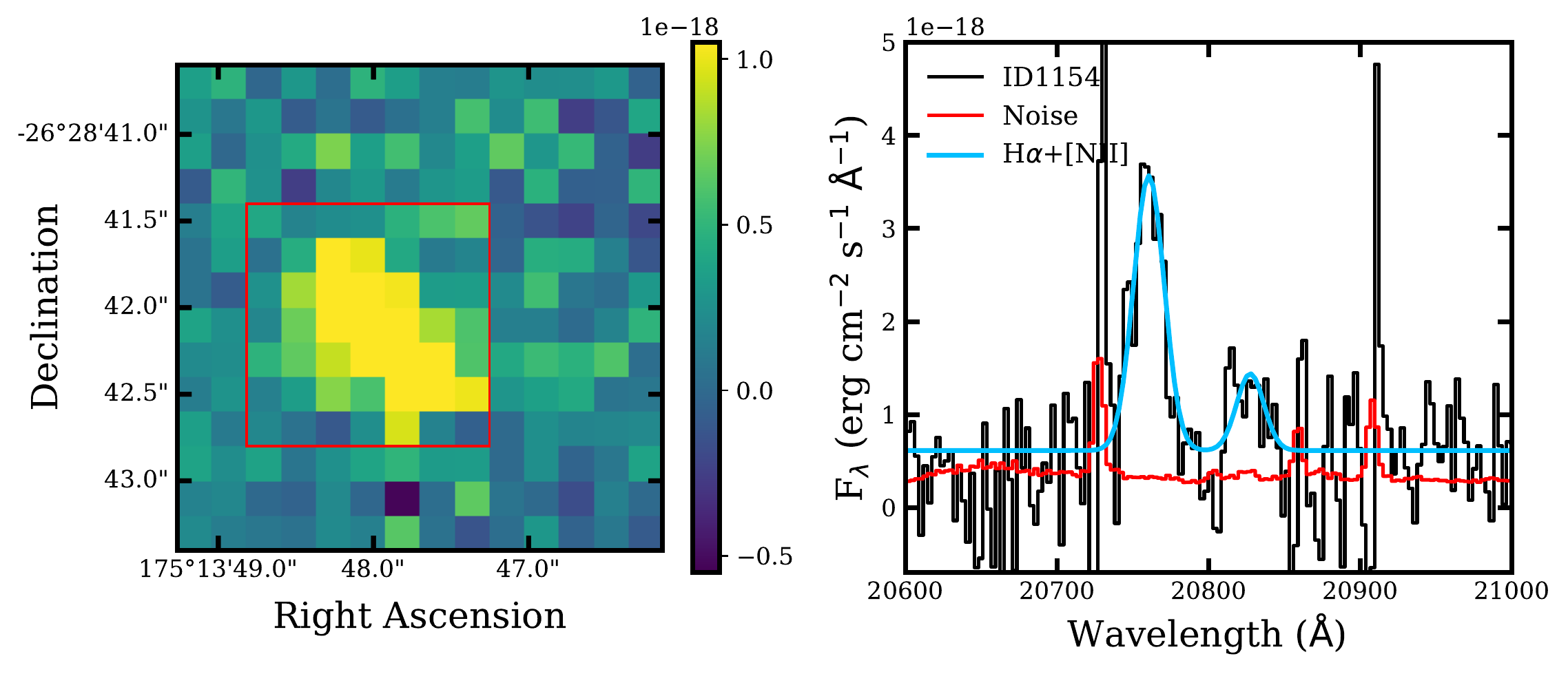}\par
      \includegraphics[width=\linewidth]{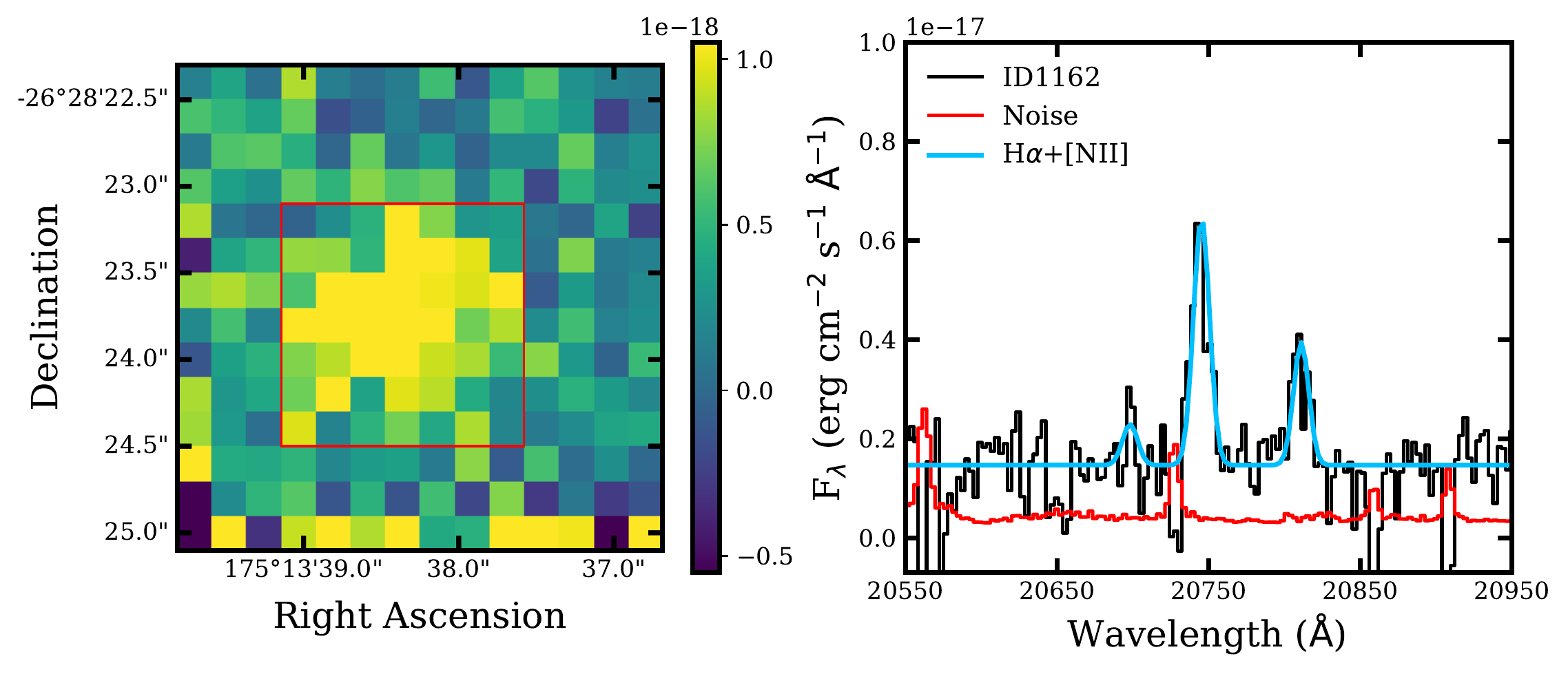}\par
      \includegraphics[width=\linewidth]{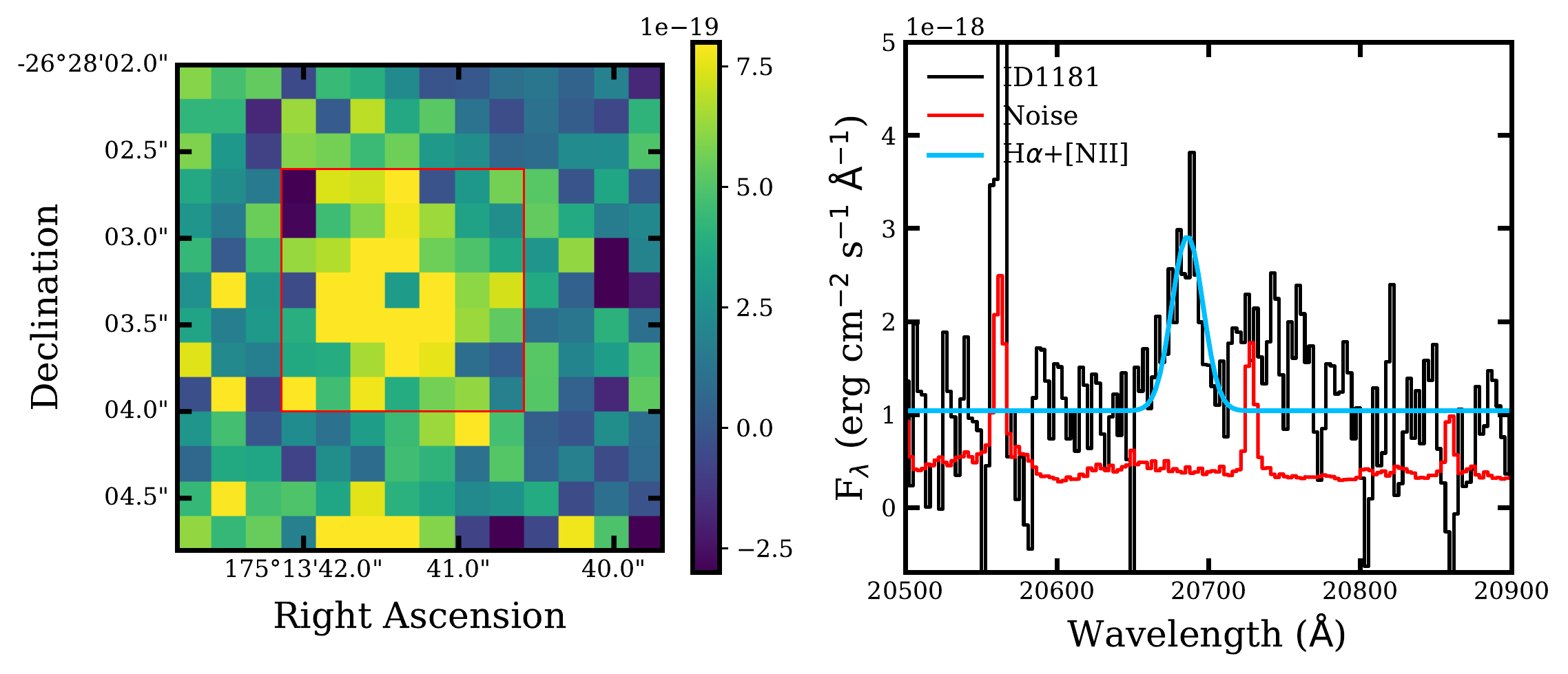}\par
      \includegraphics[width=\linewidth]{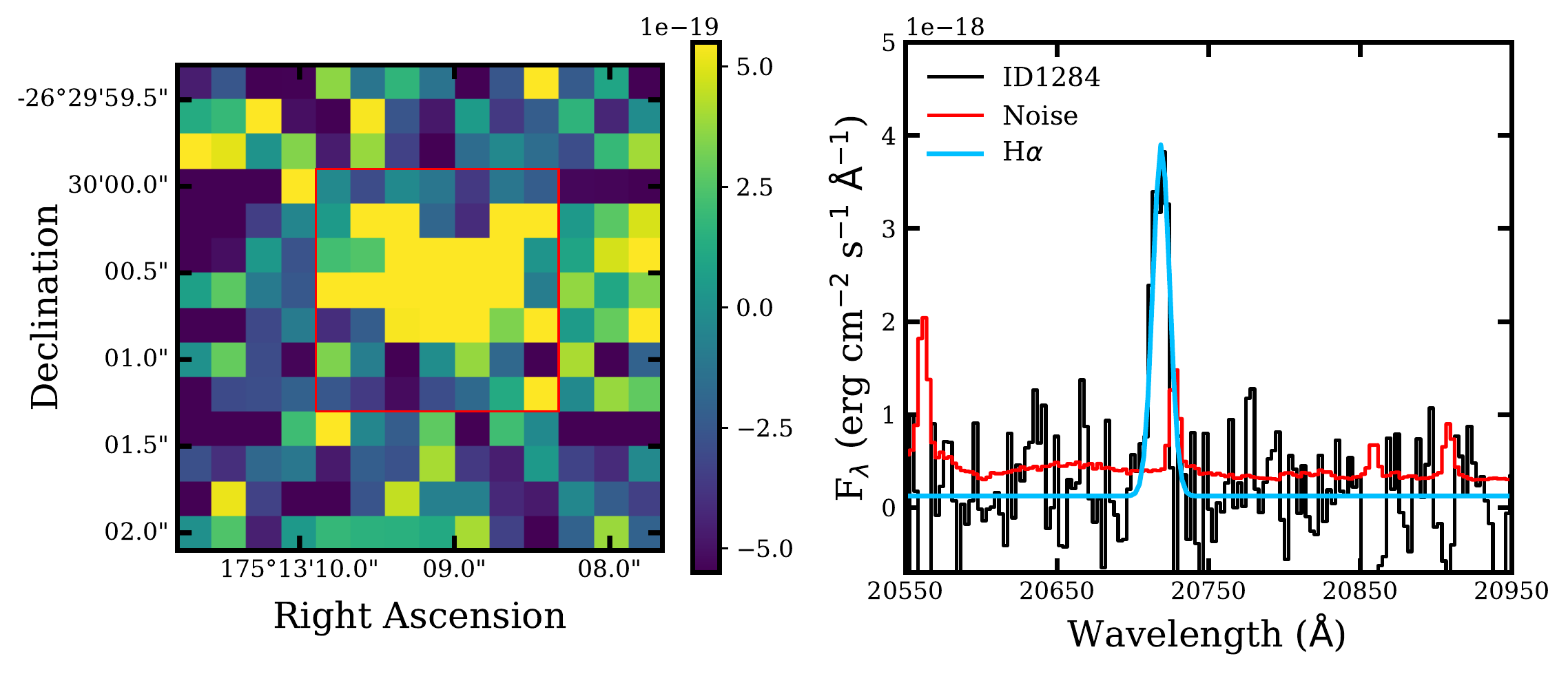}\par 
      \includegraphics[width=\linewidth]{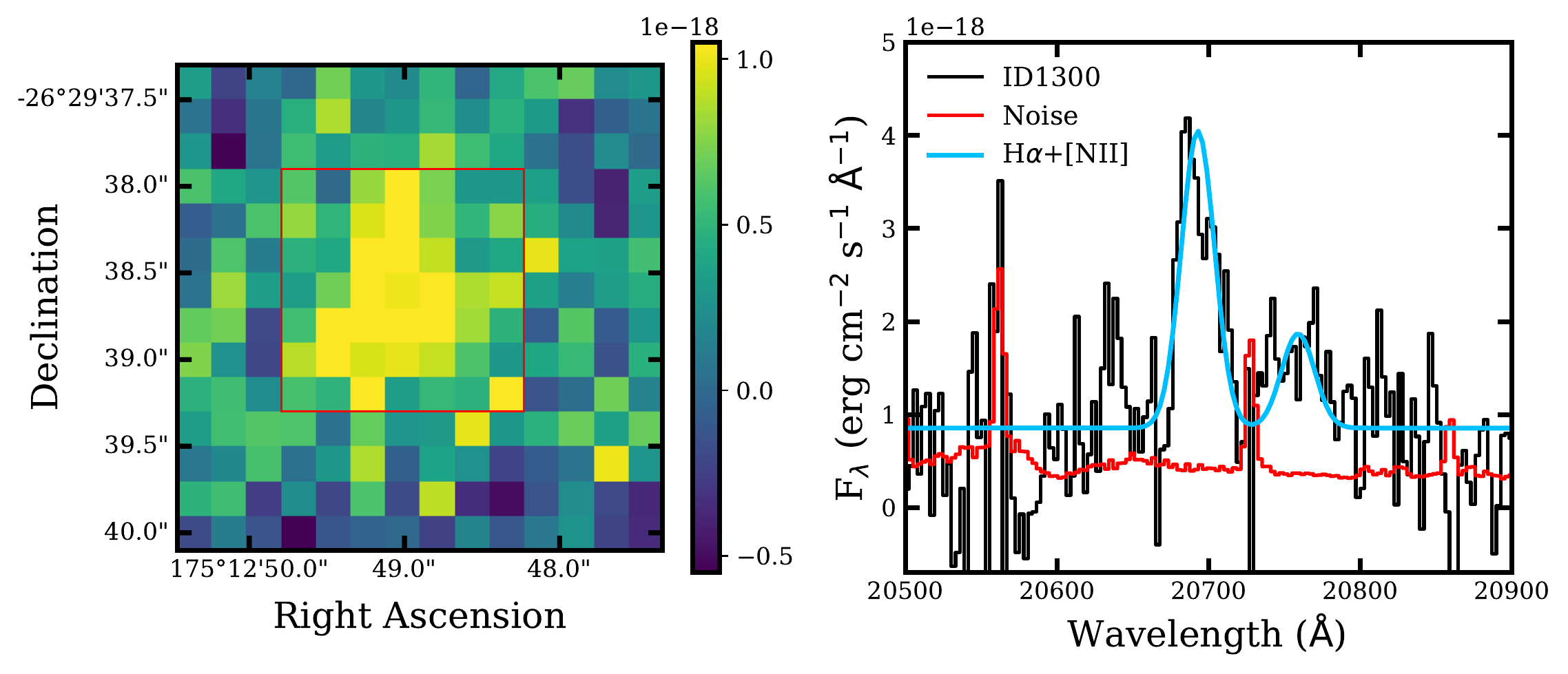}\par
      \includegraphics[width=\linewidth]{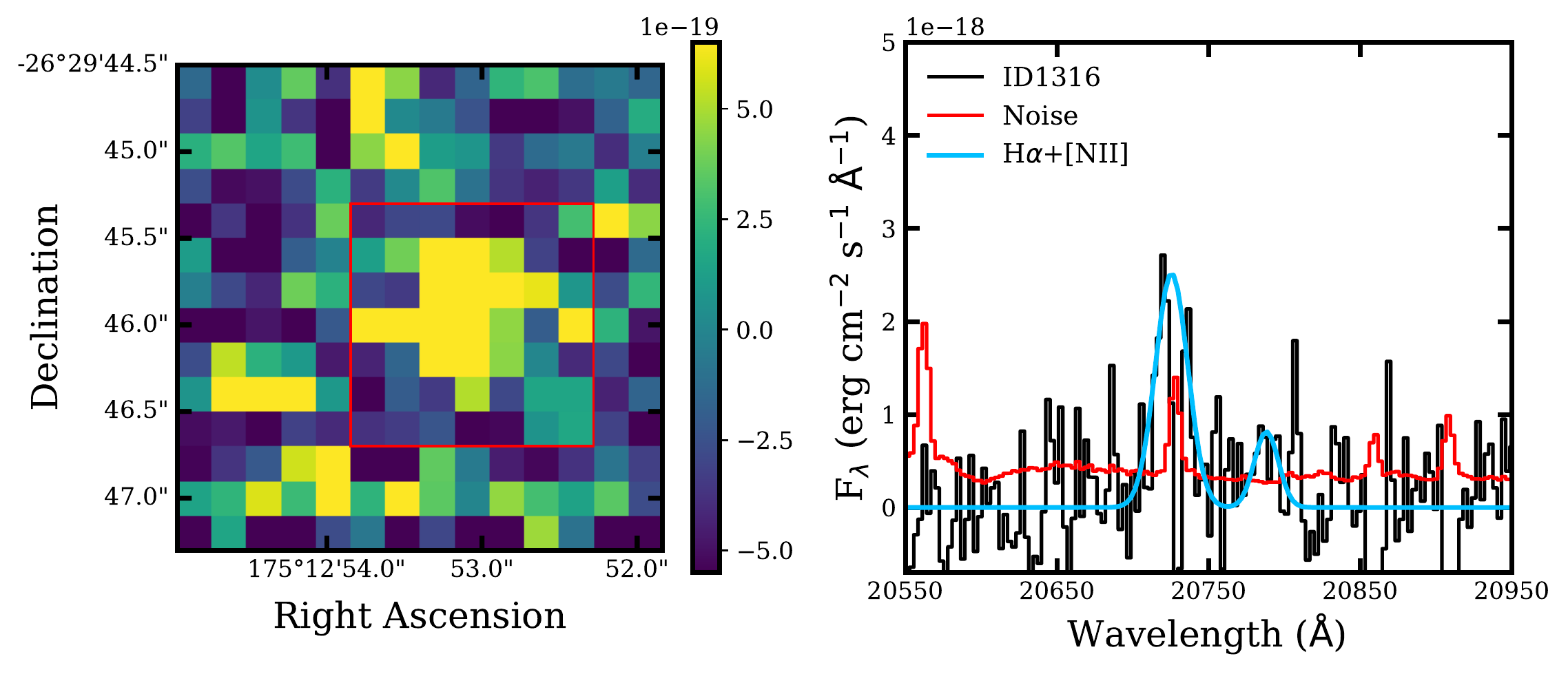}\par
      \end{multicols}
      \contcaption{}
         \label{F:SFR}
\end{figure*}

\begin{figure*}
 \centering
 \begin{multicols}{2}
      \includegraphics[width=\linewidth]{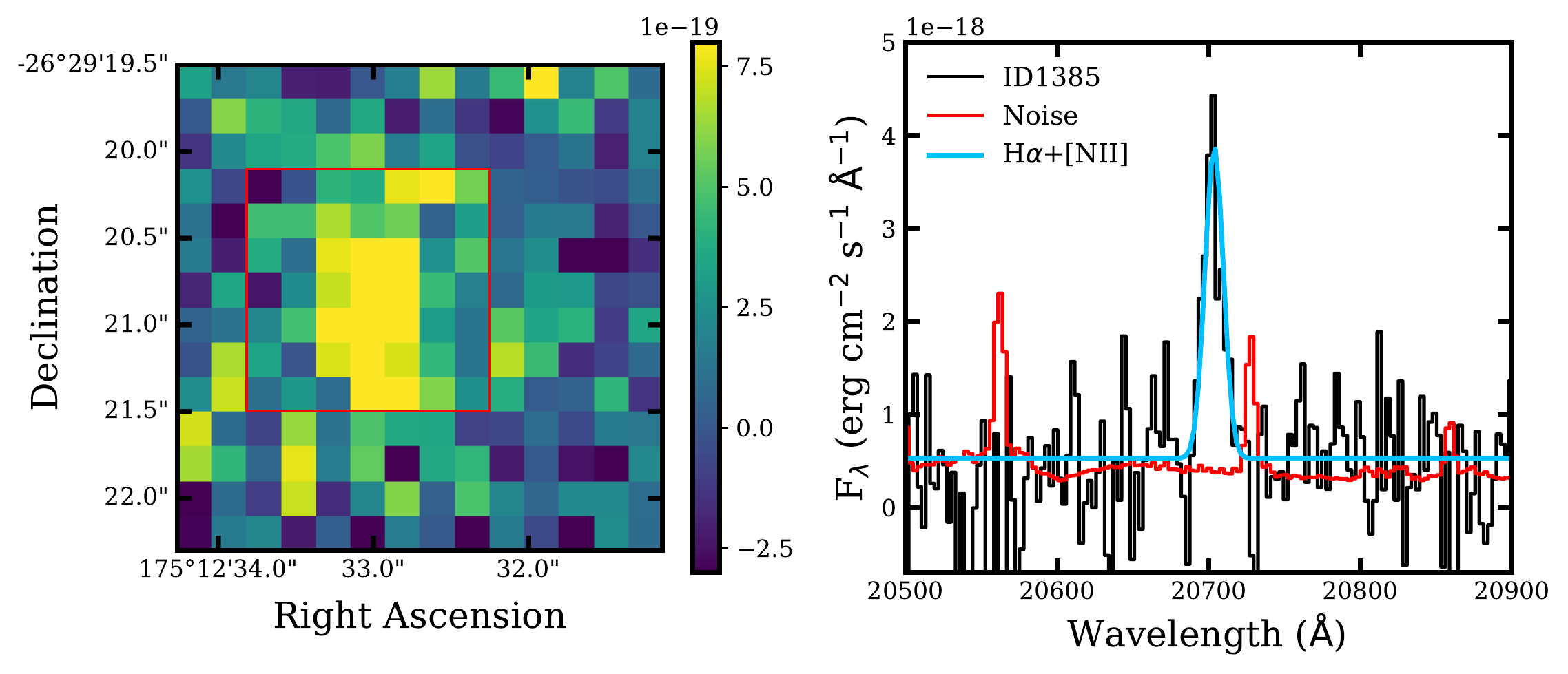}\par
      \includegraphics[width=\linewidth]{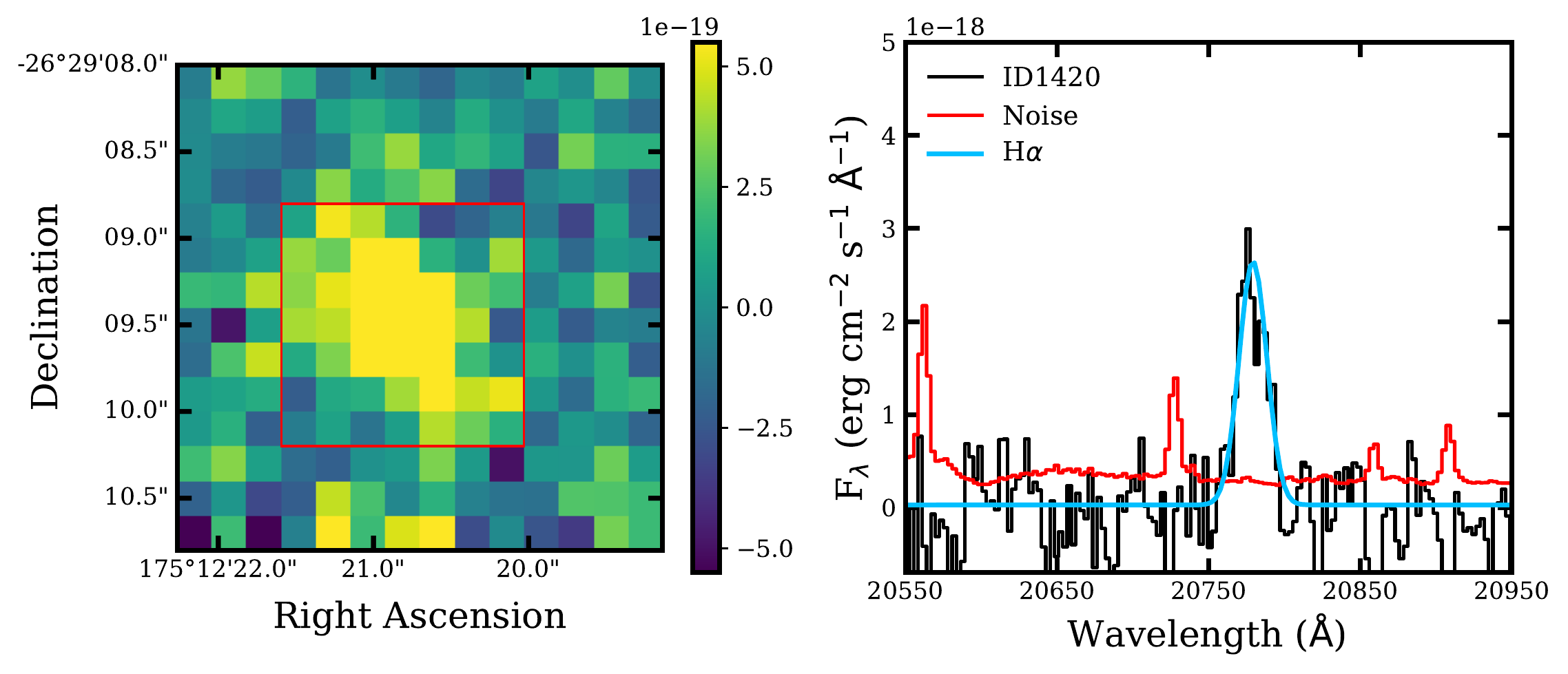}\par
      \includegraphics[width=\linewidth]{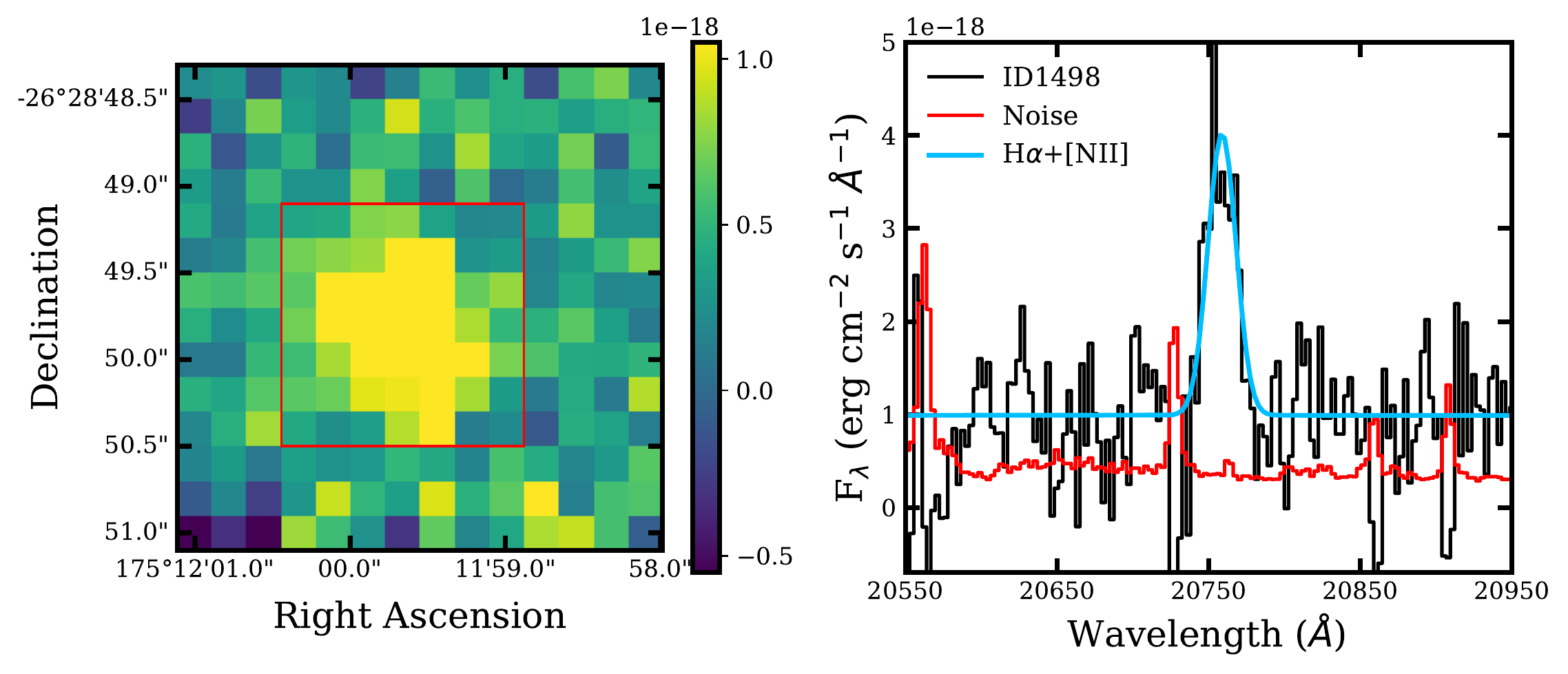}\par
      \includegraphics[width=\linewidth]{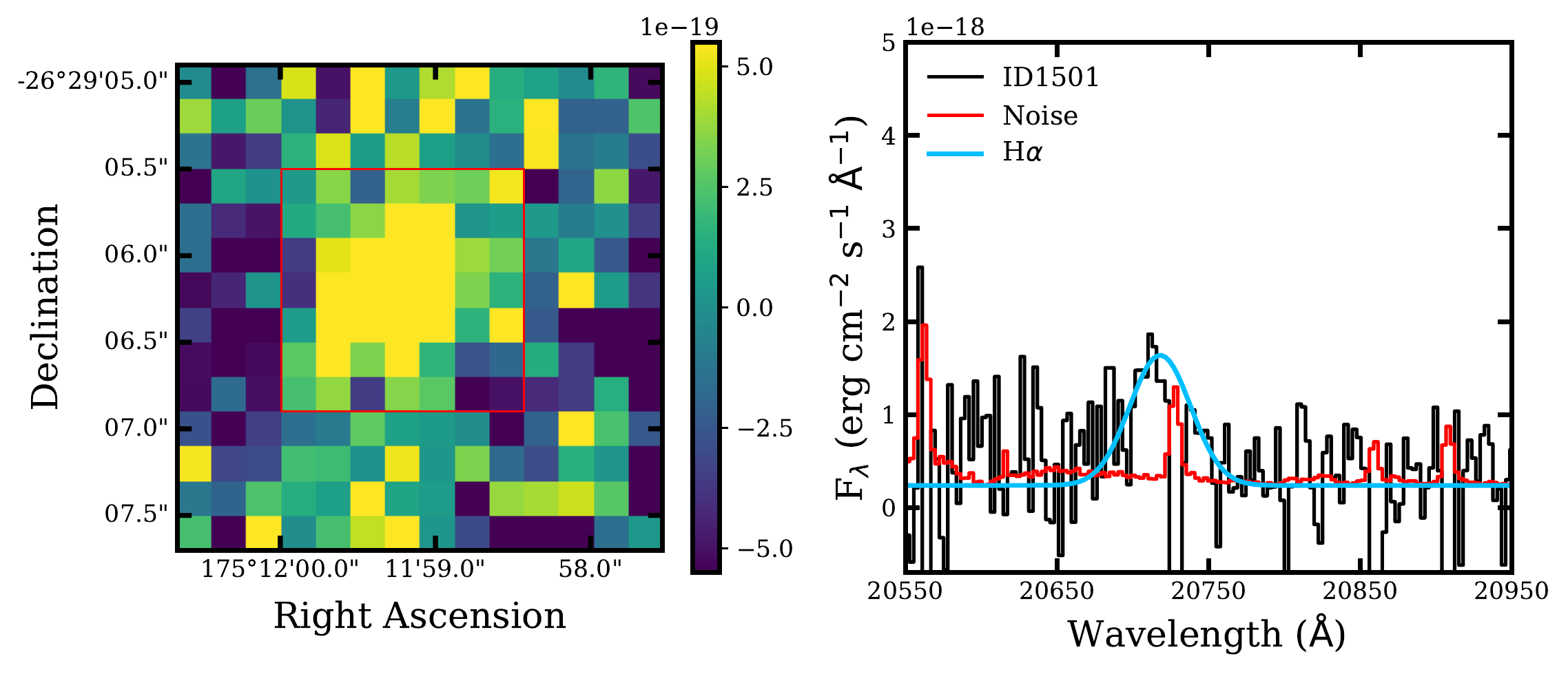}\par
      \end{multicols}
      \contcaption{}
         \label{F:SFR}
\end{figure*}

\subsection{SED fitting}
\label{SS:SED}

The field of the Spiderweb protocluster counts with extensive and deep photometry (see Table \ref{T:imaging}) comprising the rest-frame UV to NIR wavelength range for galaxies at $z=2.16$. We seek to obtain reliable stellar masses and rest-frame magnitudes for our targets based on the observed fluxes using the SED fitting technique. Due to the variety of seeing conditions and instrumental pixel sizes, we follow two different approaches to extract the observed magnitudes in different bands: First, we take as a reference the MOIRCS/Subaru NB2071 image and degrade all the other bands with PSF$<1"$ to its PSF value and pixel size. Then, we run \textit{SExtractor} (\citealt{Bertin96}) in dual-image mode using the NB2071 image for source detection while measuring the observed magnitudes (MAG\_AUTO) over the PSF and pixel size matched images. This way we ensure that we are measuring the fluxes over the same area of the targets for each band in a consistent way. Second, we carry out simple photometry (MAG\_APER) over the images with the worst spatial resolution (i.e. Suprime-Cam B-band and IRAC 3.6\,$\micron$ and 4.5\,$\micron$). 

After constructing our multi-band photometric catalog, we perform the SED fitting with the Code Investigating GALaxy Emission (CIGALE, \citealt{Boquien19}). CIGALE combines the modelling of composite stellar populations with nebular emission and dust attenuation while conserving the energy balance between the energy emitted by massive stars and its partial absorption and re-emission by dust grains. 
CIGALE follows a Bayesian fitting approach which deviates from simple $\chi^2$ minimization algorithms that provide physical properties based on the best match of templates. Instead, the estimated properties are evaluated by weighting all the models depending on their statistical agreement with respect to the best-fit, which has the heaviest weight. This naturally takes into account the uncertainties on the observations while also including the effect of intrinsic degeneracies between physical parameters. 
Finally, the physical properties and their uncertainties are estimated as the likelihood-weighted means and standard deviations. 

\begin{figure}
      \includegraphics[width=\linewidth]{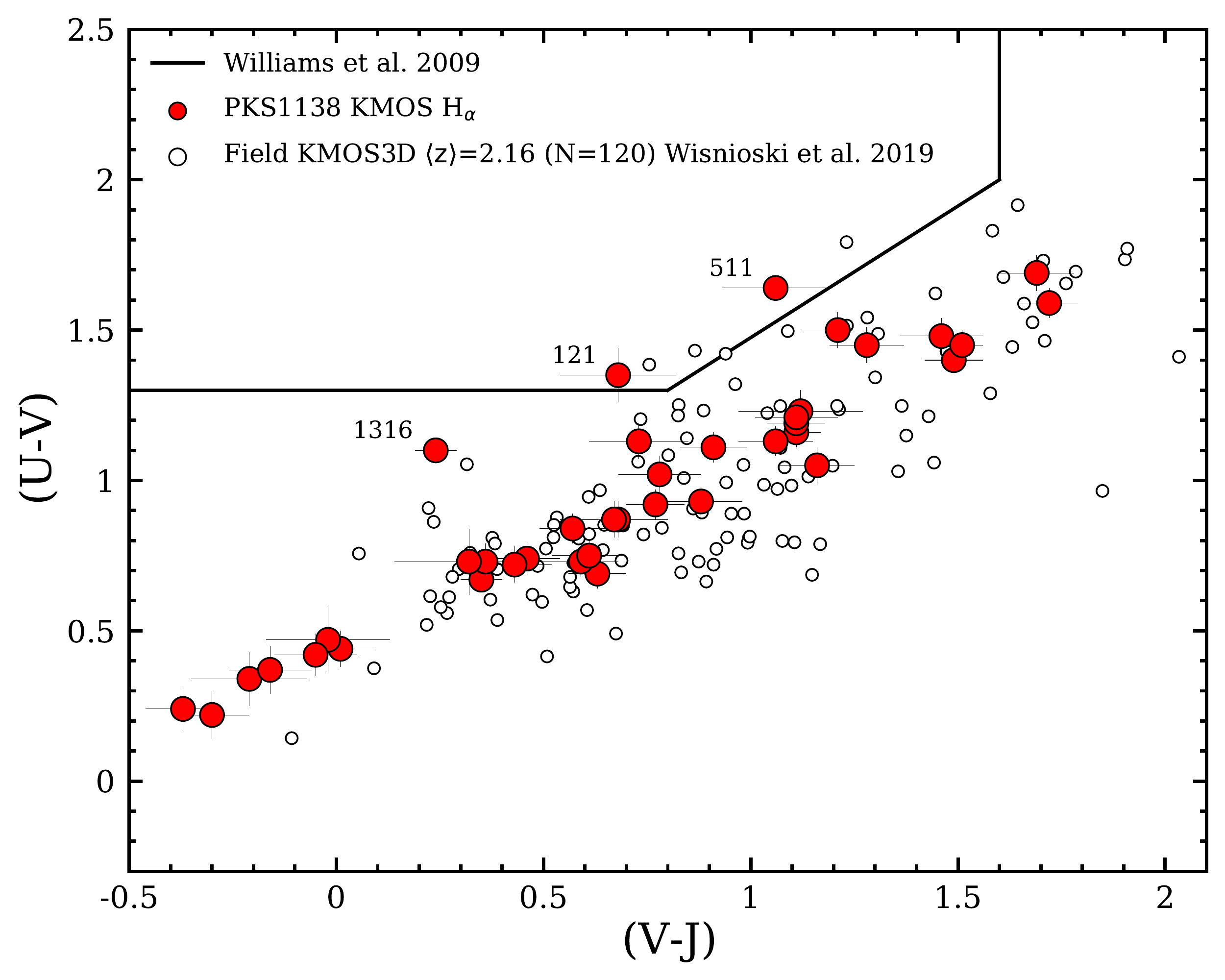}\par
      \caption{Rest-frame UVJ diagram. The limits separating the quiescent and star-forming regions follow the prescription given by \protect\cite{Whitaker13}. Red circles show the color distribution of our spectroscopically confirmed HAEs. By comparison, white circles display the color distribution of the KMOS3D sample of field galaxies at similar redshift (\protect\citealt{Wisnioski19}). Three objects display UVJ colors departing from the main sample and are labeled according to their HAE IDs (\protect\citealt{Koyama13}) for additional discussion in the main text.}
      \label{F:UVJ}
      \end{figure}

Before running CIGALE over the photometry of our targets, we create a grid based on the stellar population synthesis models of Bruzual \& Charlot (\citealt{BC03}) assuming an exponentially delayed star formation history with possible e-folding times between 1 and 8 Gyr. In addition, we constrain the possible ages of stellar populations to be younger than the age of the Universe at $z=2.16$ (i.e., $\sim3$ Gyr). Given the interacting nature of protoclusters, we also consider the possibility of a recent minor star-forming burst accounting for up to 1, 5, or 10\% of the mass fraction and with an age not older than 300 Myrs. We assume a Chabrier IMF (\citealt{Chabrier03}) and subsolar metallicity (i.e $Z = 0.004$). Nebular emission is also included in our grid of models with the ionization parameter ($U$) ranging $-2.4<\log(U)<-2.8$, which describes typical values for star-forming galaxies at $z\sim2$ (\citealt{Cullen16}). Finally, we apply Calzetti’s attenuation law (\citealt{Calzetti2000}) with extinction values ranging $\mathrm{E\left( B-V\right)_s=0-1}$ mag in steps of 0.1 mag.  We estimate the rest frame magnitudes to an average uncertainty of 0.1 mag for all bands and 0.13 dex for the stellar masses. We show the CIGALE fitted SEDs for each protocluster member within the KMOS sample in the Appendix (see Fig. \ref{F:SED}). The reduced $\chi^2$ distribution of our targets encompass the range $0.2\leq\chi^2\leq4.9$ and yields an average value of $\chi^2$=1.4, with two thirds of our sample (i.e. 26 objects) displaying $0.5\leq\chi^2\leq2$. To put our sample in context we plotted our galaxies into the rest-frame UVJ diagram (Fig.\,\ref{F:UVJ}). This diagram splits the galaxies into two different groups, an old-age sequence of quiescent galaxies (upper left corner) and a star-forming sequence of galaxies. Most of our targets lie within the star-forming region, which is consistent with their previous H$\alpha$ emitter classification (\citealt{Koyama13}).

\subsection{Size determination}
\label{SS:Structural}  

While Space-based HST observations are ideal to measure the sizes of galaxies due to its high spatial resolution and lack of atmospheric effects, the HST mosaics ($3\arcmin\times6\arcmin$) using the F475W and F814W filters cover only 31 out of the 39 KMOS spectroscopically confirmed protocluster members. Furthermore, given the redshift of our targets the HST imaging would only trace the rest-frame UV and FUV wavelength range, i.e. the very bursty and young star-forming regions, while being blind to the underlying older stellar populations which account for the bulk of the galaxies' stellar mass. Thus, we carry out the size measurement of our targets using the VLT/HAWK-I $\mathrm{K_s}$-band mosaic ($8.5\arcmin\times12\arcmin$) whose central wavelength traces the rest-frame emission around 6800\,\AA at $z=2.16$. Unfortunately, the nearby H$\alpha$ emission line may contaminate the continuum surface brightness profile originated from the stellar component for very active star-forming galaxies and type 1 AGNs. To solve this issue, we estimate the H$\alpha$ contribution to the total broad band flux within our sample, finding that it is constrained to $\lesssim15\%$ except for two cases. Therefore, we conclude that our $\mathrm{K_s}$-band size measurements are not significantly affected by the H$\alpha$ component. Then, we measure and extract an average PSF size of $\sim0.4\arcsec$ using PSFEx (\citealt{Bertin11}). This size is equivalent to a physical diameter of 3.32 kpc assuming the cosmological parameters outlined in Sect.\,\ref{S:Intro}. 

We model the surface brightness profile of our targets and measure their structural parameters (e.g. Sèrsic index, $R_e$, axis ratio and position angle) by using GALFIT (\citealt{Peng02}) over squared windows of $10\arcsec\times10\arcsec$ with our targets at its center. The models are computed following a single-component approach and allowing for free variation of all structural parameters. We fit additional components to nearby objects within the inspected window in order to remove possible light contamination on the main target. Finally, we visually check the result after subtracting the model to the original $\mathrm{K_s}$-band image and determine if the inspected object shows signs of strong residuals (e.g. bulge presence, asymmetric disc or tidal tails). 
However, we find a good agreement between the modeled single-component surface brightness profiles and the $\mathrm{K_s}$-band images in most cases. This argues against a secondary bulge component or visible interactions affecting the surface light profile of our targets. Finally, we discard those objects whose $R_e$ uncertainties are greater than $50\%$ and those displaying $R_e<1.66$ kpc (e.g. half of the PSF size) due to the difficulties to reliably resolve them. As a result, we obtained size measurements for 27 galaxies out of the 39 spectroscopically confirmed cluster members within our sample. The remaining objects did no reach enough S/N for the GALFIT models to converge into a solution or displayed too large uncertainties. The observed and modeled $\mathrm{K_s}$-band surface brightness profiles for our targets, as well as their residuals after subtraction can be found in the Appendix (see Fig.\,\ref{F:Galfit}).

\subsection{Star-formation activity}  
\label{SS:SFR_method}

Our KMOS spectroscopy campaign provides us with access to the H$\alpha$ emission line for the protocluster members at $z=2.16$. We apply the star-formation rate (SFR) calibration developed by \cite{Kennicutt98} modified for a Chabrier IMF. This calibration has proven to be one of the most reliable both at local and high redshift (e.g. \citealt{Moustakas06}, \citealt{Wisnioski19}):
\begin{equation}
\mathrm{SFR(H\alpha)}=4.65\times10^{-42}L(\mathrm{H\alpha})
\label{EQ:SFR}
\end{equation}
where $L(H\alpha)$ is the luminosity of the H$\alpha$ emission-line. We estimate the value of this quantity by measuring the H$\alpha$ spectroscopic fluxes of our targets and assuming a Calzetti's extinction law ($R_v$=4.05, \citealt{Calzetti2000}) to account for the dust attenuation. We use the extinction $A_v$ values obtained from the SED fitting to account for the diffuse dust attenuation in the galaxy's continuum following $\mathrm{A_{cont}=0.82A_{v,SED}}$ (\citealt{Wuyts13}; \citealt{Wisnioski19}). However, the nebular contribution produced in the active star-forming regions of the galaxy remains unaccounted for at this stage. To tackle this problem, \cite{Wuyts13} add an extra extinction term ($\mathrm{A_{extra}}$) that can be parametrized as $\mathrm{A_{extra}=0.9A_{cont}-0.15A_{cont}^{2}}$ and is in good agreement with the previous extinction estimates made by \cite{Calzetti2000} in the local universe. Thus, the total extinction applied to the measured H$\alpha$ fluxes is $\mathrm{A(H\alpha)=A_{cont}+A_{extra}}$. This method has been recently tested by the KMOS3D team in \cite{Wisnioski19}, which we will use as our main field comparison sample in the following sections. As it was explained in Sect. \ref{SS:EL}, two of our objects (IDs 647 and 911) display extreme H$\alpha$ line widths suggesting the presence of type 1 AGNs. In these two cases, the SFR values were computed by using only the H$\alpha$ narrow component in Eq. \ref{EQ:SFR}. The final SFR and reddening values of our protocluster sample can be found at the end of this work in Table \ref{T:BigTable}.

\subsection{Gas phase metallicities}  
\label{SS:metallicity}
\begin{figure*}
    \begin{multicols}{3}
      \includegraphics[width=\linewidth]{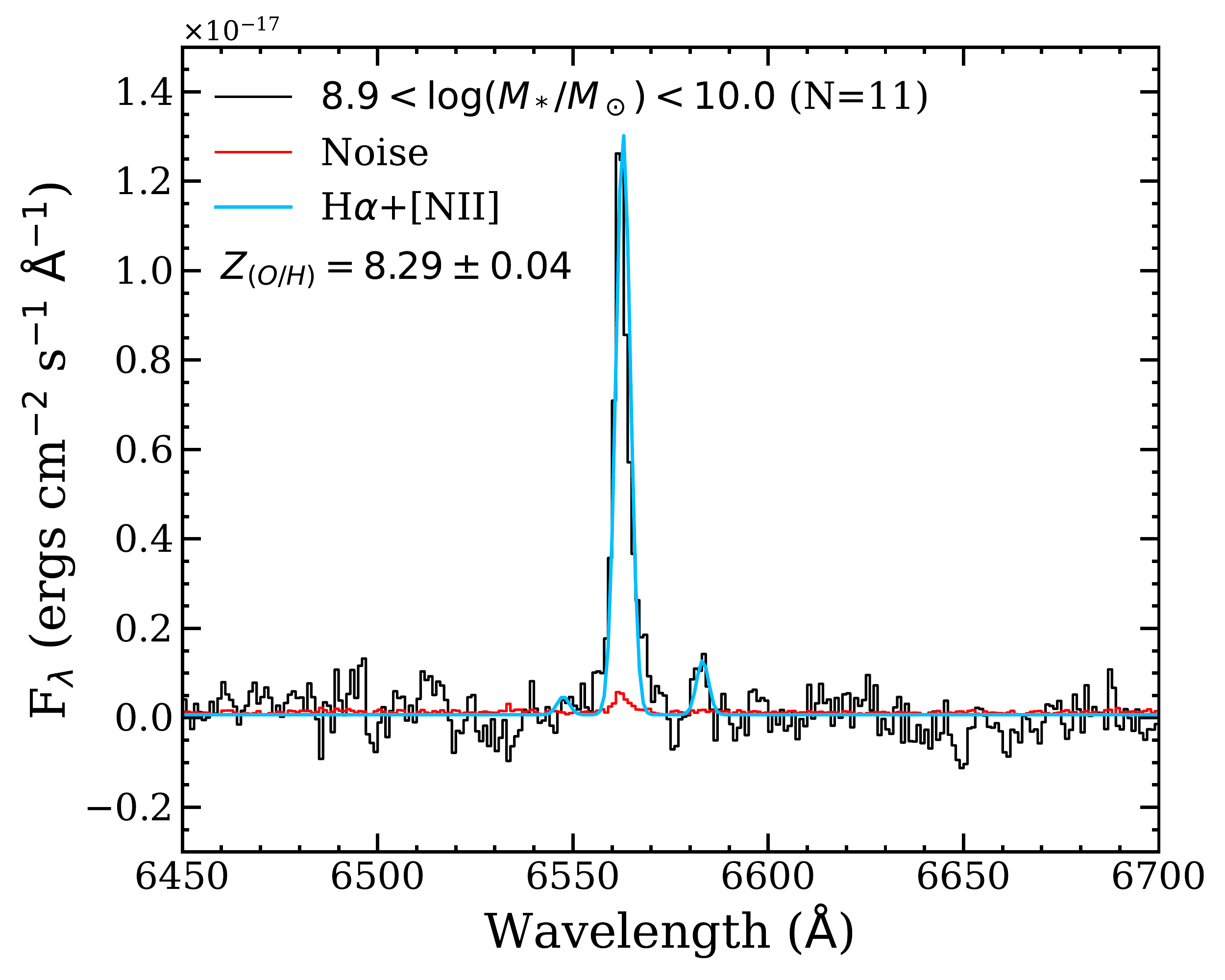}\par 
      \includegraphics[width=\linewidth]{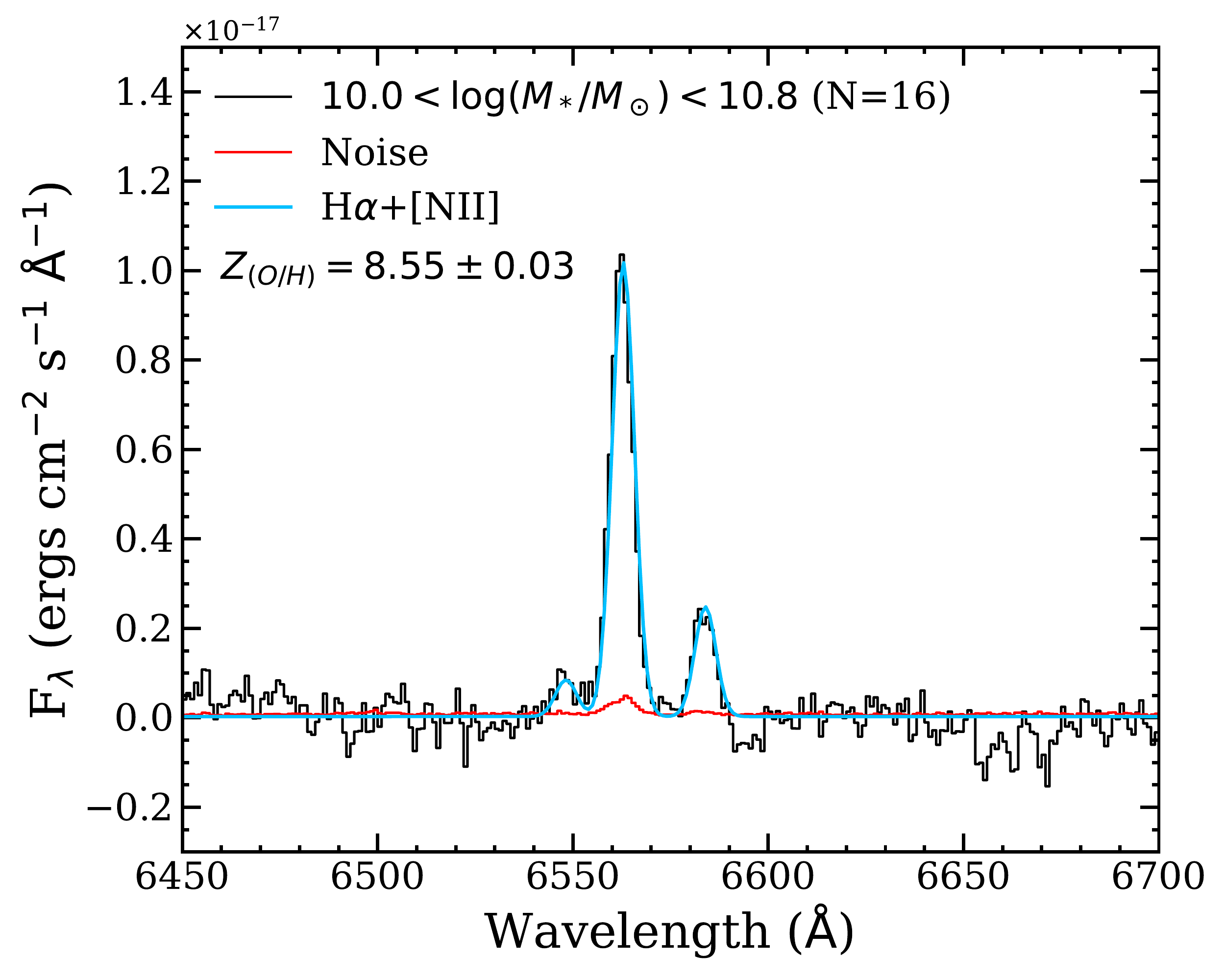}\par
      \includegraphics[width=\linewidth]{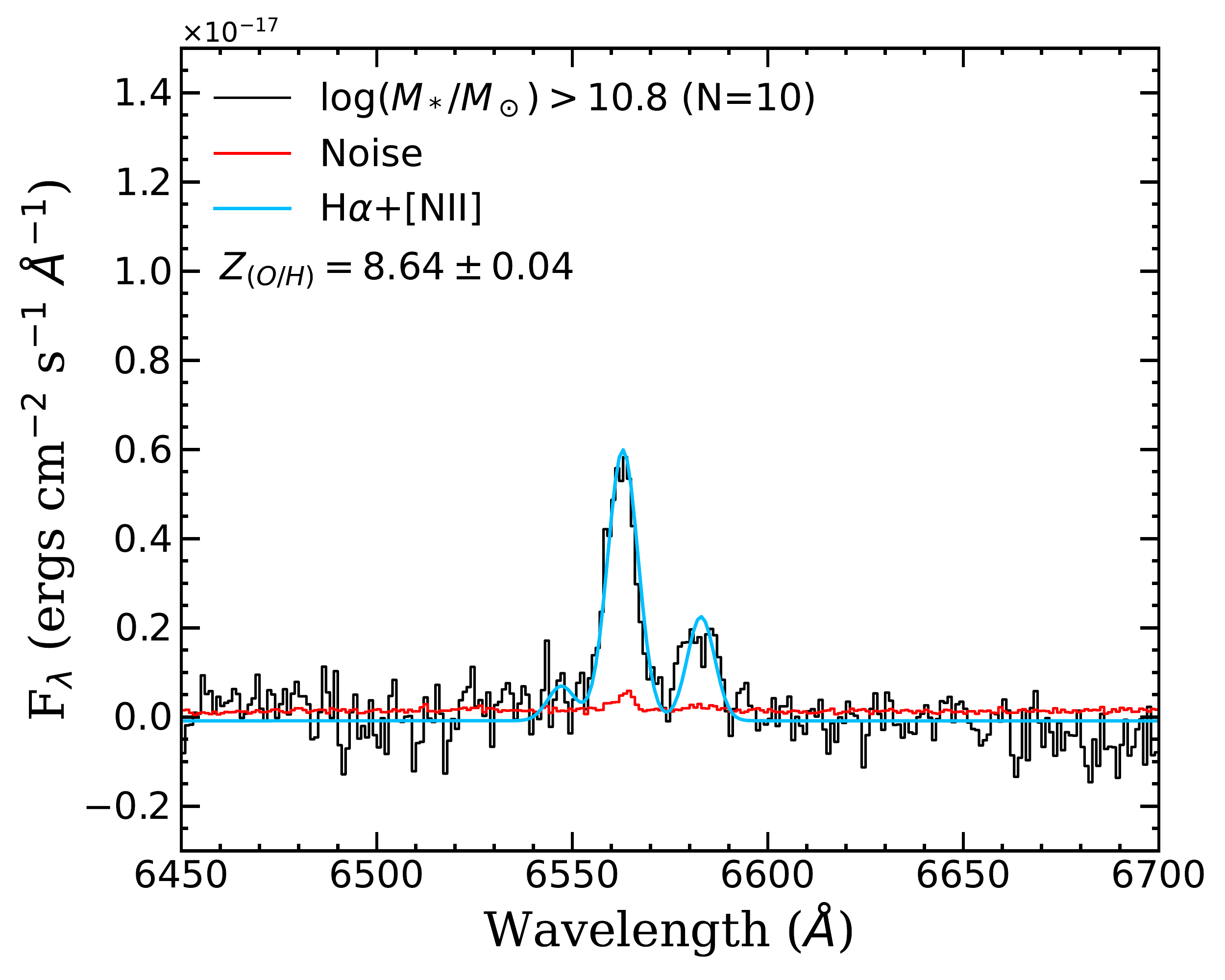}\par
      \end{multicols}
      \caption{Stacked spectra and fit around H$\alpha$ and [N{\sc{ii}}]$\lambda$6584 for the three inspected stellar-mass bins.}
         \label{F:Stacked}
      \end{figure*}

Over the last decades, several optical emission-line diagnostics have been developed to estimate the gas-phase metallicity of galaxies. Typically, these diagnostics rely on the flux ratio between ionized oxygen or nitrogen, and the hydrogen emission lines (see \citealt{Kewley08} for a review). At $z>1$ the access to these line diagnostics shifts to the NIR and simultaneous measurements of multiple emission lines become difficult due to the stretching of the spectra. In this work, we apply the N2 calibration developed by \cite{Pettini04} which involves the [N{\sc{ii}}]$\lambda$6584/H$\alpha$ ratio: 
%
\begin{equation}
   \mathrm{12+\log(O/H)=8.90+0.57\times N2}
\label{EQ:N2}
\end{equation}
where N2 is equivalent to $\log$(F{[N{\sc{ii}}]}/F{(H$\alpha$})). This method relies on a local calibration, although it has been successfully applied for field galaxies up to $z\sim3$ (\citealt{Erb06}, \citealt{Wuyts16}, \citealt{Sanders21}). However, the accuracy of its absolute values at high redshift is still under debate (\citealt{Steidel14}). Nevertheless, this work aims to investigate relative metallicity differences between the protocluster and general field populations and thus, it is not affected by calibration uncertainties as long as the chosen comparison samples are analyzed following the same approach and belong to the same cosmic epoch. The detection of [N{\sc{ii}}]$\lambda$6584 is the limiting factor when using this method to derive gas-phase metallicities. We achieve this in 24 out of 39 confirmed HAEs. In order to investigate the average properties of our full sample we decide to stack the individual spectra in three stellar-mass bins. We exclude two objects from our analysis due to their extreme H$\alpha$ line widths ($\sigma>700$ km\,s$^{-1}$), indicative of AGN activity. The stacking analysis is performed using the approach outlined in \cite{Shimakawa15}: 
\begin{equation}
\mathrm{F_{stack}}=\sum_{i}^{n}\frac{F_{i}(\lambda)}{\sigma_{i}(\lambda)^2}\Bigg/\sum_{i}^{n} \frac{1}{\sigma_{i}(\lambda)^2}
\label{EQ:stack}
\end{equation}
where $F_{i}(\lambda)$ is the flux density of an individual spectrum and $\sigma_{i}(\lambda)$ is the noise as a function of wavelength. The sums within a bin apply a median sigma clipping algorithm to reduce the contamination by sky residuals. Then, the N2 ratio is measured for the stacked spectra by simultaneously fitting two gaussian curves for H$\alpha$ and [N{\sc{ii}}]$\lambda$6584. The two objects with extremely broad H$\alpha$ components are excluded of this analysis. Finally, we estimate the average measured uncertainties to derive gas-phase abundances to be of the order of $\sim0.1$ dex for the individual measurements across our stellar mas range and $\leq0.04$ dex for the stacking analysis bins. The results of our stacking process can be seen in Fig. \ref{F:Stacked}.

\subsection{Molecular gas masses}
\label{SS:MOL}

In this section we will make use of previous works reporting radio measurements to estimate the molecular gas masses of a subsample of our HAEs. In particular, we will focus on the results recently published by \cite{Jin21} investigating the CO(1-0) luminosity function of protocluster galaxies with ATCA. In total, we share 8 overlapping targets with 6 of them being detected in CO(1-0) at more than $5\sigma$ level. The CO(1-0) luminosity ($L'_{\mathrm{CO(1-0)}}$) can be converted into a molecular gas mass estimate in a simple way assuming a conversion factor ($\alpha_{\mathrm{CO(1-0)}}$) that traces the amount of molecular gas from optically thick virialized clouds (\citealt{Dickman86}, \citealt{Solomon87}):
\begin{equation}
M_{\mathrm{mol}}=\alpha_{\mathrm{CO(1-0)}}\times L'_{\mathrm{CO(1-0)}}
\label{EQ:mol}
\end{equation}
However, $\alpha_{\mathrm{CO(1-0)}}$ is found to be approximately constant only for nearby galaxies with solar gas-phase metallicities ($\alpha_{\mathrm{CO(1-0)}}=4.36\pm0.90$, \citealt{Bolatto13}). This conversion factor increases for low metallicity regions due to the contraction of the CO-emitting surface relative to the area where the gas is H$_2$ for a fixed cloud size. Therefore, we compute our molecular gas masses by assuming the metallicity dependent conversion factor $\alpha(Z)$ outlined in \cite{Genzel15} and later revisited by \cite{Tacconi18}:
\begin{equation}
\alpha(Z)=4.36\,\sqrt{0.67\,\exp{(0.36\times 10^{(8.67-Z)}\times 10^{1.27\times(8.67-Z)})}}
\label{EQ:alpha}
\end{equation}
where $Z$=12+log(O/H) assuming the \cite{Pettini04} metallicity calibration. This conversion factor is based on the geometric sum of the metallicity corrections proposed by \cite{Genzel12} and \cite{Bolatto13}. However, these corrections strongly diverge in the low metallicity regime. Thus, we constrain the range of application of this prescription to galaxies with $\mathrm{12+log(O/H)}>8.44$, which correspond to a maximum disagreement of 0.2 dex between methods. Only three objects within our sample display lower metallicity values but none of them are affected by this constrain as they do not count with previous $L'_{\mathrm{CO(1-0)}}$ measurements. Given the distribution of our galaxies in the mass-metallicity relation (see Sect. \ref{SS:MZR}), this approach allow us to constrain the molecular gas mass of most HAEs with secure metallicity measurements. 
Finally, we will use those HAEs with available CO(1-0) information within our sample to investigate the relation between the molecular gas fraction, i.e. $f_{\mathrm{gas}}=M_{\mathrm{mol}}/(M_{*}+M_{\mathrm{mol}})$, and the gas metallicity in different environments at the cosmic noon (Sect. \ref{SS:Molprop}). 

\subsection{Environment quantification}
\label{SS:Environment}

In general, cluster membership is defined as an interval in redshift space around the value given for the whole cluster structure, which usually coincides with its BCG. Even though this may be sufficient to qualitatively disentangle the general field population of galaxies from objects residing in denser environments, we need a quantitative way to measure the environment in order to study its influence of the physical properties of galaxies within these dense regions. 

Most studies quantify the environment of their samples in two different ways: locally as a number density of cluster members (e.g. \citealt{Dressler80}), and globally by taking into account the general properties of the cluster ($M_{200}$, $R_{200}$, and $\sigma_{cl}$) which define its phase-space (\citealt{Carlberg97}). In both cases, it is required a high number of confirmed cluster members to reliably map the cluster structure and local density peaks. \cite{Koyama13} used a narrow-band MOIRCS/Subaru imaging mosaic ($7'x\ 8'$) to identify H$\alpha$ emitters (HAEs) in this field at the redshift of the Spiderweb protocluster. These HAEs form the bulk of the known galaxy populations known in this protocluster up to date ($\sim90$ objects) and they comprise the parent sample of our KMOS spectroscopic campaign, which has confirmed the membership of the narrow-band selected HAES with a $\sim93\%$ success rate (39 out 42 targets). This is comparable to previous H$\alpha$ narrow-band surveys in the field at $z\sim2$, which also find a contamination rate lower than 10\% (e.g. HiZELS, \citealt{Sobral13}). Furthermore, this contamination rate is expected to be even lower in protoclusters as they host a high number of emitters in a relatively narrow redshift window while a given field of view while the density of field contaminants (i.e. background or foreground emitters) remains constant. Based on this result, we assume that the majority of the remaining narrow-band detected HAEs can be considered cluster members with very high probability. In addition, we also include spectroscopically confirmed cluster members from other studies tracing different galaxy populations: Lyman-$\alpha$ emitters (\citealt{Pentericci00} and \citealt{Croft05}), additional HAEs cluster members (\citealt{Kurk04}) and CO emitters (\citealt{Jin21}). Combining both the spectroscopic and the narrow-band selected sources, we gather a cluster sample comprised of 125 independent objects. These sources are projected into a plane where distances between them are measured by their sky angular separation and transformed to the physical scale using a fixed cosmology with the scale factor ($a$) corresponding with the redshift of the cluster (e.g. $a=8.288$ kpc/\arcsec at $z=2.16$). Thus, it is possible to define a number surface density of objects within a given radius in the following way: 
\begin{equation}
   \Sigma_N=\frac{N}{\pi R^2_{N-1}}
\label{LocalDensity}
\end{equation}
In particular, we will use local densities defined by the minimum radius required to enclose three neighboring galaxies (i.e. $\Sigma_3$). The reason behind this choice is that high local density peaks produced by small but rather compact galaxy groups may trace the places where gravitational interactions between galaxies, such as mergers and close encounters, are more frequent. Furthermore, during the early stages of cluster assembly it is common that galaxies infall towards the protocluster not only through the surrounding filamentary structure of the cosmic web but also as part of small groups of galaxies that are accreted as a whole (\citealt{Shimakawa18a}). Therefore, this approach provides us with an opportunity to study the properties of galaxies residing in local density peaks during the protocluster assembly.

However, clusters of galaxies are by definition large-scale structures and thus, local density peaks may not always correlate with the overall cluster-mass distribution. For example, some of the most characteristic clusters properties such as the presence of the ICM or the splashback radius are related to the cluster-core itself. These components are linked to hydrodynamical cluster-specific interactions such as starvation and ram-pressure stripping, and their effects are gradually felt by the galaxies during their infalling phase. Thus, a second environmental parameter that takes into account the radial and velocity distribution of our targets with respect to the core of the cluster (i.e. the Spiderweb galaxy) is required. Our approach to trace the global environment relies on the projected clustercentric distance ($R_{\mathrm{proj}}$) of each object and its relative line-of-sight velocity with respect to the systemic velocity of the cluster ($\Delta v$), which can be measured through each object's redshift. We follow the procedure outlined by \cite{Noble13} who used a parameter ($\eta$) that defines caustic profiles in a phase-space diagram in the following way:
\begin{equation}
    \eta=(R_{\mathrm{proj}}/R_{200})\times(\left | \Delta v \right |/\sigma)
\label{GlobalDensity}
\end{equation}
where $\left | \Delta v \right |= \left |(z-z_{cl})\ c/(1+z_{cl})\right |$ and $z_{cl}$ is the redshift of the cluster. Attending to this parameter, \cite{Noble13} defined three separate regions: $\eta<0.4$ for galaxies that were accreted into the cluster core long time ago, $0.4<\eta<2$ for galaxies that have been recently accreted, and $\eta>2$ for galaxies infalling into the cluster but not yet associated with its main component. Given the heterogeneous structure of a cluster in formation such as this, we decided to use $\eta$ as a continuous parameter that models the environmental relation of a given galaxy to a large overdensity, without imposing an upper limit on $\eta$. Thus, we define the parameter space given by $\eta>2$ as the outskirts of the protocluster and $\eta<2$ as the accreted region. Our combined approach using local and global environmental parameters will allow us to investigate different environmental effects in the following sections.

\begin{figure*}
\centering
      \includegraphics[width=14cm]{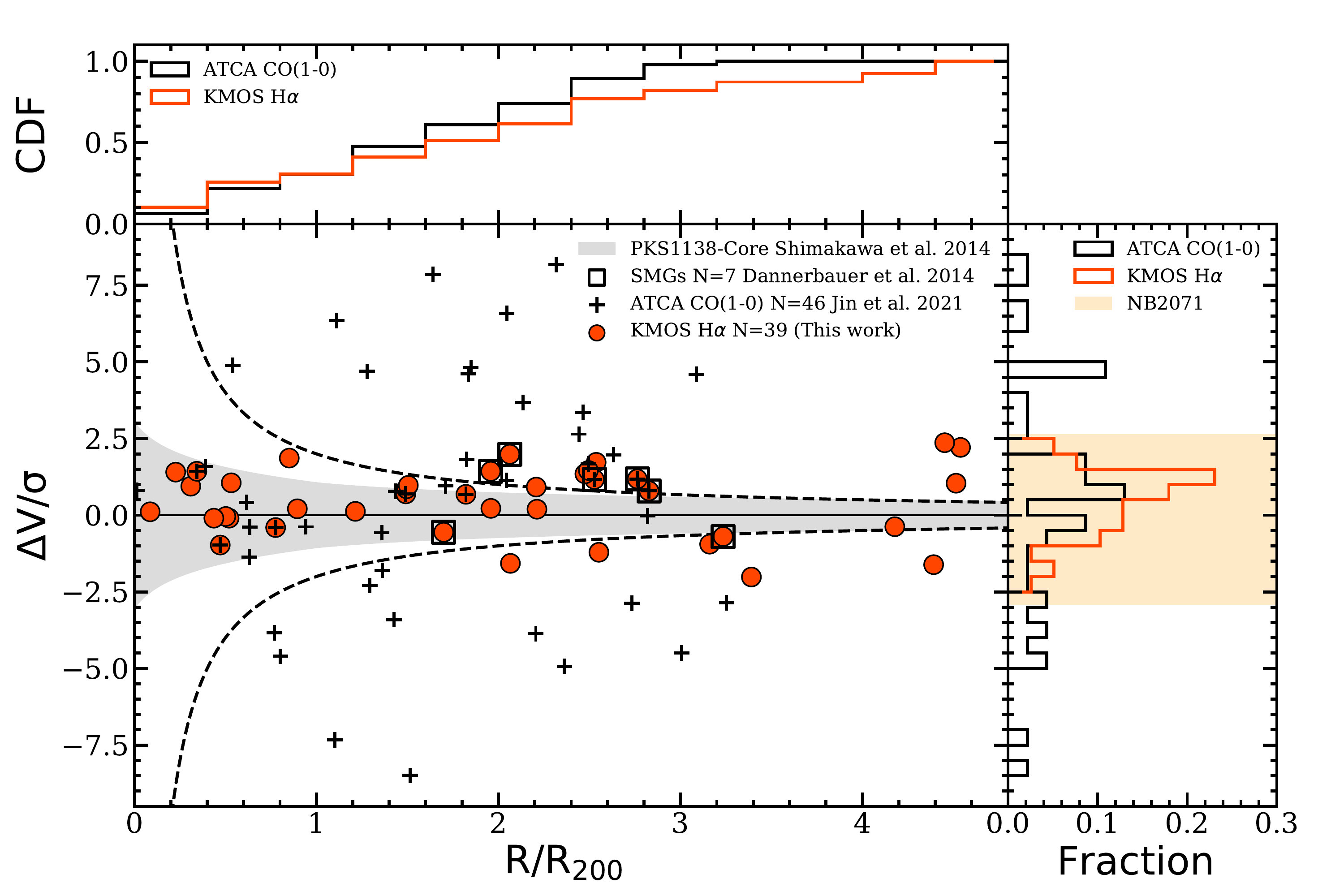}\par 
     \caption{Phase-space diagram for PKS1138 at $z=2.16$. Individual symbols represent the same samples than in Fig.\,\ref{F:Map}. The grey area represents the gravitationally bound region of the diagram assuming the R$_{200}$ and M$_{200}$ values estimated by \protect\cite{Shimakawa14} for the protocluster core. The dashed line shows the $\mathrm{\eta=2}$ countours following Eq.\,\ref{GlobalDensity}. The right side panel display the fraction of KMOS H$\alpha$ emitters and CO(1-0) emitters (\protect\citealt{Jin21}) as function of the velocity (i.e. redshift) space. The orange area in the right-hand panel mimics the width of the narrow-band filter used to identify the HAEs of the parent sample (\protect\citealt{Koyama13}). The top panel display the cumulative distribution function (CDF) of KMOS H$\alpha$ emitters and CO(1-0) emitters as function of cluster-centric radius.} 
         \label{F:phase-space}
\end{figure*}

\section{Results}
\label{S:Results}

In this section we investigate the main physical properties of our spectroscopic sample and their connection with the environment. In particular, we will first the distribution of our targets with respect to the Spiderweb galaxy at the center of the protocluster. Then, we will compare the physical properties (e.g., SFR, metallicity, size and molecular gas fractions) of protocluster members with coeval field samples and make use of local and global environmental indicators to explore possible effects across the protocluster structure. The measured values of the physical quantities quoted in this section for every galaxy can be found at the end of this work in Table \ref{T:BigTable}.

\subsection{Phase-space distribution}

We present the distribution of our 39 spectroscopically confirmed protocluster members with respect to the center of the Spiderweb protocluster in terms of clustercentric distance (i.e., as function of $\mathrm{R_{200}}$) and velocity space (i.e., redshift) in Fig.\,\ref{F:phase-space}. The center of the protocluster is defined by the position of the radio galaxy MRC 1138-262 at $z=2.156$ (also known as Spiderweb galaxy, \citealt{Miley06}). Our spectroscopic sample was selected through narrow-band techniques over a large field of view (see Sect.\,\ref{S:Data}) and thus it spreads to clustercentric distances of up to $\mathrm{5R_{200}}$ while encompassing a relatively narrow range of systemic velocities with respect to the Spiderweb galaxy. This is explicitly shown in the right-hand panel of Fig.\,\ref{F:phase-space} where we compare the narrow-band filter width (orange area) used for the target selection with the distribution of our spectroscopically confirmed protocluster members (red histogram). In contrast, the blind CO(1-0) survey carried out by \cite{Jin21} displays a more extended distribution in redshift space, suggesting that the Spiderweb protocluster may be a multicomponent system enclosing a co-moving volume similar to the Hyperion super-protocluster at $z=2.45$ (\citealt{Cucciati18}).

Finally we investigate the phase-space distribution of our sample with respect to two common environmental descriptors used in lower redshift clusters. First, we display
the gravitationally bound region of the diagram (grey area) assuming the virialization of the protocluster core with R$_{200}$=0.53\,Mpc and M$_{200}$=1.71$\times$10$^{14}$\,M$_\odot$ (\citealt{Shimakawa14}), and the escape velocity prescription given by \cite{Jaffe15} and \cite{Rhee17}. In addition, we show the $\mathrm{\eta=2}$ countours (dashed lines) marking the separation between the outskirts of the protocluster and the accreted region ($\eta<2$) according to our definition in Sect.\,\ref{SS:Environment}. Both approaches encompass a very similar fraction of our targets with 22 objects lying at $\eta<2$ and 17 at $\eta>2$.Interestingly, none of the SMGs we have in common with \citealt{Dannerbauer14} lie within $\mathrm{R_{200}}$ and only one source can be found at $\mathrm{\eta<2}$, implying that these objects are more frequent in the filamentary structure surrounding the Spiderweb protocluster and suggesting they may change their nature before reaching the inner core.

\begin{figure*}
 \centering
 \begin{multicols}{2}
      \includegraphics[width=\linewidth]{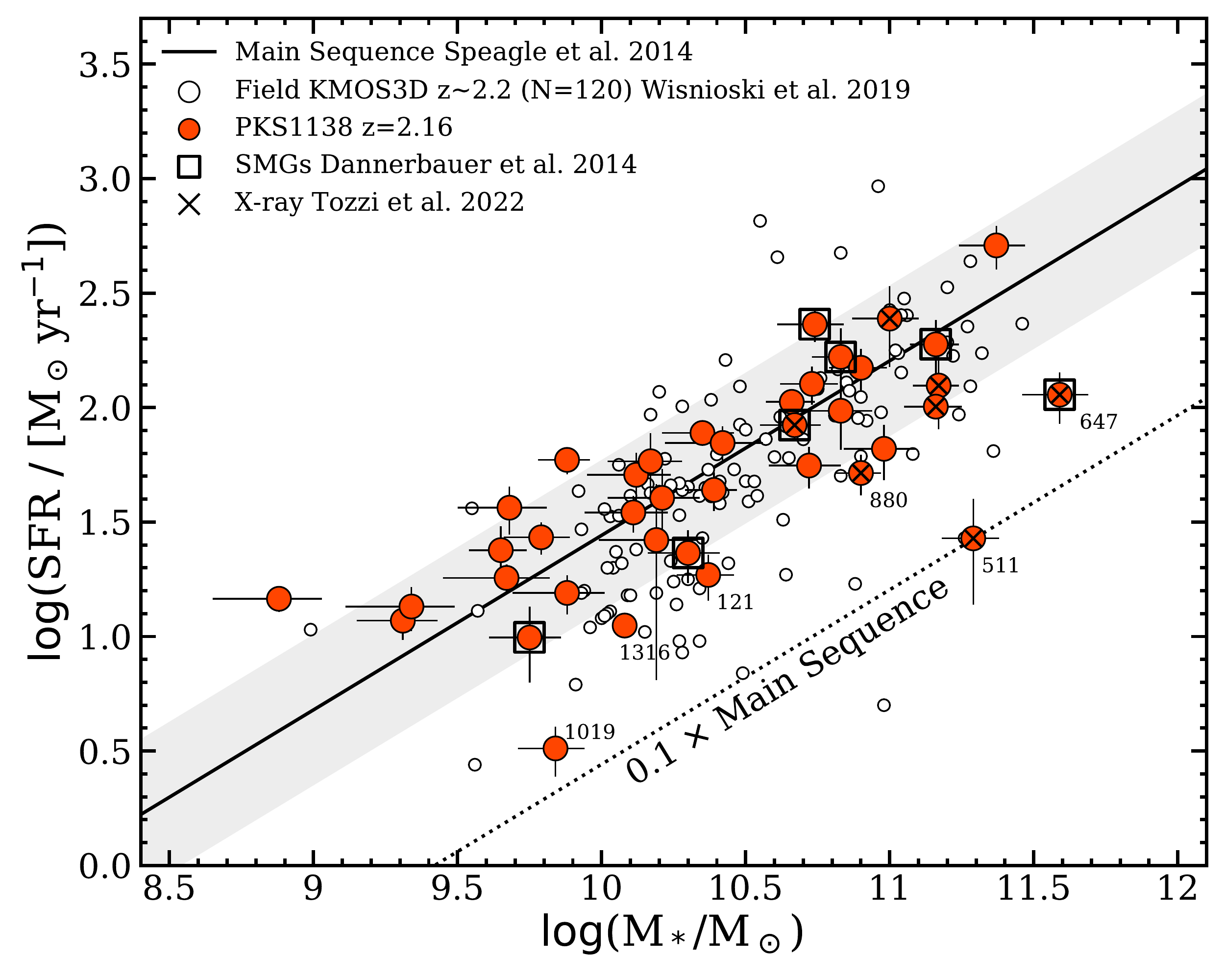}\par
      \includegraphics[width=\linewidth]{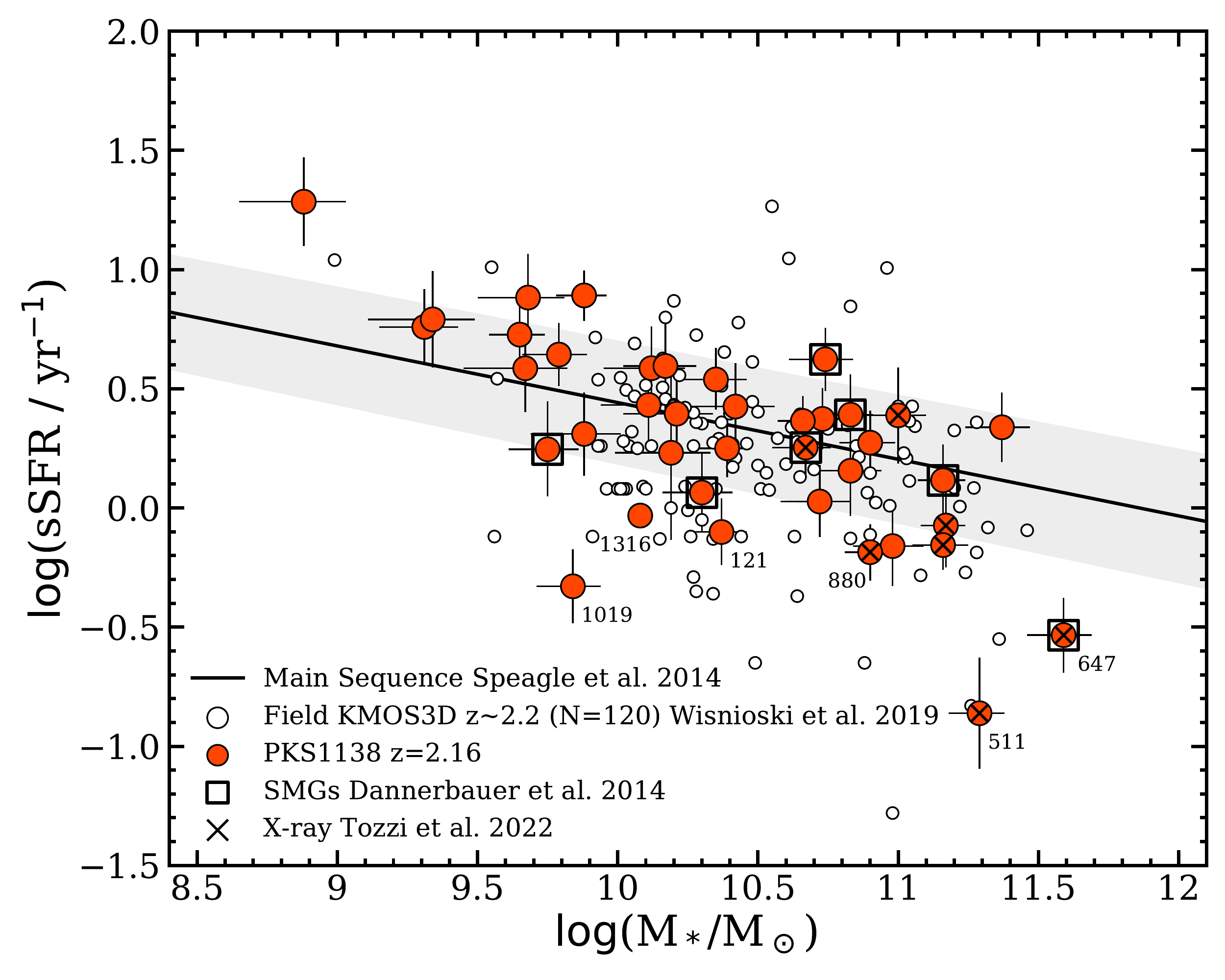}\par
      \end{multicols}
      \caption{Left: Star-forming main sequence diagram. Red circles show the distribution of our spectroscopic sample of protocluster members. Empty circles display the field comparison sample from the KMOS3D survey (\citealt{Wisnioski19}). This dataset consist of 120 field galaxies at $z\sim2.2$. The black solid line represents the star-forming main sequence parametrized for $z=2.16$ (\citealt{Speagle14}), while the grey shaded region mark the 3$\sigma$ scatter around it. The dotted line depict a sequence for quenched or nearly quenched galaxies with one tenth of the main sequence SFRs. Right: Star-forming main sequence using the specific star-formation rate (sSFR). Symbols and colors remain the same than in the left-hand panel. Six objects display SFR values below the main sequence and have been labeled according to their HAE IDs (\citealt{Koyama13}) for further discussion in the main text.}
         \label{F:SFR}
\end{figure*}
 \begin{figure*}
    \begin{multicols}{2}
      \includegraphics[width=\linewidth]{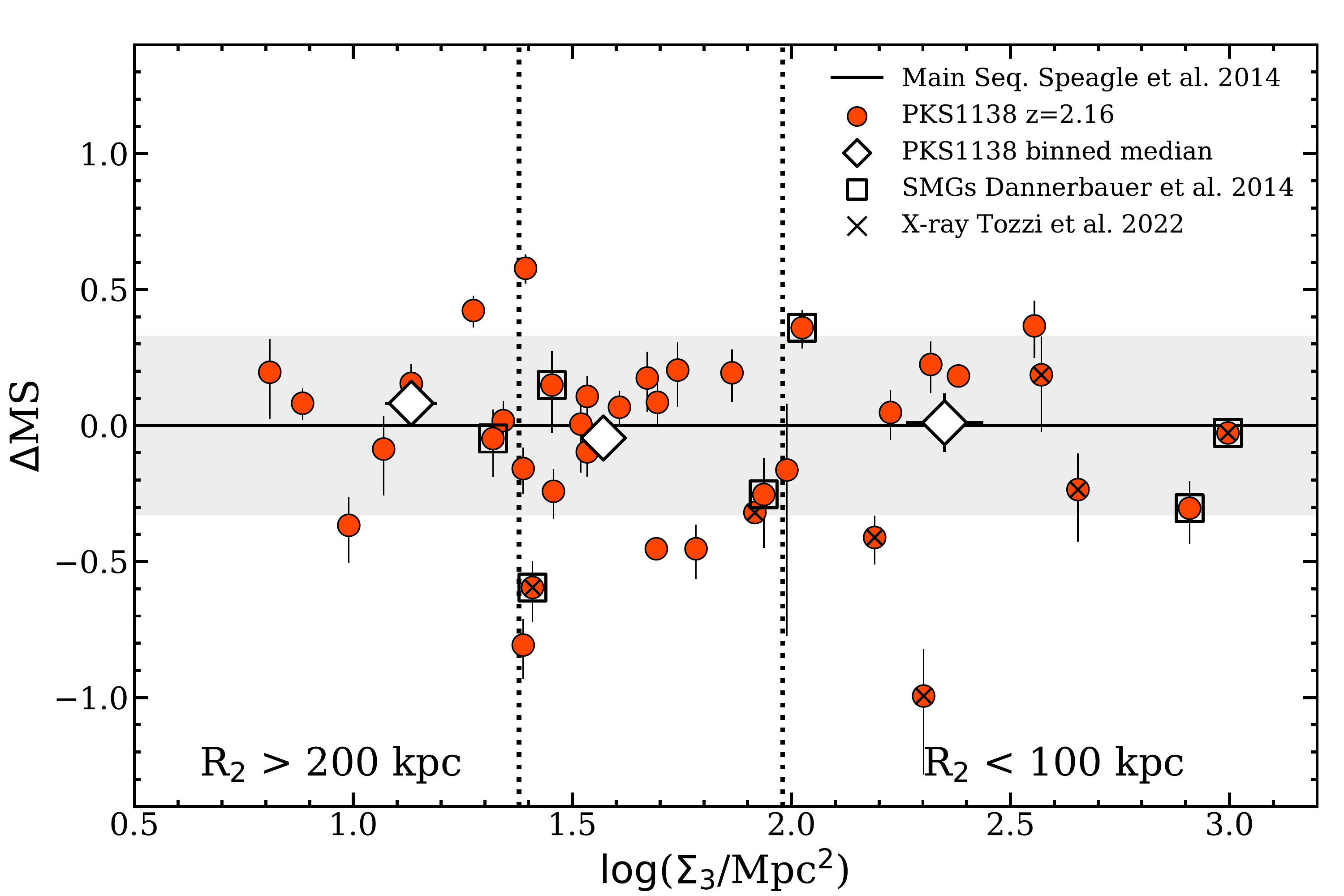}\par 
      \includegraphics[width=\linewidth]{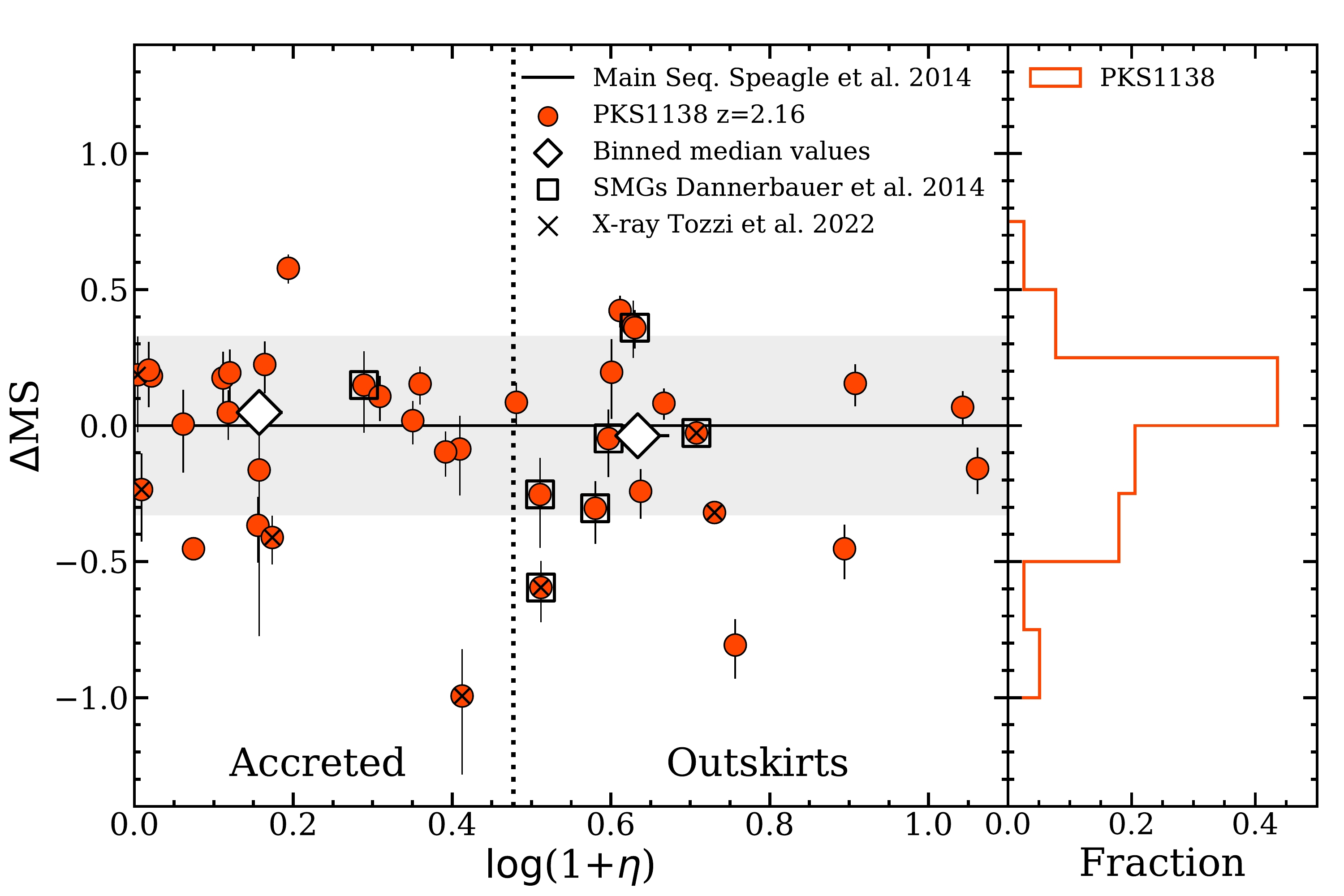}\par
      \end{multicols}
      \caption{Left: Offsets from the main sequence of galaxies as function of the local density. Right: Offsets from the main sequence of galaxies as function of the general environmental parameter $\eta$. The horizontal solid line shows the position of the main sequence of star formation (\citealt{Speagle14}) in both diagrams. Vertical dotted lines separate the density regimes outlined in Sect. \ref{SS:Environment}. Colors and symbols follow the same scheme as in Fig.\,\ref{F:SFR}.}
         \label{F:Density_SFR}
      \end{figure*}

\subsection{Star formation across different environments}
\label{SS:SFR}
In this section we investigate the influence of the environment over the star-forming properties of our sample of galaxies. In Fig.\,\ref{F:SFR}, we present the distribution of our galaxies over the SFR-$\mathrm{M_*}$ diagram. Field star-forming galaxies follow a tight correlation between these two parameters which is usually refer to as the "Main Sequence" (\citealt{Speagle14}). This correlation is represented by a solid line in Fig.\,\ref{F:SFR}, with its 3$\sigma$ limits depicted by the gray area around it. Overall, our sample is consistent with the expectations of the main sequence, with most of our galaxies scattering within the 3$\sigma$ region. Seven of our most massive targets ($\mathrm{\log M_*/M_{\odot}>10.9}$) display X-ray emission (\citealt{Tozzi22}), hinting at the presence of AGNs. 

Only six objects lie below the scatter of the main sequence even when taking into account their error bars: three of them (IDs 121, 511, 1316) show UVJ colors that detach them from the blue sequence, with the first two lying within the quiescent region and the latter displaying colors compatible with a poststarburst phase in Fig \ref{F:UVJ}. In addition, two of these objects (IDs 511 and 1316) lie within the inner core of the protocluster ($r<R_{200}$) and may be representative of its nascent quenched population. Another object (ID 647) is likely a type 1 AGN given its extreme H$\alpha$ broad component (see Fig.\,\ref{F:IFU}) and its X-ray emitting nature, which is also the case of IDs 511 and 880. The remaining objects (ID 1019) is classified as blue star-forming galaxies in the UVJ diagram and show no signs of AGN activity in their emission line profiles. On the other hand, two galaxies lie clearly above the main sequence scatter, though one of them has very low stellar mass ($\mathrm{\log M_*/M_{\odot}<9.0}$). Seven of our objects have previously been identified as submillimeter galaxies (SMGs) by LABOCA 870$\mu$m observations (\citealt{Dannerbauer14}) displaying $\mathrm{SFR_{FIR}\sim1000\,M_{\odot}/yr}$. However, their VLT/KMOS SFR(H$\alpha$) range between 10 and 300 $\mathrm{M_{\odot}/yr}$, making them compatible with the main sequence of star-forming galaxies except for the case of ID 647 (which likely host an AGN). This indicates the dust-extinction of these sources is highly underestimated as was suggested by \cite{Dannerbauer14}. In general, HAEs in the Spiderweb protocluster also occupy a similar locus than the KMOS3D field sample of \cite{Wisnioski15} at $z\sim2.2$. These results suggest that most of our sample show no significant differences in terms of star-formation compared to their field counterparts at similar redshift. 
      
This can be seen easily by inspecting the right-hand panel of Fig.\,\ref{F:SFR}, where we investigate the main sequence using the specific star-formation diagram (i.e. $\mathrm{sSFR=SFR/M_*}$). The sSFR allow us to investigate the star-forming activity of a given galaxy normalized by its present stellar-mass, which ease the process of identifying galaxies that are more efficient forming new generations of stars than expected per unit of stellar-mass. The scatter of our sample increases with respect to the previous figure though we find similar results. We detect a group of galaxies lying below the "Main Sequence" at $\mathrm{\log M_*/M_{\odot}\geq10.0}$ with sSFR values between 0.1 and 0.8 dex lower than expected, while at lower masses a few galaxies seem to display enhanced sSFR. 

We also investigate the possible influence of the environment over these results by combining both local and global environmental indicators. The local density is measured as the projected surface density enclosing 3 neighboring galaxies (i.e. $\Sigma_3$). This allows us to inspect the influence of local density peaks where small groups of galaxies are clustered and could be interacting in a relatively small area of the sky, although they may not necessarily lie close to the cluster-core. Fig.\,\ref{F:Density_SFR} shows the distribution of our galaxies in local density space as function of the offsets with respect to the main sequence (i.e. $\Delta\mathrm{MS=SFR-SFR_{MS}}$). We bin our sample into three density regimes defined by the minimum radius required to enclose three neighboring galaxies within a circle. Following this approach, our high-density regime is defined by $R_3<100$ kpc, while the intermediate density regime encompasses $200>R_3>100$ kpc, and the low density regime trace galaxies with measured $R_3>200$ kpc. As it can be seen in Fig.\,\ref{F:Density_SFR}, the median values within each bin display a flat trend across almost 3 orders of magnitude in $\Sigma_3$. A similar result is shown when applying a higher number of neighboring objects such as $\Sigma_5$. These results suggest that local density peaks made of small groups of galaxies do not have a strong influence on the SFR of its galaxies, regardless of how dense they are. Intriguingly, five out of the seven X-ray emitters lie within the densest bin while the other two reside at intermediate densities. This suggests that local density peaks where galaxy-galaxy interactions are more frequent may promote the triggering of AGN activity. On the other hand, the sample of seven SMGs spread almost evenly between the three local density regimes. A possible caveat to this approach is, however, the fact that the number surface density of galaxies in a given patch of the sky does not take into account the projection effects, i.e. the possible distance and velocities in the line of sight of the observations. This could potentially present some groups of galaxies as high density peaks while its members lie at great distances in reality.

Finally, we use the global environmental indicator described by Eq. \ref{GlobalDensity} and introduced by \cite{Noble13} to check the potential influence of the protocluster's structure over its members. This indicator relies on the combined knowledge of the clustercentric distance of every galaxy and their redshift (or velocity) offset with respect to the systemic redshift of the cluster. We took the position and redshift of the Spiderweb radio galaxy (MRC $1138-262$ at $z=2.156$, \citealt{Miley06}) as reference for our measurements. Our results are shown in the right panel of Fig.\,\ref{F:Density_SFR}. The median value of galaxies within the accreted region is slightly above the main sequence, while galaxies residing in the outskirts show a small SFR deficit. Nonetheless, all median values are within the $3\sigma$ scatter of the main sequence, similarly to our previous findings using the local density environmental indicator. We will further investigate these results in Sect. \ref{S:Discussion}.

\subsection{Gas-phase metallicities}
\label{SS:MZR}

\begin{figure}    
  \centering
  \includegraphics[width=\linewidth]{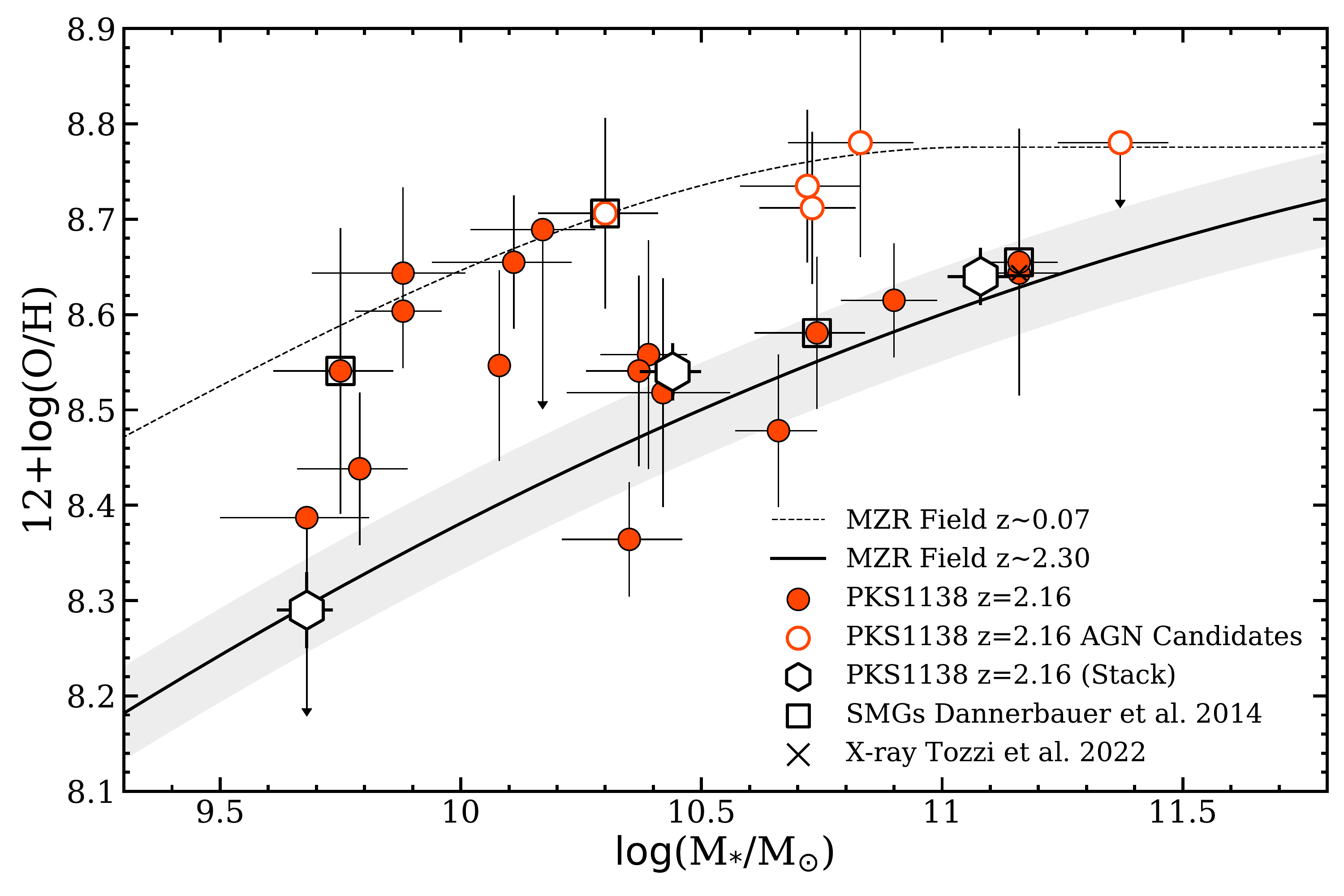}\par
  \caption{Mass-metallicity relation (MZR) diagram. Gas-phase metallicity measurements for individual objects in PKS1138 are shown by filled and empty circles, with the latter being a population of AGN candidates according to their [N{\sc{ii}}]/H$\alpha$ ratios (see Sect. \ref{SS:MZR}). The empty hexagons display the results of our stacking analysis in three stellar-mass bins. The solid line depicts the field MZR using a second order polynomial fit over the combined results of the \protect\cite{Erb06} sample, the KMOS3D sample of \protect\cite{Wuyts16} and the MOSDEF sample of \protect\cite{Sanders21} at $z\sim2.3$. The grey area shows the mean metallicity uncertainty for the samples used to define the field MZR at $z\sim2.3$. The dashed line mark the position of the local MZR (\protect\citealt{Kewley08}) assuming the \protect\cite{Pettini04} metallicity calibration but adjusted to the \protect\cite{Chabrier03} IMF.}
    \label{F:MOH}
\end{figure} 
\begin{figure*}
    \begin{multicols}{2}
      \includegraphics[width=\linewidth]{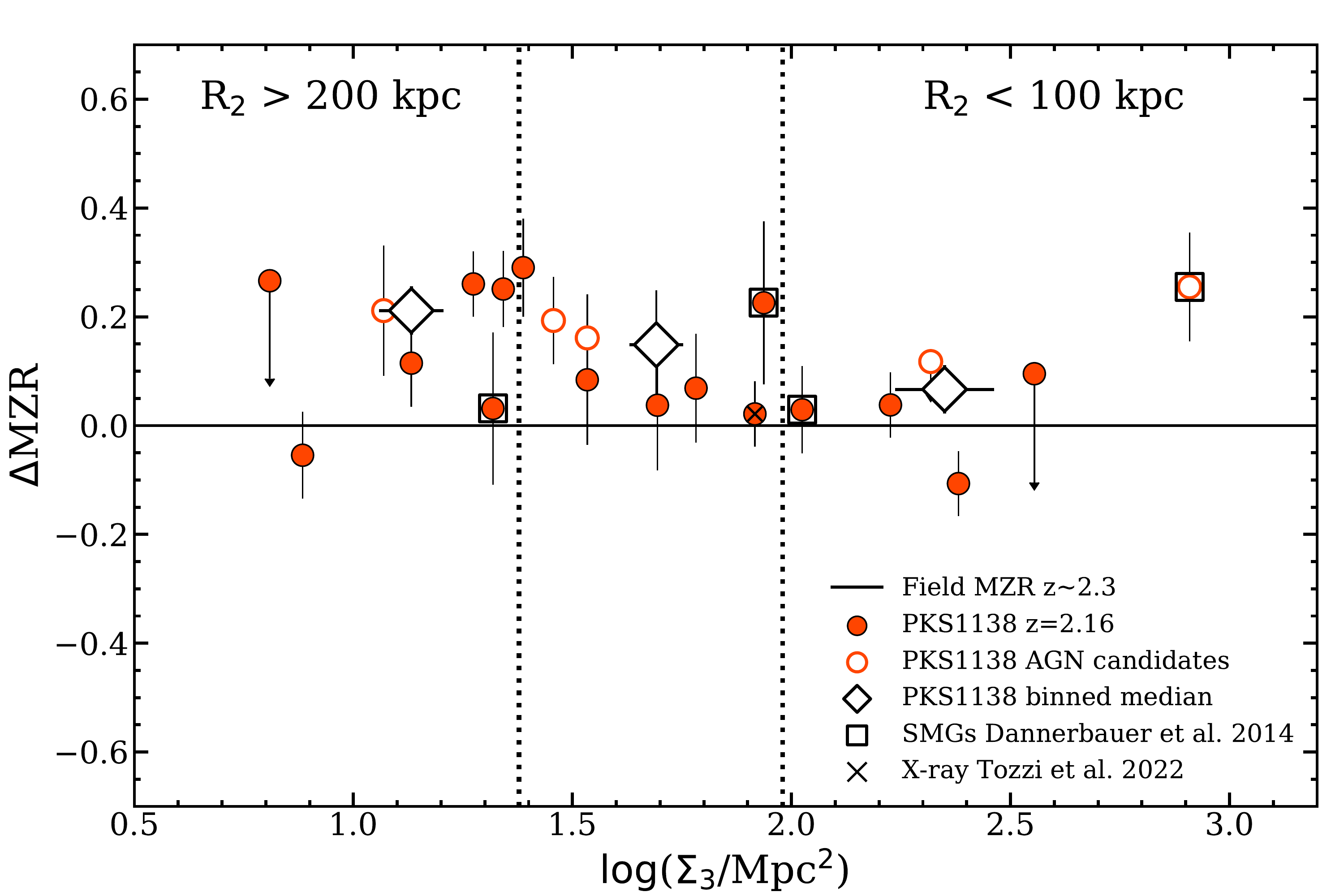}\par 
      \includegraphics[width=\linewidth]{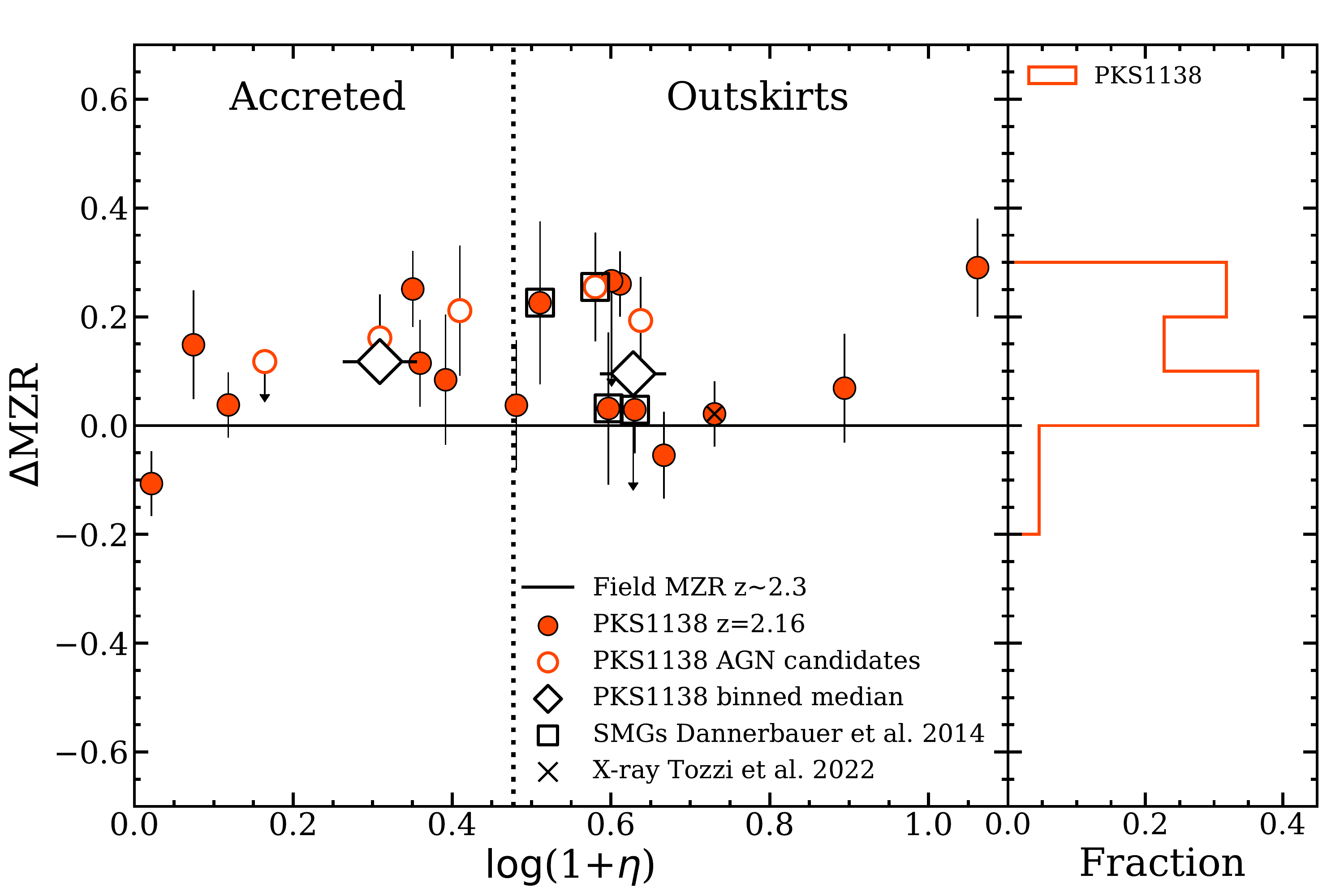}\par
      \end{multicols}
      \caption{Left: Offsets from the field MZR as function of local density values. Right: Offsets from the field MZR as function of the general environmental parameter $\eta$. The horizontal solid line shows the position of the field MZR from Fig.\,\ref{F:MOH}. Vertical dotted lines separate the density regimes outlined in Sect. \ref{SS:Environment}}
         \label{F:Density_OH}
\end{figure*}

In this section we investigate the impact of the environment over the metal enrichment of the ISM of protocluster galaxies. We follow the approach outlined in Sect. \ref{SS:metallicity} to compute the gas-phase metallicities of our targets based on the \cite{Pettini04} calibration. Our results are shown in Fig.\,\ref{F:MOH}, where we inspect the distribution of our galaxies with respect to the field mass-metallicity relation (MZR) at $z\sim2.3$. We derive individual metallicity values for 24 out of 39 HAEs in our KMOS sample. Out of these 24 objects, two are excluded due to their broad H$\alpha$ profiles ($\sigma>700$ km\,s$^{-1}$). Another three galaxies display $\mathrm{S/N}<2$ in the [N{\sc{ii}}] line and are shown as upper limits in Fig.\,\ref{F:MOH}. The remaining 15 galaxies do not achieve enough S/N in the [N{\sc{ii}}] emission line to be measured, and thus they are only included as part of our stacking analysis. 

Among the difficulties to study the environmental dependence of the MZR, the construction of an unbiased field comparison sample is one of the most prominent problems (\citealt{Stott13b}; \citealt{Namiki19}). According to the fundamental metallicity relation (FMR, \citealt{Mannucci10}), it is expected that high SFR samples are naturally biased towards lower oxygen metallicity values. Similarly, field samples that are selected using UV to optical bands (e.g. \citealt{Erb06}) may overlook dusty galaxies and thus, underestimate their mean metallicity, especially at the low-mass end. In order to mitigate these effects, we build our field MZR upon the combined results of the KMOS3D survey (\citealt{Wuyts16}), the MOSDEF survey (\citealt{Sanders21}) and the ancillary sample \cite{Erb06}, which share a similar SFR lower limit ($\sim10\,\mathrm{M_{\odot}/yr}$, see Fig.\,\ref{F:SFR}) and reddening values with our protocluster spectroscopic sample. In Fig.\,\ref{F:MOH}, the field MZR (solid line) is built by fitting an error-weighted second order polynomial of the form y=a$x^2$+b$x$+c with $\mathrm{y=12+\log(O/H)}$ and $\mathrm{x=\log(M_*/M_\odot)}$ to the mass-binned results of the three samples of $z\sim2.3$ field galaxies. The resulting polynomial coefficients of the fitted relation (solid line) in Fig.\,\ref{F:MOH} are a=-0.03827, b=1.02301 and c=1.97744, with the typical uncertainty (0.05 dex) being shown by the grey area. For reference, we also add the local MZR from \cite{Kewley08} (dashed line in Fig.\,\ref{F:MOH}) assuming the same metallicity calibration but adjusted to the \cite{Chabrier03} IMF. The local relation of \cite{Kewley08} saturates at $\mathrm{\log M_*/M_{\odot}=11.06}$. Thus, we assume a constant metallicity value beyond this limit.

Furthermore, we consider the effect of AGN contamination by splitting our sample in objects below and above $\log$([N{\sc{ii}}]/H$\alpha$)$\geq-0.35$ in the same figure (filled and empty circles respectively). This threshold has been recently proposed by \cite{Agostino21} as a modification of the \cite{Kauffmann03} line to separate the purely star-forming from the composite and AGN regions in the BPT diagram (\citealt{Baldwin81}) at relatively low [O{\sc{iii}}]/H$\beta$ ratios (e.g. $\log$([O{\sc{iii}}]/H$\beta$)$\,\leq-0.2$). This threshold can be used as a [N{\sc{ii}}]/H$\alpha$ upper limit beyond which the ionization source of most objects is either composite or dominated by the AGN contribution. Thus, we will hereafter label objects beyond this threshold as AGN candidates. Based on this criterion, we find 5 AGN candidates within our sample (empty circles in Fig.\,\ref{F:MOH}). Finally, we also consider that X-ray emission may hint the presence of AGN activity in some objects. Among the sample of 22 galaxies that entered our metallicity analysis, we only find one object with significant X-ray emission in previous works (\citealt{Pentericci02}; \citealt{Tozzi22}).

Most of our individual measurements display metallicity values above the field MZR, even when the AGN candidates are excluded from this analysis. The need for a significant detection of the [N{\sc{ii}}] emission line (see Sect. \ref{SS:EL}) could potentially explain this result, naturally biasing our individual measurements towards objects with high metallicity values. To overcome this, we analyze the average properties of our sample by resorting to the stacking analysis (white hexagons in Fig.\,\ref{F:MOH}). Our results show that on average, our sample is compatible with the field MZR in the high mass ($\mathrm{\log M_*/M_{\odot}>10.8}$) and low mass ($\mathrm{\log M_*/M_{\odot}<10.0}$) ends. However, intermediate mass galaxies display stacked gas-phase metallicities $0.06\pm0.03$ dex above the field MZR. This result is at odds with recent findings in overdense regions at $z\geq2$ (\citealt{Valentino15}; \citealt{Chartab21}), which find a metallicity deficit in protocluster galaxies. However, others have reported metallicity enhancements (up to 0.2 dex) in massive protoclusters at a similar epoch (\citealt{Kulas13}; \citealt{Shimakawa15}) or no significant differences with respect to the field (\citealt{Kacprzak15}). In Sect. \ref{S:Discussion}, we will discuss physically motivated scenarios that may be responsible for these contradicting results in protoclusters at the cosmic noon. 

One of these works (\citealt{Shimakawa15}), however, is partially based on a HAEs belonging to the Spiderweb protocluster. The discrepancy between our result and theirs may be related to several sample selection factors: First, their comparison sample is solely based on the \cite{Erb06} UV selected galaxies which tend to be biased towards lower metallicities. Furthermore, their spectroscopic campaign is based on slit spectra, which in some cases could miss a significant fraction of the emission coming from the galaxy disk. Even though the slit losses can be roughly estimated and thus corrected, this process can also be highly uncertain depending on the orientation and shape of the object with respect to the slit. In addition, the KMOS $2.8\arcsec\times2.8\arcsec$ IFU allows to extract the flux from a given aperture without missing flux in the process. Finally, our spectroscopic program is based on a larger sample of sources across a wider area, improving the overall accuracy of the stacked measurements. 

We also examine the relation between the metallicity enhancement measured in our sample and the environment these HAEs live in by following the two-fold approach outlined in Sect.\,\ref{SS:Environment}. First, we consider the impact of the local environment in the left panel of Fig.\,\ref{F:Density_OH}, where we show the metallicity offsets with respect to the field MZR as a function of local density indicators ($\Sigma_3$). After splitting our sample in three density bins we find a median binned metallicity enhancement of $\sim0.05-0.2$ dex across the density regimes examined. Interestingly, there is a mild decreasing metallicity trend towards higher local density values. By analyzing the median values of our sample with respect to the global environment indicator $\eta$ (Fig.\,\ref{F:Density_OH} right), we find an almost flat metallicity trend with a slightly higher median value within the accreted region of the protocluster. In summary, while the measurements coming from the highest local density peaks (i.e. $\Sigma_3$) within our sample show the lowest median metallicity values, we do not observe a clear trend from the core to the outskirts of the protocluster (in terms of $\eta$), suggesting that local and global environmental indicators trace different structures within the protocluster. Finally, we measure individual [N{\sc{ii}}]-based gas metallicities for four out of the seven SMGs within our sample. Two of them are consistent with the expectations in the field while another two show $\sim0.2$ dex enhancement (Fig.\,\ref{F:Density_OH}). Nonetheless, the three objects without [N{\sc{ii}}] detection likely have lower metallicity than the field, indicating that SMGs may share a similar distribution to the HAEs in the MZR. We will further explore the possible causes of these results in Sect.\,\ref{S:Discussion}.

\subsection{Mass-size relation}
\label{SS:Mass-size}

\begin{figure}
 \centering
      \includegraphics[width=\linewidth]{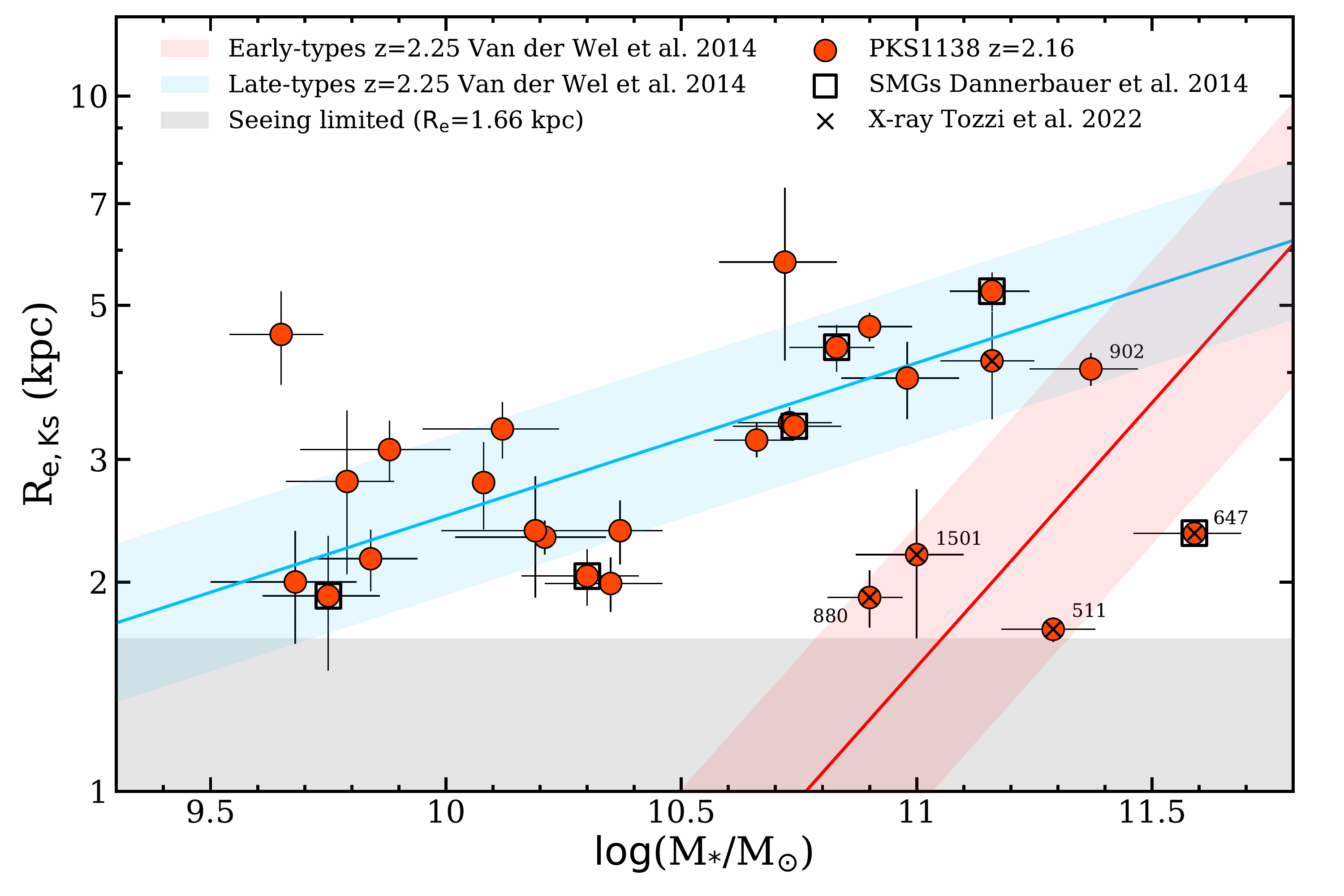}\par
      \caption{Mass-size relation. Red circles show the distribution of the KMOS H$\alpha$ emitters of this work. Empty squares display the sample of SMGs studied in \protect\cite{Dannerbauer14}. The blue and red sequences respectively depict the position of the the late and early-type galaxies in the mass-size plane at $z=2.25$ (\protect\citealt{Vanderwel14}). The gray area display the limiting seeing size in the physical scale. We label a few objects with their IDs for specific discussion in the main text.}
         \label{F:mass-size}
      \end{figure}

We inspect the distribution of our KMOS-detected HAEs with respect to the stellar mass-size relation of field galaxies at similar redshift (\citealt{Vanderwel14}). The effective radius ($\mathrm{R_e}$) of our targets is estimated using the VLT/HAWKI $\mathrm{K_s}$-band images of this protocluster. Due to the excellent seeing conditions and depth of these images ($\mathrm{FWHM}\sim0.4\arcsec$), we are able to measure the effective radius of 27 of our KMOS HAEs. However three of these sources lie below the seeing limit (gray area). Most of our targets show surface brightness profiles compatible with the presence of extended disks (Fig.\,\ref{F:mass-size}) in line with the predictions for late-type galaxies in the field at $z\geq2$ (light blue band). Other works exclusively investigating HAEs find a flatter trend than \cite{Vanderwel14} at the high-mass end of the mass-size relation (e.g. \citealt{Stott13a}; \citealt{Paulino-Afonso17}). This disagreement may arise from the different selection techniques (UVJ colors and H$\alpha$ emission respectively) employed to identify star-forming galaxies. Nevertheless, we find that the mean effective radius of our sample ($\mathrm{R_e=3.14\pm0.21}$) and the one from \cite{Paulino-Afonso17} agrees within the errors ($\mathrm{R_{e,PA17}=2.86\pm0.14}$), though our sample may be biased towards higher sizes due to the seeing limitations and the higher fraction of massive galaxies.

On the other hand, five objects lie within or close to the early type sequence in the mass-size relation, meaning that their morphology is considerably more compact than the rest of sources within our sample. Four of these objects are confirmed X-ray emitters (\citealt{Tozzi22}) suggesting that they are likely hosting AGNs (IDs 511, 647, 880 and 1501). For ID 647, its $\mathrm{K_s}$-band emission may be dominated by the H$\alpha$ broad component coming from the AGN torus, explaining its compact effective radius. In addition, IDs 511 and 880 display SFR values bellow the 1$\sigma$ scatter of the main sequence in Fig.\,\ref{F:SFR}. These objects may be transitioning from their late-type star-forming phase towards the early-type quiescent regime. The object (ID 902) which lies in the overlapping region between the early and late-type sequences in the mass-size diagram is also a massive AGN candidate according to its [N{\sc{ii}}]/H$\alpha$ ratio.

 \begin{figure*}
    \begin{multicols}{2}
      \includegraphics[width=\linewidth]{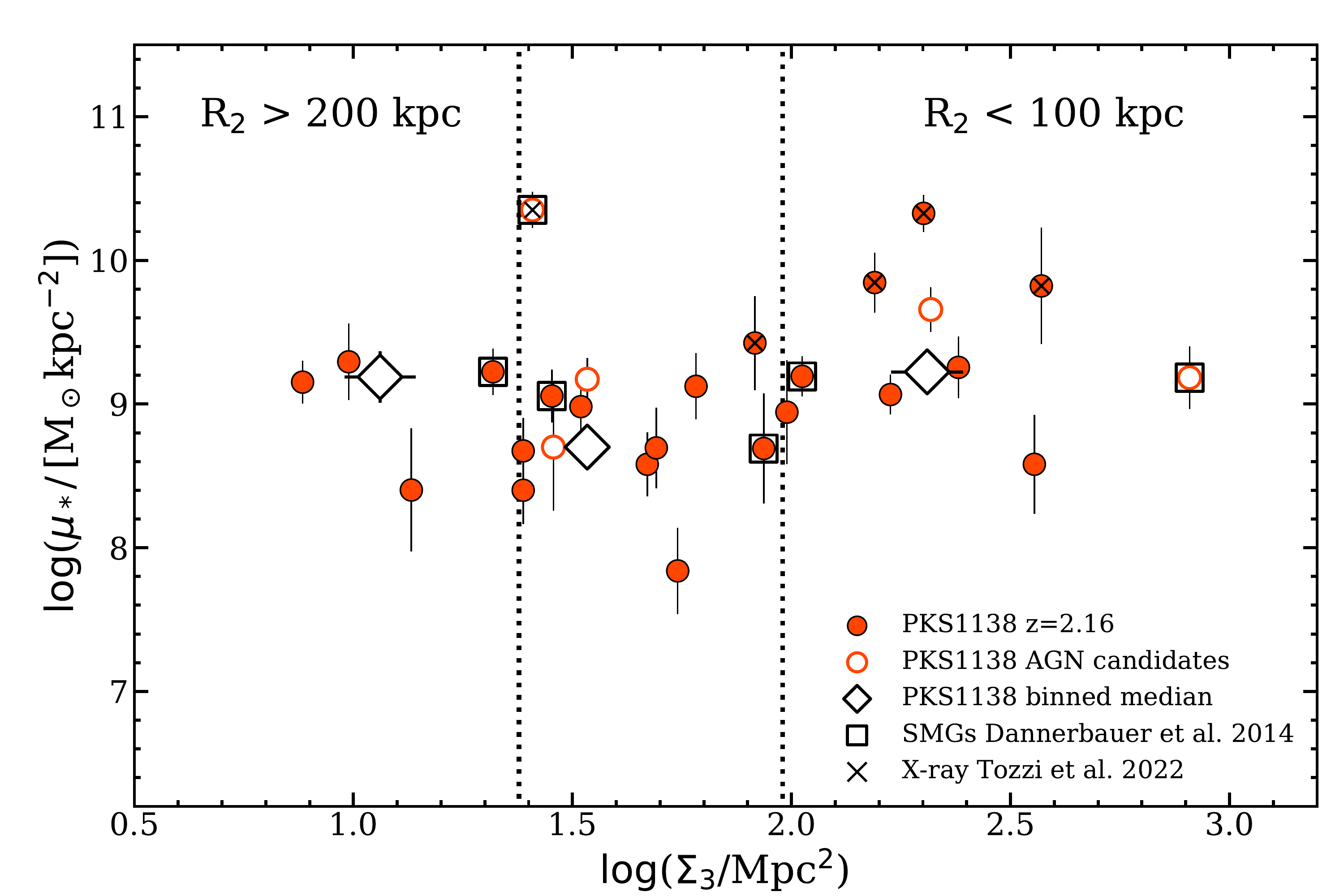}\par
      \includegraphics[width=\linewidth]{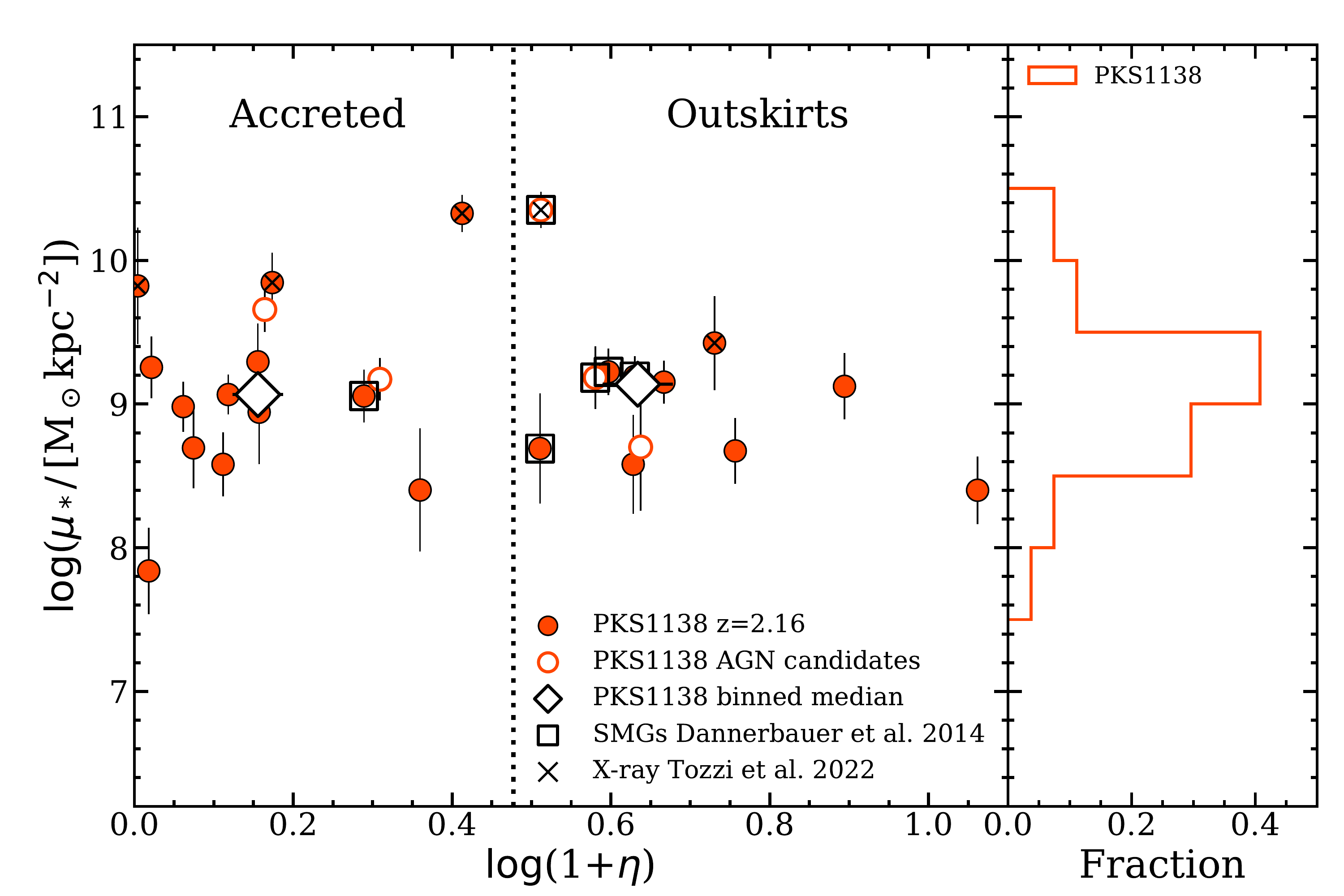}\par
      \end{multicols}
      \caption{Left: Local density distribution of our sample as a function of mass surface density defined as $\mathrm{\mu_*=M_*/(\pi R_e^2)}$. Right: general environment distribution ($\eta$) of our targets as a function of the mass surface density defined above. Vertical dotted lines separate the density regimes outlined in Sect. \ref{SS:Environment}. Colors and symbols follow the same scheme as in Fig.\,\ref{F:Density_OH}.}
         \label{F:Density_Size}
      \end{figure*}

In overdense regions, environmentally driven processes such as mergers and close encounters may temporarily distort the stellar morphology of interacting galaxies or create tidal tails (e.g. \citealt{Moore96}), both increasing or reducing the stellar-mass surface density depending on the geometry of the interaction and the physical properties of the galaxies involved (e.g. mass and gas reservoir, \citealt{Lagos18a}). On the other hand, such events may as well channel large amounts of gas towards their central regions, triggering starburst events or AGN activity in some cases, and favoring the creation of prominent bulges. In order to test these effects we inspect the stellar-mass surface density ($\mathrm{\mu_*=M_*/(\pi\,R_e^2)}$) of our objects as function of both local density and general environment indicators in Fig.\,\ref{F:Density_Size}. Our spectroscopic sample display a rather flat trend ($\sim10^9-10^{10}\,M_\odot\,$kpc$^{-2}$) across its stellar-mass range ($\log M_*=9.5-11.5$) and three orders of magnitude in local density ($\Sigma_3$). This can be understood as a consequence of the tight distribution that most of our targets display along the late-type sequence in the mass-size relation (Fig.\,\ref{F:mass-size}). Similar results are achieved after repeating this analysis by comparing the distribution of our targets as function of the global environment parameter instead ($\eta$, right panel). The small variations between the binned median stellar-mass surface values across different environmental regimes in Fig.\,\ref{F:Density_Size} can be partly explained by a combination of factors: the heterogeneous nature of our sample (including HAEs, AGN candidates and SMGs) and the limited number of object per density bin.

\subsection{Molecular gas properties.}
\label{SS:Molprop}

In this section, we explore the relation between the molecular gas fraction ($f_{\mathrm{gas}}=\mathrm{M_{mol}/(M_*+M_{mol})}$) and gas phase metalicity of 7 HAEs within our KMOS sample. Our objects were selected after crossmatching the CO(1-0) line flux catalog published by \cite{Jin21} with our spectroscopic sample. Out of 8 overlapping targets, we obtained metallicity measurements based on the [N{\sc{ii}}]/H$\alpha$ ratio for 7 of them. The molecular gas mass of every object is derived following the scaling relations published by \cite{Tacconi18} and the procedure outlined in Sect\,\ref{SS:MOL}.

\cite{Bothwell13a} and \cite{Hunt15} showed that the gas-phase metallicity gradually increases as the gas reservoir of the galaxy is consumed (i.e. towards lower gas fractions) and the SFR slowly decreases. Such combination of processes were put into context by \cite{Peng14} in their gas regulator model, which also considers the influence of gas inflows and outflows in the evolution of the aforementioned physical properties. This model relies on four main input parameters governing the evolution of galaxies: the star formation efficiency, the gas inflow rate, the mass-loading factor ($\lambda$=outflows/SFR) and the return mass fraction. In this model, the inflowing gas is assumed to scale with the growth rate of the dark matter halo while the outflow rate correlates with SFR and the return fraction is a constant parameter fixed for a given IMF. For a more in depth discussion of the model we refer to \cite{Peng14}. 

\begin{figure}
      \includegraphics[width=\linewidth]{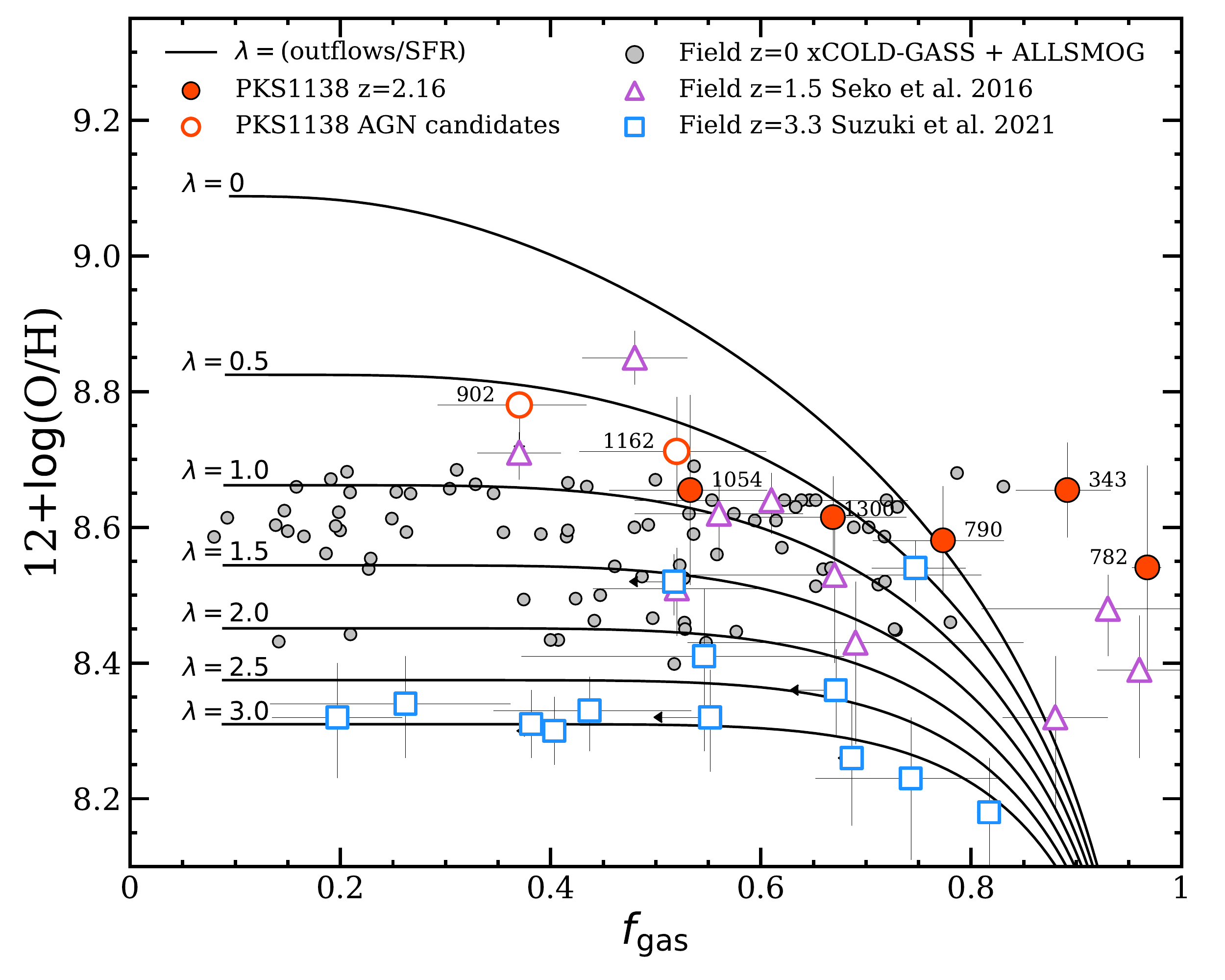}\par 
      \caption{Molecular gas fraction versus gas phase metallicity. The solid lines show the tracks of constant mass-loading factor (outfows/SFR) assuming equilibrium in the context of the gas regulator model (\protect\citealt{Peng10}). Filled and empty circles display our subsample of 7 HAEs with measured gas fractions from CO(1-0). Blue edged squares show the distribution of field galaxies at $z=3.3$ from \protect\cite{Suzuki21}. Violet edged triangles depict the field sample of \protect\cite{Seko16} at $z=1.5$. The local field comparison sample is composed by galaxies from the XCOLD GASS (\protect\citealt{Saintonge17}; \protect\citealt{Catinella18})  and the ALLSMOG (\protect\citealt{Bothwell14}) surveys. We label our objects according to their HAE IDs (\protect\citealt{Koyama13}) for discussion in the main text.}
         \label{F:massload}
      \end{figure}

In this work, we follow the approach outlined in \cite{Suzuki21}, who also relies on the gas regulator model to investigate the gas fraction gas-phase metallicity relation in the field at $z\sim3.3$. For reference, we overplot the lines of constant mass-loading factor ($\lambda$) for $\lambda=0-2.5$ in Fig.\,\ref{F:massload}. Our 7 protocluster members show a mild negative trend between the gas metallicity and gas fraction. However, they are shifted towards lower mass-loading factor values compared to the field samples of \cite{Suzuki21} at $z=3.3$ and \cite{Seko16} at $z=1.5$, whose values in Fig.\,\ref{F:massload} have been recomputed to match the calibration and IMF of this work. This means that at our objects display higher metallicity values at a fixed gas fraction. In the context of the gas regulator model and assuming that our galaxies are representative of the main sequence of star-forming galaxies at $z=2.2$ (see Fig.\,\ref{F:SFR}), the shift in $\lambda$ may be imply that the enrichment of the ISM may be (at least partly) driven by the suppression of outflows in our targets. This effect could be the result of the influence of the protocluster environment, where a denser IGM compared to the field would exert external pressure over the gas halo of the protocluster members, preventing the outflows from star-formation to leave the gravitational potential of the galaxies and forcing them to recycle the already enriched gas. However, the small size of our current sample and the influence of AGN candidates prevent us from discarding other possibilities at this stage. We will discuss some of these possibilities in Sect.\,\ref{S:Discussion}.

\section{Discussion}
\label{S:Discussion}

In this work we have searched for environmental imprints on the star-formation, gas-phase metallicity, morphology and gas fraction of a sample of 39 spectroscopically confirmed protocluster galaxies at $z\sim2.2$. In the following subsections we will discuss the implication of our results on different galaxy evolution scenarios.

\subsection{Star forming galaxies in protoclusters}
\label{SS:SFR_DIS}

Strong gravitational interactions such as mergers should be more frequent in high-z protoclusters than in local universe clusters according to N-body and hydrodynamical simulations (\citealt{Gottlober01}; \citealt{Genel14}). This can be understood as a natural consequence of the assembling phase these structures are experiencing, with small groups infalling towards the highest density peak in the field, and the low relative velocity between the galaxies therein (\citealt{Hine16}). Therefore, it is also expected that such interactions increase the fraction of starbursting galaxies within protoclusters. However, different works on protoclusters at $z=2-3$ have found mixed results in terms of the star-formation activity of their members. While \cite{Koyama13}, \cite{Shimakawa18b}, \cite{Sattari21} and this work find no significant differences between the protocluster and field population, others such as \cite{Shimakawa18a} and \cite{Wang21} find that protocluster galaxies display enhanced SFRs (specially in the low stellar mass range), and that the amplitude of such enhancement positively correlates with higher density regions within the protocluster structure. 

Two considerations should be taken into account when analyzing these conflicting results. First, most overdense structures that are classified as protoclusters at $2<z<3$ undergo a major though rapid phase of mass accretion and mass build-up until $z\sim1.5$, when some of the most massive structures start showing a well developed red sequence (e.g. XMMU J2235-2557 at $z=1.4$, \citealt{Rosati09} or XLSSC 122 at $z=1.99$ \citealt{Willis20}). This implies that protoclusters belonging to the same epoch ($z=2-3$) may be going through different evolutionary stages and thus, their galaxy population could be affected in different ways by the environment. In fact, recent studies at low redshift have shown that the dynamical properties of clusters do influence the evolution of their galaxies both in terms of star-formation and AGN activity (\citealt{Stroe21}). In that sense, the Spiderweb protocluster is one of the most massive structures at $z\sim2$ and show signs of virialization within its inner core ($\mathrm{<0.5 Mpc}$, \citealt{Shimakawa14}), while other protoclusters such as USS1558 at $z=2.53$ (\citealt{Hayashi16}) or BOSS1244 at $z=2.2$ (\citealt{Zheng21}) are formed by several less massive clumps ($<10^{14}\,\mathrm{M_\odot}$) hosting galaxies with low proper velocities, thus favoring galaxy-galaxy interactions driven starbursts. 
This could explain why we find no significant differences in terms of star-formation between the Spiderweb protocluster and the field population, in contrast with results in younger and less massive structures. Second, the detection of high-z protoclusters and the analysis of their galaxy populations are carried out through a wide variety of techniques (e.g. SED, narrow-band, spectroscopic surveys, etc.) and focusing on different wavelength ranges. This could potentially bias the selection of the samples, making difficult to establish a fair comparison between the galaxy populations of different protoclusters. Finally, the signs of SFR enhancement in protocluster galaxies are oftenly restricted to the low stellar mass regime (e.g. at $\mathrm{\log M_*/M_{\odot}<9.5}$ in \citealt{Hayashi16}), adding further uncertainties since most studies are not able to achieve a high completeness level in this mass range due to observational depth limitations, as it is the case of this work.

Recently, \cite{Tozzi22} reported fourteen protocluster members with significant X-ray emission hinting at a possible enhancement of the AGN fraction with respect to the field. Our KMOS sample contains seven of these objects. They are massive HAEs ($\mathrm{\log M_*\geq10.8}$) which tend to lie below the main sequence of star formation, with four of them displaying $\mathrm{\Delta MS\leq-0.3}$ dex (Fig.\,\ref{F:SFR}). We find that the overlapping X-ray emitters are sparsely distributed from the outskirts to the core of the protocluster in terms of $\eta$ (Fig.\,\ref{F:Density_SFR}). However, five of them reside within the densest bin in terms of $\Sigma_3$ while the other two lie at intermediate local densities. These results suggest that AGNs may preferentially be located at local density peaks but not necessarily within the innermost regions of the protocluster. The impact of the X-ray emitters can also be spotted in the mass-size relation (Fig:\,\ref{F:mass-size}), where four out of the five objects with measured $\mathrm{R_e}$ display rather compact morphologies. In contrast, the majority of the HAEs that entered our analysis show sizes compatible with the presence of extended disks. We must note, however, that our $\mathrm{R_e}$ measurements are based on $\mathrm{K_s}$-band which overlaps with the H$\alpha$ emission line at $z=2.16$. After checking the H$\alpha$ contribution to the total $\mathrm{K_s}$-band flux we find that it accounts for $\lesssim15\%$ except for two cases. Therefore, we conclude that the size measurements are not driven by contamination from star-forming gas emission. For AGNs, a significant fraction of the $\mathrm{K_s}$ emission could be originated by the H$\alpha$ broad component at their center, making them look more compact. Furthermore, three of these objects also display low SFRs (see Sect.\,\ref{SS:Mass-size}) or UVJ colors similar to those of early-type galaxies, suggesting that they may be transitioning towards a quiescent state. However, it is not possible to discern between these two options with the current depth and seeing limitations of our data.

Finally, we also inspect the properties and spatial distribution of a sample of seven HAEs identified as SMGs by \cite{Dannerbauer14}. We find a factor 3-10 lower H$\alpha$-based SFRs than the FIR-based ones, confirming previous reports of unaccounted dust obscuration in the rest-frame optical regime of this starbursting objects (\citealt{Dannerbauer14}). Furthermore, these seven SMGs are found at projected clustercentric distances larger than $\mathrm{1.6\,R_{200}}$ (Fig.\,\ref{F:phase-space}), with six of them residing at the protocluster's outskirts according to their global environment indicator ($\eta>2$, Fig.\,\ref{F:Density_SFR}). While the spatial distribution of our SMGs is not significantly offset with respect to the HAE population as in \cite{Zhang22}, it is clear that these objects are not yet part the innermost regions of the Spiderweb protocluster (Fig.\,\ref{F:Map}, see also \citealt{Dannerbauer14}). Several authors have established links between the violent dust-obscured star-forming activity of SMGs at $z>2$ and the surge of red ellipticals at later epochs (e.g. \citealt{Swinbank06}; \citealt{Michalowski10}; \citealt{Toft14}; \citealt{Simpson14}). In fact, \cite{Smail14} reported a similar spatial mismatch for SMGs in a cluster at $z\sim1.6$, hypothesizing that this infalling population may constitute the progenitors of some of the fainter ellipticals dominating cluster cores by $z\sim0$, whilst the pre-existing passive cluster members may have formed following the same channel at even earlier epochs. Several studies have spectroscopically identified a few members of the nascent passive population of the Spiderweb protocluster (e.g. this work, but see also \citealt{Doherty10}; \citealt{Tanaka13}). Assuming that these galaxies were formed through a similar SMG phase in the past, and that the duration of the starburst causing the such phase lasted a few hundred Myrs (e.g. \citealt{Riechers11}; \citealt{Hickox12}), we estimate their formation period at $z\lesssim2.5$. On the other hand, the current SMGs in this protocluster may deplete their gas reservoirs and start their quenching process by $z\approx2$, concurrently to their gradual infall towards the protocluster core.






\subsection{ISM enrichment at the cosmic noon}
\label{SS:OH_DIS}

We have analyzed the gas-phase metallicity properties of 22 individual protocluster members (Fig.\,\ref{F:MOH}) based on their [N{\sc{ii}}]/H$\alpha$ ratios. We obtained enhanced individual values (up to 0.3 dex) with respect to the field MZR at $z\sim2.3$. However, our stacking analysis yields metalliciy values only slightly above the field MZR for intermediate mass galaxies ($\mathrm{\Delta MZR=0.06\pm0.03}$), while no significant metallicity enhancement is observed in the high and low-mass regime. We find that there exist a population of AGN candidates according to their high [N{\sc{ii}}]/H$\alpha$ ratios (see Sect. \ref{SS:MZR}). However, observation by \cite{Tozzi22} reported X-ray emission for only one of these candidates. Even though the X-ray emission originated through AGN activity can be obscured for several reasons (\citealt{Hickox18}), the low number of X-ray detections between the targets with high [N{\sc{ii}}]/H$\alpha$ ratios suggests that our results are not severely affected by AGN contamination and thus, we may be witnessing a population of highly metal-enriched objects.

Previous studies on the MZR evolution in protoclusters at the cosmic noon have shown mixed results. For example, \cite{Kulas13} found a significant metallicity excess of $0.1-0.2$ dex with respect to the field, in particular for galaxies in the low-mass end in a protocluster with a virial mass well above $\mathrm{M_{vir}\geq10^{14}\,M_\odot}$ (\citealt{Steidel05}). A possible scenario to explain this result involves the recycling of outflowing chemically enriched gas due to the external pressure exerted by the surrounding IGM in clusters. This effect would predominantly affect galaxies in the intermediate to low-mass regime, as the gravitational potential of the most massive systems is strong enough to retain their metals regardless of the environment (\citealt{Oppenheimer08}). Interestingly, this scenario is expected to be more efficient in massive virialized clusters, as the density of the IGM correlates with the total mass of the cluster and the accretion of pristine gas which could potentially dilute the metallicity could have stopped some time ago due to the shock heating of their main halo (\citealt{Dekel06}). Nevertheless, other studies in less massive structures ($\mathrm{M_{vir}\sim10^{13}\,M_\odot}$) report a relatively constant metallicity deficit of similar magnitude across the entire galaxy stellar-mass range, which can be understood as a consequence of the cold gas accretion into these relatively young assembling protoclusters (\citealt{Valentino15}; \citealt{Chartab21}). Finally, several authors have reported a diversity of behaviors within their samples, with massive galaxies being less metal enriched than the low-mass ones but with absolute values close to the field MZR (\citealt{Sattari21}; \citealt{Wang21}), or no clear environmental dependence at all (\citealt{Kacprzak15}). 

Using hydrodynamical simulations, \cite{Dekel09a} proposed that protoclusters at the cosmic noon experience a gradual but rapid transformation from an early assembly phase where the cold gas accretion dominates in the halo, to a short transition period where cold streams can still penetrate an otherwise shock heated medium, to a final stage where the cold streams are completely supressed and the cluster achieves a virialized state. These three sequential scenarios could potentially explain the variety of metallicity results in the literature as part of an evolutionary path that involves the co-evolution of the galaxies and their host protocluster halo. In this context, the mild metallicity enhacenment that we detect in the Spiderweb protocluster ($\mathrm{M_{200}\sim2\times10^{14}\,M_\odot}$, \citealt{Shimakawa14}) suggest that the protocluster halo may have recently become shock heated to the point where inflows are no longer efficient to supply cold gas to the member galaxies.

The environmental analysis of our individual measurements in Fig.\,\ref{F:Density_OH} reveals that the accreted region of the spiderweb protocluster display similar median metallicity values than the outskirts in terms of the global environment indicator ($\eta$). On the other hand, a declining metallicity trend is found towards high local density peaks ($\Sigma_3$). This trend, however, may be predominately driven by massive galaxies dominating the highest $\Sigma_3$ bin ($\mathrm{\log M_*/M_\odot=10.66\pm0.21}$) while lower mass galaxies populate the intermediate regime ($\mathrm{\log M_*/M_\odot=10.36\pm0.14}$) and the least dense regions ($\mathrm{\log M_*/M_\odot=10.17\pm0.18}$). This mass segregation is not present when examining the metallicity variations as function of $\eta$, with the accreted and outskirts regions sharing very similar stellar mass median values ($\mathrm{\log M_*/M_\odot=10.39\pm0.15}$ and $\mathrm{10.36\pm0.13}$ respectively). Nonetheless, larger number statistics and deeper observations, specially for the intermediate to low-mass star-forming population ($\mathrm{\log M_*/M_\odot\lesssim10.0}$), are required to determine if significant metallicity differences between environmental regimes exist within this protocluster.

\subsection{The role of the molecular gas fractions}
\label{SS:MOL_DIS}

The gas regulator model (\citealt{Peng14}) provides us with a practical frame to investigate the relation between the possible suppresion of outflows via the external pressure of the IGM and the gas-phase metallicity. As we discussed in Sect.\,\ref{SS:Molprop}, this model relies on a few input parameters, being one of them the mass-loading factor ($\lambda$=outflows/SFR). In Fig.\,\ref{F:massload} we explore the relation of this parameter with both the molecular gas fraction and the gas-phase metallicity of our targets in comparison with field samples at several cosmic epochs. It is expected that galaxies move towards higher mass-loading factors from the local to the high-z universe (\citealt{Suzuki21}). This can be explained by the different star-formation histories that galaxies at different redshifts experience, with those at high-z experimenting vigorous but short burst of star-formation (increasing the feedback too), while the stellar-mass growth proceeds in a more gradual manner at lower redshifts. Indeed, the majority of the $z=3.3$ field galaxy sample discussed by \cite{Suzuki21} is consistent with $\lambda=2.5-3$, while results at $z=1.5$ display $\lambda=1-2$ (\citealt{Seko16}) and local universe samples lie around $\lambda\sim1-1.5$ (ALLSMOG from \citealt{Bothwell14}; xCOLD-GASS from \citealt{Saintonge17} and \citealt{Catinella18}). In contrast, a small subsample of 7 HAEs embedded in the Spiderweb protocluster display unusually low mass-loading factors for galaxies at $z\gtrsim2$. This subsample is composed of two objects (IDs 343 and 782) with relatively low stellar mass ($\mathrm{\log M_*/M_\odot=10.11}$ and 9.75 respectively) and strong MZR offsets ($\gtrsim0.2$ dex) with respect to the field. Another three objects (IDs 790, 1054 and 1300) with high stellar masses ($\mathrm{10.7<\log M_*/M_\odot<11.2}$) but small offsets ($\mathrm{\Delta MZR\lesssim0.05}$ dex) and two additional galaxies (IDs 902 and 1162) which share similar characteristics to the previous ones in terms of stellar mass but show $\mathrm{0.11<\Delta MZR<0.16}$ and they are labeled as AGN candidates due to their high [N{\sc{ii}}]/H$\alpha$ ratios. 

As we have extensively discussed in previous sections, most of our spectroscopic sample display SFRs consistent with those of the main sequence of star-formation. In principle, this would suggests that the lower $\lambda$ values are mostly driven by a lower outflow contribution in these protocluster galaxies, supporting the scenario where galaxies are forced to recycle their gas due to the external pressure of the IGM and thus, yielding higher metallicity values at a fixed molecular gas fraction. Following the accretion mode evolutionary sequence depicted in \citealt{Dekel09a}, this scenario would also suggest that cold streams are no longer efficient on supplying pristine gas to some members of the Spiderweb protocluster, in agreement with the theoretical predictions of \citealt{Dekel06} for a protocluster of $\mathrm{M_{vir}\geq10^{14}\,M_\odot}$ at $z=2.16$.

However, we should also consider the current statistical limitations and biases of our sample. For example, these seven galaxies are drawn from those with [N{\sc{ii}}] detection within our sample, which predominately trace the metal rich end of the protocluster distribution (see Sect.\,\ref{SS:MZR}). This would naturally bias these objects towards higher metallicity values in Fig.\,\ref{F:massload} and thus, they may not be representative of the full population of HAEs in the Spiderweb protocluster if metallicity and gas fraction information were available for all them. In the following, we will continue discussing our current results while keeping in mind this limitation. In particular, the two galaxies with the lowest stellar mass values are at the same time the ones with the strongest $\mathrm{\Delta MZR}$ and highest gas fraction (IDs 343 and 782), indicating that their ISM has been significantly enriched while their gas reservoir remains nearly intact. It is tempting to think of these two objects as candidates to support environmentally driven gas recycling, specially by taking into account their low stellar mass (see Sect.\,\ref{SS:OH_DIS}). However, we would require significantly larger number statistics to confirm such scenario. As for the remaining five objects, their high stellar mass ($\mathrm{\log M_*/M_\odot>10.7}$) makes them less prone to suffer this kind of environmental effect (\citealt{Oppenheimer08}). Nonetheless, it is still interesting to find relatively high gas fractions for such metal enriched galaxies when the universe was only 3 Gyrs old. The rapid build-up of metals in these massive objects, if not mediated by any environmental effect, would require extreme episodes of star formation in the recent past, thus transitioning from high to low $\lambda$ in a short time scale.  

\section{Conclusions}
\label{S:Conclusions}

In this work, we have investigated the star-formation, galaxy size, gas-phase metallicity, and molecular gas fraction of a sample of HAEs within the Spiderweb protocluster at $z=2.16$. We based our results on new multi-object near-infrared spectroscopy with VLT/KMOS accompanied by previously reported spectroscopic and photometric observations covering from the UV to the submillimeter wavelength regime. This wealth of data allowed us to accurately trace the environment of the protocluster using both local and global environment indicators after confirming the cluster membership of over a hundred galaxies. We have examined the physical properties of 39 HAEs as a function of environment, and discussed the implications of these results on different scenarios of environmentally driven galaxy evolution during the early stages of massive cluster assembly. In the following paragraphs we summarize the main conclusions of this work:
\begin{enumerate}
\item We analized a sample of 42 narrow-band selected H$\alpha$ emission-line candidates with VLT/KMOS in the field of the Spiderweb protocluster at $z=2.16$. We spectroscopically confirm the protocluster membership of 39 HAEs and detect one foreground [O{\sc{iii}}] emitter at $z>3$. Thus, we achieved a $\sim93\%$ success rate on the cluster membership confirmation based on narrow-band selected H$\alpha$ emitters.
\item We measure the star formation activity of our spectroscopically confirmed HAEs finding that most of these objects are consistent with the field main sequence of star-formation at z$\sim$2.16 (\citealt{Speagle14}; \citealt{Wisnioski15}). Furthermore, these results show no correlation with several environmental indicators (e.g. local density and global environment), suggesting that the star-formation of HAEs has not been strongly affected by the environment of the Spiderweb protocluster.
\item Our morphological analysis reported that our HAEs sample is predominantly composed of late-type galaxies with observed $\mathrm{K_s}$-band effective radius consistent with those of field galaxies at z$\sim$2.2 (\citealt{Vanderwel14}). Only 5 HAEs show compact morphologies compatible with those of early-type galaxies at this redshift. However, no environmental correlation was found when exploring the distribution of stellar-mass surface density ($\mu_*$) of our targets across different density regimes. 
\item We extract [N{\sc{ii}}]-based metallicities for a subsample of 22 HAEs. We find that 5 of them have [N{\sc{ii}}]/H$\alpha$ ratios consistent with those of AGN candidates according to \cite{Agostino21}. Most of the remaining protocluster members still show individual enhanced metallicities by up to $\sim0.25$ dex with respect to the field mass-metallicity relation at similar redshift (\citealt{Erb06}; \citealt{Wisnioski19}; \citealt{Sanders21}). However, our stacking analysis reveal just a mild metallicity enhancement at intermediate masses ($\mathrm{\Delta MZR=0.06\pm0.03}$) and no significant differences at the high or low mass ends. 
\item We also studied a subsample of seven SMGs in this protocluster from \cite{Dannerbauer14}. Even though their distribution in the star-forming main sequence is similar to the parent sample of HAEs when using SFR(H$\alpha$), the comparison with FIR-based SFRs from \cite{Dannerbauer14} suggests that these sources are highly dust-obscured. SMGs are also consistent with the MZR traced by HAEs of similar stellar mass, and their $\mathrm{R_{e,K_s}}$ are comparable to those of late-type galaxies at $z\sim2.2$. However, SMGs are preferentially found in the outskirts (six) of this protocluster rather than in their accreted region (one), in agreement with previous works in (proto-)clusters at $1.5<z<2.5$ (\citealt{Dannerbauer14}; \citealt{Smail14}; \citealt{Zhang22}).
\item After exploring the gas metallicity gas fraction relation, we find that the 7 HAEs with available metallicities and molecular gas fractions are distributed along mass-loading factor tracks comparable to those of the local universe but in contrast with previous studies at $z>2$. Taking into account that these are normal star-forming galaxies, we could interpret this as a sign of suppressed outflow activity, indicative of environmental effects in the form of external pressure over the galaxies' halo by a denser IGM in the Spiderweb protocluster. However, a systematic analysis of the metal enrichment and gas fraction across the protocluster structure would be required to confirm or discard such scenario.
\end{enumerate}

\section*{Acknowledgements}

We would like to thank Christian Herenz for his thorough feedback during the revision process of this manuscript. This research is based in part on observations collected at the European Organisation for Astronomical Research in the Southern Hemisphere under ESO programme 095.A-0500(B). This research is based in part on data collected at the Subaru Telescope, which is operated by the National Astronomical Observatory of Japan (NAOJ). We are honored and grateful for the opportunity of observing the Universe from Maunakea, which has cultural, historical, and natural significance in Hawaii.
This work is based in part on observations made with the Spitzer Space Telescope, which was operated by the Jet Propulsion Laboratory, California Institute of Technology under a contract with NASA. This research is based in part on observations made with the NASA/ESA Hubble Space Telescope obtained from the Space Telescope Science Institute, which is operated by the Association of Universities for Research in Astronomy, Inc., under NASA contract NAS 5–26555. This research made use of Astropy,\footnote{http://www.astropy.org} a community-developed core Python package for Astronomy \citep{Astropy13, Astropy18}. JMP acknowledges the funding support from the Marietta Blau Grant, financed by the Austrian Science Ministry, as well as the support from the Instituto de Astrof\'isica de Canarias for allowing him to develop part of this work in their headquarters at the Canary Islands, and the support from Tohoku University. HD acknowledges financial support from the Spanish Ministry of Science, Innovation and Universities (MICIU) under the 2014 Ramón y Cajal program RYC-2014-15686 and under the AYA2017-84061-P, co-financed by FEDER (European Regional,Development Funds) and support from the ACIISI, Consejería de Economía, Conocimiento y Empleo del Gobierno de Canarias and the European Regional Development Fund (ERDF) under grant with reference PROID2020010107. HD, RC, ZC and ALR acknowledge financial support from the Agencia Estatal de Investigación del Ministerio de Ciencia e Innovación (AEI-MCINN) under grant (La evolución de los cíumulos de galaxias desde el amanecer hasta el mediodía cósmico) with reference (PID2019-105776GB-I00/DOI:10.13039/501100011033). TK acknowledges the support by Grant-in-Aid for Scientific Research (A) (KAKENHI \#18H03717). NAH acknowledges support from UKRI STFC grant ST/T000171/1.

\section*{Data Availability}

The VLT/KMOS spectroscopic data used in this work is publicly available through the ESO Science Archive Facility (program ID 095.A-0500, PI: Y. Koyama). The ATCA CO(1-0) map belongs to the COALAS project (large program ID: C3181, PI: H. Dannerbauer). The photometric data used in this work is publicly available through the Subaru Mitaka Okayama Kiso Archive (SMOKA) system, the Hubble Legacy Archive (proposal ID 10327, PI: H. Ford), the NASA/IPAC Infrared Science Archive (campaign ID 1634, PI: D. Stern) and the ESO Science Archive Facility (program IDs 088.A-0754, 091.A-0106, 094.A-0104, PI: A. Kurk). Any other dataset within this article will be shared on reasonable request to the corresponding author.



\bibliographystyle{mnras}
\bibliography{references.bib} 




\appendix

\section{Additional material}
      
      

\begin{figure*}
 \centering
 \begin{multicols}{4}
      \includegraphics[width=\linewidth]{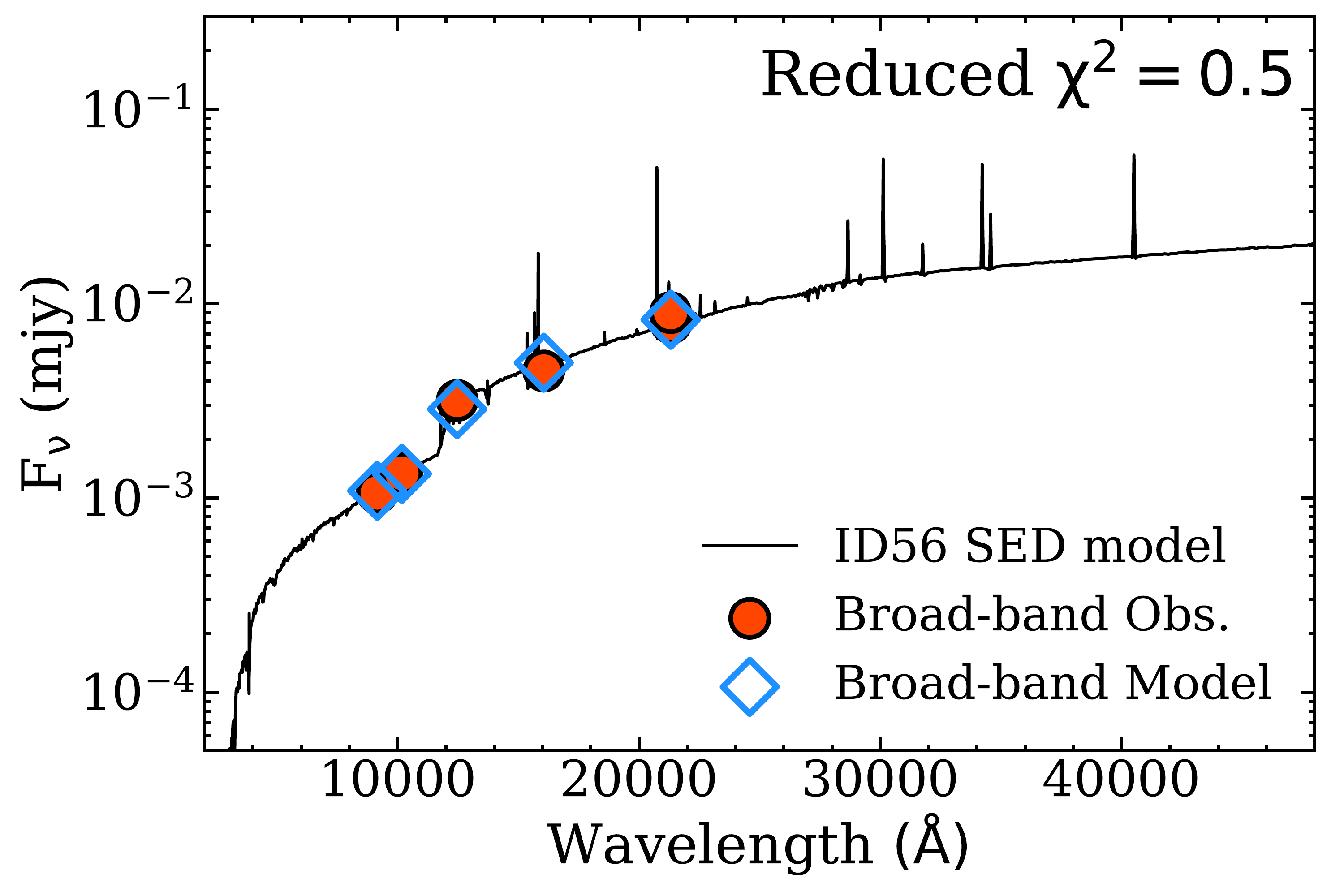}\par
      \includegraphics[width=\linewidth]{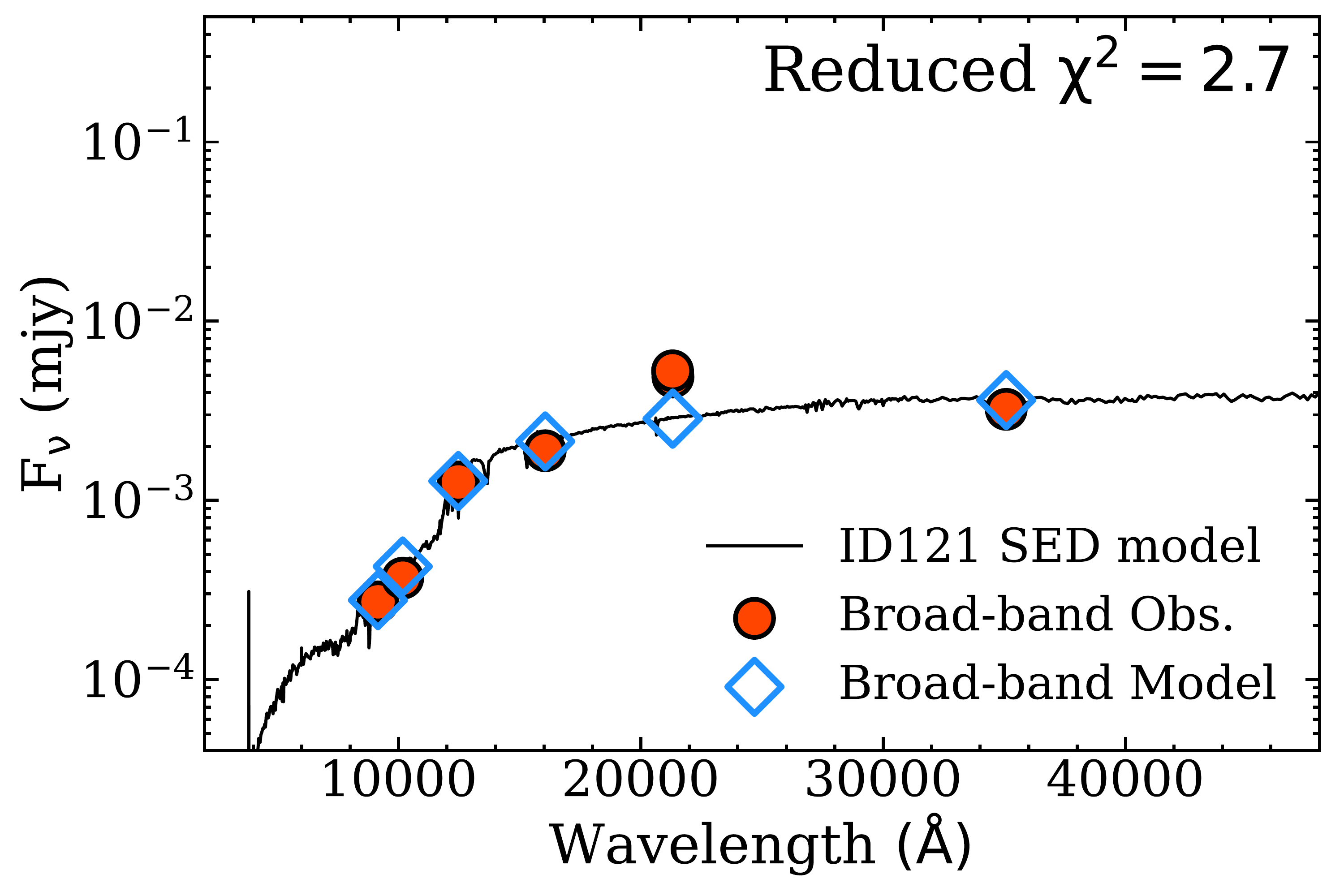}\par
      \includegraphics[width=\linewidth]{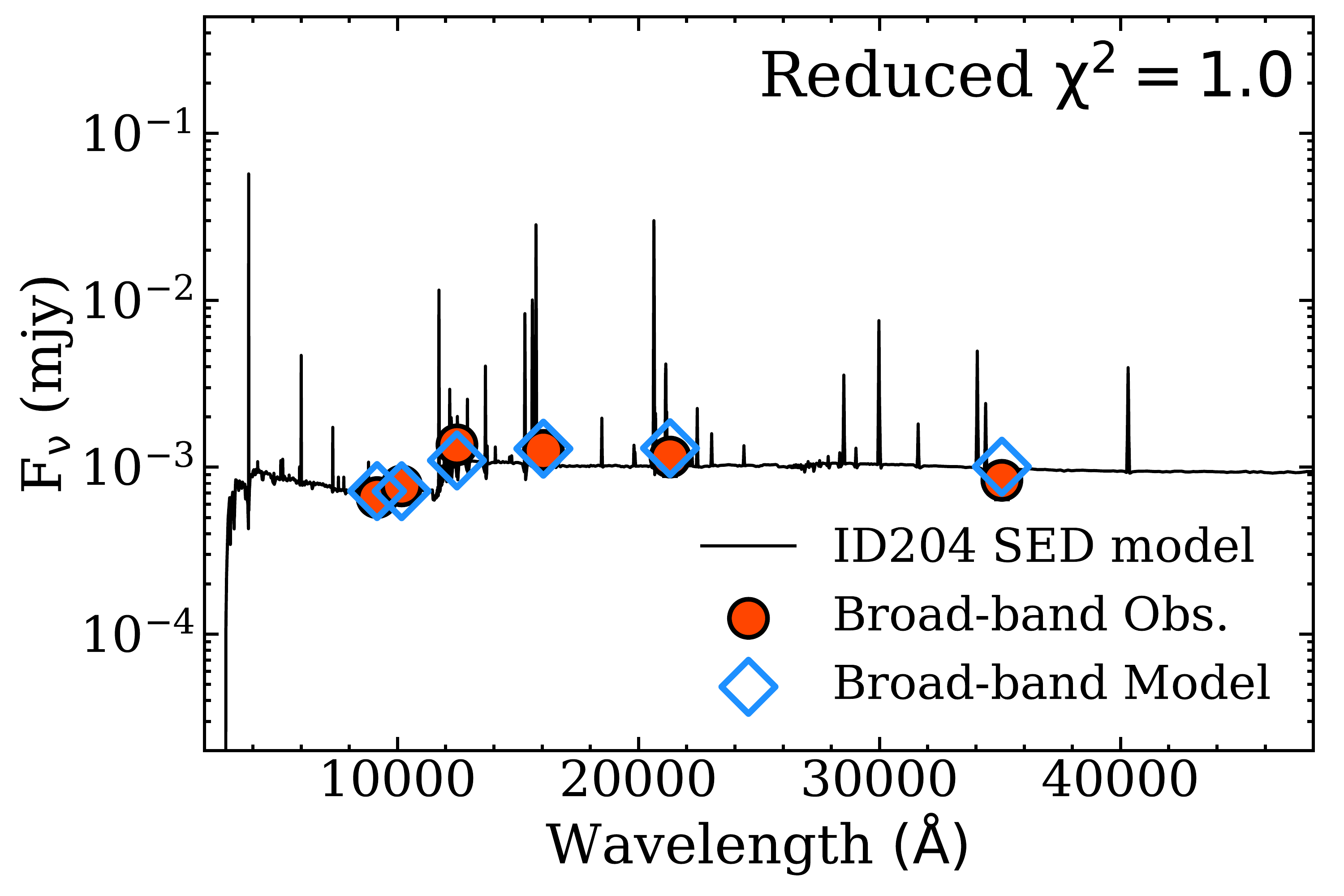}\par
      \includegraphics[width=\linewidth]{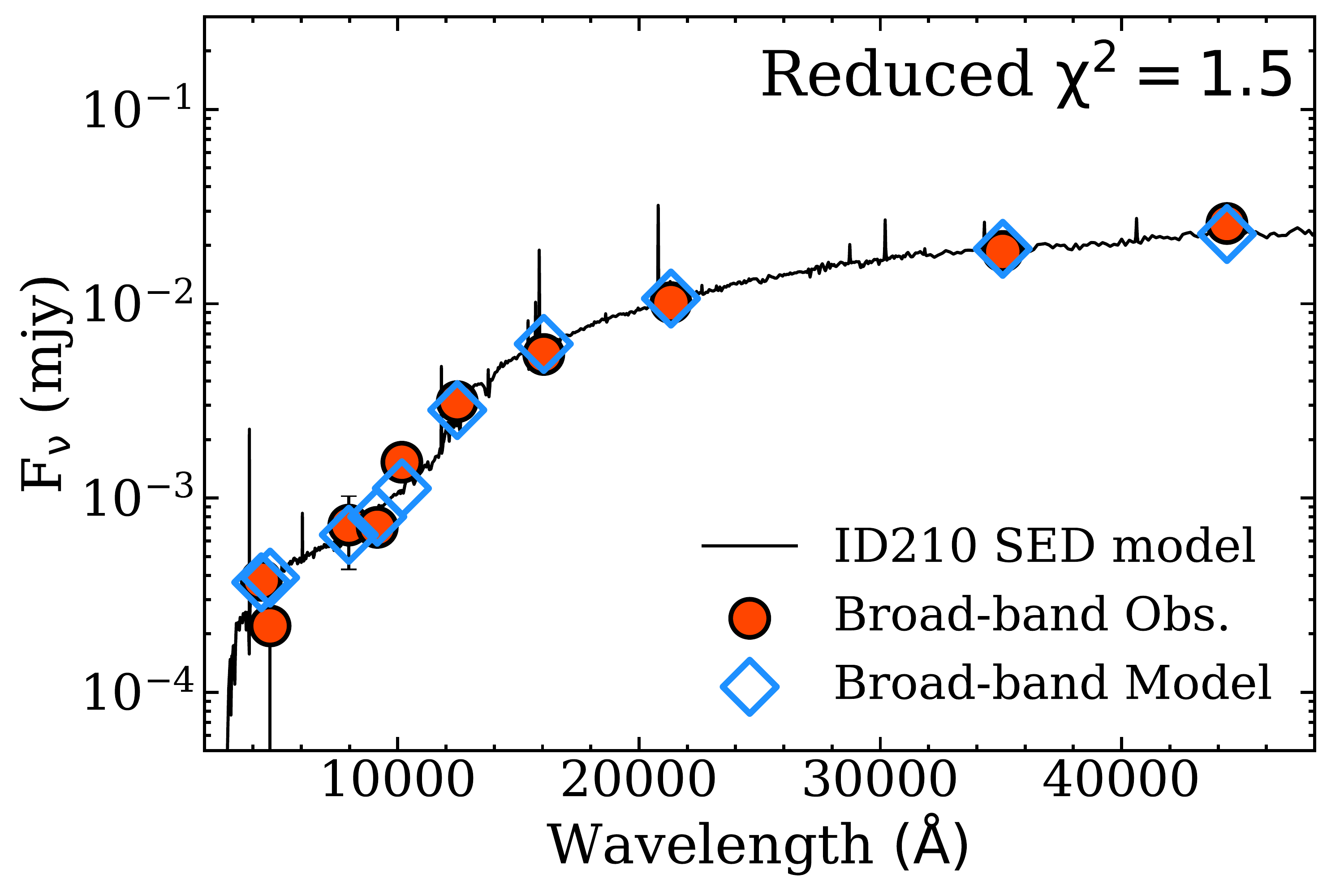}\par
      \includegraphics[width=\linewidth]{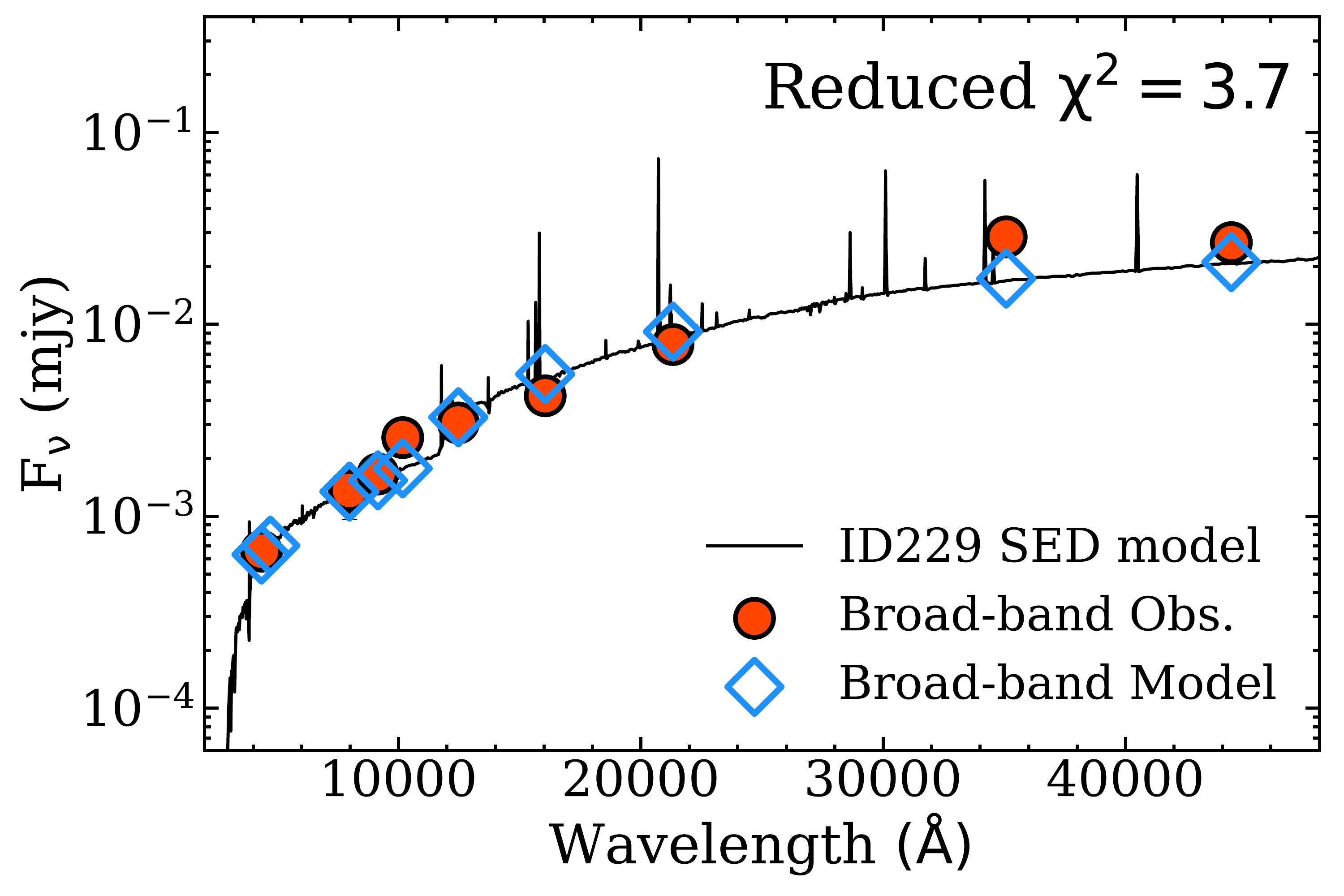}\par
      \includegraphics[width=\linewidth]{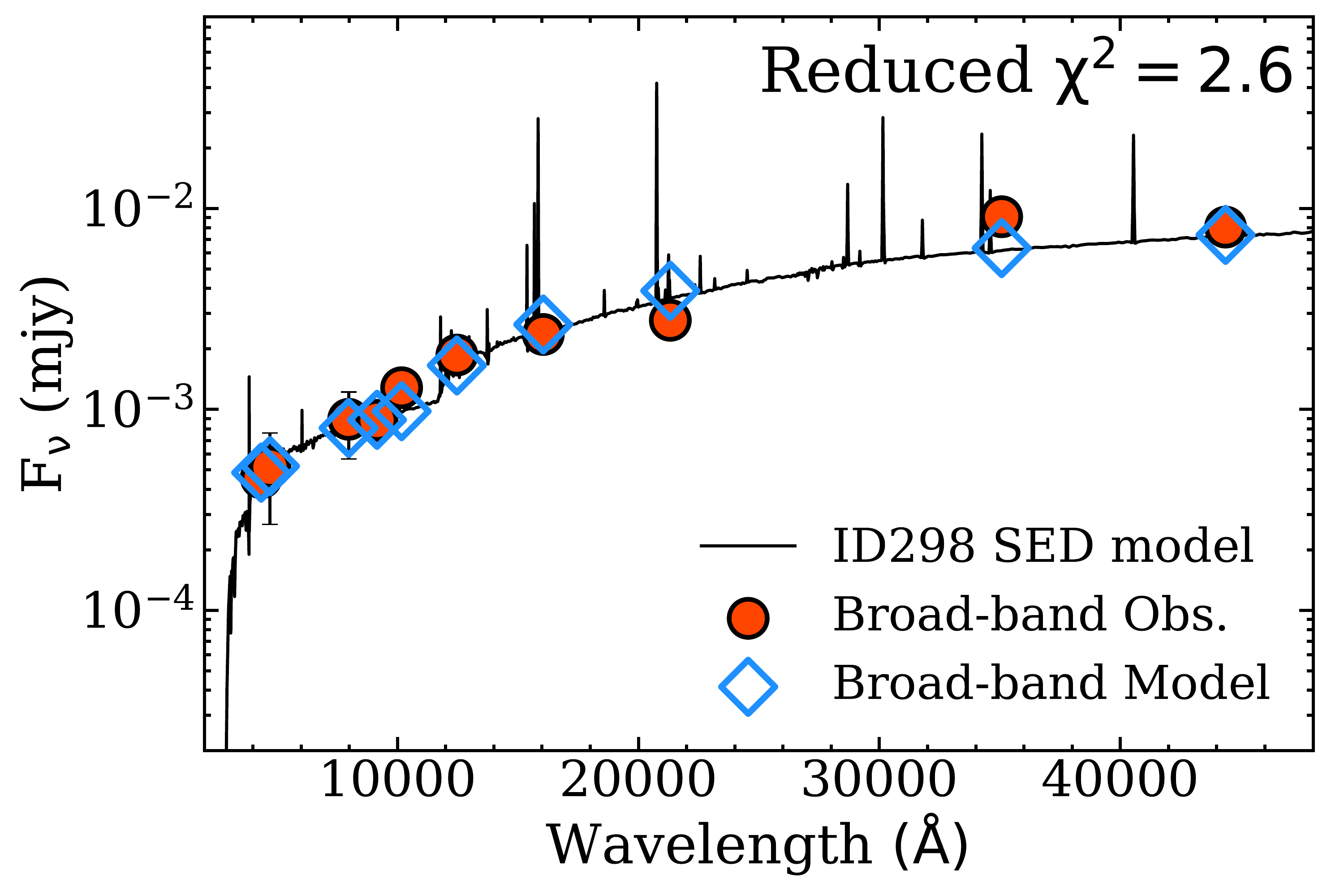}\par
      \includegraphics[width=\linewidth]{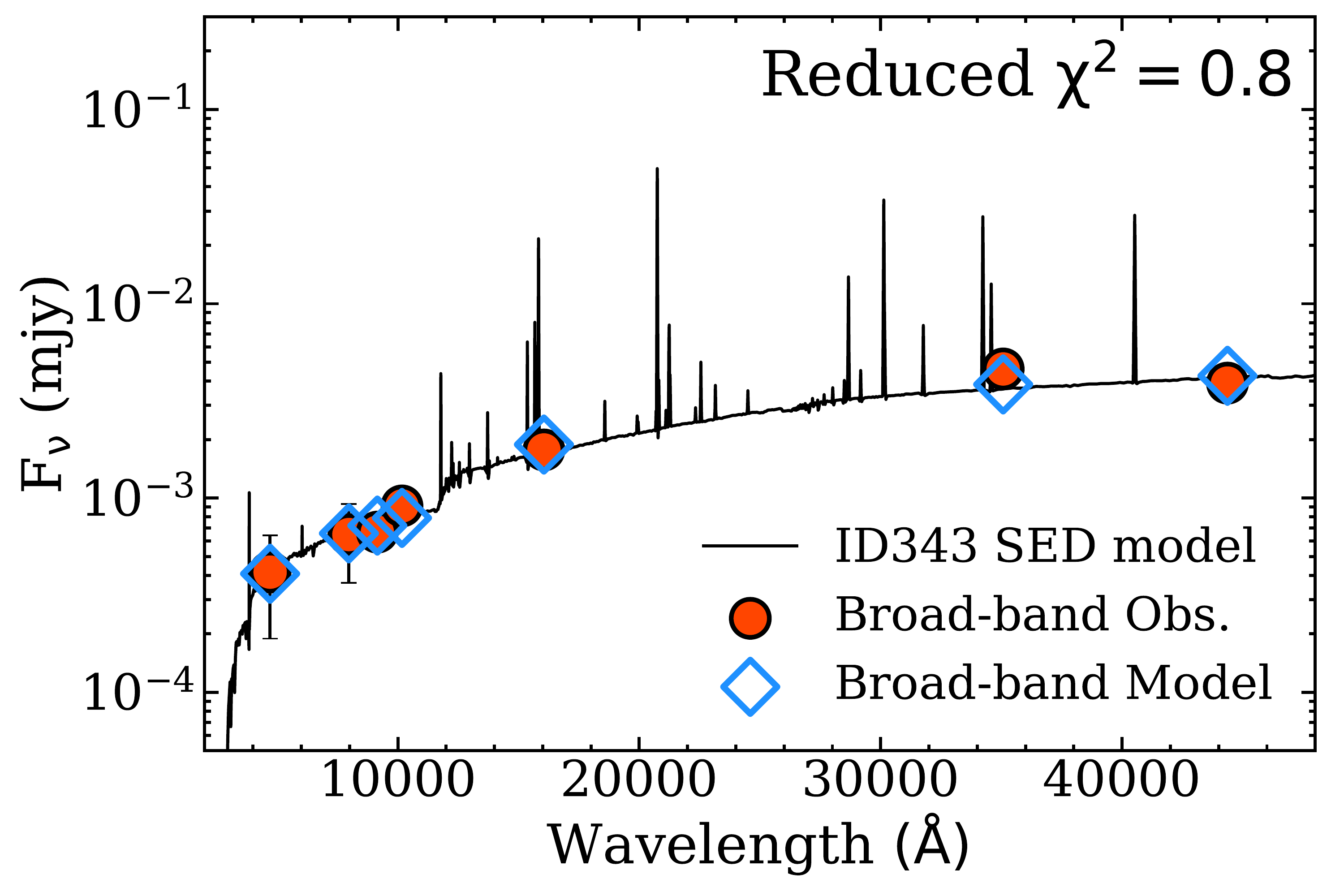}\par
      \includegraphics[width=\linewidth]{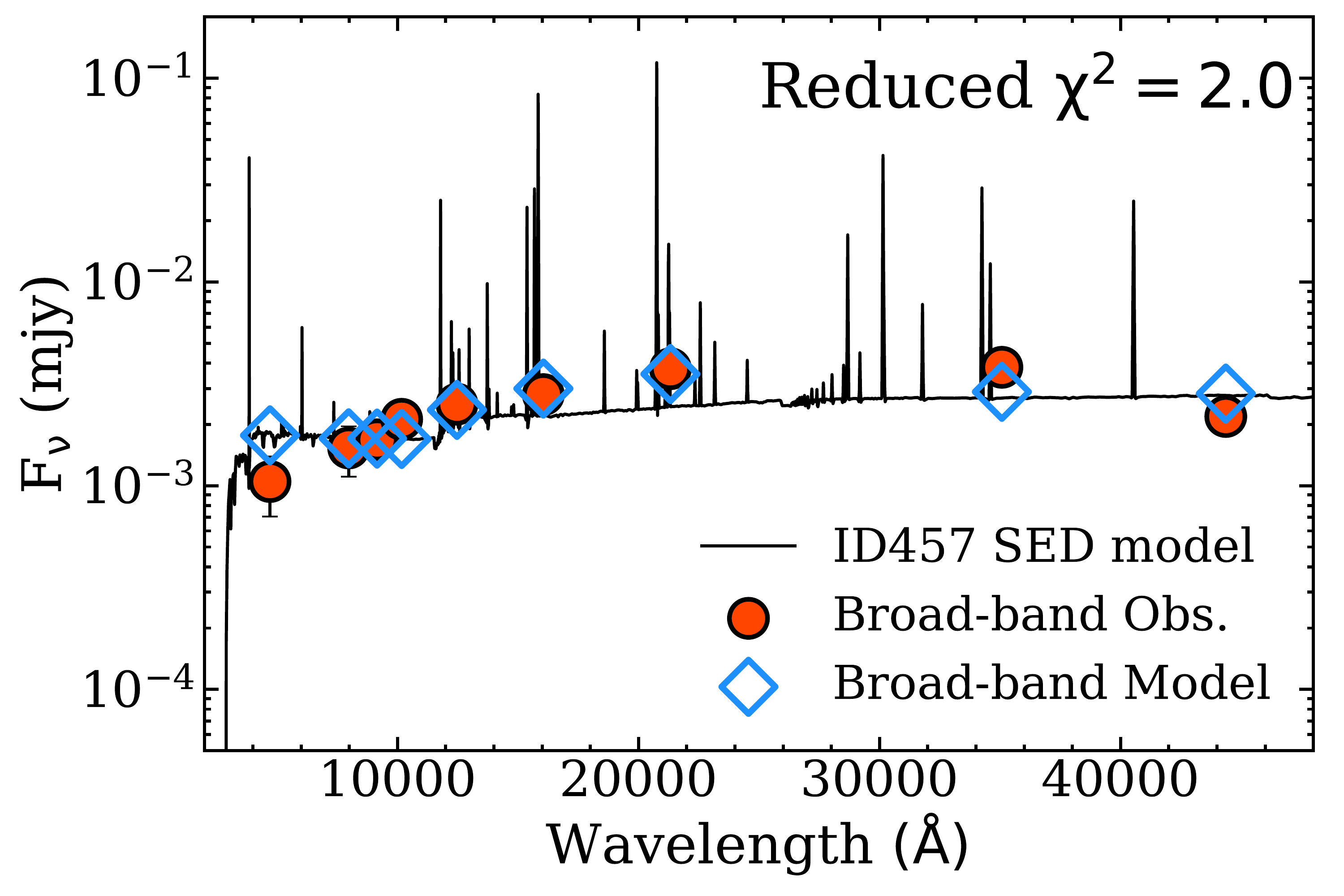}\par
      \includegraphics[width=\linewidth]{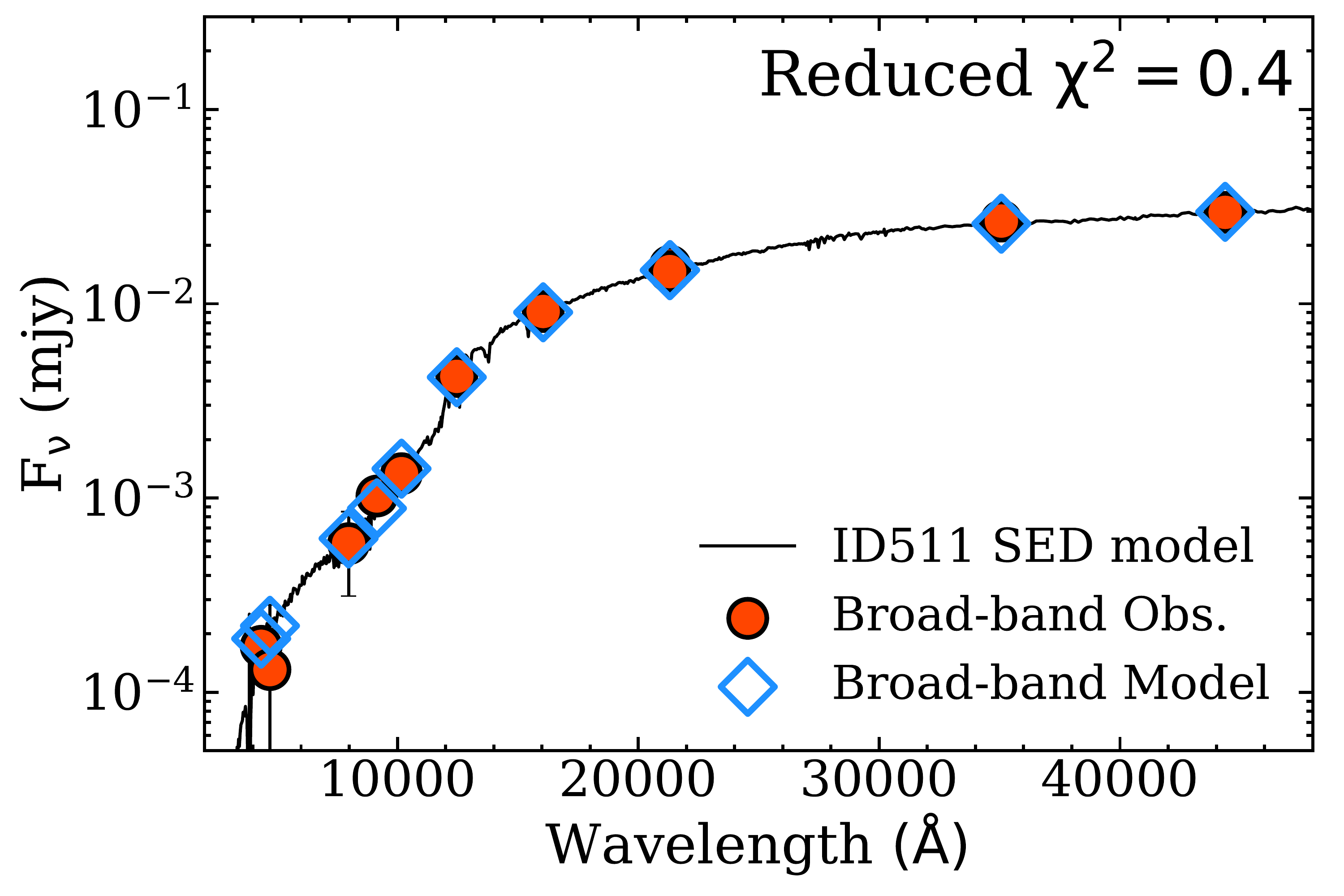}\par
      \includegraphics[width=\linewidth]{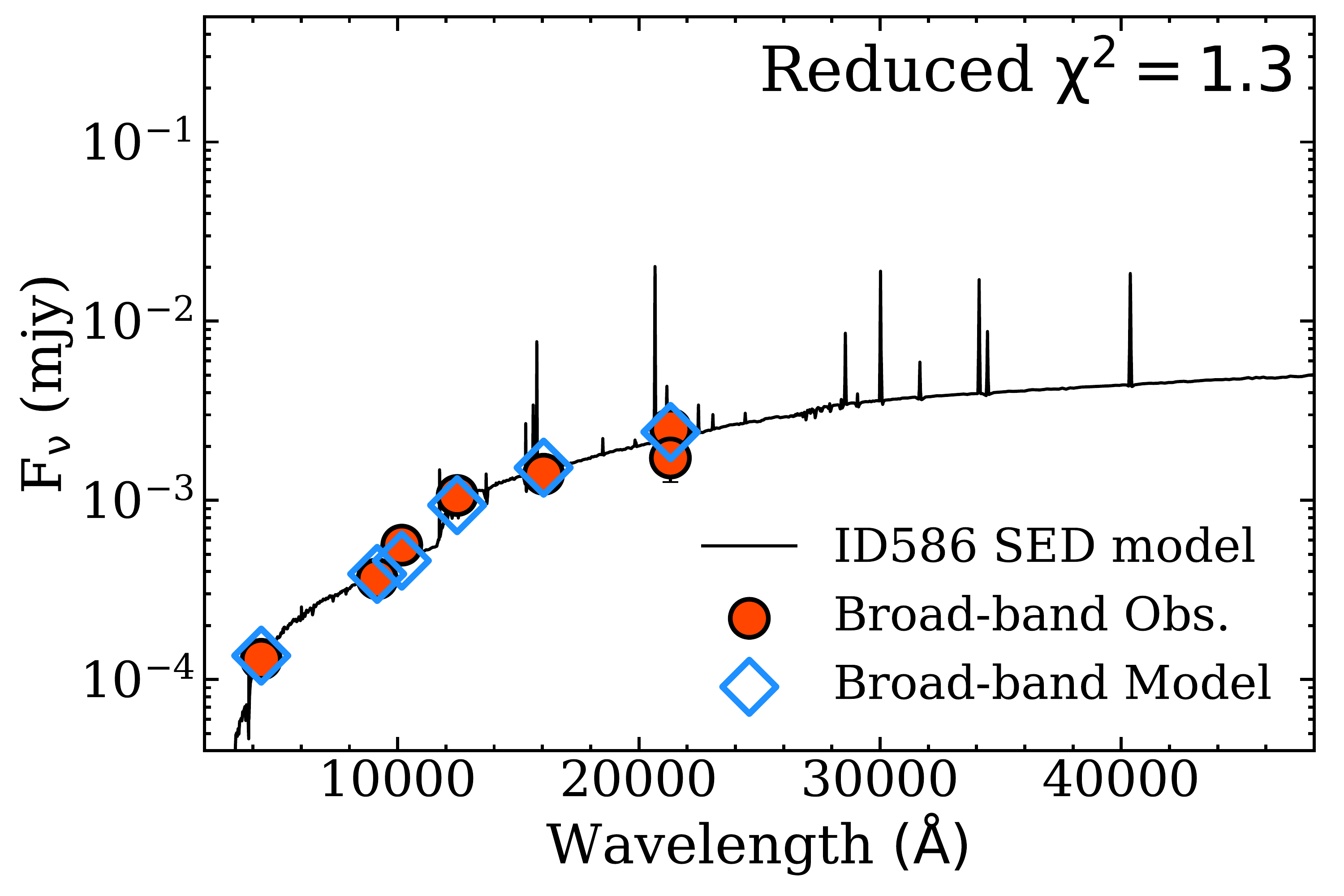}\par
      \includegraphics[width=\linewidth]{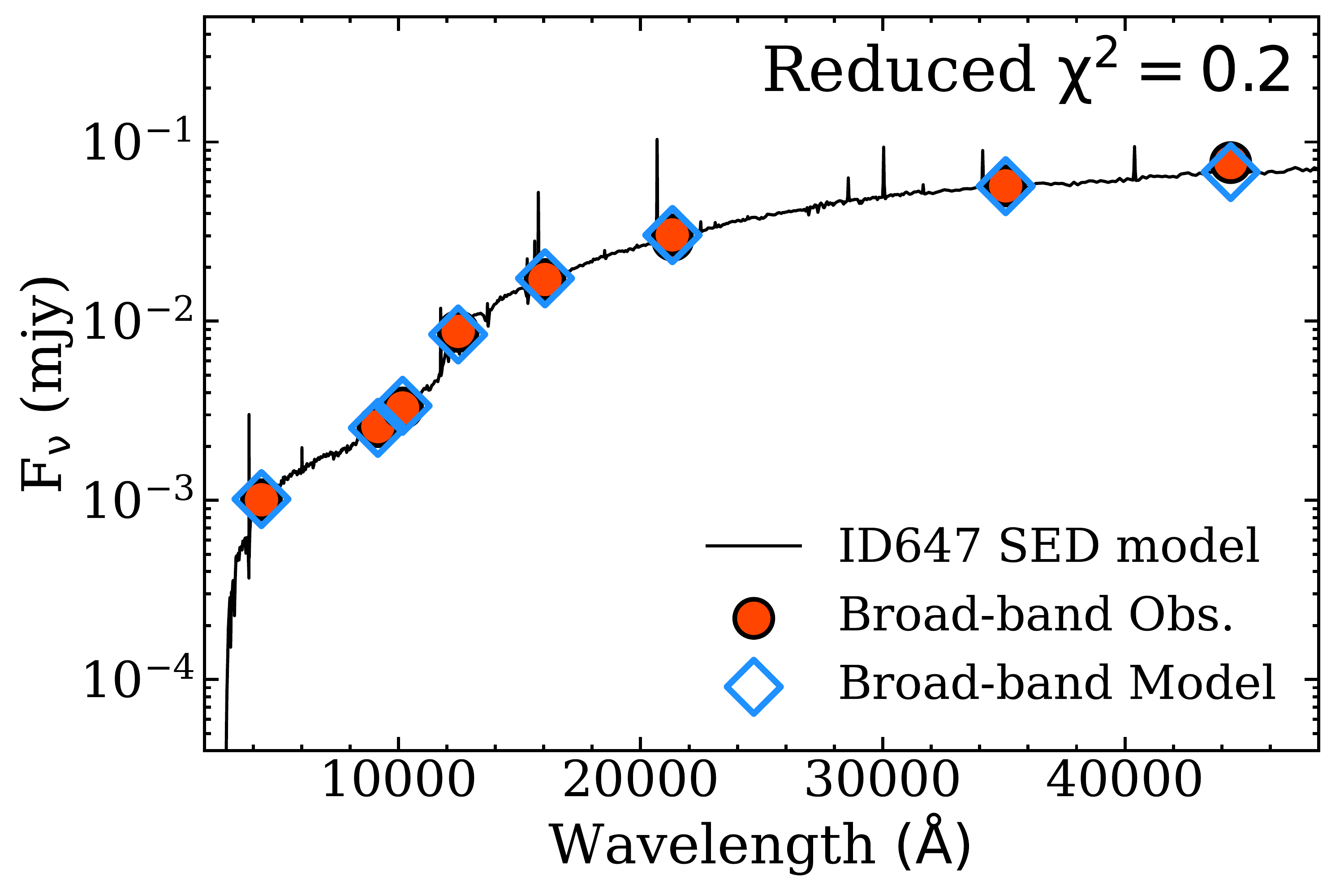}\par
      \includegraphics[width=\linewidth]{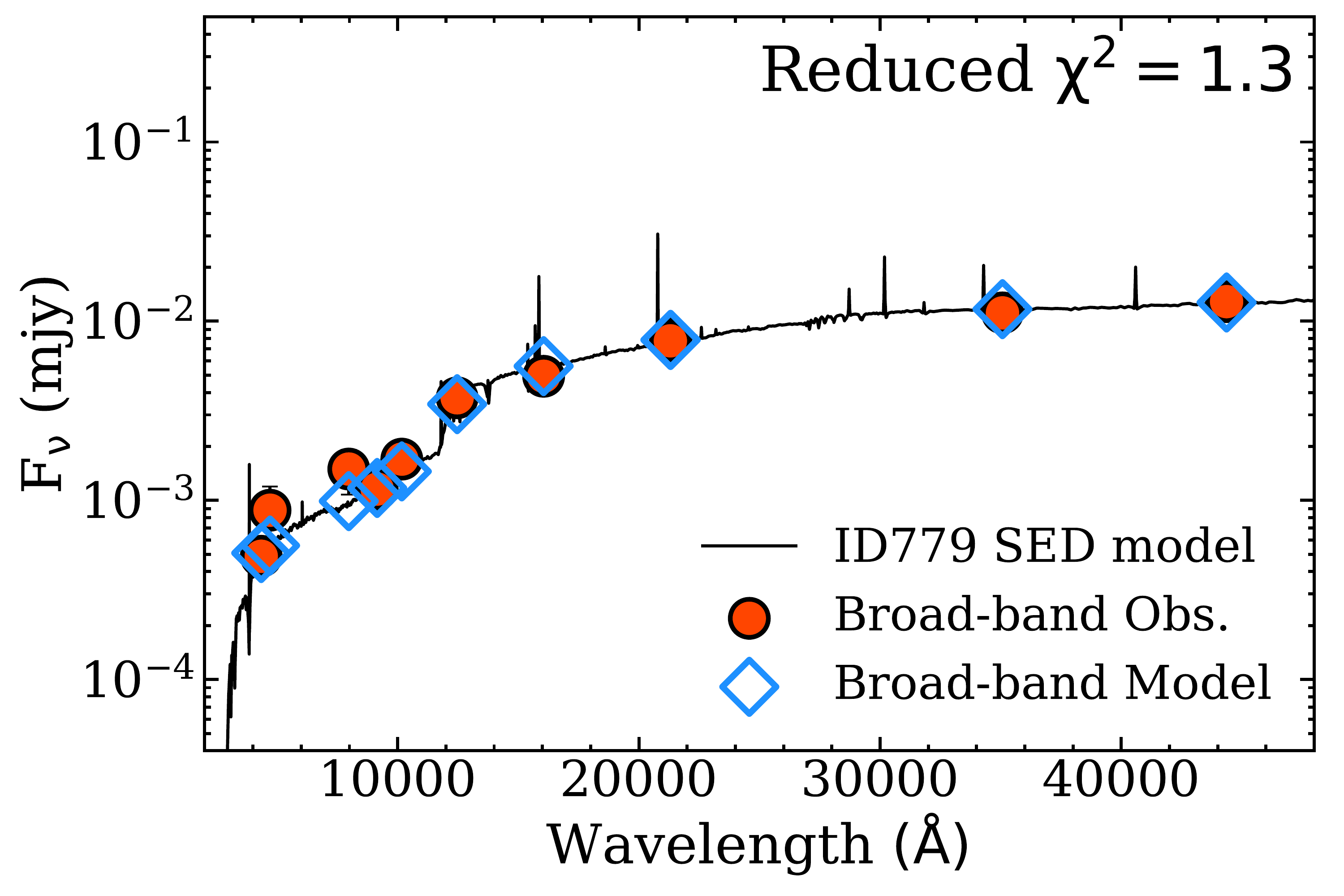}\par
      \includegraphics[width=\linewidth]{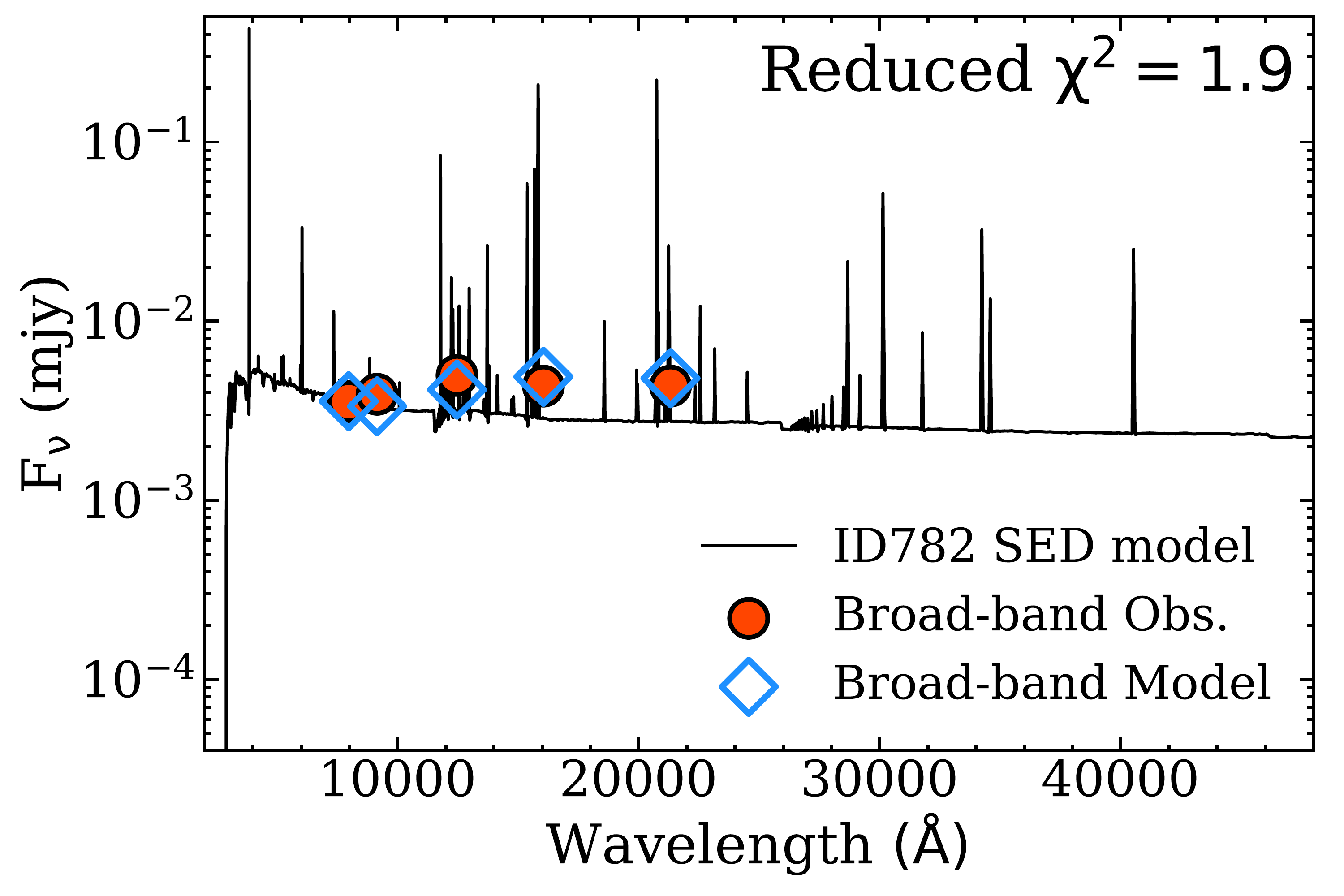}\par
      \includegraphics[width=\linewidth]{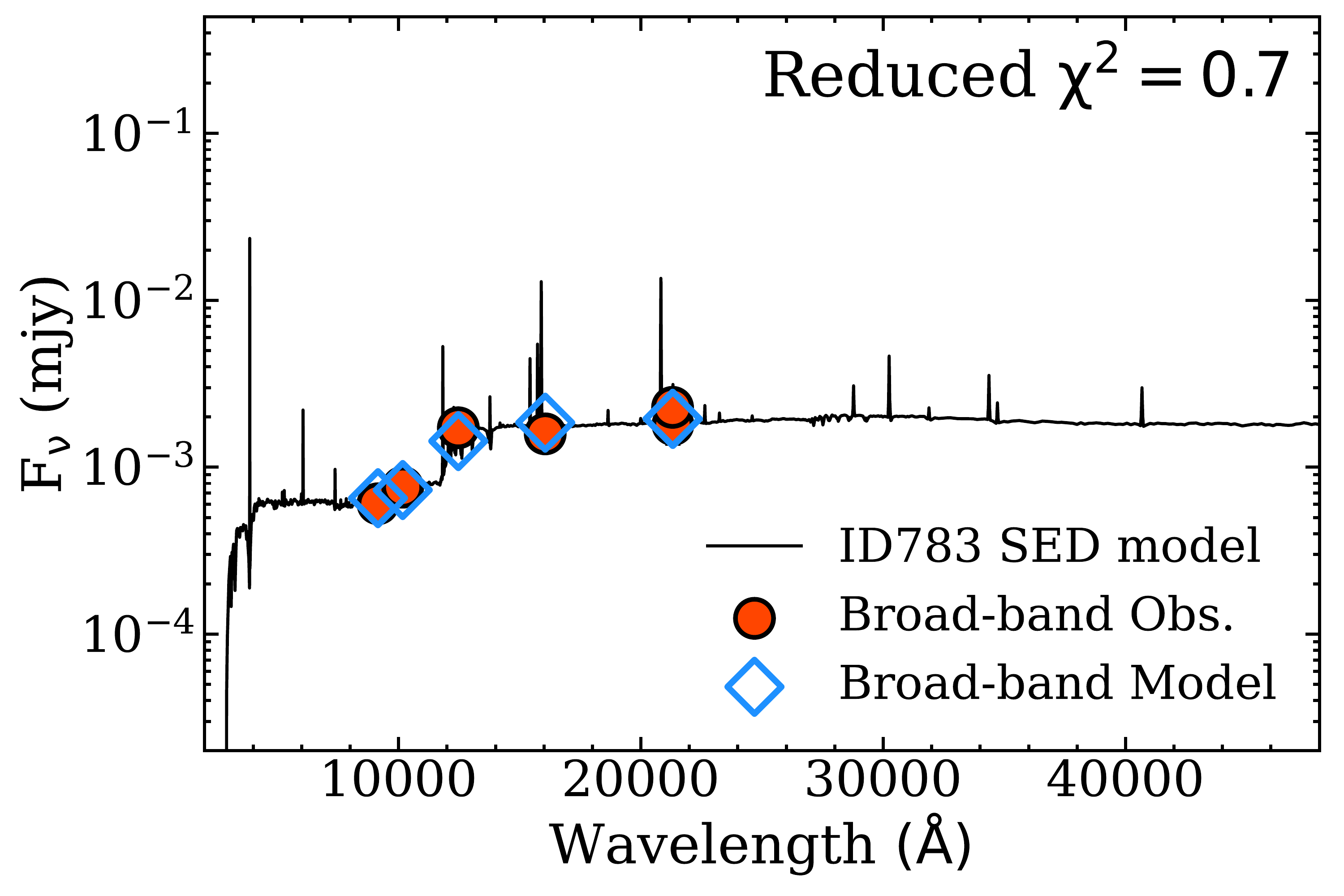}\par
      \includegraphics[width=\linewidth]{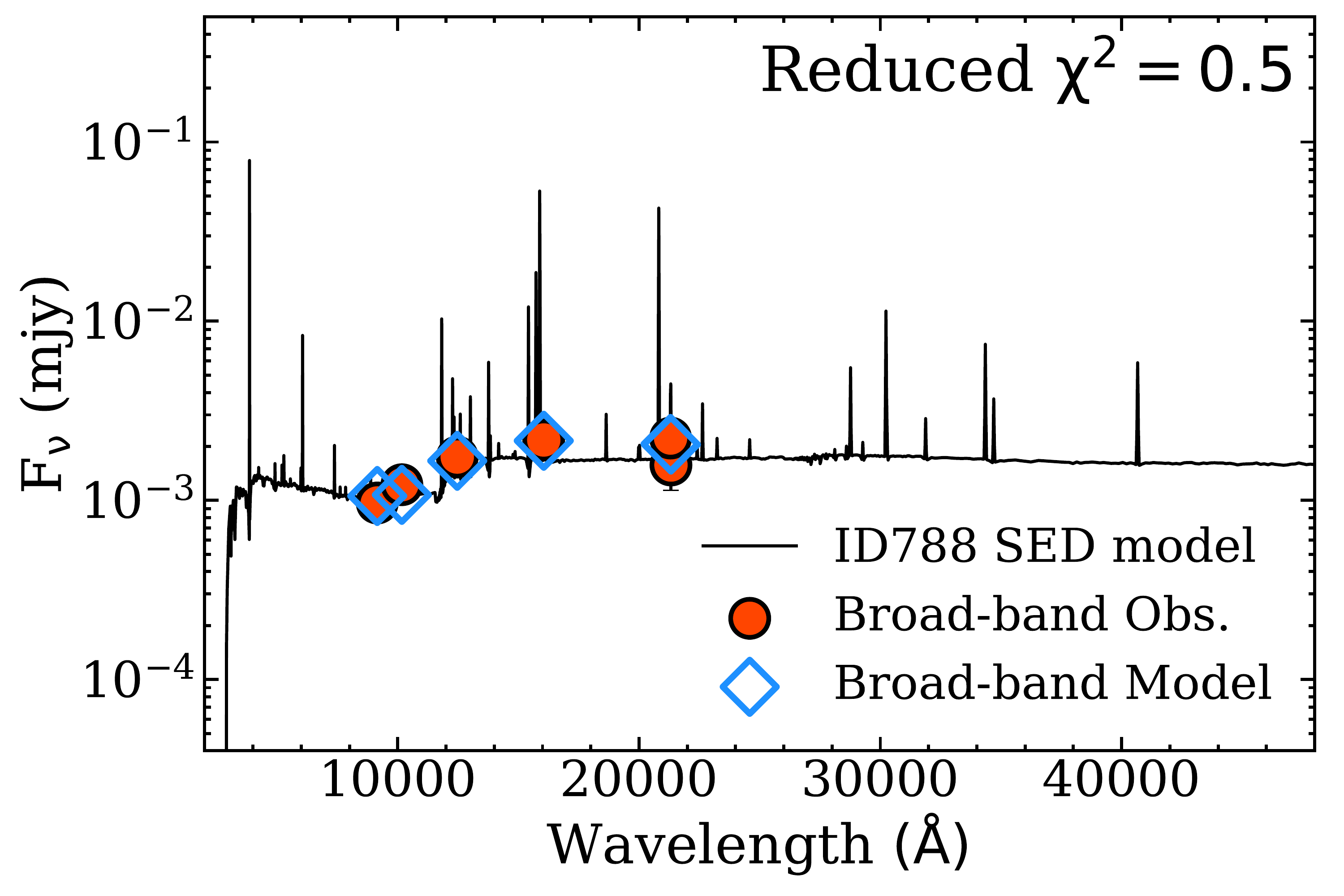}\par
      \includegraphics[width=\linewidth]{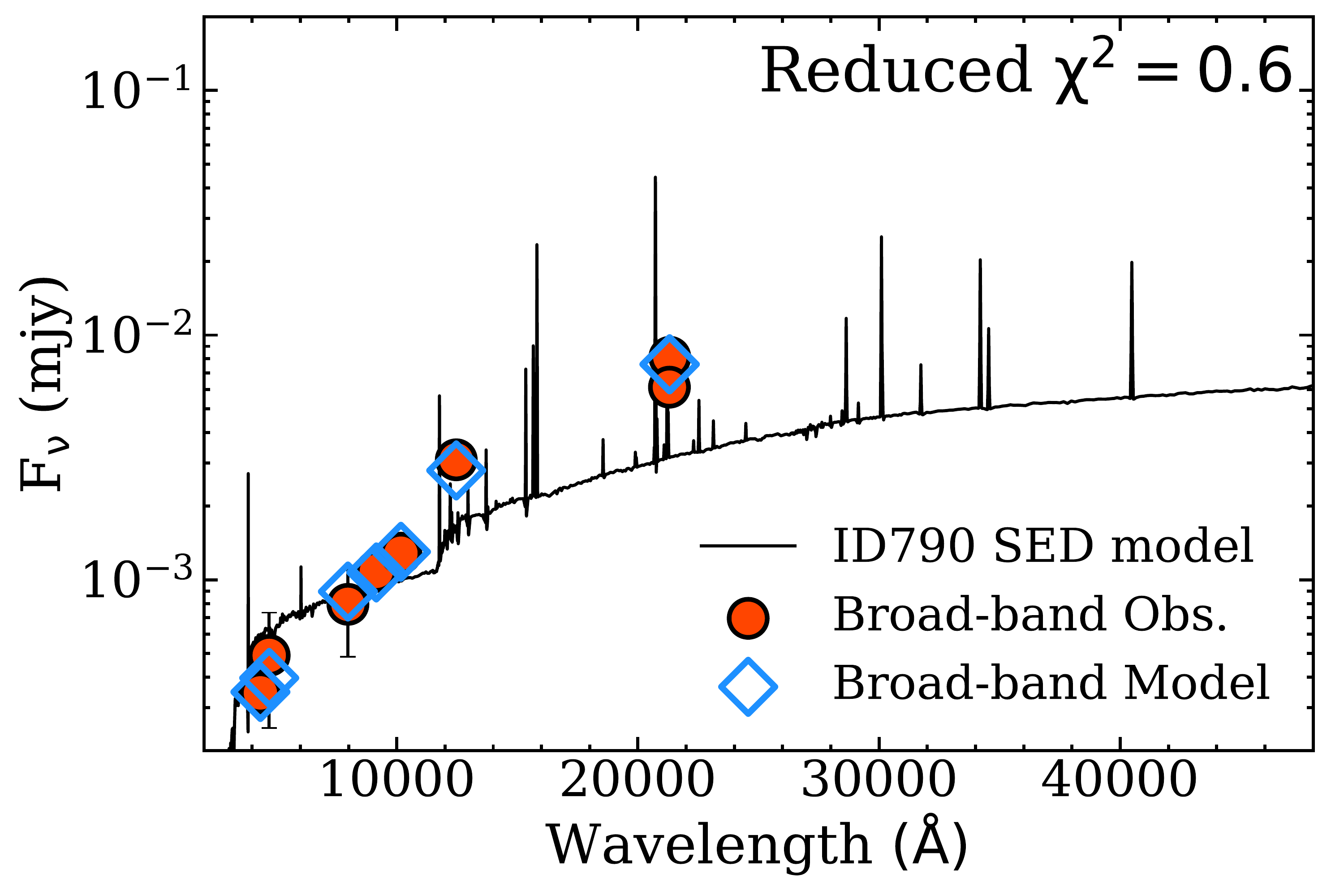}\par
      \includegraphics[width=\linewidth]{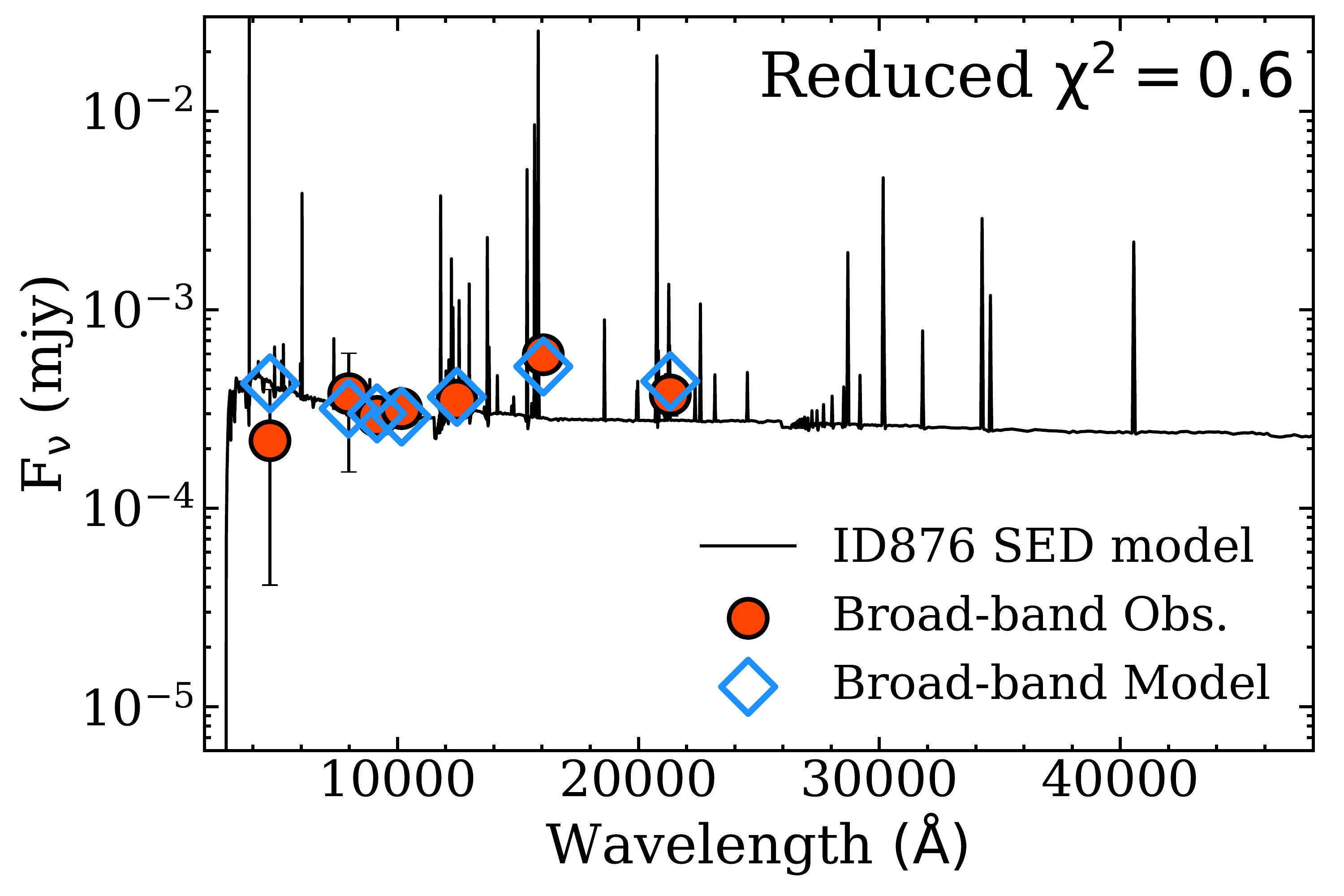}\par
      \includegraphics[width=\linewidth]{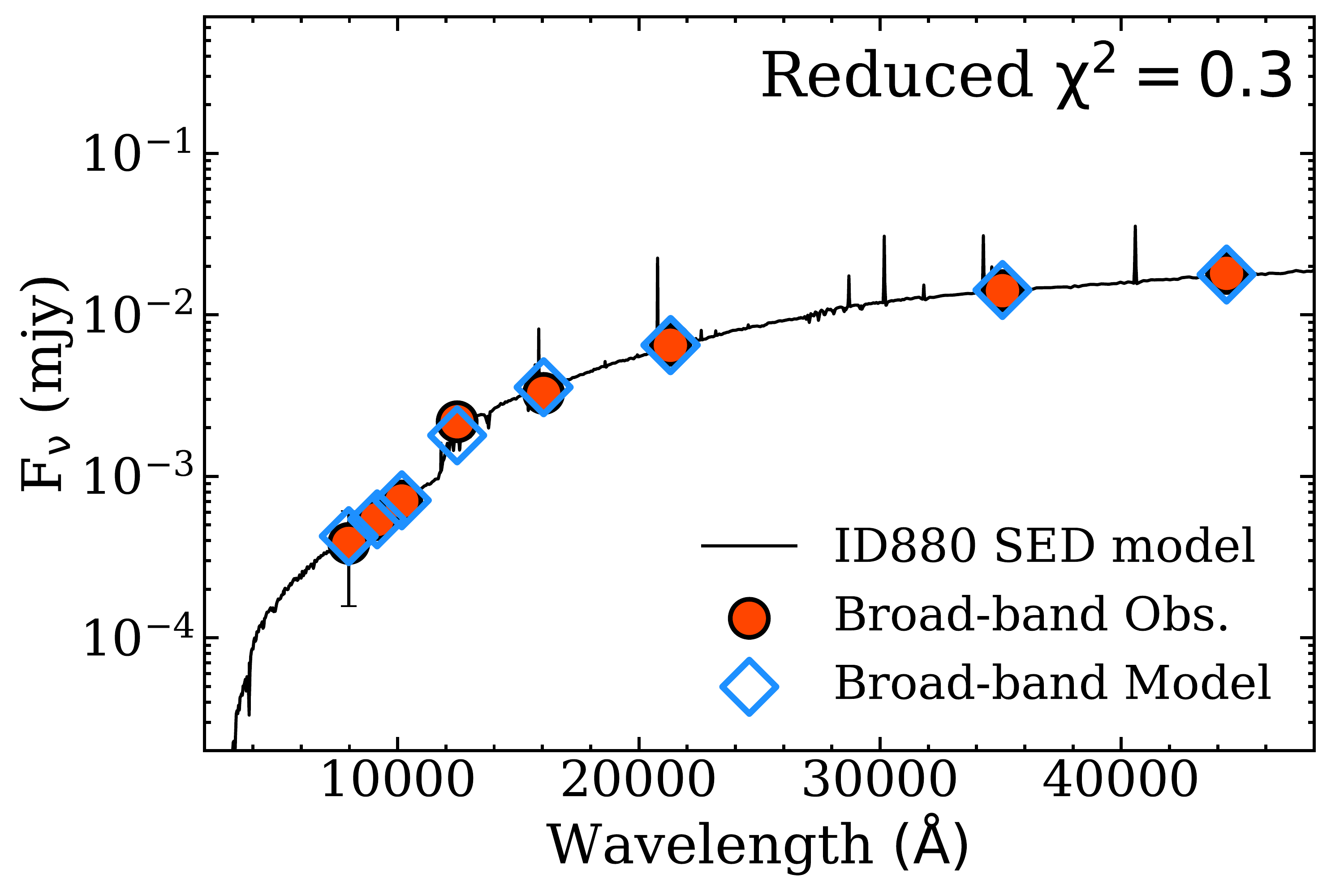}\par
      \includegraphics[width=\linewidth]{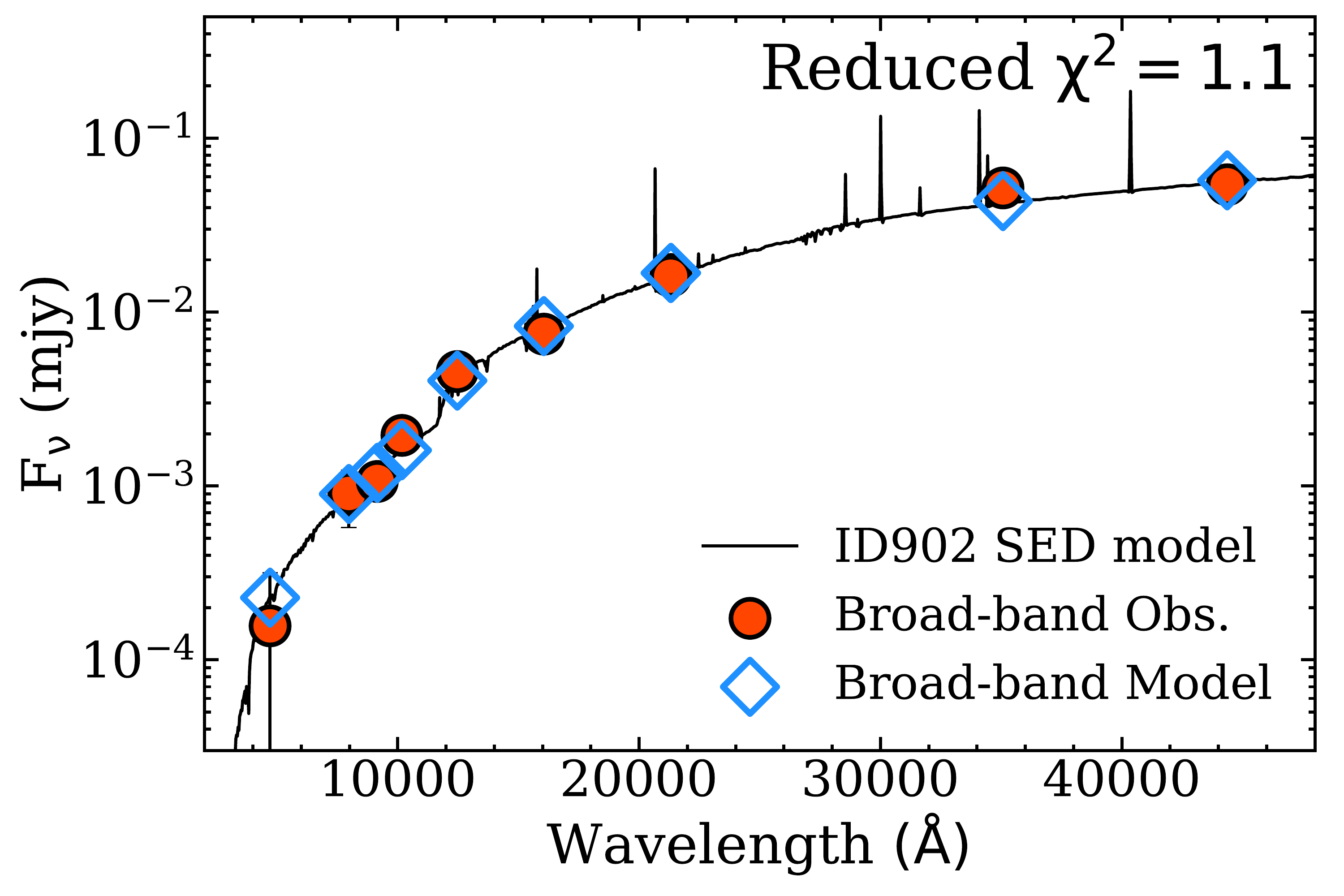}\par
      \includegraphics[width=\linewidth]{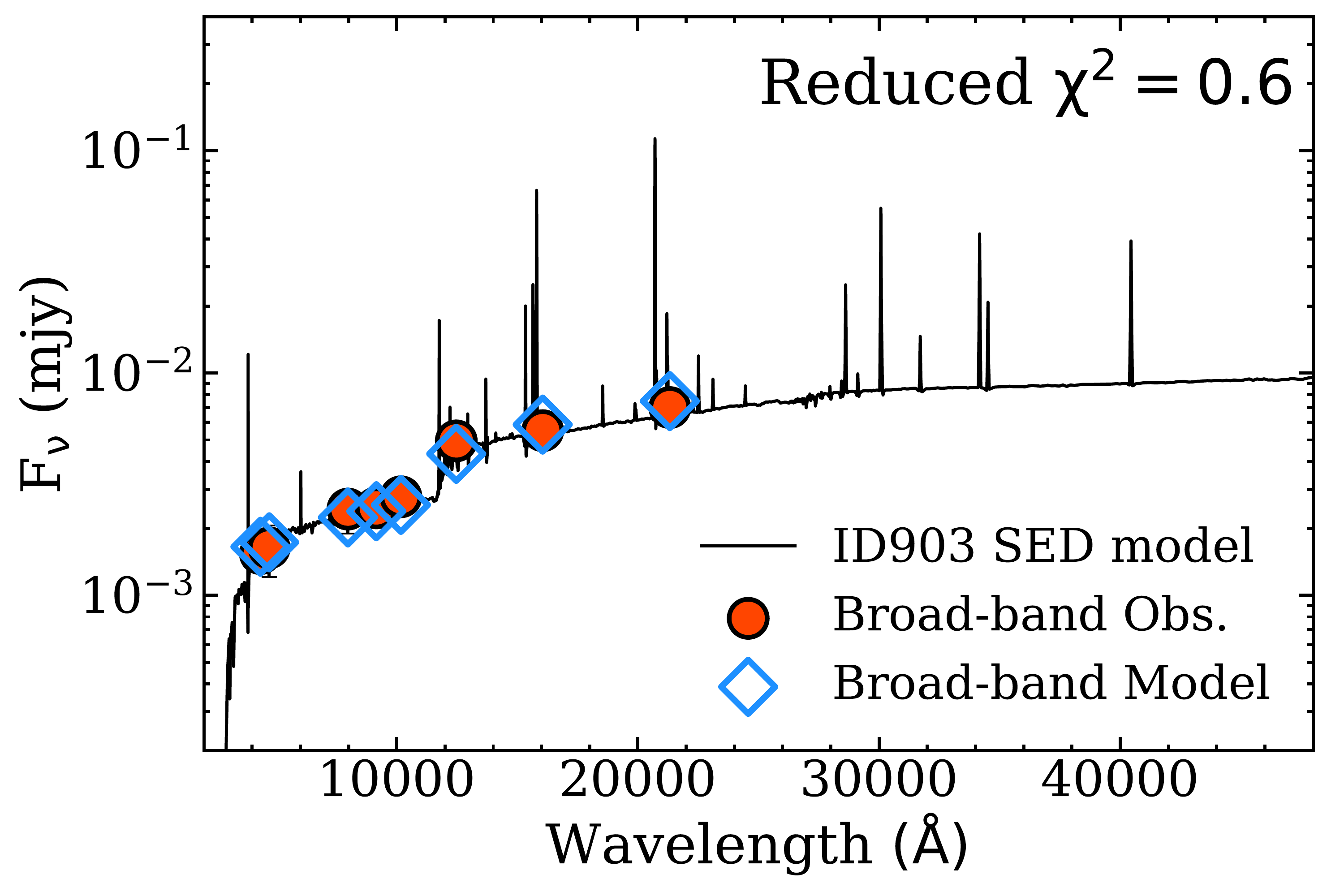}\par
      \includegraphics[width=\linewidth]{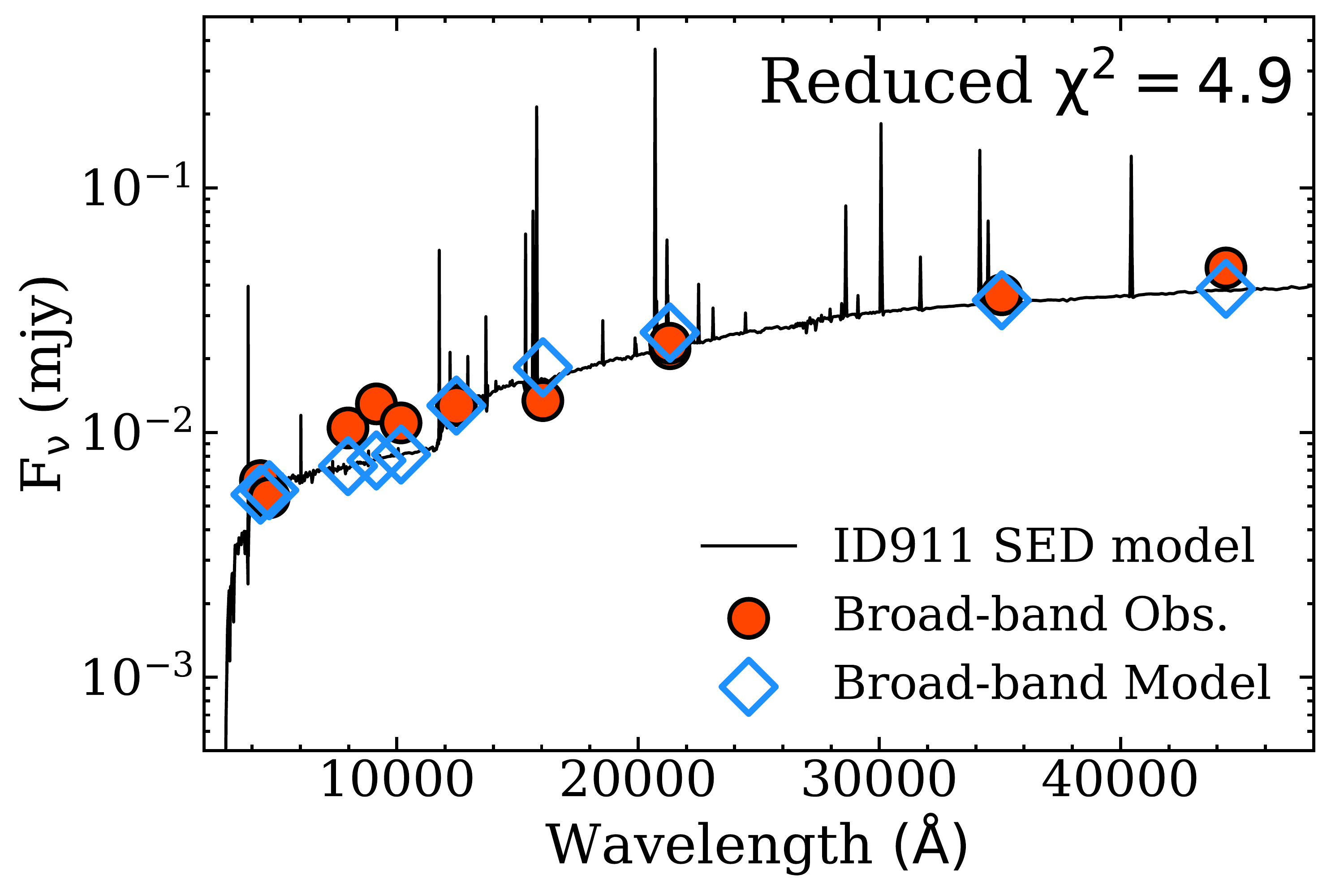}\par
      \includegraphics[width=\linewidth]{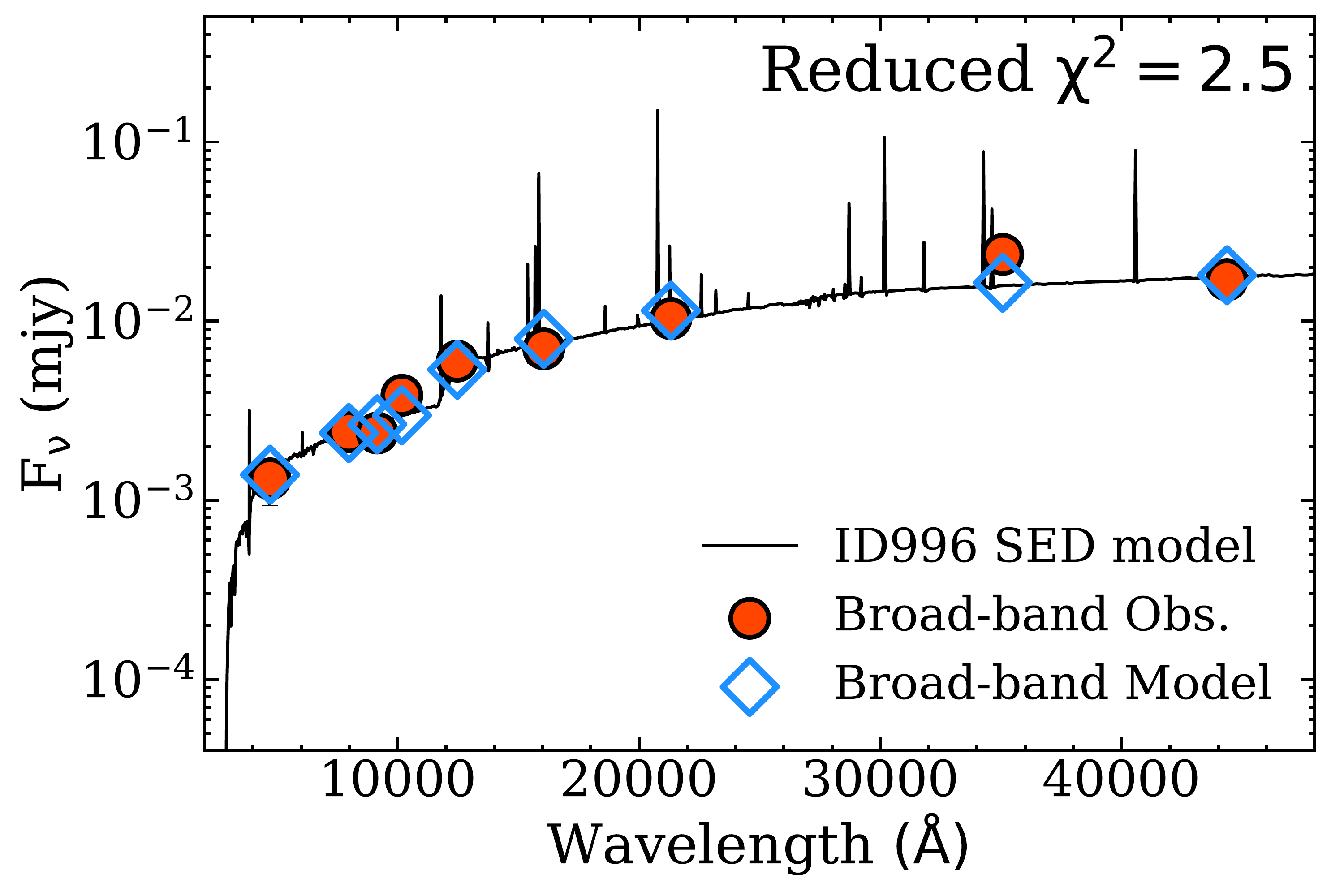}\par
      \includegraphics[width=\linewidth]{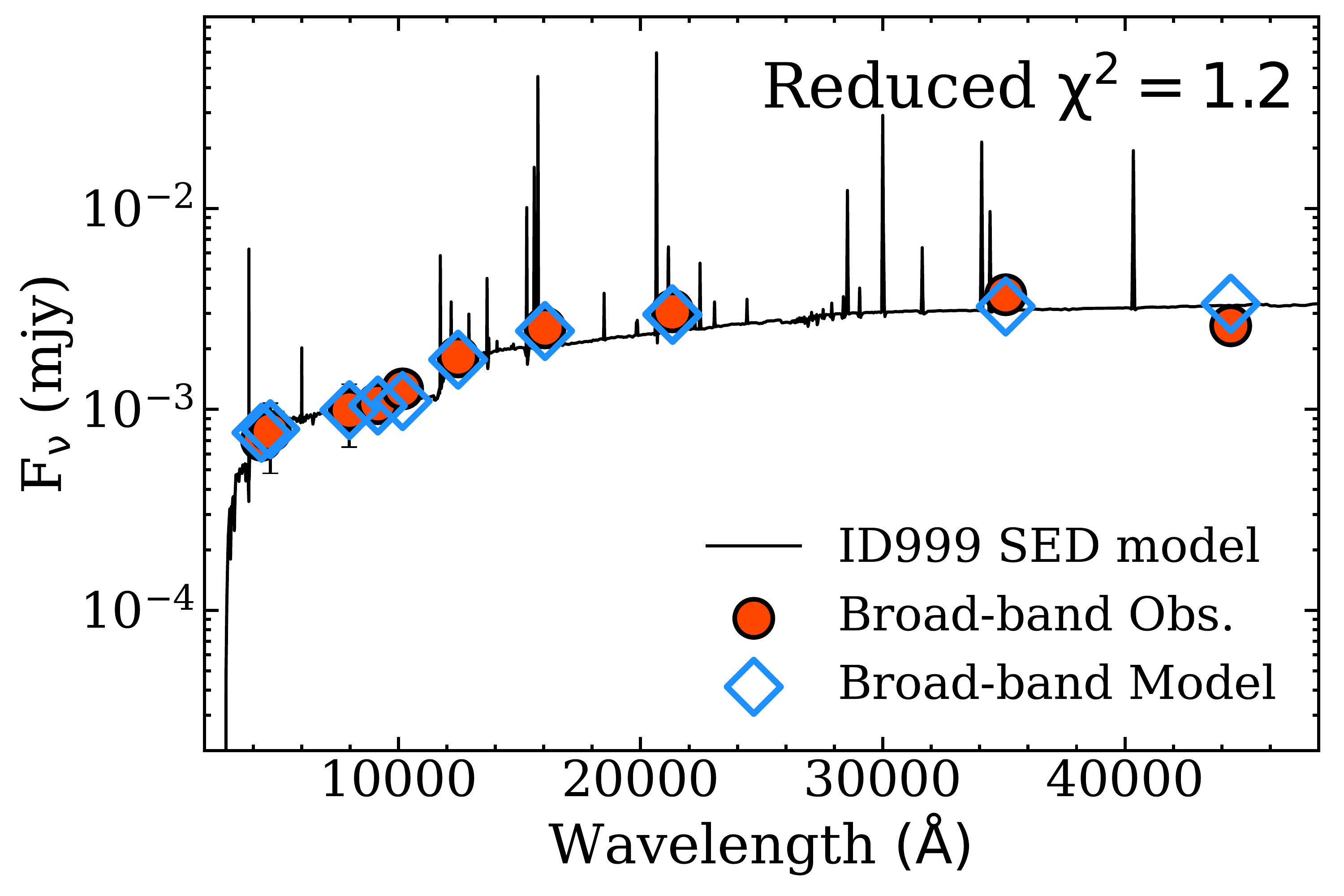}\par
      \includegraphics[width=\linewidth]{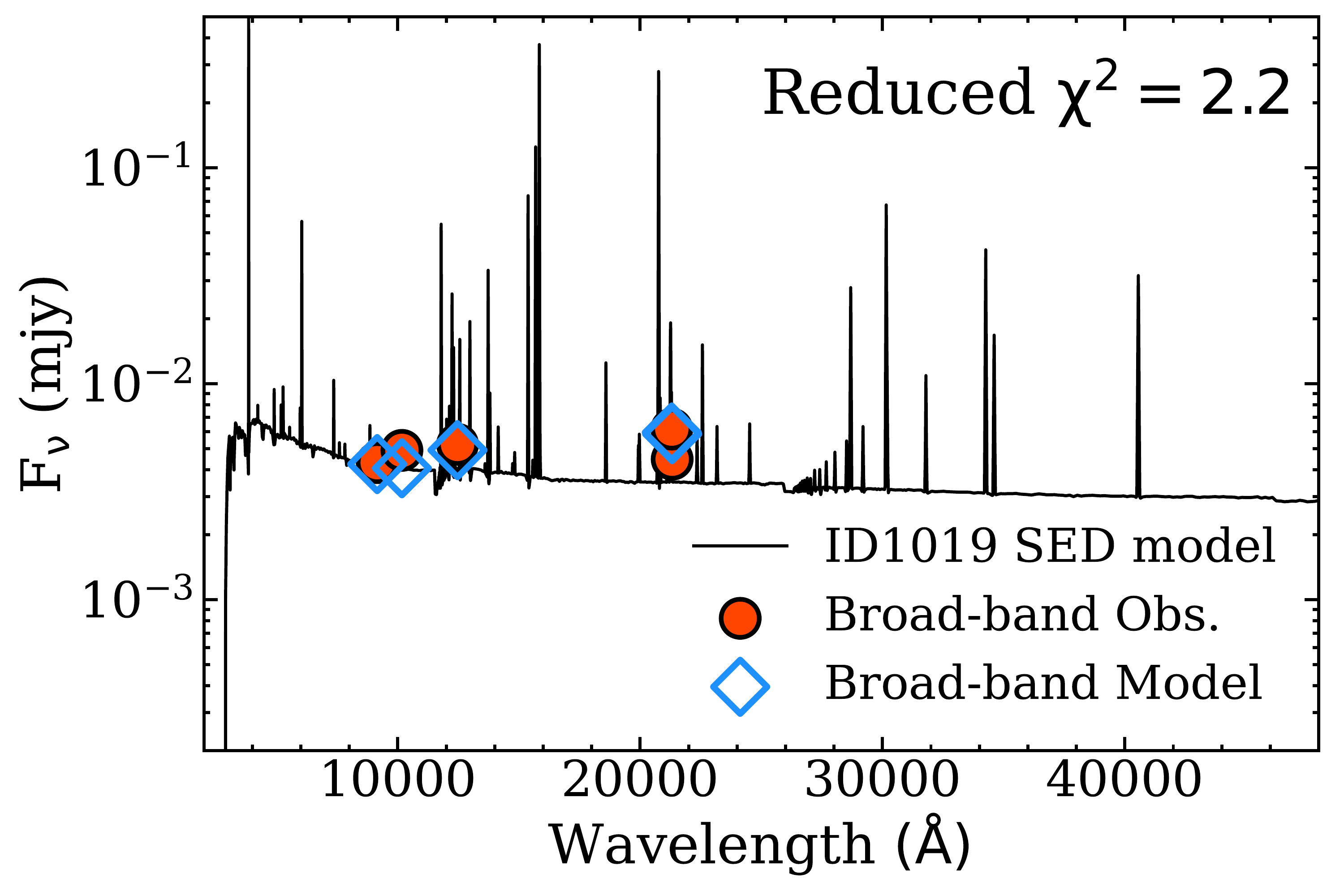}\par
      \includegraphics[width=\linewidth]{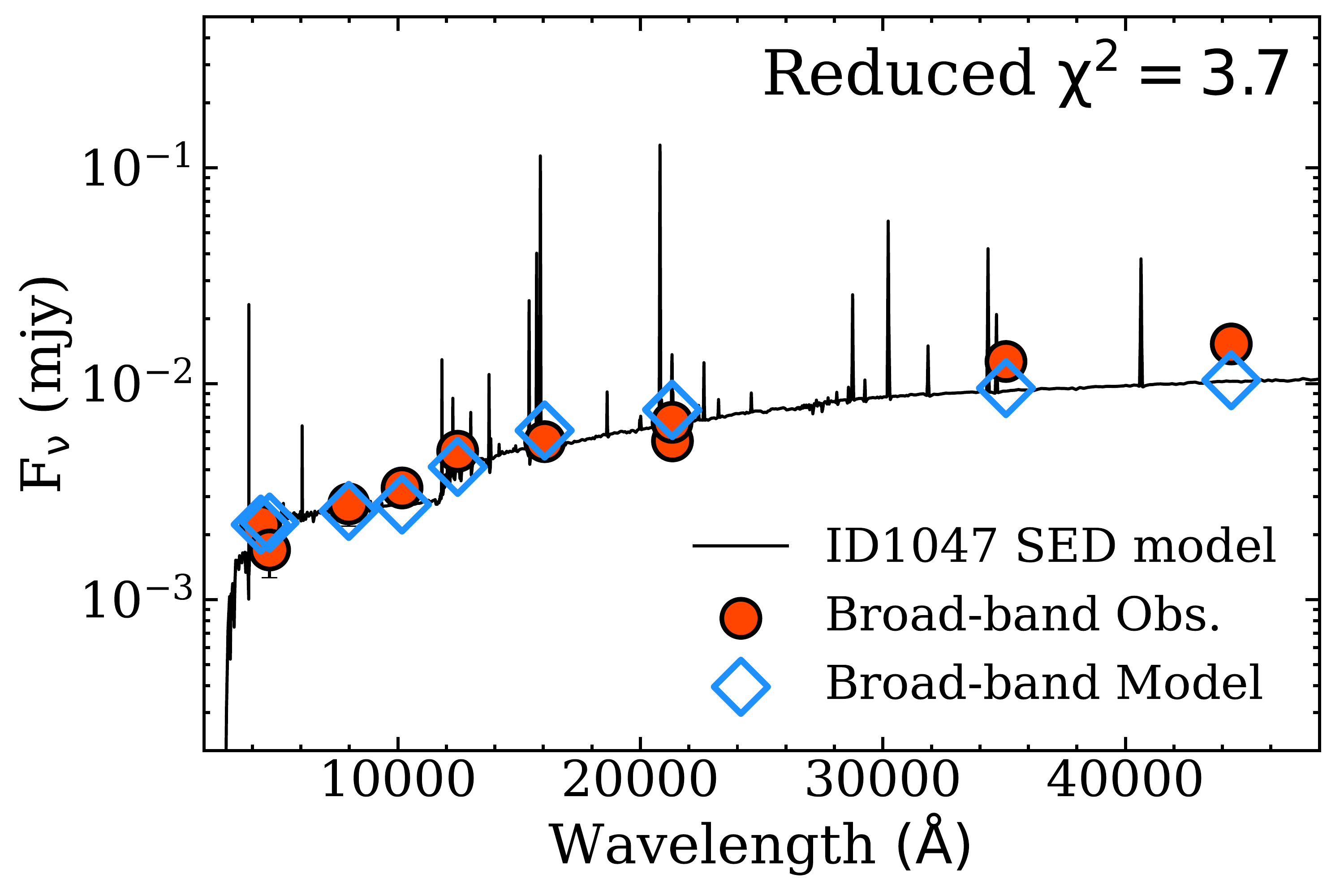}\par
      \includegraphics[width=\linewidth]{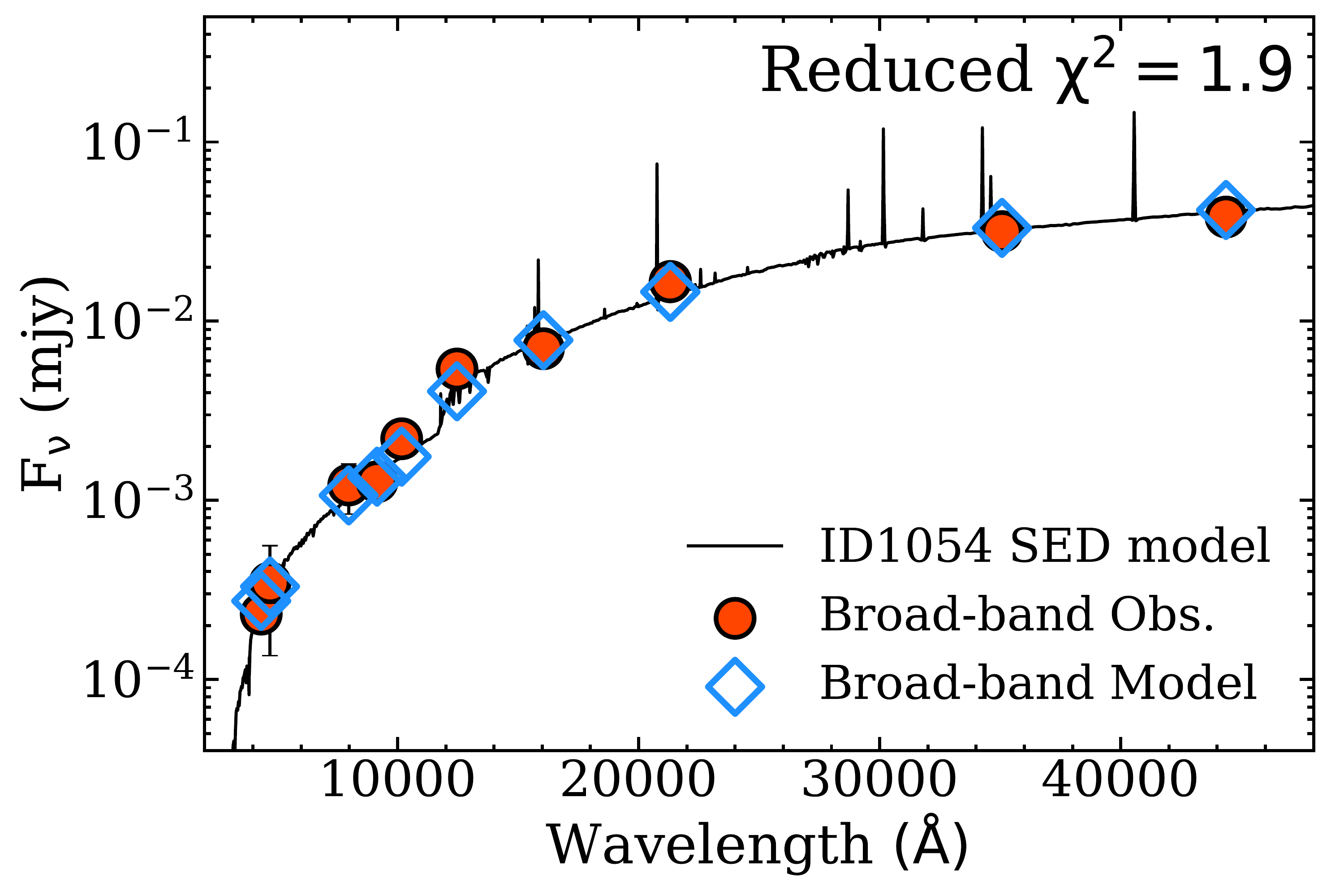}\par
      \includegraphics[width=\linewidth]{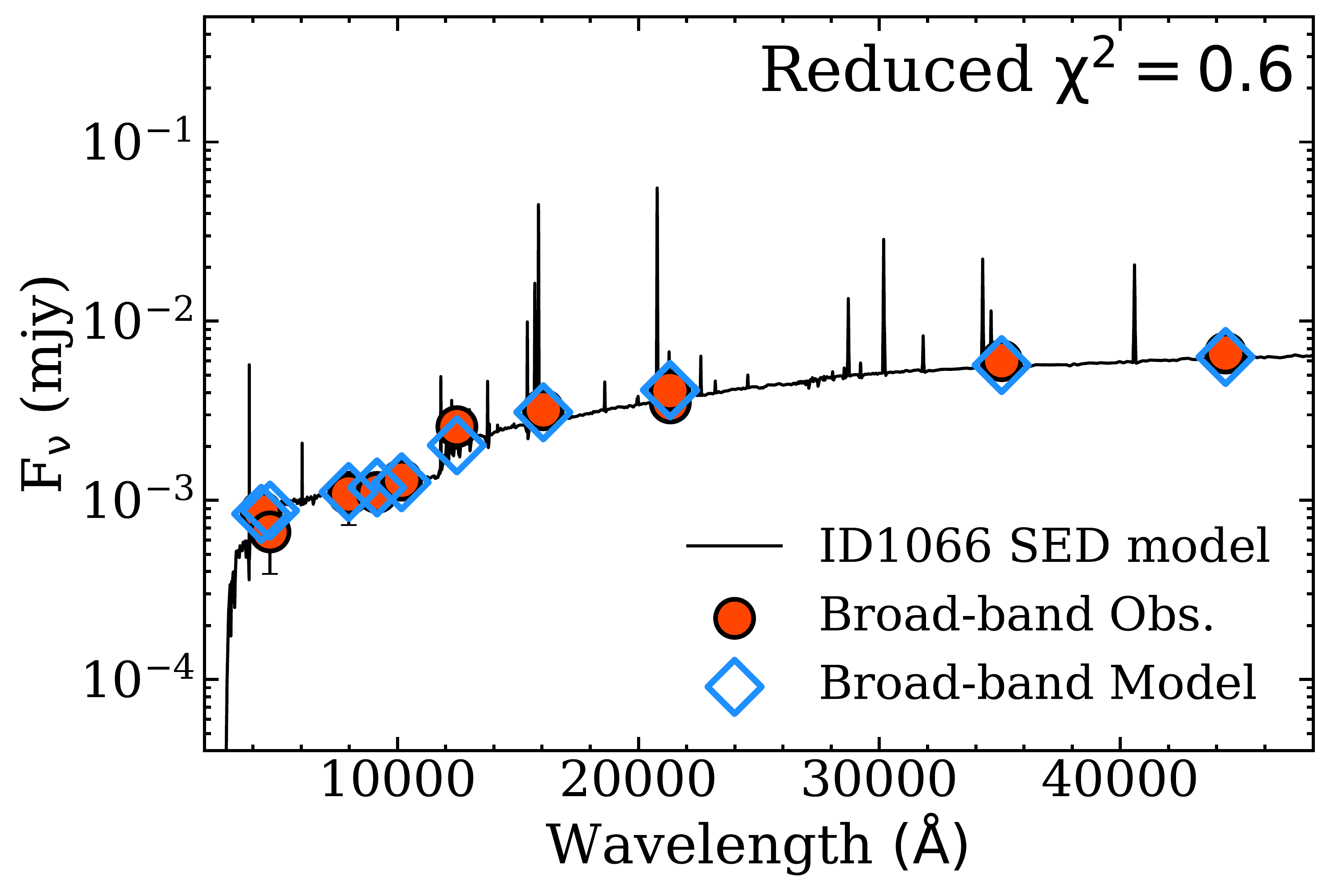}\par
      \includegraphics[width=\linewidth]{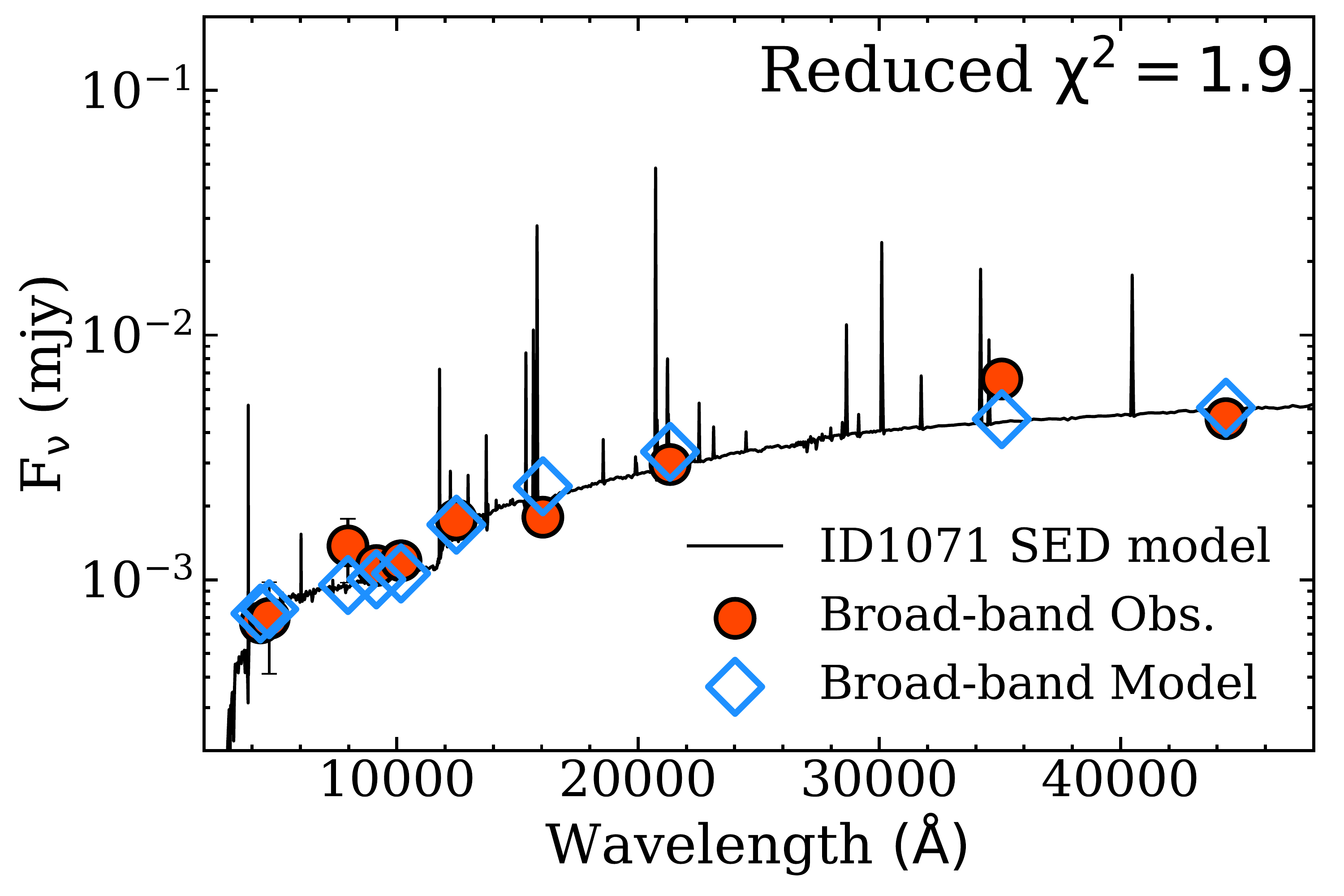}\par
      \includegraphics[width=\linewidth]{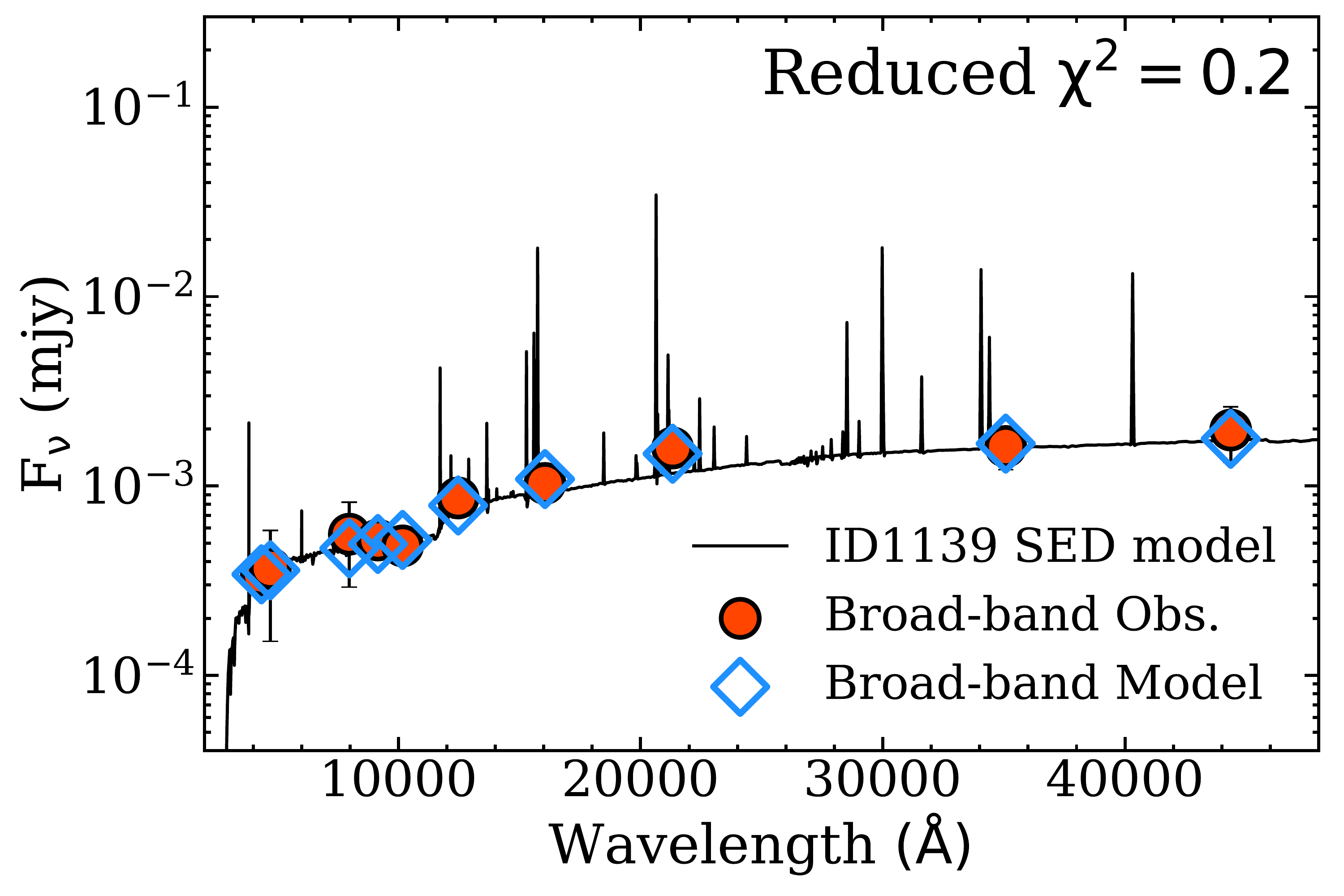}\par
      \includegraphics[width=\linewidth]{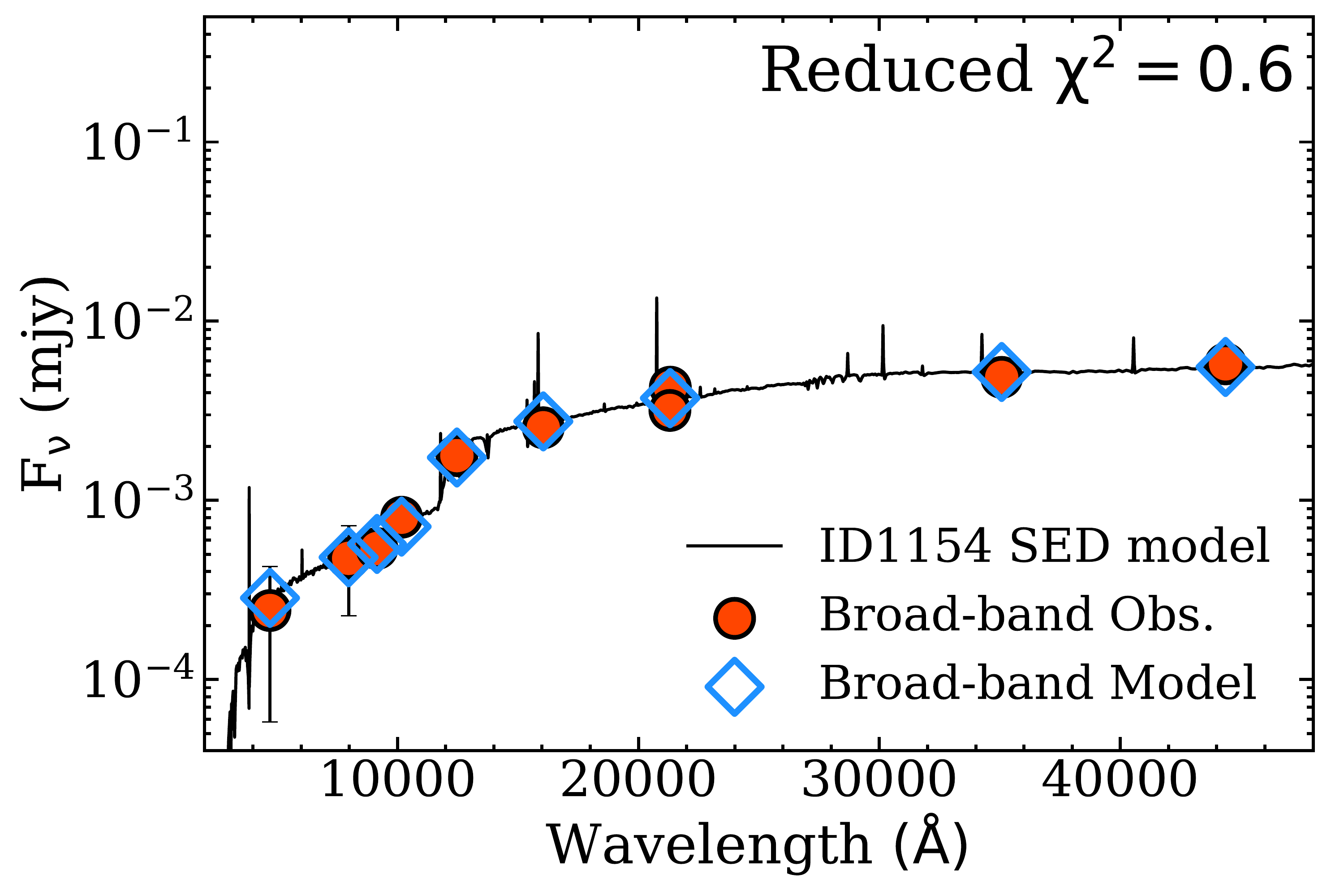}\par
      \includegraphics[width=\linewidth]{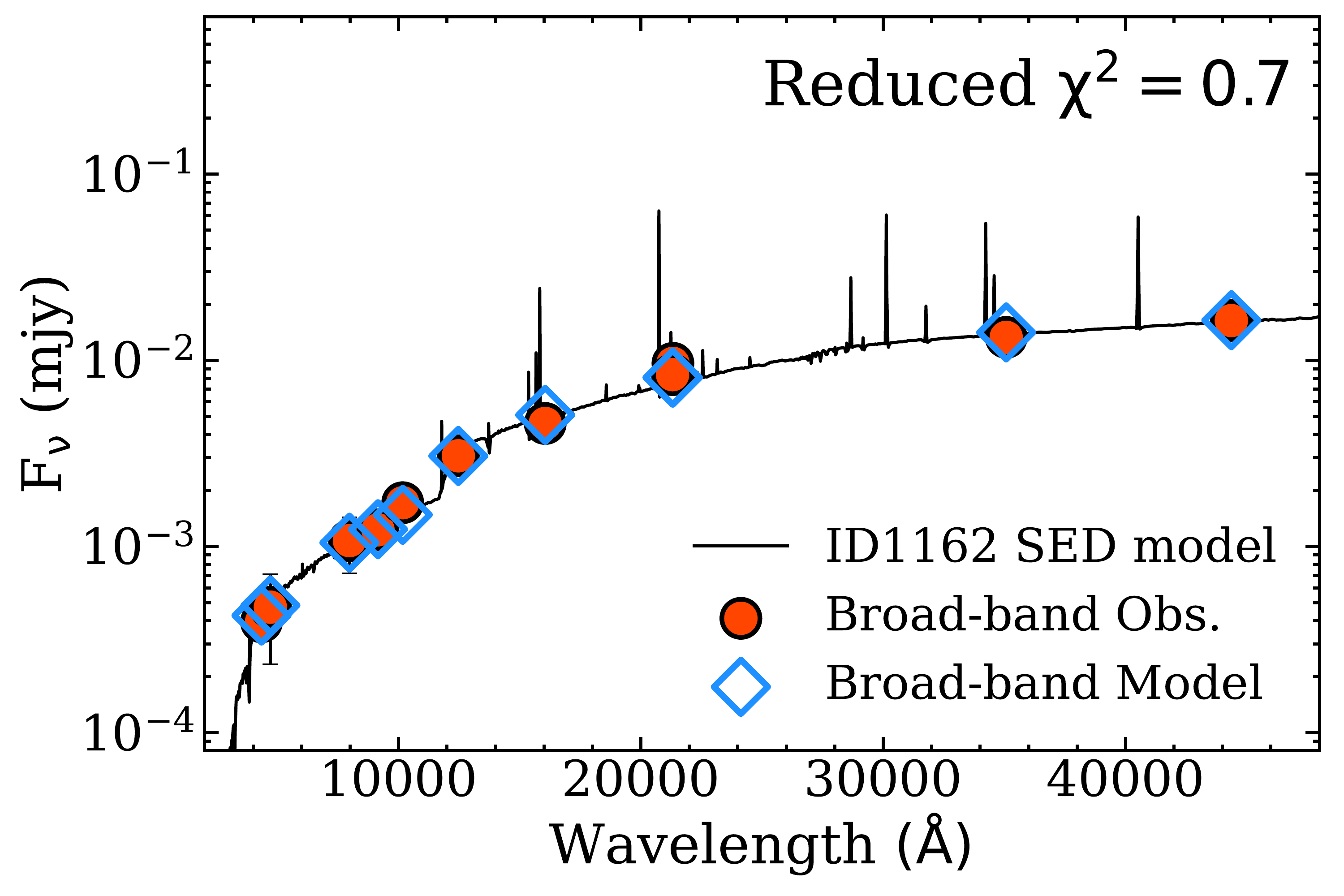}\par
      \includegraphics[width=\linewidth]{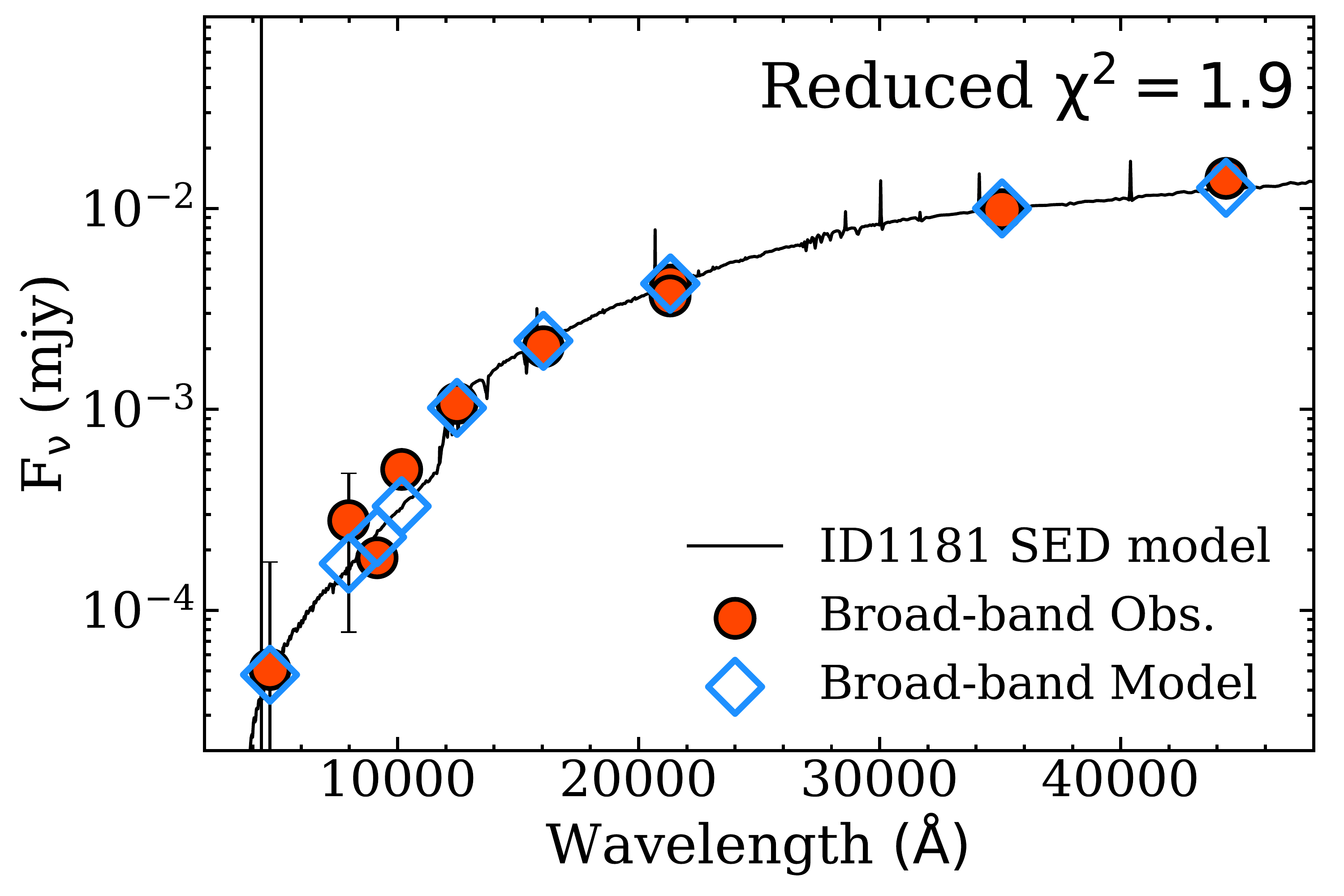}\par
      \end{multicols}
      \caption{Cigale SED best model and broad-band observations for every member of the protocluster sample.}
         \label{F:SED}
\end{figure*}

\begin{figure*}
 \centering
 \begin{multicols}{4}
      \includegraphics[width=\linewidth]{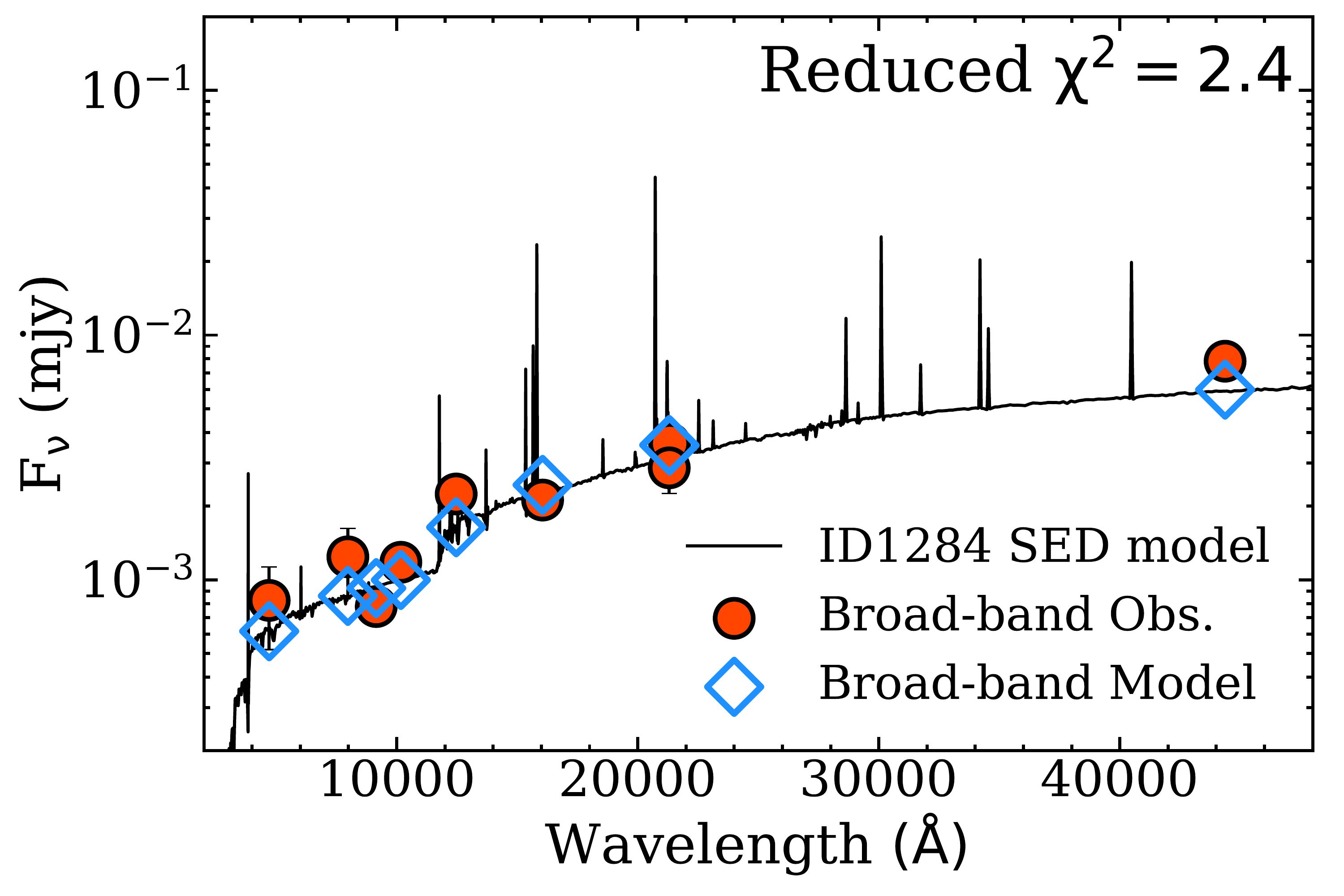}\par
      \includegraphics[width=\linewidth]{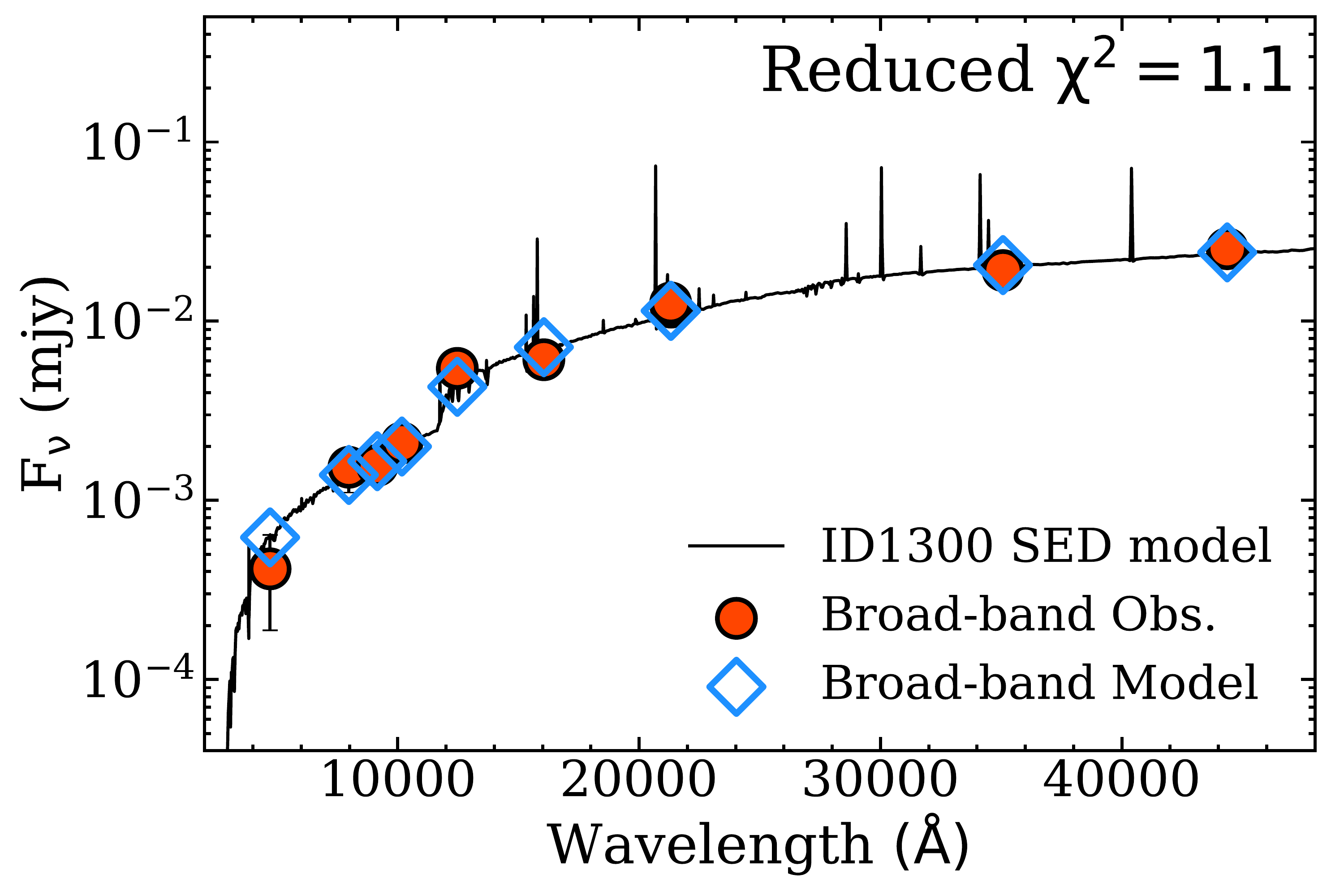}\par
      \includegraphics[width=\linewidth]{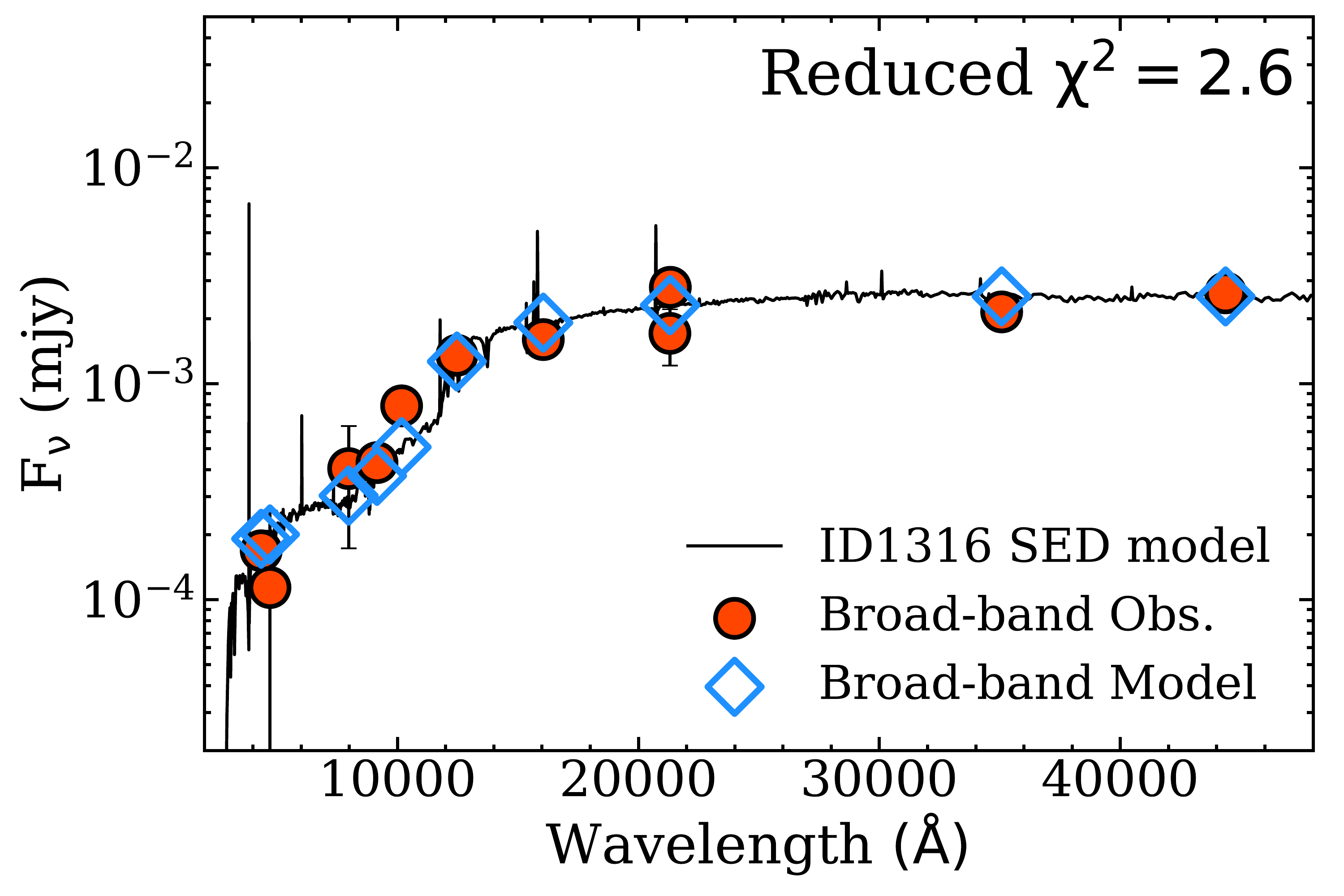}\par
      \includegraphics[width=\linewidth]{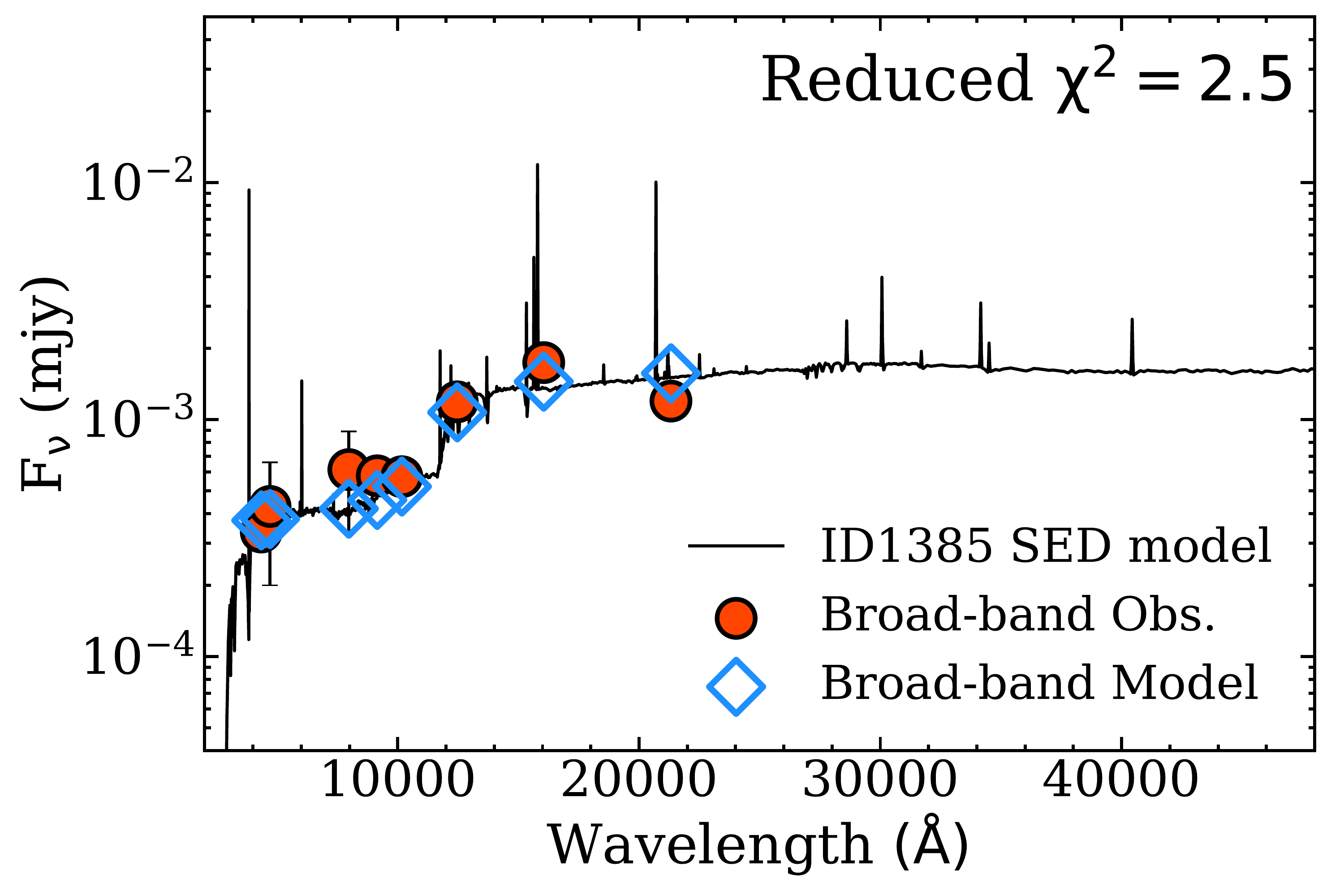}\par
      \includegraphics[width=\linewidth]{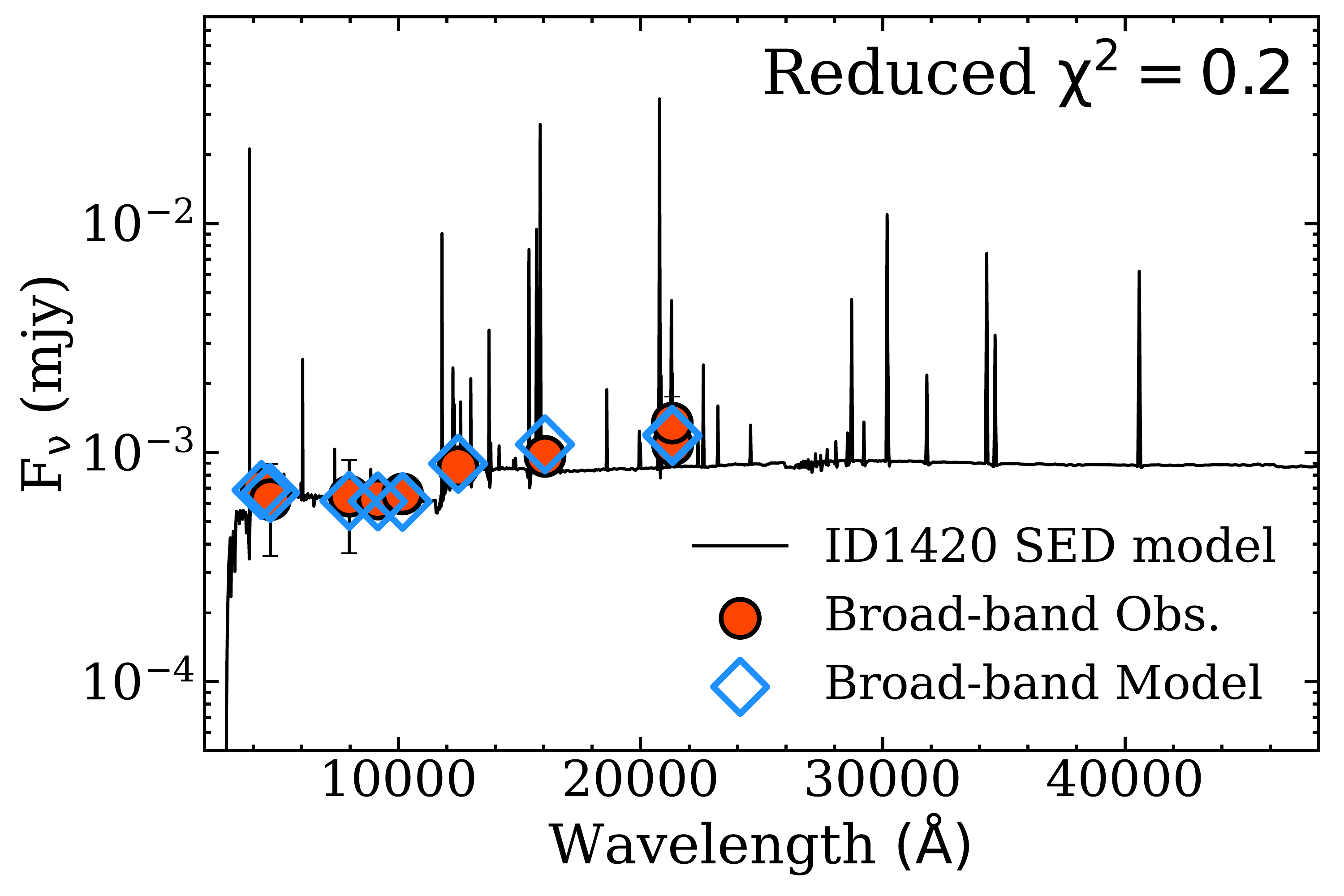}\par
      \includegraphics[width=\linewidth]{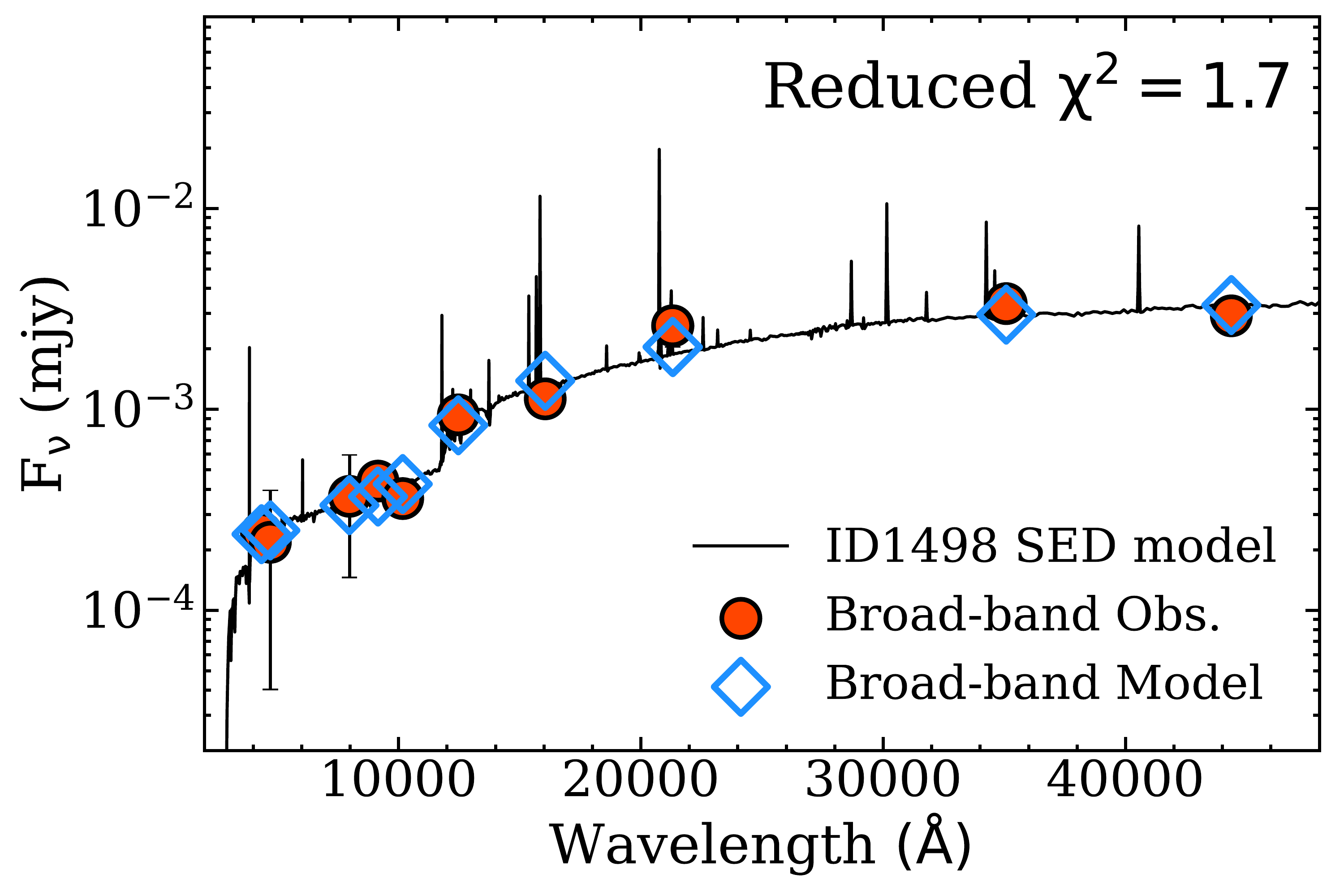}\par
      \includegraphics[width=\linewidth]{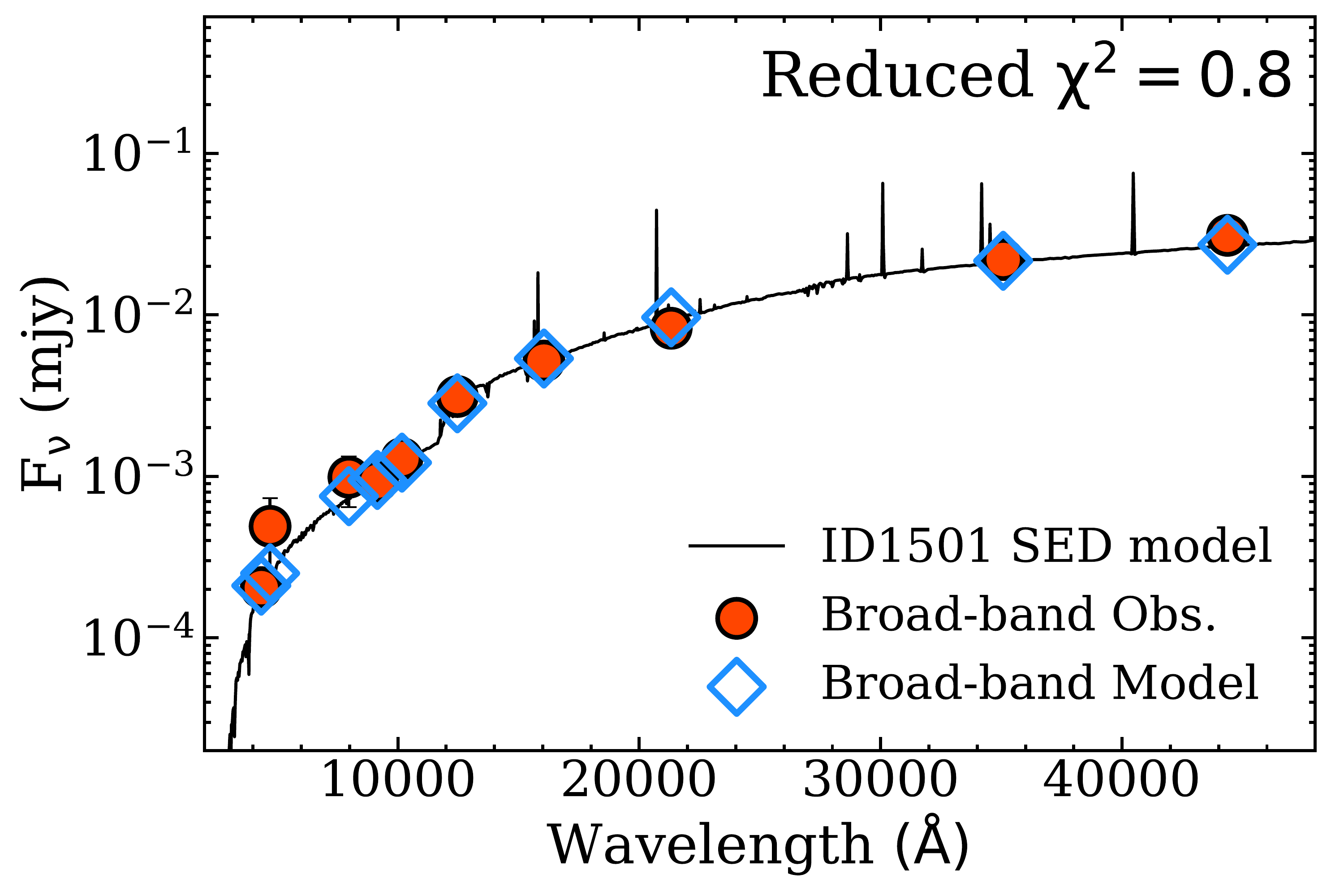}\par
      \end{multicols}
      \contcaption{}
\end{figure*}

\begin{figure*}
 \centering
 \begin{multicols}{2}
      \includegraphics[width=\linewidth]{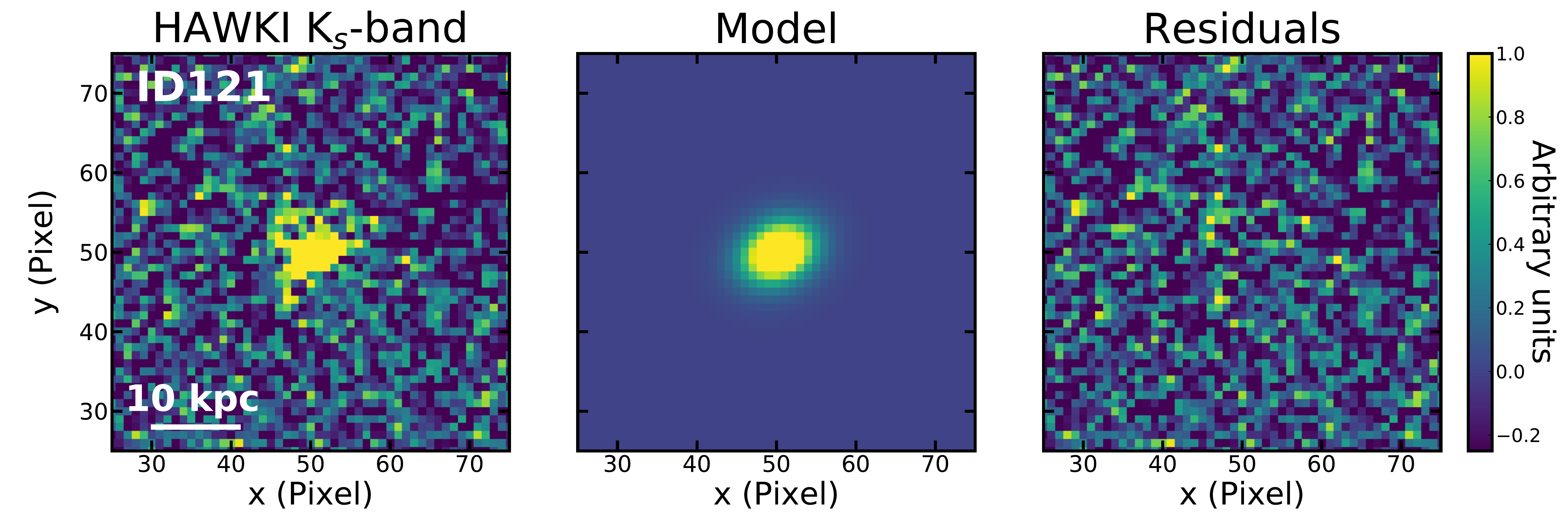}\par
      \includegraphics[width=\linewidth]{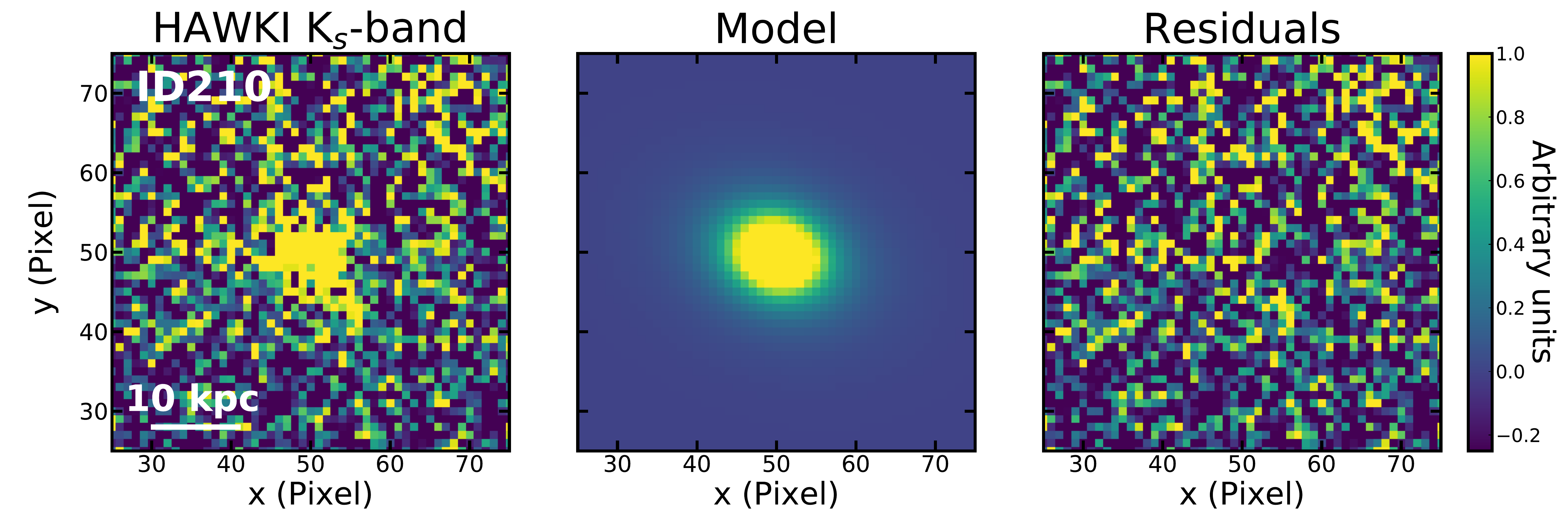}\par
      \includegraphics[width=\linewidth]{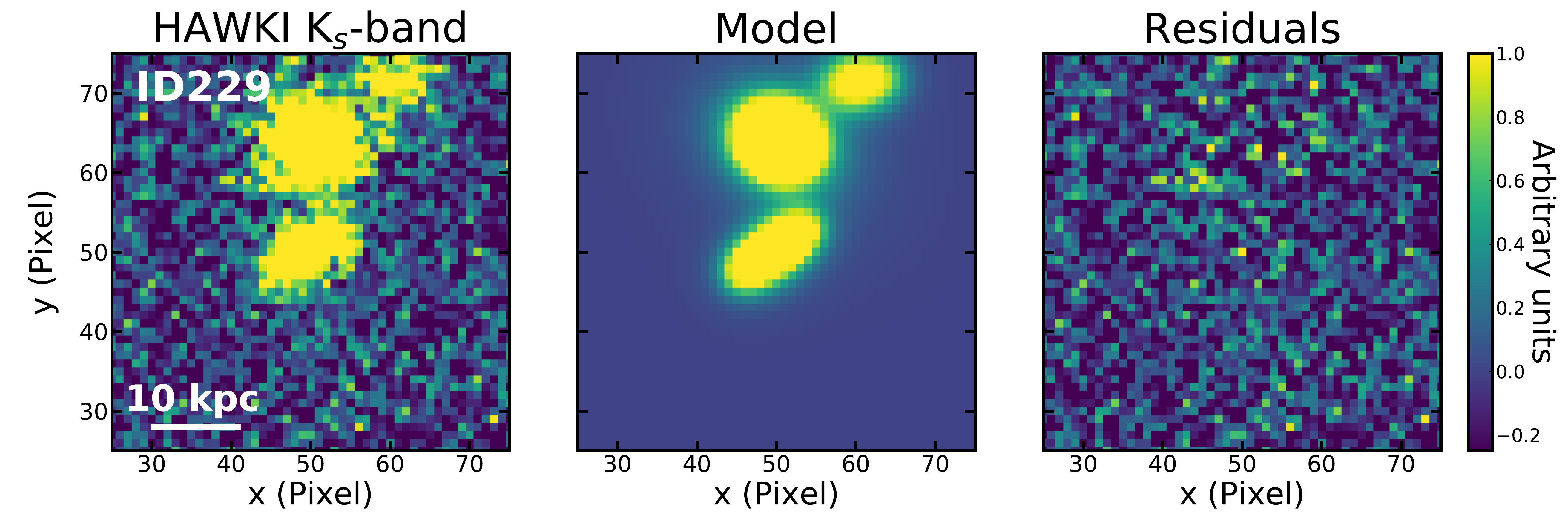}\par
      \includegraphics[width=\linewidth]{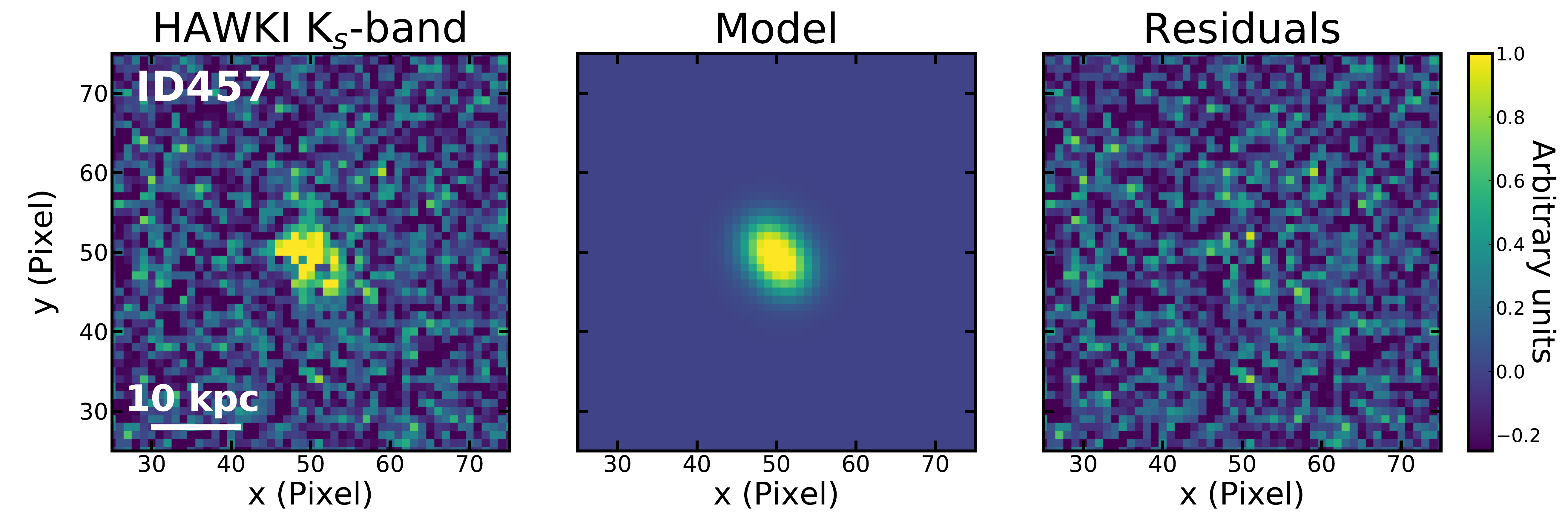}\par
      \includegraphics[width=\linewidth]{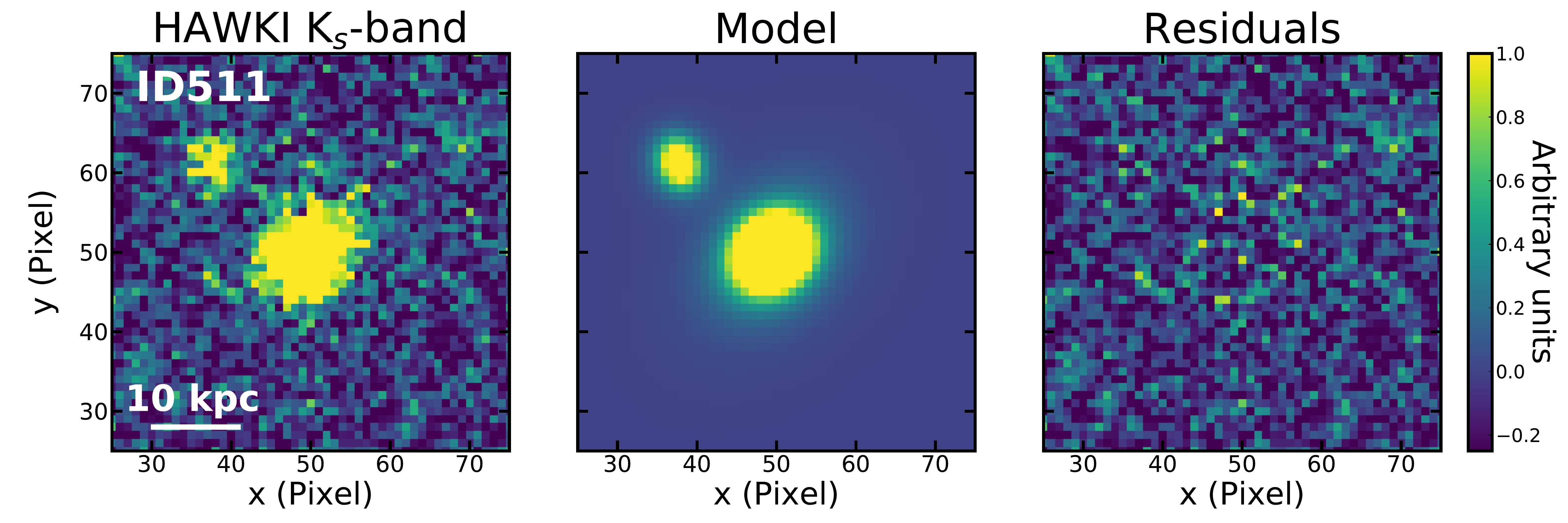}\par
      \includegraphics[width=\linewidth]{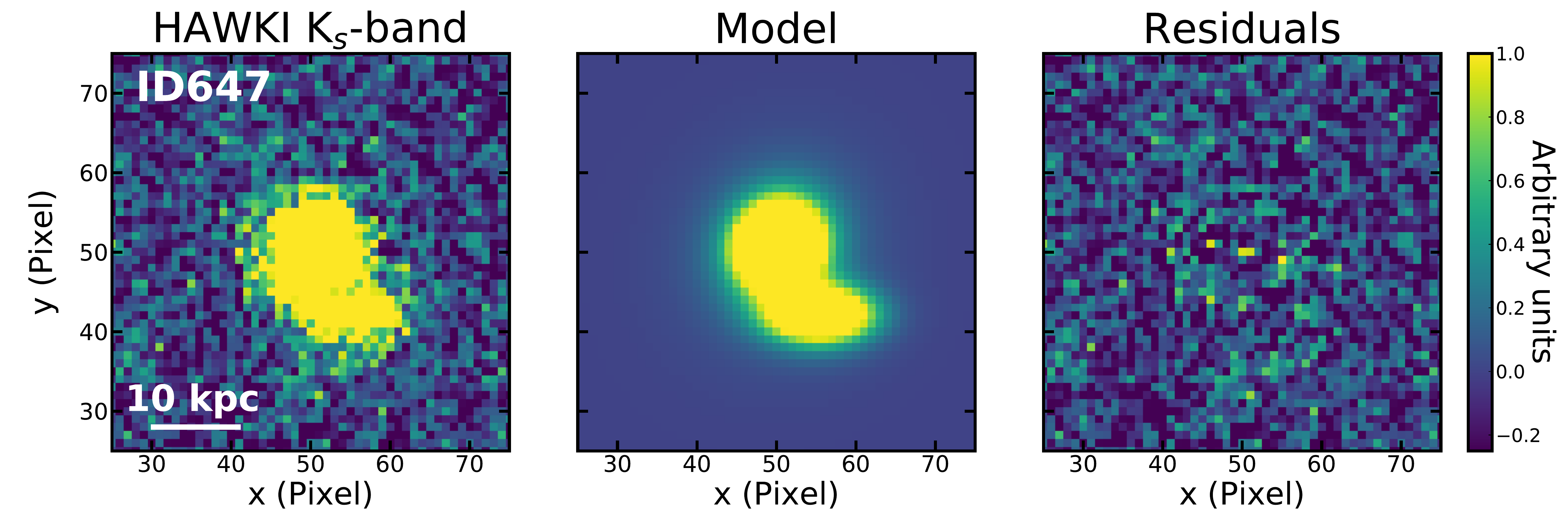}\par
      \includegraphics[width=\linewidth]{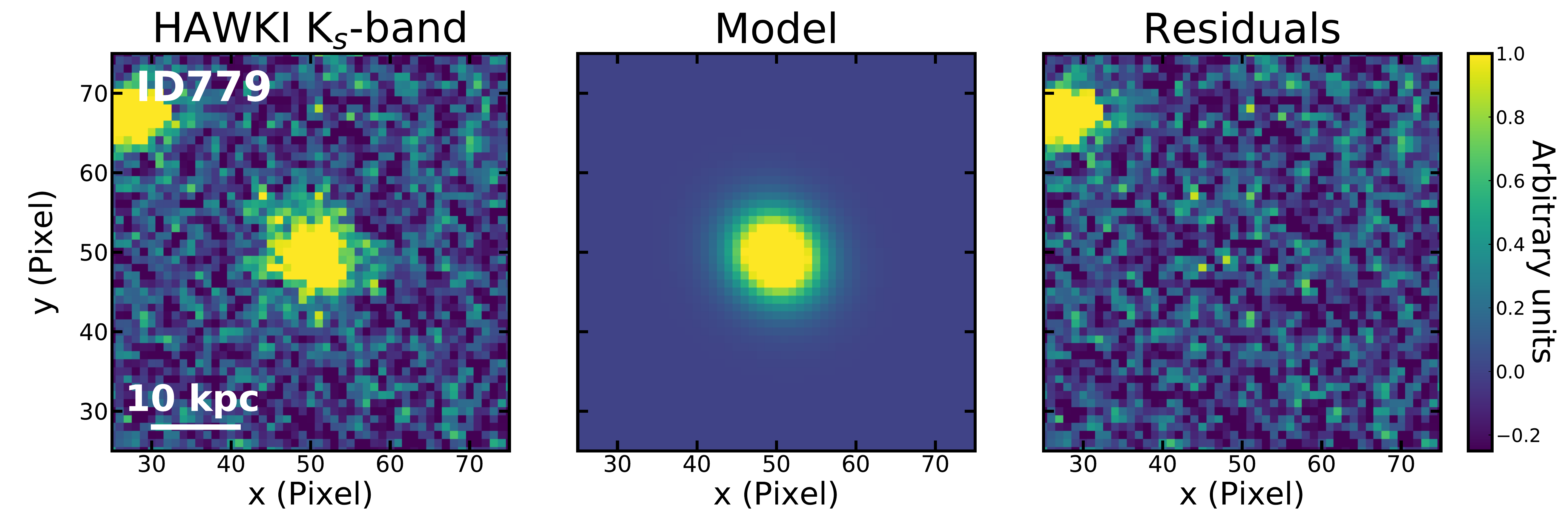}\par
      \includegraphics[width=\linewidth]{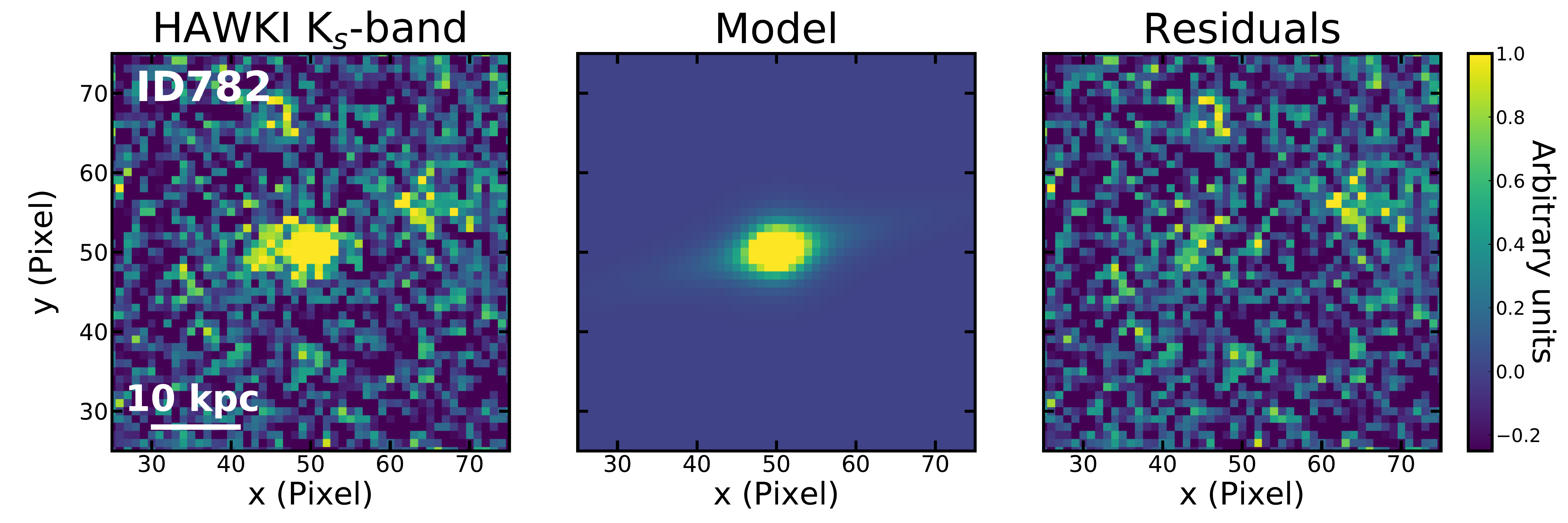}\par
      \includegraphics[width=\linewidth]{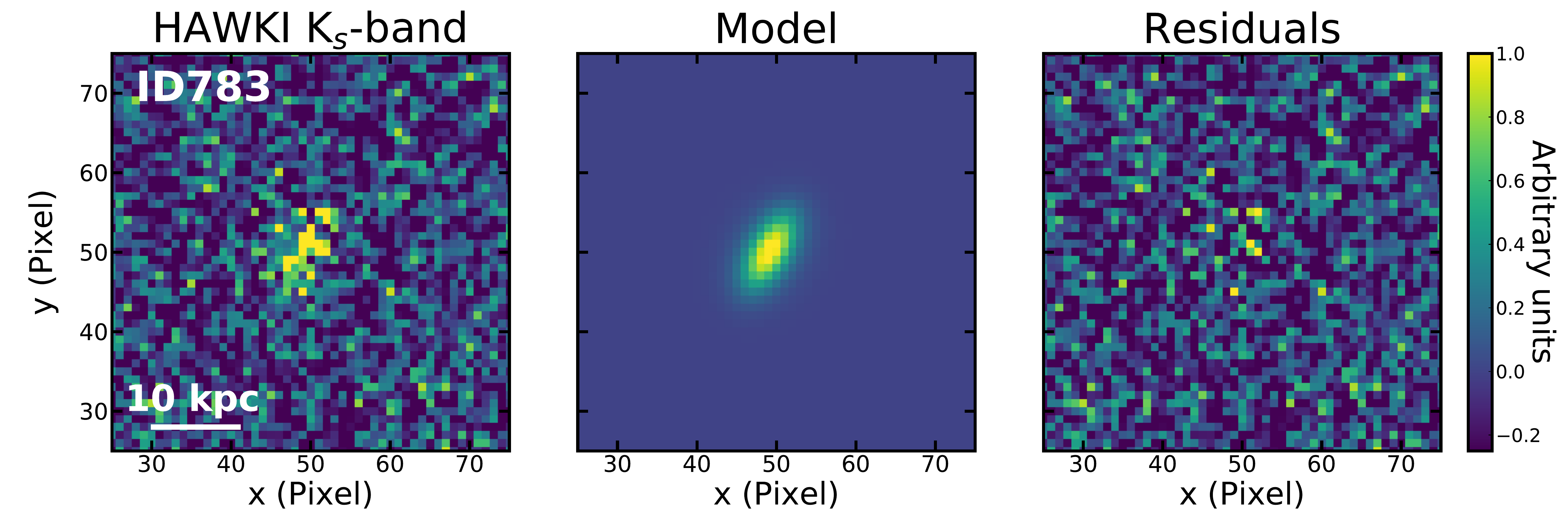}\par
      \includegraphics[width=\linewidth]{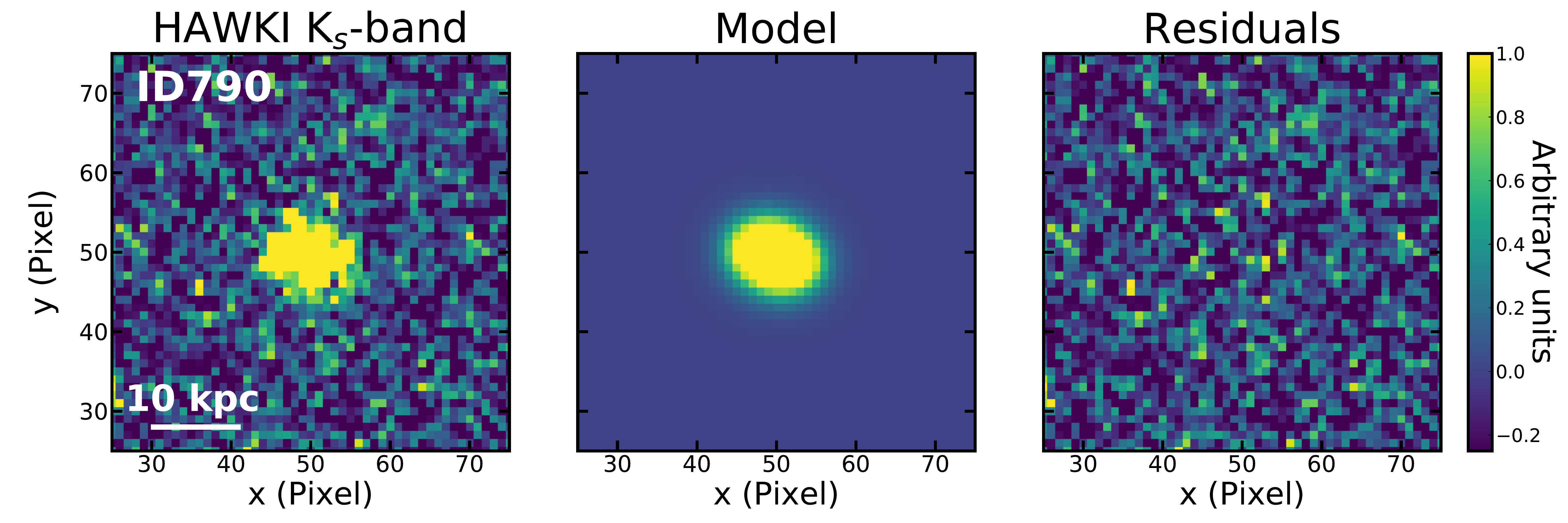}\par

      \end{multicols}
      \caption{VLT/HAWKI $\mathrm{K_s}$-band images, Galfit models, and residuals for the objects included in our Mass-Size relation  (Fig.\,\ref{F:mass-size}).}
         \label{F:Galfit}
\end{figure*}

\begin{figure*}
 \centering
 \begin{multicols}{2}
      \includegraphics[width=\linewidth]{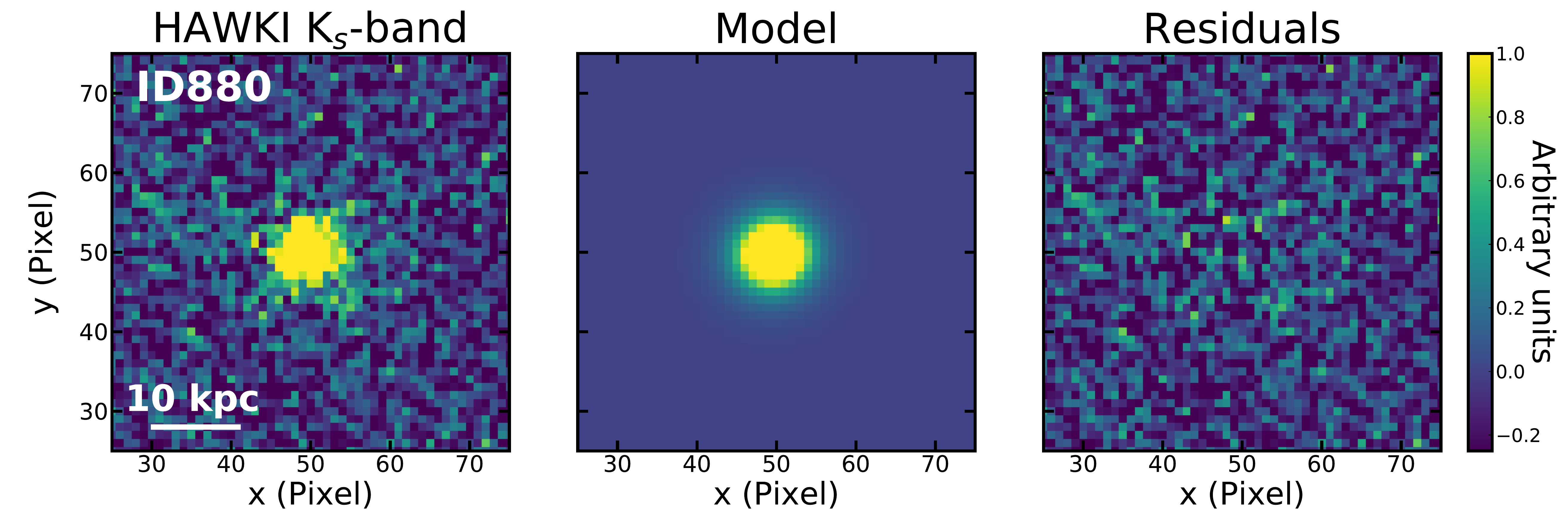}\par
      \includegraphics[width=\linewidth]{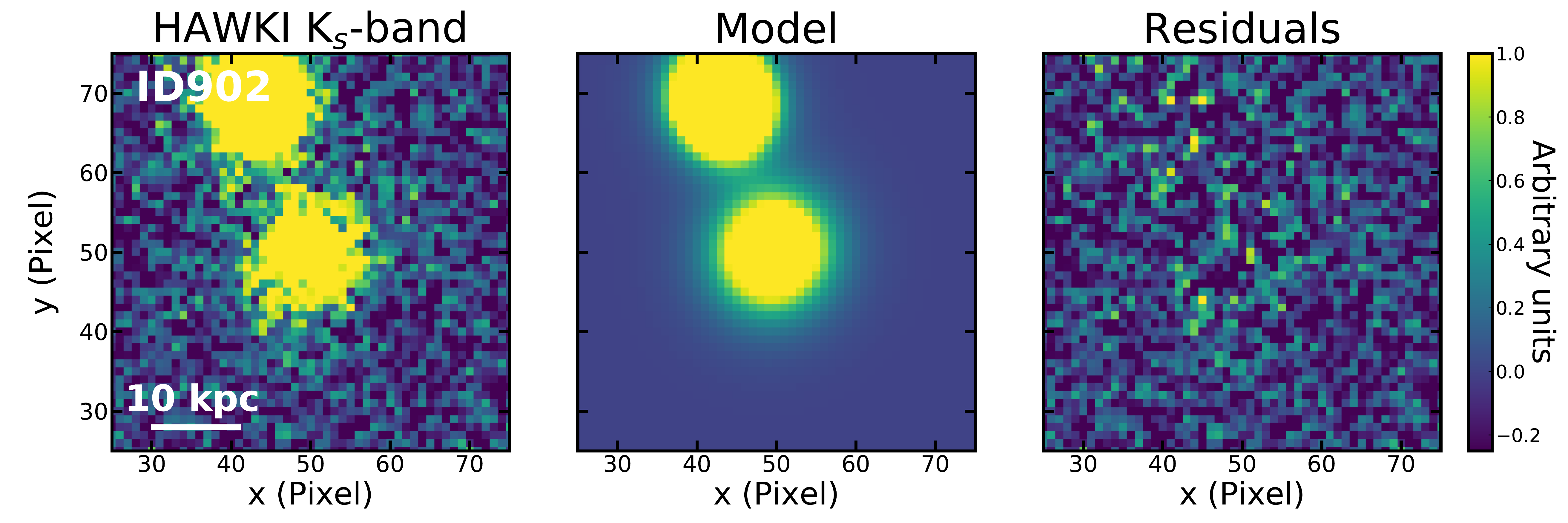}\par
      \includegraphics[width=\linewidth]{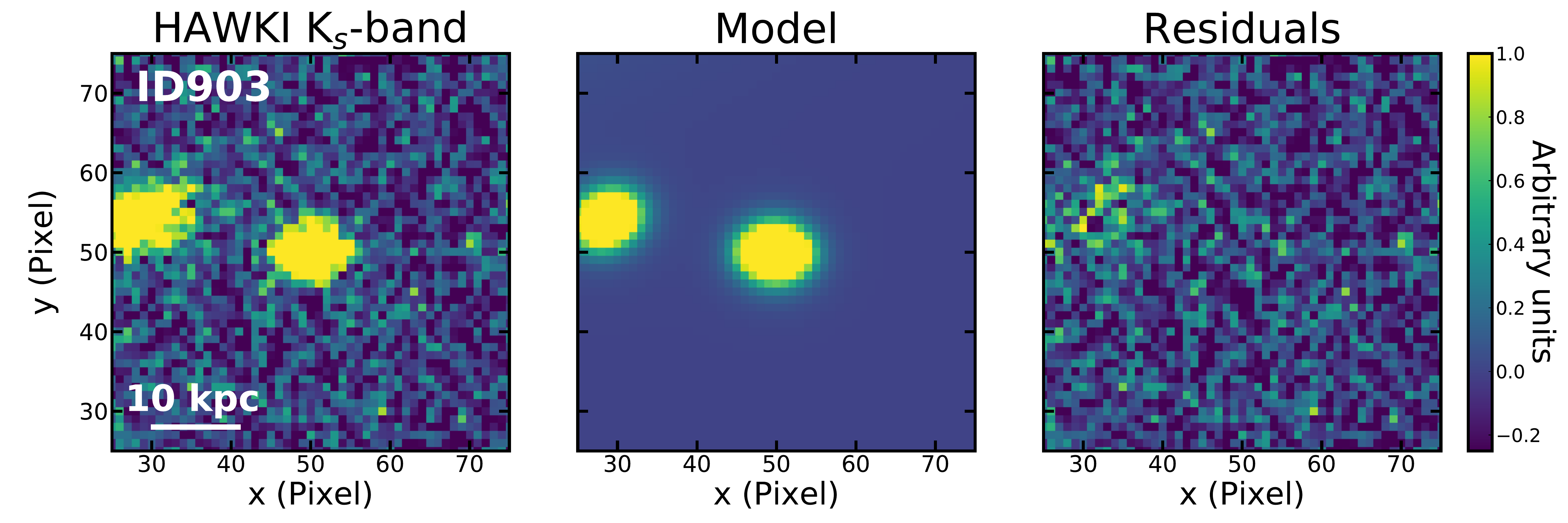}\par
      \includegraphics[width=\linewidth]{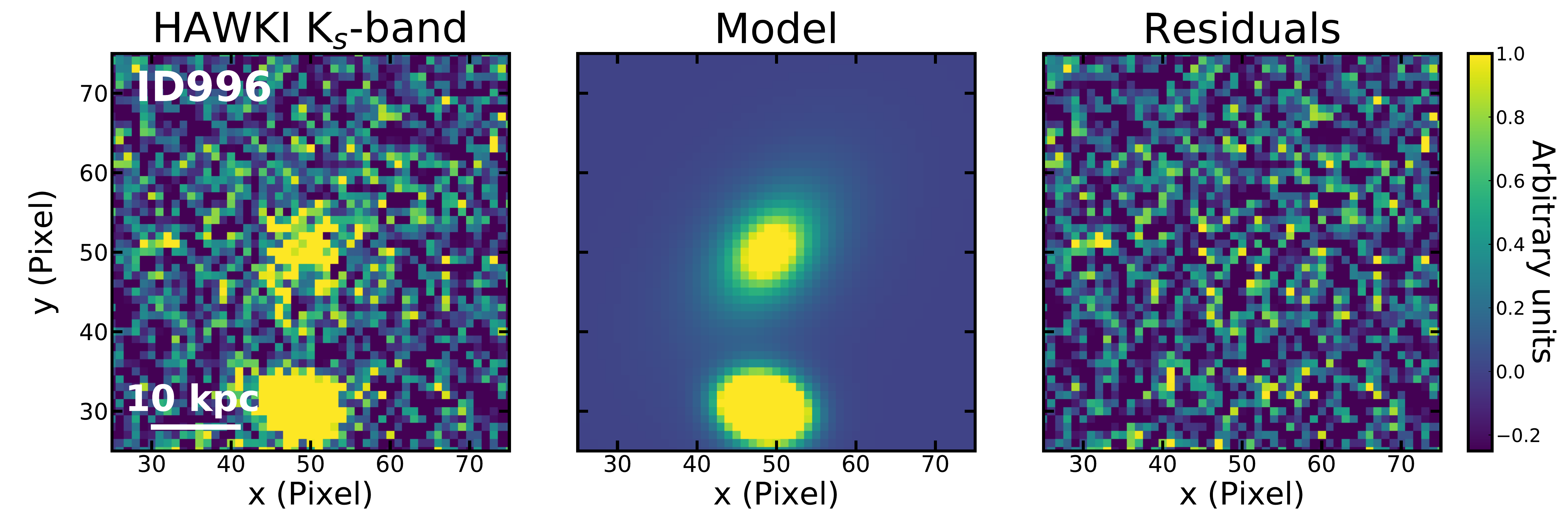}\par
      \includegraphics[width=\linewidth]{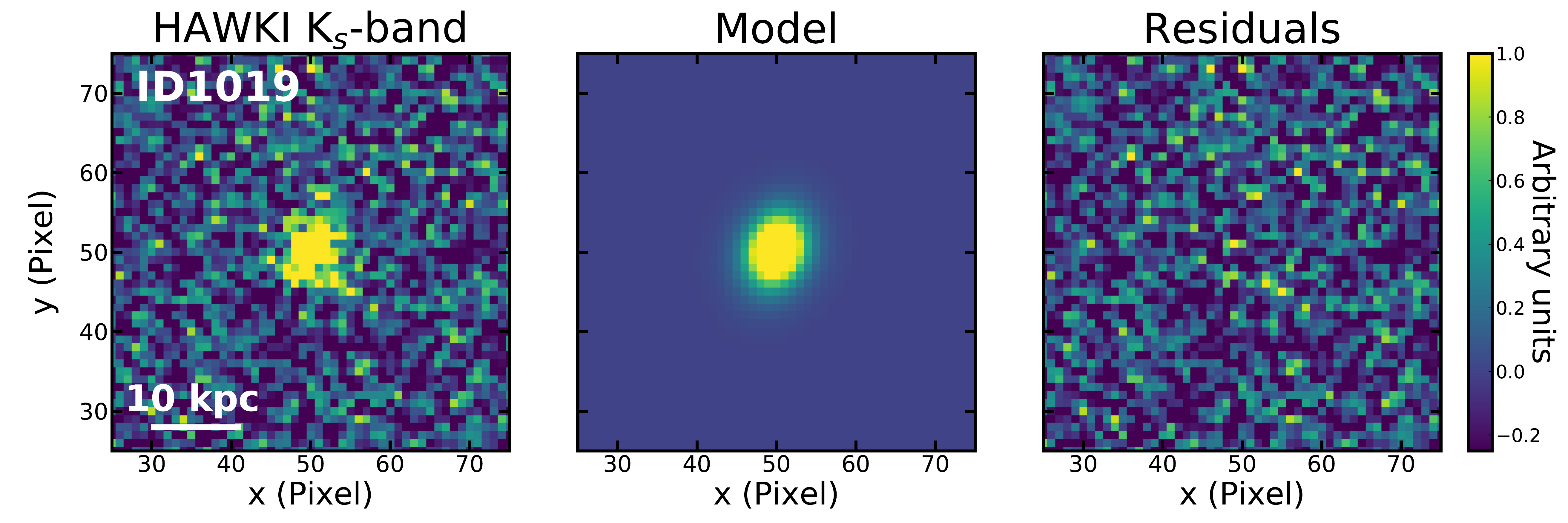}\par
      \includegraphics[width=\linewidth]{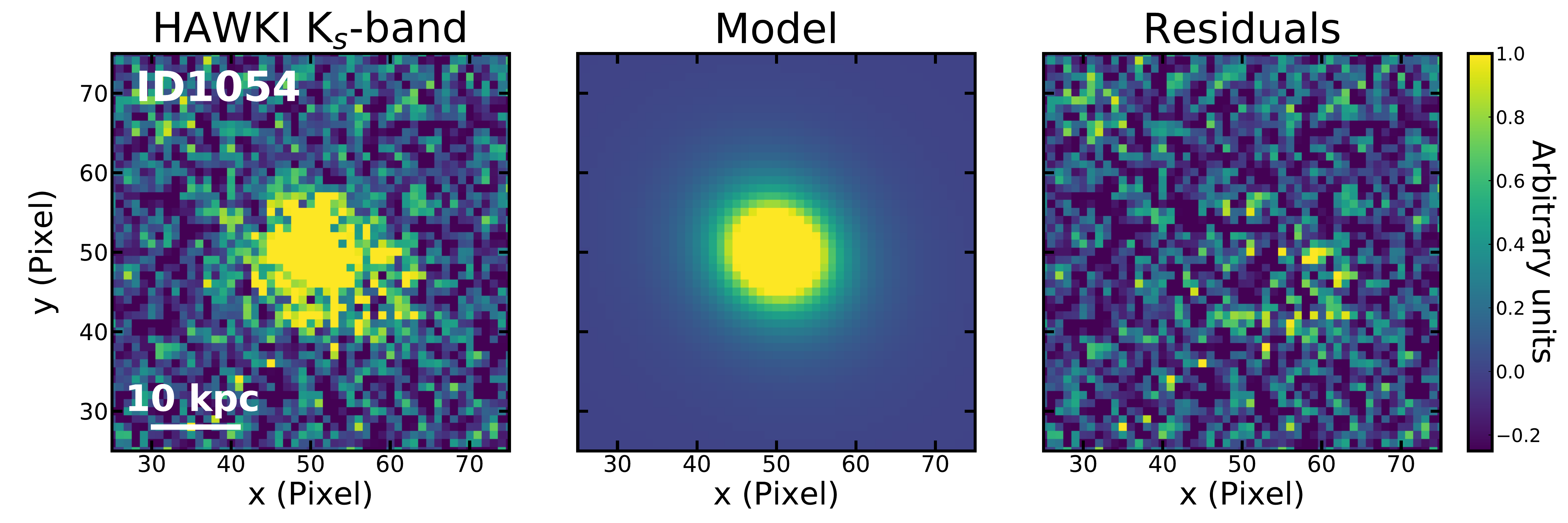}\par
      \includegraphics[width=\linewidth]{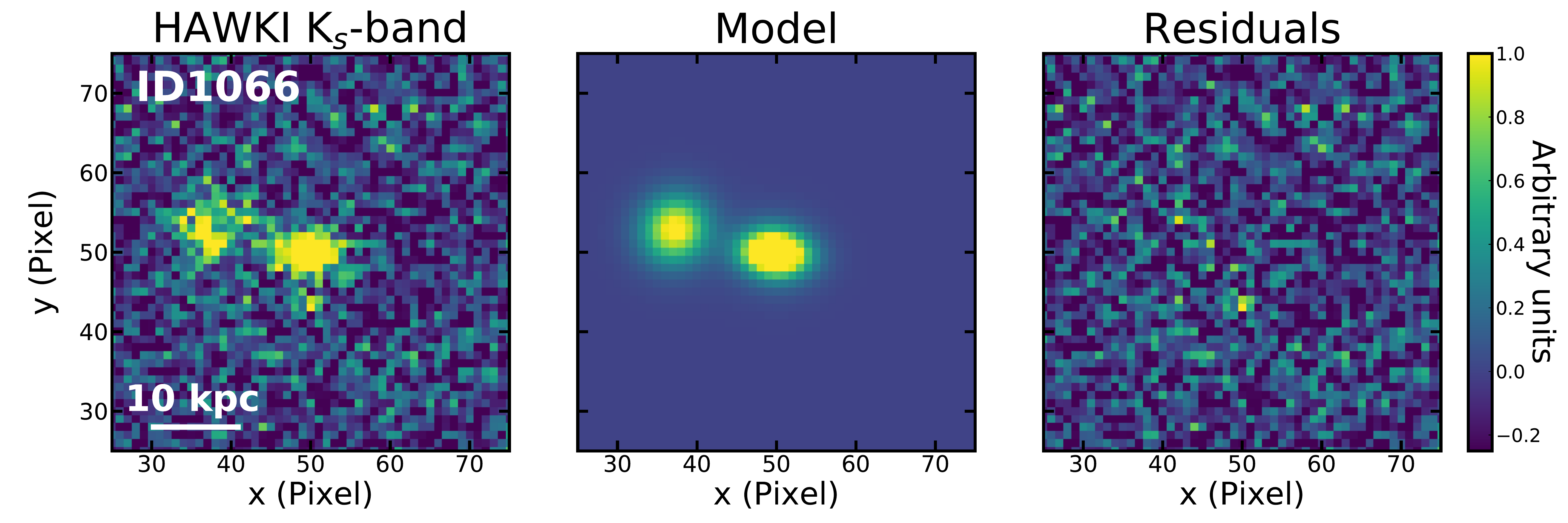}\par
      \includegraphics[width=\linewidth]{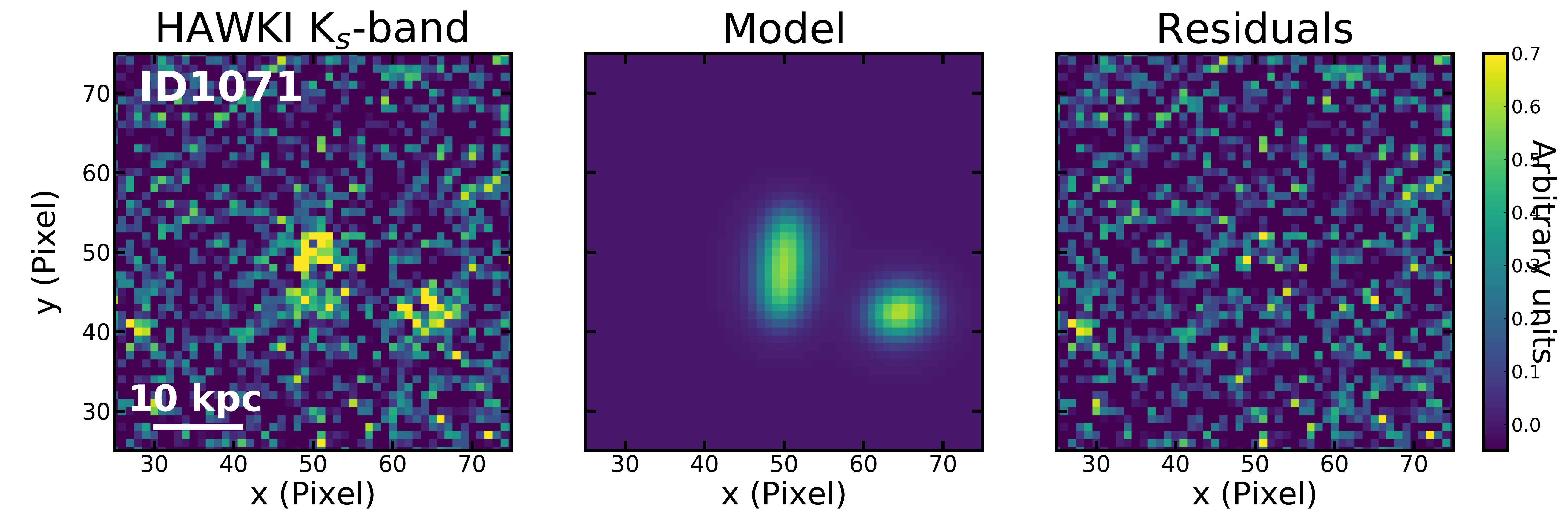}\par
      \includegraphics[width=\linewidth]{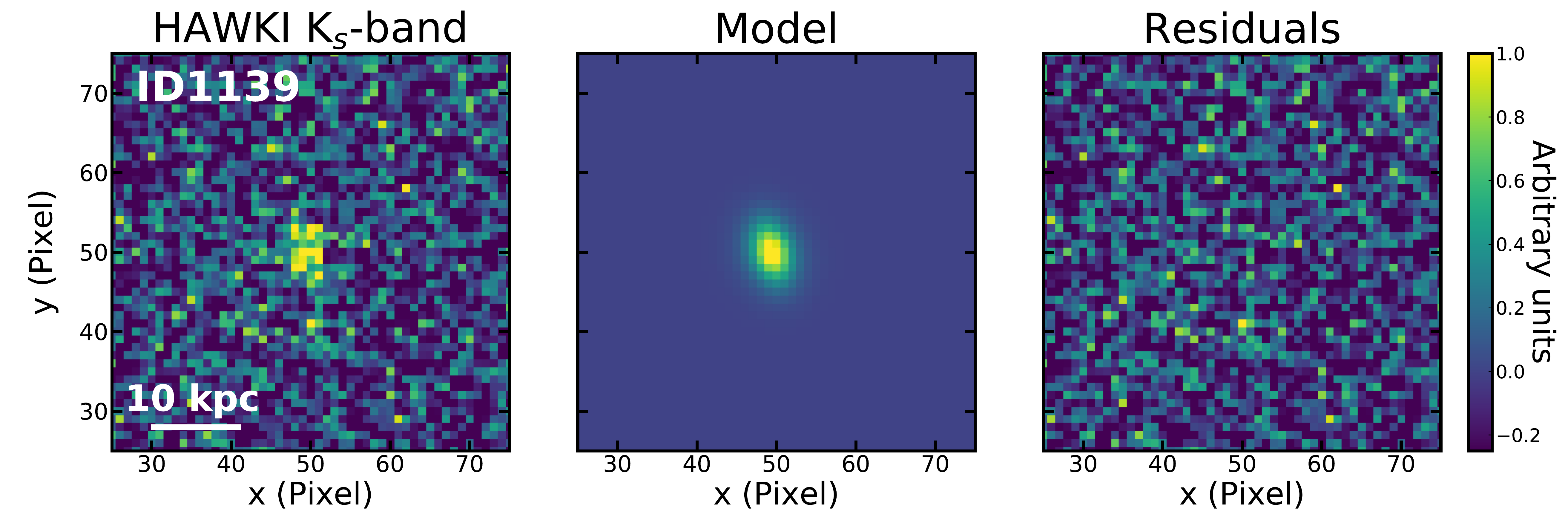}\par
      \includegraphics[width=\linewidth]{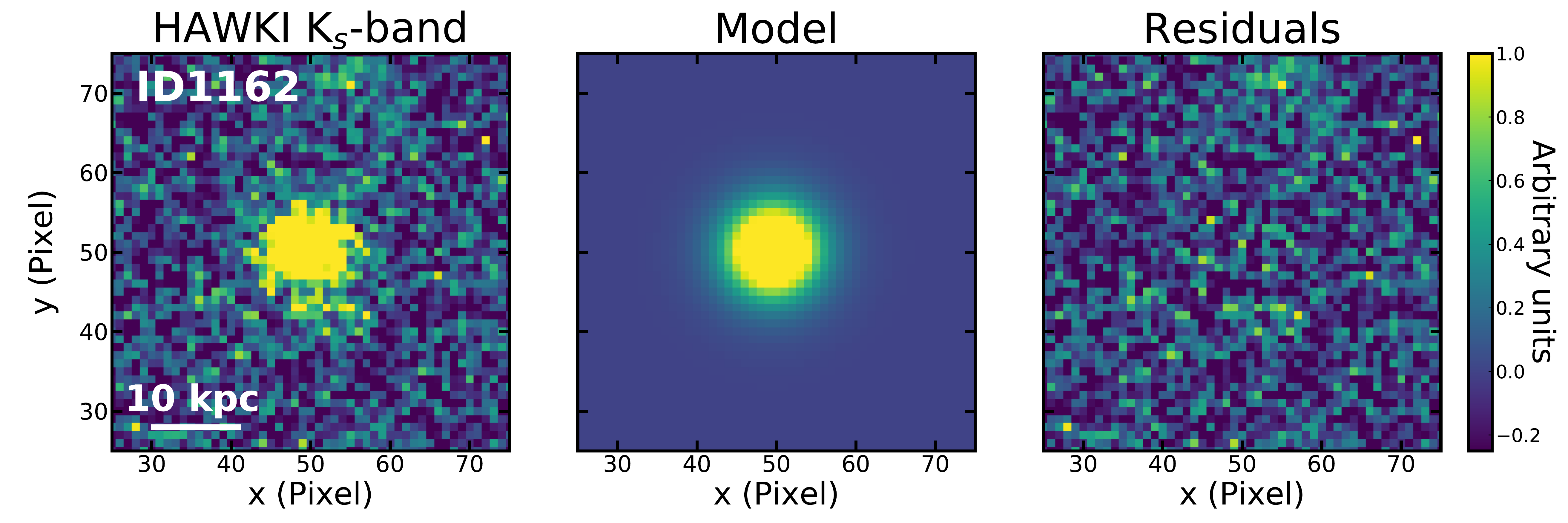}\par
      \includegraphics[width=\linewidth]{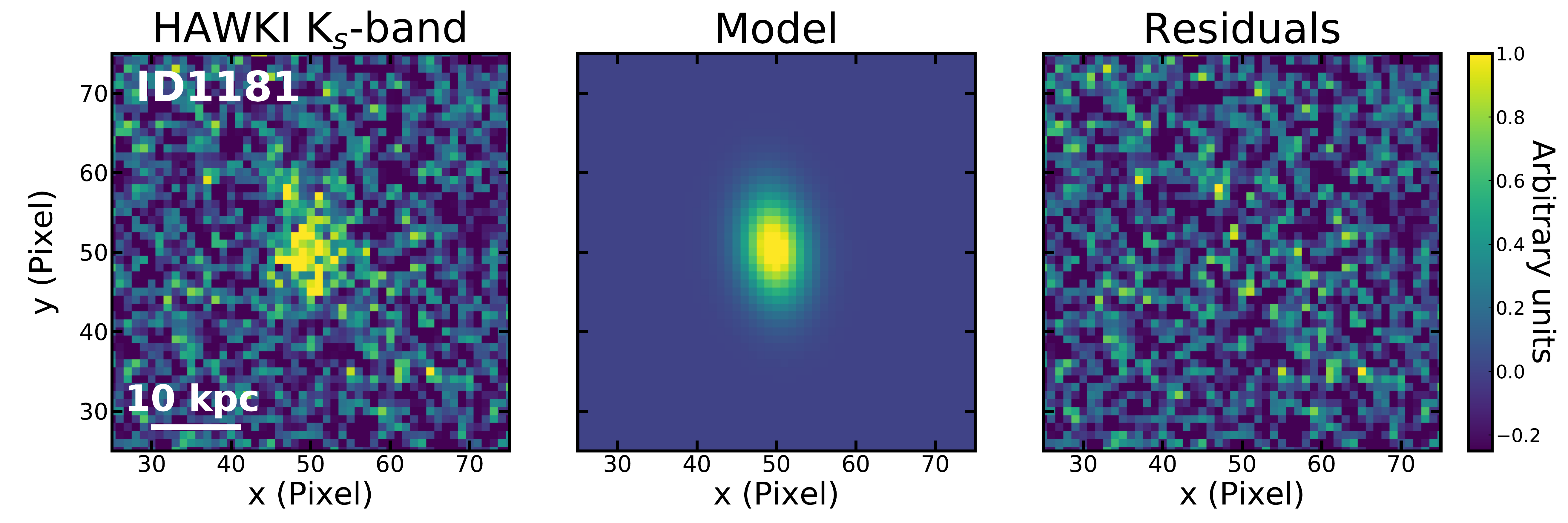}\par
      \includegraphics[width=\linewidth]{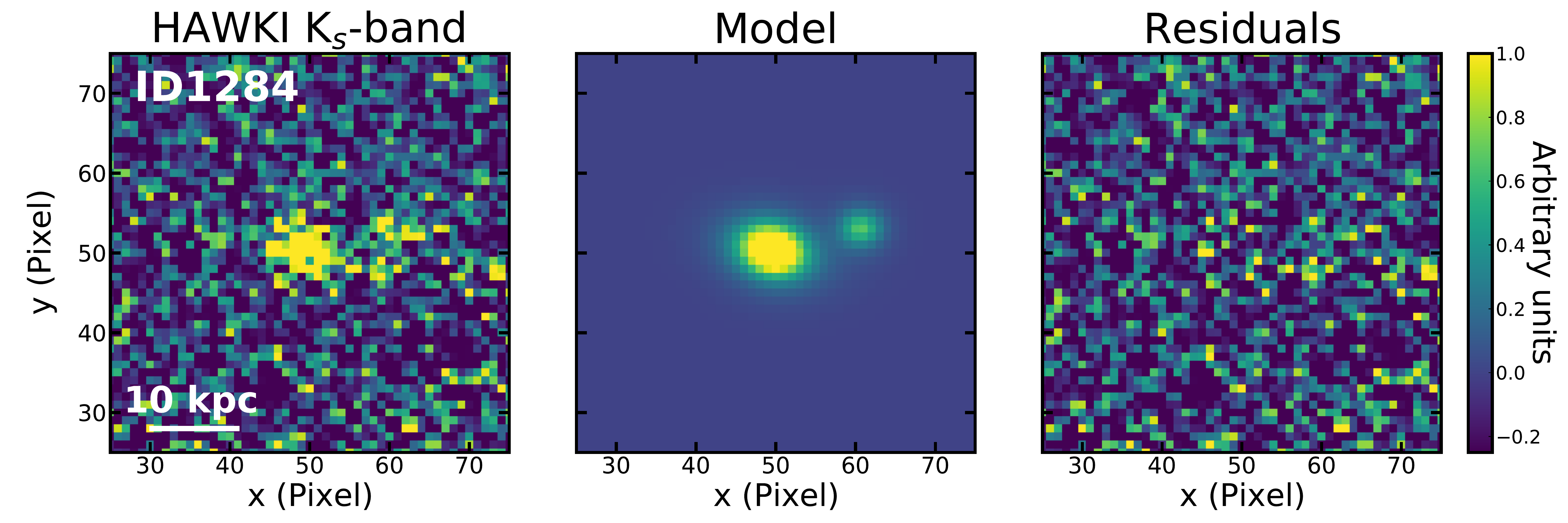}\par
      \includegraphics[width=\linewidth]{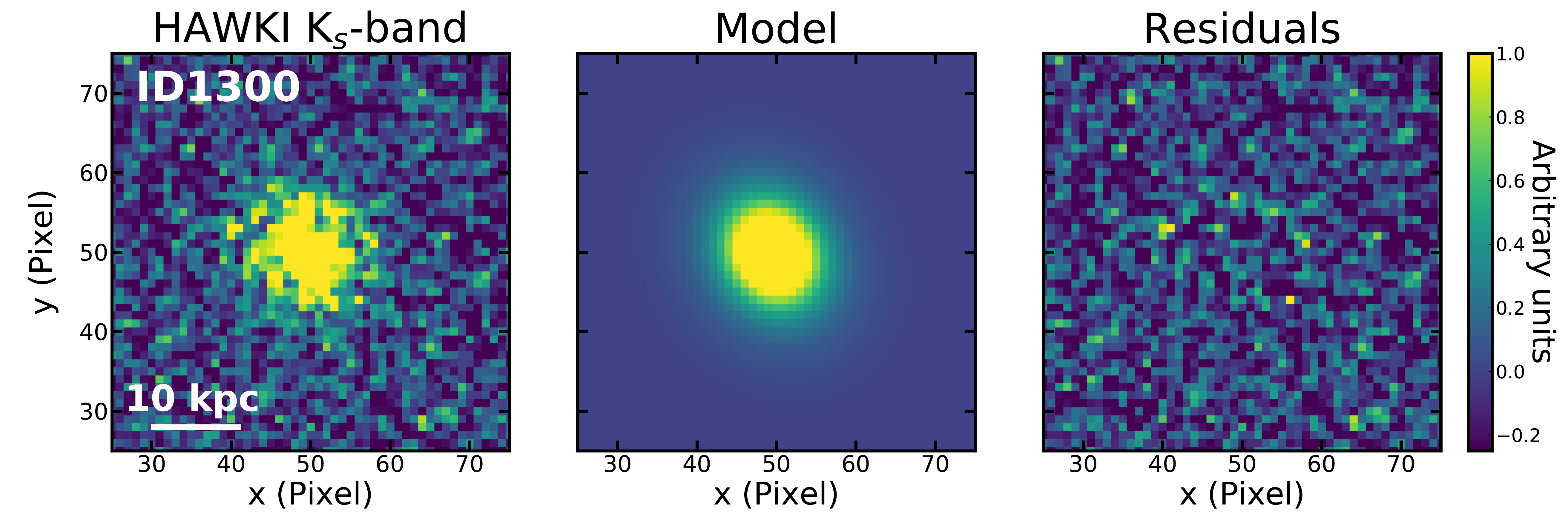}\par
      \includegraphics[width=\linewidth]{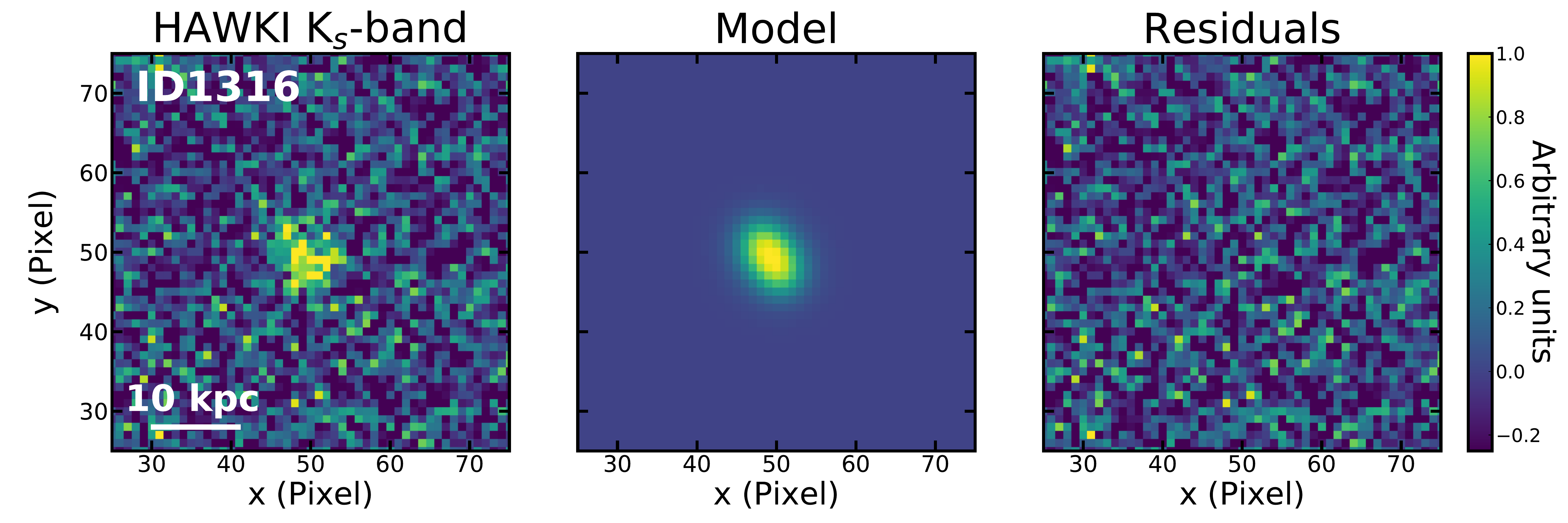}\par
      \includegraphics[width=\linewidth]{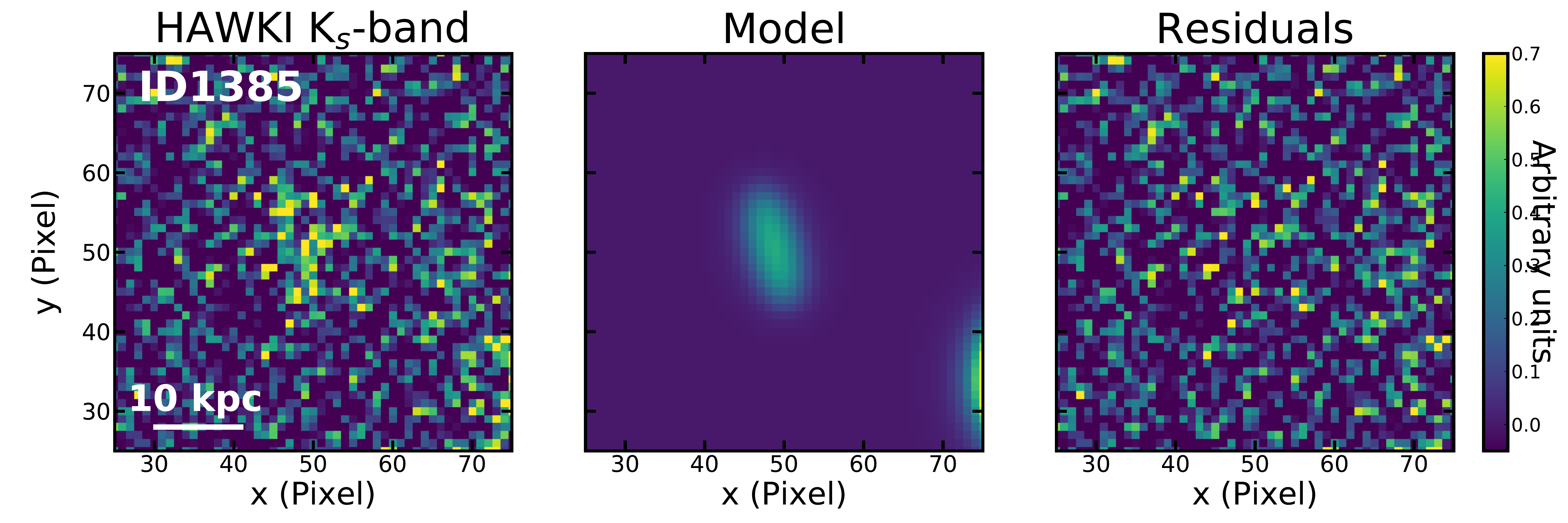}\par
      \includegraphics[width=\linewidth]{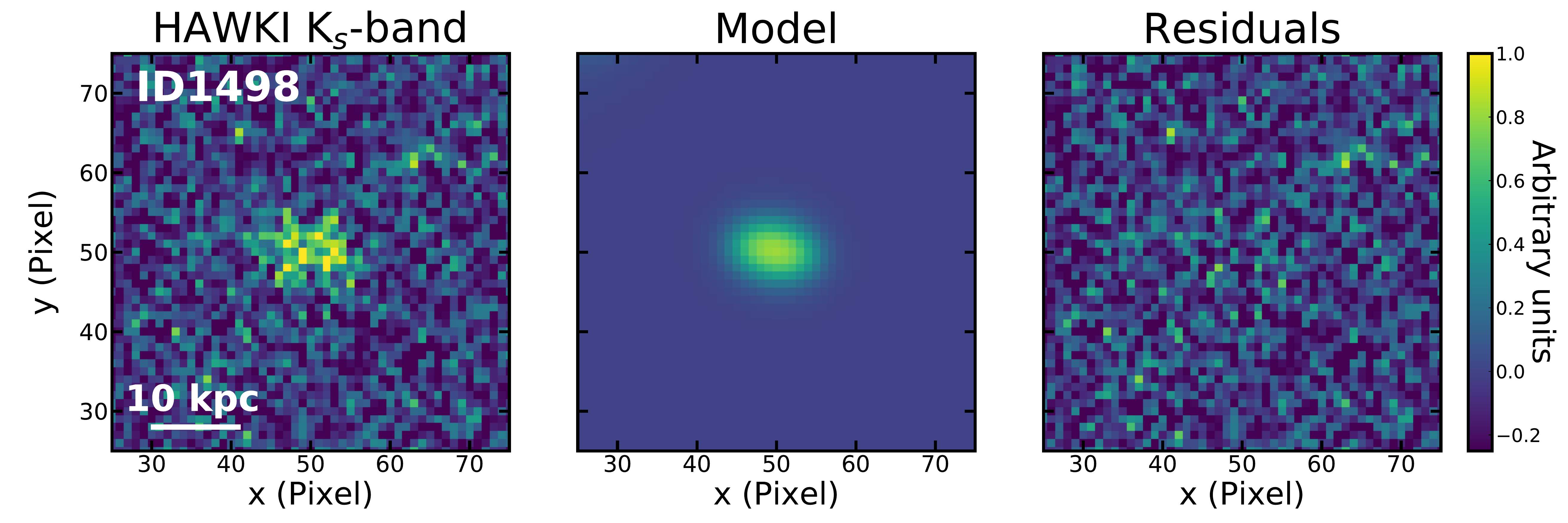}\par 
      \end{multicols}
      \contcaption{}
\end{figure*}

\begin{figure*}
 \centering
 \begin{multicols}{2}
      \includegraphics[width=\linewidth]{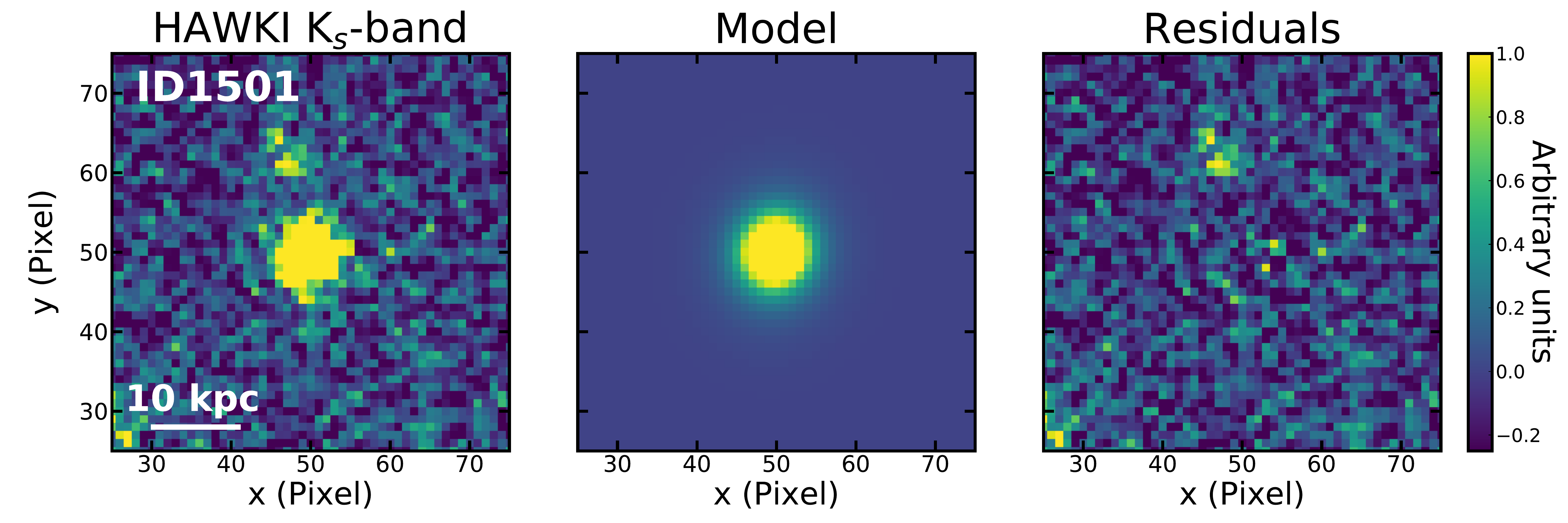}\par
      \end{multicols}
      \contcaption{}
\end{figure*}

\renewcommand{\arraystretch}{1.5}
\begin{landscape}
\begin{table}
\caption{Physical properties of the galaxies studied in this work. Coordinates (RA and DEC) are shown in degrees. The redshifts (z) are based on the H$\alpha$ measurements of this work. The molecular masses have been computed following the metallicity dependent calibration by \protect\cite{Tacconi18} and making use of the ATCA CO(1-0) information published by \protect\cite{Jin21}. See Sect.\,\ref{SS:MOL} for more details. $\mathrm{E(B-V)_s}$ represents the stellar contribution to the reddening assuming the \protect\cite{Calzetti2000} extinction law (see Sect.\,\ref{SS:SFR_method}). In the last two columns we highlight the presence of X-ray emission according to \protect\cite{Tozzi22} and the identification of sources as SMGs according to \protect\cite{Dannerbauer14}.}

\begin{threeparttable}
\begin{tabular}{cccccccccccccccc}
\hline
\noalign{\vskip 0.00cm}
ID & RA  & DEC   & z  &  $\mathrm{\log\,M_*/M_\odot}$  & SFR & $\mathrm{R_{e,K_s}}$ & $\mathrm{12+\log(O/H)}$ & $\mathrm{\log\,M_{mol}/M_\odot}$ & U-V & V-J & $\mathrm{E(B-V)_s}$ & $\Sigma_3$ & $\eta$ & X-ray & SMG\\
  & (J2000) & (J2000) & & & ($\mathrm{M_\odot/yr}$) & (kpc) & & & & & & ($\mathrm{Mpc^{-2}}$) & & & \\
\noalign{\vskip 0.00cm}
\hline 
\hline 
\noalign{\vskip 0.00cm}
56 & 175.1411667 & -26.5369722 & 2.1533 & $10.83^{+0.11}_{-0.15}$ & $97\pm31$ & - & $8.78\pm0.12$ & - & 1.23 & 1.12 & 0.44 & 11.72 & 1.57 & - & - \\
121 & 175.1466667 & -26.5210278 & 2.1415 & $10.37^{+0.09}_{-0.11}$ & $19\pm4$ & $2.4\pm0.3$ & $8.54\pm0.10$ & - & 1.35 & 0.68 & 0.12 & 60.56 & 6.84 & - & - \\
204 & 175.1541250 & -26.5514444 & 2.1444 & $9.31^{+0.12}_{-0.16}$ & $12\pm2$ & - & - & - & 0.37 & -0.16 & 0.02 & 13.55 & 7.09 & - & - \\ 
210 & 175.1555833 & -26.5048056 & 2.1684 & $11.16^{+0.09}_{-0.11}$ & $101\pm8$ & $4.2\pm0.7$ & $8.65\pm0.06$ & - & 1.5 & 1.21 & 0.34 & 82.52 & 4.38 & Yes & - \\
229 & 175.1573750 & -26.4867500 & 2.1574 & $10.98^{+0.11}_{-0.14}$ & $66\pm18$ & $3.9\pm0.5$ & - & - & 1.05 & 1.16 & 0.49 & 9.76 & 0.43 & - & - \\
298 & 175.1642500 & -26.5068333 & 2.1626 & $10.42^{+0.14}_{-0.20}$ & $70\pm13$ & - & $8.52\pm0.12$ & - & 0.93 & 0.88 & 0.38 & 49.44 & 2.03 & - & - \\
343 & 175.1670833 & -26.4963889 & 2.1609 & $10.11^{+0.12}_{-0.17}$ & $35\pm6$ & - & $8.66\pm0.07$ & $11.02\pm0.12$ & 0.87 & 0.68 & 0.31 & 21.98 & 1.24 & - & - \\
457 & 175.1785833 & -26.4686944 & 2.1622 & $9.79^{+0.10}_{-0.13}$ & $27\pm4$ & $2.8\pm0.7$ & $8.44\pm0.08$ & - & 0.44 & 0.01 & 0.09 & 13.56 & 1.29 & - & - \\
511 & 175.1843750 & -26.4853056 & 2.1694 & $11.29^{+0.09}_{-0.11}$ & $27\pm13$ & $1.7\pm0.1$ & - & - & 1.64 & 1.06 & 0.25 & 200.32 & 1.59 & Yes & - \\
586 & 175.2637083 & -26.4807778 & 2.1492 & $10.17^{+0.11}_{-0.15}$ & $58\pm19$ & - & $<8.69$ & - & 1.13 & 1.06 & 0.47 & 6.44 & 2.99 & - & - \\ 
647 & 175.2599167 & -26.4625278 & 2.1510 & $11.59^{+0.10}_{-0.13}$ & $114\pm29$  \tnote{a} & $2.4\pm0.1$ & $8.80\pm0.07$ \tnote{a} & - & 1.45 & 1.28 & 0.42 & 25.63 & 2.25 & Yes & Yes \\
779 & 175.2505833 & -26.4823056 & 2.1665 & $10.66^{+0.08}_{-0.09}$ & $106\pm14$ & $3.2\pm0.2$ & $8.48\pm0.08$ & - & 1.11 & 0.91 & 0.37 & 7.66 & 3.64 & - & - \\
782 & 175.2492500 & -26.5118333 & 2.1617 & $9.75^{+0.11}_{-0.14}$ & $10\pm3$ & $1.9\pm0.4$ & $8.54\pm0.15$ & $11.22\pm0.13$ & 0.22 & -0.3 & 0.00 & 86.45 & 2.24 & - & Yes\\
783 & 175.2489583 & -26.5523889 & 2.1730 & $9.88^{+0.13}_{-0.19}$ & $16\pm3$ & $3.1\pm0.3$ & $8.64\pm0.09$ & - & 0.73 & 0.32 & 0.14 & 24.41 & 10.53 & - & - \\
788 & 175.2485000 & -26.5544722 & 2.1719 & $9.67^{+0.15}_{-0.22}$ & $18\pm3$ & - & - & - & 0.47 & -0.02 & 0.05 & 40.48 & 10.04 & - & - \\
790 & 175.2484167 & -26.5108611 & 2.1645 & $10.74^{+0.10}_{-0.13}$ & $231\pm38$ & $3.4\pm0.2$ & $8.58\pm0.08$ & $11.27\pm0.09$ & 1.21 & 1.11 & 0.46 & 105.80 & 3.27 & - & Yes \\
876 & 175.1952083 & -26.4780833 & 2.1636 & $8.88^{+0.15}_{-0.23}$ & $15\pm2$ & - & - & - & 0.34 & -0.21 & 0.02 & 24.72 & 0.56 & - & - \\
880 & 175.1944583 & -26.4862222 & 2.1663 & $10.90^{+0.10}_{-0.12}$ & $52\pm11$ & $1.9\pm0.2$ & - & - & 1.48 & 1.46 & 0.57 & 154.84 & 0.49 & Yes & - \\
902 & 175.1919167 & -26.4864722 & 2.1490 & $11.37^{+0.10}_{-0.13}$ & $511\pm110$ & $4.1\pm0.2$ & $<8.78$ & $11.14\pm0.06$ & 1.59 & 1.72 & 0.70 & 207.93 & 0.46 & - & - \\
903 & 175.1921667 & -26.4902222 & 2.1553 & $10.35^{+0.11}_{-0.14}$ & $77\pm7$ & $2.0\pm0.2$ & $8.36\pm0.06$ & - & 0.74 & 0.46 & 0.25 & 240.56 & 0.05 & - & - \\
911 & 175.1915833 & -26.4880000 & 2.1557 & $11.17^{+0.07}_{-0.09}$ & $125\pm44$  \tnote{a} & - & $8.38\pm0.08$ \tnote{a} & - & 0.69 & 0.63 & 0.25 & 450.88 & 0.02 & Yes & - \\
996 & 175.2447500 & -26.5062500 & 2.1657 & $10.72^{+0.11}_{-0.14}$ & $56\pm12$ & $5.8\pm1.6$ & $8.74\pm0.08$ & - & 0.92 & 0.77 & 0.35 & 28.62 & 3.34 & - & - \\
999 & 175.2465833 & -26.4656389 & 2.1473 & $9.88^{+0.08}_{-0.10}$ & $59\pm8$ & - & $8.61\pm0.06$ & - & 0.67 & 0.35 & 0.21 & 18.79 & 3.09 & - & - \\
1019 & 175.2432917 & -26.5566389 & 2.1635 & $9.84^{+0.10}_{-0.13}$ & $3\pm1$ & $2.2\pm0.2$ & - & - & 0.24 & -0.37 & 0.00 & 24.41 & 4.71 & - & - \\
1047 & 175.2412917 & -26.4934167 & 2.1703 & $10.67^{+0.09}_{-0.12}$ & $84\pm6$ & - & - & - & 0.73 & 0.59 & 0.21 & 992.00 & 4.10 & Yes & Yes\\
1054 & 175.2408750 & -26.5133611 & 2.1644 & $11.16^{+0.08}_{-0.09}$ & $189\pm53$ & $5.2\pm0.3$ & $8.66\pm0.14$ & $11.21\pm0.09$ & 1.45 & 1.51 & 0.63 & 20.82 & 2.95 & - & Yes \\
1066 & 175.2390833 & -26.4937500 & 2.1663 & $10.30^{+0.11}_{-0.14}$ & $23\pm6$ & $2.0\pm0.2$ & $8.71\pm0.10$ & - & 0.84 & 0.57 & 0.23 & 810.69 & 2.81 & - & Yes \\
\noalign{\vskip 0.0cm}
\hline
\noalign{\vskip 0.0cm}
\end{tabular}
\begin{tablenotes}
\item[a] These objects display broad H$\alpha$ profiles ($\sigma>700$ km\,s$^{-1}$). SFRs and gas phase metallicities have been computed using only the narrow-band component of their emission lines.
\end{tablenotes}
\end{threeparttable}
\label{T:BigTable}
\end{table}
\end{landscape}


\renewcommand{\arraystretch}{1.5}
\begin{landscape}
\begin{table}
\contcaption{}
\begin{threeparttable}
\begin{tabular}{cccccccccccccccc}
\hline
\noalign{\vskip 0.0cm}
ID & RA  & DEC   & z  &  $\mathrm{\log\,M_*/M_\odot}$  & SFR & $\mathrm{R_{e,K_s}}$ & $\mathrm{12+\log(O/H)}$ & $\mathrm{\log\,M_{mol}/M_\odot}$ & U-V & V-J & $\mathrm{E(B-V)_s}$ & $\Sigma_3$ & $\eta$ & X-ray & SMG\\
  & (J2000) & (J2000) & & & ($\mathrm{M_\odot/yr}$) & (kpc) & & & & & & ($\mathrm{Mpc^{-2}}$) & & & \\
\noalign{\vskip 0.0cm}
\hline 
\hline 
\noalign{\vskip 0.0cm}
1071 & 175.2379583 & -26.4744167 & 2.1576 & $10.19^{+0.14}_{-0.20}$ & $26\pm20$ & $2.4\pm0.5$ & - & - & 0.75 & 0.61 & 0.30 & 97.64 & 0.44 & - & - \\
1139 & 175.2300833 & -26.5119167 & 2.1447 & $9.68^{+0.13}_{-0.18}$ & $37\pm9$ & $2.0\pm0.4$ & $<8.39$ & - & 0.72 & 0.43 & 0.22 & 358.48 & 3.25 & - & - \\
1154 & 175.2299167 & -26.4783333 & 2.1630 & $10.39^{+0.08}_{-0.10}$ & $44\pm8$ & - & $8.56\pm0.12$ & - & 1.13 & 0.73 & 0.24 & 34.18 & 1.47 & - & - \\
1162 & 175.2272917 & -26.4732500 & 2.1610 & $10.73^{+0.09}_{-0.11}$ & $127\pm24$ & $3.4\pm0.2$ & $8.71\pm0.08$ & $10.76\pm0.10$ & 1.16 & 1.11 & 0.49 & 34.18 & 1.04 & - & - \\
1181 & 175.2281250 & -26.4676111 & 2.1520 & $10.83^{+0.08}_{-0.10}$ & $166\pm55$ & $4.4\pm0.3$ & - & - & 1.69 & 1.69 & 0.63 & 28.39 & 0.95 & - & Yes \\
1284 & 175.2191667 & -26.5002500 & 2.1569 & $10.21^{+0.13}_{-0.19}$ & $40\pm14$ & $3.2\pm0.4$ & - & - & 0.87 & 0.67 & 0.29 & 33.06 & 0.15 & - & - \\
1300 & 175.2135833 & -26.4940833 & 2.1531 & $10.90^{+0.09}_{-0.11}$ & $149\pm31$ & $4.7\pm0.2$ & $8.65\pm0.14$ & $11.19\pm0.10$ & 1.19 & 1.11 & 0.46 & 168.27 & 0.31 & - & - \\
1316 & 175.2148333 & -26.4960833 & 2.1575 & $10.08^{+0.02}_{-0.02}$ & $11\pm1$ & $2.8\pm0.4$ & $8.55\pm0.10$ & - & 1.1 & 0.24 & 0.00 & 49.14 & 0.19 & - & - \\
1385 & 175.2090833 & -26.4891389 & 2.1553 & $9.65^{+0.09}_{-0.11}$ & $24\pm6$ & $4.5\pm0.7$ & - & - & 0.73 & 0.36 & 0.19 & 54.99 & 0.04 & - & - \\
1420 & 175.2057500 & -26.4858889 & 2.1661 & $9.34^{+0.15}_{-0.23}$ & $14\pm3$ & - & - & - & 0.42 & -0.05 & 0.06 & 73.15 & 0.32 & - & - \\
1498 & 175.1998750 & -26.4804722 & 2.1628 & $10.12^{+0.12}_{-0.17}$ & $51\pm13$ & $3.3\pm0.3$ & - & - & 1.02 & 0.78 & 0.28 & 46.86 & 0.29 & - & - \\
1501 & 175.1997500 & -26.4850833 & 2.1568 & $11.00^{+0.10}_{-0.13}$ & $245\pm94$ & $2.2\pm0.5$ & - & - & 1.4 & 1.49 & 0.63 & 371.92 & 0.01 & Yes & - \\ 
\noalign{\vskip 0.0cm}
\hline
\noalign{\vskip 0.0cm}
\end{tabular}
\end{threeparttable}
\end{table}
\end{landscape}


\bsp	
\label{lastpage}
\end{document}